\documentclass[preprint2]{aastex}
\usepackage{epsfig,psfig}
\usepackage{amsmath}
\usepackage{graphics}
\usepackage{longtable}
\usepackage{geometry}
      \newcommand{\ps}{\,{\rm s}$^{-1}$}
    
\newcommand{\cm}{\,{\rm cm}}    \newcommand{\km}{\,{\rm km}}

\newcommand{\HI}{H~{\sc i}}
\newcommand{\HII}{H~{\sc ii}}
\newcommand{\vlsr}{$V_{\rm LSR}$}       
\newcommand{\twCO}{$^{12}$CO}  \newcommand{\thCO}{$^{13}$CO}
\newcommand{\CeiO}{C$^{18}$O}

\shorttitle{SNR-MC association and their distribution statistics}
\shortauthors{Zhou et al.}

\begin{document}
%\begin{CJK*}{UTF8}{gbsn}

\title{A Systematic Study of Associations between Supernova Remnants and Molecular Clouds}

\author{Xin Zhou\altaffilmark{1}, Yang Su\altaffilmark{1}, Ji Yang\altaffilmark{1}, Xuepeng Chen\altaffilmark{1}, Yan Sun\altaffilmark{1}, Zhibo Jiang\altaffilmark{1}, Min Wang\altaffilmark{1}, Hongchi Wang\altaffilmark{1}, Shaobo Zhang\altaffilmark{1}, Ye Xu\altaffilmark{1}, Qingzeng Yan\altaffilmark{1}, Lixia Yuan\altaffilmark{1}, Zhiwei Chen\altaffilmark{1}, Yiping Ao\altaffilmark{1}, Yuehui Ma\altaffilmark{1}
%\author{Xin Zhou (周鑫)\altaffilmark{1,2}, Yang Su (苏扬)\altaffilmark{1,2}, Ji Yang (杨戟)\altaffilmark{1,2}, Xuepeng Chen (陈学鹏)\altaffilmark{1,2}, Yan Sun (孙燕)\altaffilmark{1,2}, Zhibo Jiang (江治波)\altaffilmark{1,2}, Min Wang (王敏)\altaffilmark{1,2}, Hongchi Wang (王红池)\altaffilmark{1,2}, Shaobo Zhang (张少博)\altaffilmark{1,2}, Ye Xu (徐烨)\altaffilmark{1,2}, Qingzeng Yan (闫庆增)\altaffilmark{1,2}, Lixia Yuan (苑利霞)\altaffilmark{1,2}, Zhiwei Chen (陈志维)\altaffilmark{1,2}, Yiping Ao (敖宜平)\altaffilmark{1,2}, Yuehui Ma (马月辉)\altaffilmark{1,2}
}
\affil{$^1$Purple Mountain Observatory and Key Laboratory of Radio Astronomy, Chinese Academy of Sciences, 10 Yuanhua Road, Nanjing 210033, People's Republic of China; xinzhou@pmo.ac.cn
%$^4$Key Laboratory of Modern Astronomy and Astrophysics, Nanjing University, Ministry of Education, Nanjing 210023, China
}

%\email{xinzhou@pmo.ac.cn}

\begin{abstract}
%snr-mc association based on morphology
%8/8 and 1/1 circular integrated radius vs. flux profile, with self-adaptive bin
% may also apply to moment2 map as supplementary evidence
% final table: snr name (other name), (type,) l, b, coverage: 1'*1', vlsr, bml: v(npoint), MA: v(coeff), distance, reference

We universally search for evidence of kinematic and spatial correlation of supernova remnant (SNR) and molecular cloud (MC) associations for nearly all SNRs in the coverage of the MWISP CO survey, i.e.\ 149 SNRs, 170 SNR candidates, and 18 pure pulsar wind nebulae (PWNe) in $1^\circ < l < 230^\circ$ and $-5.^\circ5 < b < 5.^\circ5$. 
Based on high-quality and unbiased \twCO/\thCO/\CeiO\ (J = 1--0) survey data, we apply automatic algorithms to identify broad lines and spatial correlations for molecular gas in each SNR region. 
%The searching method is demonstrated to be efficient.
%%Among 147 SNRs studied in this paper, 58 of them are found to be associated with MCs, and 68 of them are considered to be possibly associated with MCs. We find 48 SNRCs to be associated with MCs, and 88 SNRCs to be possibly associated with MCs. We also find candidates of associated MCs for 14 pure PWNe.
The 91\% of SNR-MC associations detected previously are identified in this paper by CO line emission. %, and those that are not identified here are mostly based on spatial correlations with \HI\ gases.
%, indicating that CO line emission is efficient for detecting SNR-MC associations.
%%Assuming all SNR-MC associations detected in previous works are real, the accuracy of fixed SNR-MC associations suggested in this paper would be at least 72\%, and at least 58\% for possible SNR-MC associations.
%%There are 7 previously detected SNR-cloud associations but not identified here, i.e.\ about 8\%, which are mostly based on spatial correlations with \HI\ gases.
%Seven previously detected SNR-cloud associations (about 8\%) are not identified in this paper, which are mostly based on spatial correlations with \HI\ gases.
Overall, there could be as high as 80\% of SNRs associated with MCs.
The proportion of SNRs associated with MCs is high within the Galactic longitude less than $\sim50^\circ$.
%statistics
Kinematic distances of all SNRs that are associated with MCs are estimated based on systemic velocities of associated MCs.
%The radius and age distributions of SNRs associated with MCs follow lognormal distributions, which peak at $\sim$8.3 pc and $\sim$2900 yr, respectively. 
The radius of SNRs associated with MCs follows a lognormal distribution, which peaks at $\sim$8.1 pc. The progenitor initial mass of these SNRs follows a power-law distribution with an index of $\sim$$-2.3$ that is consistent with the Salpeter index of $-2.35$.
We find that SNR-MC associations are mainly distributed in a thin disk along the Galactic plane, while a small amount distributed in a thick disk. 
With the height of these SNRs from the Galactic plane below $\sim$45 pc, the distribution of the average radius relative to the height of them is roughly flat, and the average radius increases with the height when above $\sim$45 pc.

\end{abstract}


\section{Introduction}
Supernova remnants (SNRs) release large amounts of momentum, energy, and heavy elements into the interstellar medium (ISM), which modify the physical and chemical properties of the ISM and trigger star formation. %\citep{McKeeOstriker1977}
Such stellar feedback, affecting the next generation of star formation, primarily characterizes the nonlinear evolution of the ISM. 
SNRs are also prime candidates of Galactic cosmic ray (CR) sources. Some SNRs have bright $\gamma$-ray emission, which may originate from CR particles impacting on the surrounding dense medium \citep[e.g.,][]{Aharonian+2008a, Giuliani+2011, Fukui+2021}. These sources provide valuable opportunities to study CR acceleration in SNRs.
Many SNRs originate from core-collapse supernova explosions of massive stars. Because their progenitor massive stars formed in molecular clouds (MCs) and had relatively short lifetimes, these SNRs are expected to be associated with their parent MCs. In addition, type Ia supernovae may also occur near dense molecular gases \citep[e.g.,][]{Lee+2004, Zhou+2016, Chenxp+2017}. 
It is not clear that how many SNRs are associated with MCs. Supposing half of SNRs originated from core-collapse supernovae \citep{ReynosoMangum2001}, there may be more than half of SNRs associated with MCs.

%Among different evidences of SNR-MC interaction, i.e.\ shifted and broadened CO line emission \citep[e.g.,][etc.]{Denoyer1979b, Seta+1998, Kilpatrick+2016}, OH maser line emission \citep[e.g.,][etc.]{GossRobinson1968, Frail+1994, Frail+1996, Green+1997, WardleYusef-Zadeh2002, Yusef-Zadeh+2003, Hewitt+2009oh}, other shock excited molecular line emission \citep[e.g.,][etc.]{Wootten1981}, enhanced IR emission \citep[e.g.,][etc.]{ArendtRichard1989, Saken+1992, Reach+2005, Reach+2006, Hewitt+2009ir}, etc., molecular line emission is wildly used in searching associated MC, since it originates from associated molecular gas and provides its local standard-of-rest (LSR) velocity information.
Among different evidences of SNR-MC interaction, i.e.\ shifted and broadened CO line emission \citep[e.g.,][etc.]{Denoyer1979b, Seta+1998, Kilpatrick+2016}, OH maser line emission at 1720 MHz \citep[e.g.,][etc.]{GossRobinson1968, Frail+1994, Frail+1996, Green+1997, WardleYusef-Zadeh2002, Yusef-Zadeh+2003, Hewitt+2009oh}, shock excited molecular line emission \citep[e.g.,][etc.]{Wootten1981}, enhanced IR emission \citep[e.g.,][etc.]{ArendtRichard1989, Saken+1992, Reach+2005, Reach+2006, Hewitt+2009ir}, etc., molecular line emission is wildly used in searching associated MC, since it originates from associated molecular gas and provides its local standard-of-rest (LSR) velocity information.
To date, about 80 Galactic SNRs have been confirmed or suggested to be associated with MCs, out of about 300 Galactic SNRs \citep[][and references therein]{Gaensler+2008, Jiang+2010, Tian+2010, Eger+2011, Jeong+2012, Frail+2013, Chen+2014, Fukuda+2014, Su+2014, Zhou+2014, Zhu+2014, Zhang+2015, Voisin+2016, Zhoup+2016a, Zhou+2016, deWilt+2017, Lau+2017, Liu+2017, Su+2017a, Liu+2018, Maxted+2018, Su+2018, Ma+2019, Yu+2019, Green2019, RanasingheLeahy2022}.
Most of these Galactic SNR-MC associations have been confirmed or suggested through CO or OH maser line emission. 
In particular, \cite{Kilpatrick+2016} performed an effective systematic search for broad \twCO~(J=2--1) line regions toward 50 SNRs, and confirmed the detection in 19 SNRs including 9 newly identified ones.
\cite{Sofue+2021} also provided some supplementary morphological search results of CO line emission toward 63 Galactic SNRs in the $10^\circ \le l \le 50^\circ$, $|b|\le1^\circ$ region.
%Only a small percentage of SNRs with clear evidence, e.g., uncontaminated broad CO emission or OH 1720~MHz maser emission, are confirmed to be associated with MCs. 
Only a small percentage of SNRs are confirmed to be associated with MCs with clear evidence, e.g., uncontaminated broad CO line emission or OH 1720~MHz maser emission.
Many SNRs are located in crowded regions with multiple MCs overlapping each other in the line-of-sight, hence, molecular lines toward these SNRs are overlapping each other. %with conjunctive line emission, 
It is difficult to distinguish potential disturbed molecular gases in these SNRs. 
In addition, physical conditions for the formation of OH 1720 MHz masers are strict, i.e.\ in regions with moderate temperatures and densities ($T\sim 50$--125~K, $n_{\rm H_2}\sim10^{5}~\cm^{-3}$) behind slow C-type shocks \citep{Lockett+1999, Hewitt+2009oh}, and many SNRs do not have such physical conditions.
A large amount of potential SNR-MC associations prohibit general statistical studies of MC environments around SNRs.

Intending to investigate SNR-MC associations, we present in this work a general study of CO line emission toward most SNRs in the northern sky, based on the Milky Way Imaging Scroll Painting (MWISP) unbiased three CO isotope lines survey using the 13.7-meter millimeter wavelength telescope at Qinghai station. The survey data has appropriate spatial resolution, high sensitivity, and large coverage.
Both kinematic evidence and spatial correlations are examined, and further statistical analyses of SNRs associated with MCs are performed.

\section{Observations}\label{sec:obs}
%SNRs were selected from those in the coverage of the ongoing Milky Way Imaging Scroll Painting (MWISP\footnote{http://www.radioast.nsdc.cn/mwisp.php}) project, i.e.\ nearly all SNRs in $1^\circ < l < 230^\circ$ and $-5^\circ.5 < b < 5^\circ.5$. 
SNRs in the coverage of the Milky Way Imaging Scroll Painting (MWISP\footnote{http://www.radioast.nsdc.cn/mwisp.php}) project, i.e.\ nearly all SNRs in $1^\circ < l < 230^\circ$ and $-5.^\circ5 < b < 5.^\circ5$, are investigated.
The MWISP project is an unbiased survey of \twCO/\thCO/\CeiO\ (J = 1--0) emission lines \citep[see][and references therein, for details]{Su+2019,Sun+2020}, using the Purple Mountain Observatory Delingha (PMODLH) 13.7 m millimeter wavelength telescope \citep{Zuo+2011}. The three CO lines were simultaneously observed by a $3\times3$ multibeam sideband separating superconducting receiver \citep{Shan+2012}. The region is mapped via on-the-fly (OTF) observation mode, with a half-power beamwidth (HPBW) of $\sim$51$''$. Spectral resolutions of the three CO lines are $\sim$0.16~\km\ps\ for \twCO\ (J = 1--0), and $\sim$0.17~\km\ps\ for \thCO\ and \CeiO\ (J = 1--0). The typical rms noise level is about 0.5 K for \twCO\ (J = 1--0); and about 0.3~K for \thCO\ and \CeiO\ (J = 1--0), corresponding to their spectral resolutions. 
%All data are processed using GILDAS/CLASS software package\footnote{http://www.iram.fr/IRAMFR/GILDAS}. 
%All data were processed using dedicated pipelines based on C, IDL, & Python by MWISP working group and the GILDAS/CLASS package\footnote{http://www.iram.fr/IRAMFR/GILDAS}. 
All data were processed using dedicated pipelines by MWISP working group and the GILDAS/CLASS package\footnote{http://www.iram.fr/IRAMFR/GILDAS}. 
A linear fit was performed in the baseline subtraction. The OTF raw data was meshed with a grid spacing of 30$''$, and was corrected for beam efficiency using T$_{mb}$=T$_A^*/\eta_{mb}$.
The data of the full extent of each SNR was extracted, covering an area of 2 to 4 times the size of the remnant. Moreover, for pulsar wind nebulae (PWNe), areas of 16 to 32 times their sizes are covered. SNRs G82.2+5.3 and G93.3+6.9 are not fully covered, nevertheless, more than half of their extents are covered.

Radio continuum emission data at 200~MHz were obtained from the Galactic and Extragalactic All-sky Murchison Widefield Array (GLEAM) survey\footnote{https://www.mwatelescope.org/gleam} \citep{Wayth+2015, Hurley-Walker+2017, Hurley-Walker+2019a}. 1.4~GHz radio continuum emission data were obtained from the NRAO VLA Sky Survey \citep[NVSS;][]{Condon+1998} and The HI/OH/Recombination line survey \citep[THOR;][]{Beuther+2016, Wang+2020}.
4850~MHz radio continuum data were also obtained from the Green Bank 6-cm/Parkes-MIT-NRAO (GB6) survey \citep{Condon+1991, Condon+1993, Condon+1994}.
%other radio data CGPS, VGPS, MAGPIS, 
20~cm radio continuum data are from the Multi-Array Galactic Plane Imaging Survey \citep[MAGPIS\footnote{http://third.ucllnl.org/gps};][]{Helfand+2006}.
The information on SNRs in the \cite{Green2019} catalog\footnote{http://www.mrao.cam.ac.uk/surveys/snrs/snrs.info.html} and the catalog of high-energy SNRs \citep[SNRcat\footnote{http://www.physics.umanitoba.ca/snr/SNRcat};][]{FerrandSafi-Harb2012} was used. Based on the radio continuum emission, we define circular regions to designate individual SNRs by visual inspection, mostly along their outermost bright radio continuum shell.

\section{Methods}\label{sec:method}
%two aspects:
%macro-turbulence
%broadening of molecular emission line: shock-cloud interaction, macro-turbulence in advance => broad line or line wing broadening
%spatial correspondence: accumulation by shock, pre-SN progenitor's wind-blown bubble => cavity or shell
%line broadening and spatial correspondence are evidence of SNR-MC association
%Searching for these evidence, we investigate molecular line emission around SNR in two aspects.
Molecular gases impacted by SNR shock will be disturbed in two respects, i.e.\ heating and macro-turbulence. Molecular gas behind a transmitted shock in a molecular clump will be accelerated and heated, which is usually detected as broadened line wing in CO line profiles. 
In addition, molecular gas engulfed by SNR will become cold as the SNR gets old, but still maintains the injected momentum. Hence, relative motion between different parts of molecular gas in SNR, i.e.\ macro-turbulence, leads to the corresponding CO line profile being broader than that from surrounding quiescent molecular gas.
The macro-turbulence in molecular gas injected by SNR could be very general, especially in old SNRs where most of the dense shocked molecular gases become cold.
%Stellar wind of SNR's progenitor may also contribute some turbulence to local molecular gas, but it should be minors, due to presure balance between wind-blown bubble and surrounding interstellar medium.
%Therefore, shocked molecular gases usually result in broadened CO line profile, which usually presents as an extra peak unlike molecular outflows in star-forming regions.
Overall, these effects would result in broadened CO lines from shocked molecular gases. 
Unlike CO lines from molecular outflows in star-forming regions, these broadened CO lines usually present as extra peaks beside narrow CO lines from surrounding quiescent molecular gases. 
Such kinematic evidence in SNRs is very useful to confirm SNR-MC associations.
It is noteworthy that such broadened CO lines might not be present in some young SNRs that are associated with MCs, because CO molecules nearby can be dissociated by UV radiation from progenitor massive stars or strong shock heating of these young SNRs, and, in general, the kinetic energy of SNR's shock is not sufficiently converted into a molecular cloud \citep[e.g.,][and references therein]{Inoue+2012, Zhoup+2016, Celli+2019, Sano+2020}.

An SNR will also reshape nearby MCs, which can destroy less dense molecular gas and leave a cavity, or can accumulate molecular gas and form a shell \citep[e.g.,][]{Seta1998iau, Inutsuka+2015}. 
Spatial correlation is commonly used as evidence of SNR-MC association. Many known SNR-MC associations are proposed based on their spatial correlations \citep[see Table 2 in][]{Jiang+2010}.
%Therefore, spatial correlations between SNRs that could be traced by radio continuum emission and MCs are also evidence of SNR-MC associations. %can also be considered as an evidence of SNR-MC association. 
In some cases, spatially correlated MCs originate from wind-blown bubbles of the SNR's progenitor, especially for very young SNRs or PWNe, for which kinematic evidence barely exists \citep[][and references therein]{Chevalier1999, Chen+2013}.
%Note that for identifying the SNR-MC association, 
Note that the spatial correlation evidence is not as robust as the kinematic evidence, since overlapping of multiple MCs across the Galactic plane through the line of sight would complicate the spatial distribution of molecular gases.

Our goal is to search universally for evidence of kinematic and spatial correlation of SNR-MC association. 
Thereafter, we also need to settle associated spectral components, by comparing the searching results of broad line and spatial correlation, comparing them with the radio continuum emission of SNRs, and eliminating the contamination of overlaid energetic objects.
Finally, we can determine the systemic velocity of MCs associated with SNRs. %, based on automatically searching results.
%Systemic velocities of all MCs possibly associated with SNRs are also obtained in the automated search.
In this paper, the velocity of the intensity peak of the associated spectral component, which represents the local standard of rest (LSR) velocity of the main part of molecular gas, is considered as the systemic velocity of MCs associated with SNRs.
The detail of searching method is presented in Section~\ref{sec:methodbl} and \ref{sec:methodsc}.

\subsection{Broad Line Identification}\label{sec:methodbl}
%macro-turbulence
%moment analysis
Kinematic evidence of SNR-MC associations would present as broadened lines, and the goal is to search these broadened lines in \twCO~(J=1--0) data.
At the beginning, we smooth all \twCO~(J=1--0) spectra by bin nearby three channels, to get a high signal-to-noise ratio in broad line searching. The spectral resolution after smoothing is 0.48~\km\ps, which is about half of the narrowest \twCO~(J=1--0) line.
Then, we decompose these spectra into individual spectral components, of which the minimum intensity of bottoms is $1\sigma$, the minimum intensity of peak-to-valley differences is $3\sigma$, and the minimum intensity of peaks is $5\sigma$.
%To eliminate some baseline variations, the intensity variance is also greater than the minimum of the corresponding spectrum.
Broad line identifications are performed based on these spectral components.
For these spectral components, we identify broad lines first, then find broader lines in the SNR region than those in the background if exist.
Broad lines are identified by following criteria.
\begin{itemize}
\item The full width at half maximum (FWHM) of the spectral component be larger than 6~\km\ps.
\item The kurtosis of the component be larger than 1.1 times that of the best-fitting Gaussian function with the same FWHM in the same velocity range.
\item The square rooted variance of channels with intensity greater than the best-fitting Gaussian function be smaller than or equal to the square rooted variance of all channels of the whole component.
\item All required values be over three times their $\sigma$ errors. %except for the square rooted variance of channels with intensity greater than the best-fitting Gaussian function.
\end{itemize}
%These criteria already

As indicated by the Larson's line width-size relationship, the typical FWHM of \twCO~(J=1--0) lines of normal, parsec scale, quiescent MCs is below $\sim$6~\km\ps\ \citep{HeyerDame2015}. 
The \twCO~(J=1--0) line of most shocked molecular gases is with FWHM greater than 6\km\ps\ \citep[e.g., SNR HB3;][]{Zhou+2016}. 
The FWHM of the \twCO~(J=1--0) line of some small shocked molecular clumps can be lower than 6~\km\ps; however, such clumps only occupy a small part of the shocked molecular gas in the SNR in most cases.
Therefore, we adopted 6~\km\ps\ as a lower limit of the FWHM of broad lines. 
Note that, for SNRs adjacent to MCs of much larger size, a larger FWHM threshold will be applied to identify broad lines associated with SNRs in the following step, which depends on the FWHM of background spectral components.
The information on the width of background lines can also help us to eliminate some line overlapping effects.

\begin{figure*}[ptbh!]
\centerline{{\hfil\hfil
\psfig{figure=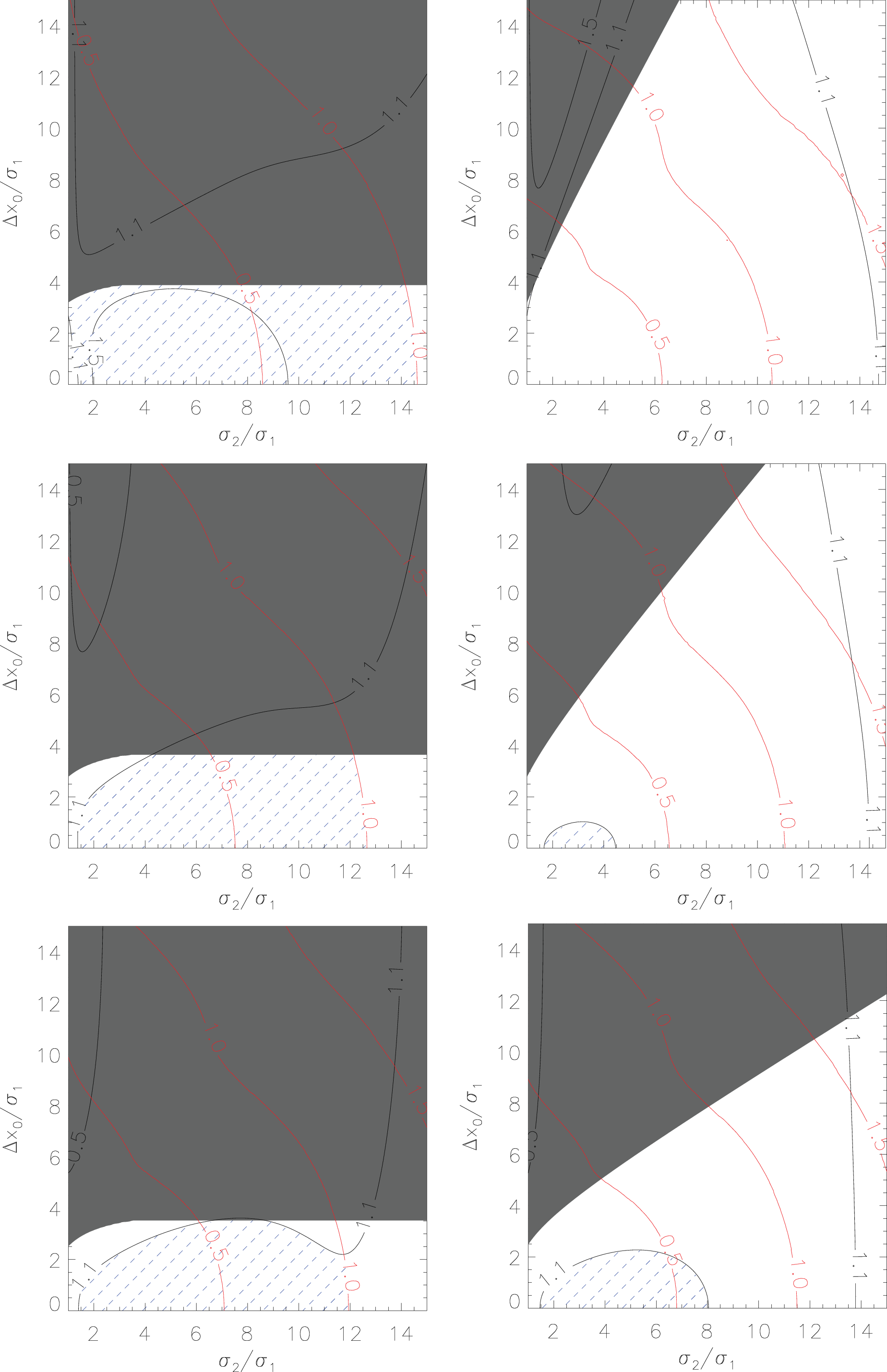,height=5in,angle=0, clip=}
\hfil\hfil}}
\caption{Test results of broad line identification. The spectrum comprising two Gaussian components is used for test, with standard deviations of $\sigma_1$ and $\sigma_2$ and the distance between their peaks of $\Delta x_0$. The FWHM of the first Gaussian component is set to 3, i.e.\ $\sigma_1$ is 1.27. Results are similar for other values of $\sigma_1$.
Black contours show the distribution of the kurtosis ratio between the spectrum and a best-fitting Gaussian function.
Red contours show the distribution of the square rooted variance ratio between the part of the spectrum larger than the best-fitting Gaussian function and the whole spectrum.  
Panels in the left column show cases that the broad Gaussian component is weaker than the narrow component, with peak ratios of 0.2, 0.5, and 0.8 from top to bottom, respectively. The right column is opposite, with the ratios of peaks between the broad and narrow components of 5, 2, and 1.2 from top to bottom, respectively.
Gray scale regions indicate where two Gaussian components can be resolved.
Blue dashed lines denote regions selected by full broad line identification criteria.
}
\label{f:2gautest}
\end{figure*}
We use the kurtosis value to characterize the deviation from a Gaussian profile, of which a larger value indicates a distribution with extra wings.
Together with the variance of channels with intensity greater than the best-fitting Gaussian function, we can eliminate some line overlapping effects.
%See figure~\ref{f:2gautest}
As shown in Figure~\ref{f:2gautest}, components comprising a strong narrow line plus a weak broad wing are mostly selected, and those comprising a strong broad line plus a weak narrow line are eliminated. 
The CO emission of shocked MC by an SNR usually presents as a strong narrow line plus a broadened line wing, where the narrow line originates from the quiescent matrix MC and the broad wing from shocked molecular gas.
%The SNR shock transmitted into MC is hardly to pass through the whole matrix MC in the lifetime of SNR.
Narrow lines of some small molecular clumps in SNRs could be weaker than broad lines, or no narrow line left at all. However, in most cases, such small molecular clumps contribute only a small portion of shocked molecular gas comparing to that from the large matrix MC \citep[e.g.,][]{Zhou+2016}, and velocities of these broad lines are usually significantly shifted from that of corresponding narrow lines, which are easily to be distinguished as separate spectral components.
Commonly, overlapping CO lines from different MCs in the line-of-sight do not have such features.
For instance, in some star-forming regions, central dense molecular gases are mostly disturbed, where broad CO lines are stronger.
These criteria are strict, which may eliminate some individual broad lines, but they provide more accurate results for SNRs in complicated backgrounds.

In the following step, we further select broader lines within the SNR region than that in the background. The size of the SNR region is 1.1 times enlarged than that indicated by its radio continuum emission. These broader lines are considered as primary kinematic evidence of the SNR-MC association.
%The final FWHM threshold adopted to select broad lines originating from SNRs is estimated by two conditions: larger than or equal to the average FWHM of background broad lines, and larger than all broad lines in at least one of equally divided sub-regions in the background, of which the area is comparable to the SNR region.
The final FWHM threshold for selecting broad lines originating from SNRs satisfies the condition that their FWHM is larger than or equal to the average FWHM of background broad lines. In practice, some SNRs still have many broad lines at different velocities selected. In this case, we add an additional condition to select broad lines for reference.
The additional condition requires that the final FWHM threshold be larger than all broad lines in at least one of equally divided sub-regions in the background, which have areas comparable to that of the SNR region. Note that, for SNRs with simple backgrounds, this additional condition does not change the broad line identification results. This additional condition is hereafter referred to as the clean subbackground region condition.
%Therefore, the final FWHM thresholds are estimated from background broad lines, which are larger than or equal to the average FWHM of background broad lines and makes no broader lines left in one equally divided subregions at least in the background, of which the area is comparable to the SNR region.
Thereby, we can better identify broad lines originating from SNRs in backgrounds where CO lines are broad, e.g., nearby the Galactic center.

As noted above, full broad line identification criteria are strict, which provide better accuracy but less completeness.
Full broad line identification criteria plus the clean subbackground region condition are even more strict.
As a complement, a second broad line identification procedure is performed, applying only partial criteria, i.e.\ FWHM and error criteria, and the final FWHM threshold being larger than or equal to the average FWHM of background broad lines.
%The SNR, containing broad lines identified by the partial criteria but not the full criteria, is considered as possible SNR-MC association.
Broad lines identified only by partial criteria are considered as broad line candidates.

At last, identified broad lines in an SNR are divided into different components, according to their intensity peaks being in each other's velocity range or not.
The systemic velocity of broad line groups is determined as the velocity of the overall intensity peak of corresponding \thCO~(J=1--0) emission. If there is no significant \thCO~(J=1--0) emission, the systemic velocity is applied as the velocity of the overall intensity peak of \twCO~(J=1--0) emission.
The case that broad lines being adjacent to the associated narrow lines is common, since the SNR shock is slow when it transmitted into a dense cloud. Broad lines are also usually strong near the observed boundary of shocked MCs, where the column density of shocked molecular gas is high, and LSR velocities in the line-of-sight direction of these broad lines are low.
The systemic velocity obtained based on broad lines would represent the LSR velocity of most quiescent molecular gases in most cases.
Nevertheless, some broad lines can be shifted to velocities far away from associated narrow lines, and are identified as separated broad line components.
Therefore, further examination is needed to settle the systemic velocity of associated MCs, e.g., spatial correlation examination that is introduced in Section~\ref{sec:methodsc}.

\subsection{Spatial Correlation Coefficient}\label{sec:methodsc}
%spearman correlation coefficience between molecular emission and radio continuum emission.
To examine spatial correlation between SNRs and MCs, we calculate spatial correlation coefficients (abbreviated as SCCs below) between a series of circular rings and channel maps of \twCO~(J=1--0) emission.
Positions and sizes of SNRs, which are used to derive circular rings, are mostly determined by radio continuum emission of SNRs. %, and data in other bands will be used if necessary.(data introduction)
For preparation, \twCO~(J=1--0) data is moment masked at first \citep[see][for reference]{Dame2011}, to eliminate noise that affects the result of weak emission.
\twCO~(J=1--0) channel maps are also linearly interpolated to an $80\times80$ pixel resolution and normalized to maximum intensity of 1. The $80\times80$ pixel resolution would simplify the calculation, and enables the smallest circular ring to be still resolved.
%The minimum inner radius of circular rings is 4 pixels, corresponding to a minimum of a quarter of the radius of SNRs, and the maximum inner radius is 20 pixels. 
The minimum inner radius of circular rings, corresponding to a quarter of the radius of SNRs, is at least 4 pixels, and the maximum inner radius is 20 pixels. 
%The step of inner radius in calculation is 1 pixel.
The increment of the inner radius of the series of circular rings is 1 pixel.
Thicknesses of circular rings are from 2 to 30 pixels at intervals of 2 pixels.
%These parameters yield effective results in experimentation.
%Results obtained with these parameters are sufficiently valid in calculation. %in experimentation.
The results obtained with these parameters are fully feasible for the following calculations. %in experimentation.
After these preparations, we calculate Spearman correlation coefficients between \twCO\ tile images and ring template images, as $\rho=\frac{\Sigma_i(x_i-\bar x)(y_i-\bar y)}{\sqrt{\Sigma_i(x_i-\bar x)^2 \Sigma_i(y_i-\bar y)^2}}$.
This SCC would be large for molecular gases with similar spatial distributions to circular rings.
Since the large number of pixels outside the SNR has a large impact on the SCC result, we only consider pixels inside ring templates or within twice the SNR radius.
In the following step, we select the largest coefficient for each channel map, and make a coefficient versus velocity channel plot. 
In the plot, velocity channels with coefficients over $3\sigma$ levels usually congregate into different groups that correspond to velocity components in CO spectra.
Groups of coefficients with peaks over $5\sigma$ confidence level, or over $3\sigma$ confidence level and larger than 0.5, are considered as candidates of correlated components. If there is no coefficient over $5\sigma$ or larger than 0.5, the maximum coefficient over $3\sigma$ will be given for reference.
In addition, SCCs that consider pixels inside ring templates or within the SNR radius are also calculated, which can better examine thin molecular shell structures around an SNR.
Note that the SCC would be underestimated in some cases, e.g., correlations between partial shells or other irregular shapes, nevertheless, they are still noticeable on the coefficient versus velocity plot in most cases. %in practice.
%The correlation coefficient for some partial shells or assymetric distributions would be undervalued, but still notable in practice. %examination in our experience.

\subsection{Demonstration}\label{sec:demon}
%demonstrate: completeness and accuracy
We perform our searching method to several known SNR-MC associations for demonstration at first, and it helps us to optimize the parameters of the algorithm.
Among these known SNR-MC associations, SNR IC~443, SNR HB~3, and SNR G16.0$-$0.5 are representative of SNRs with clear, simple, and complicate CO emission backgrounds, respectively.

\subsubsection{IC~443}\label{sec:ic443}
\begin{figure*}[ptbh!]
\centerline{{\hfil\hfil
\psfig{figure=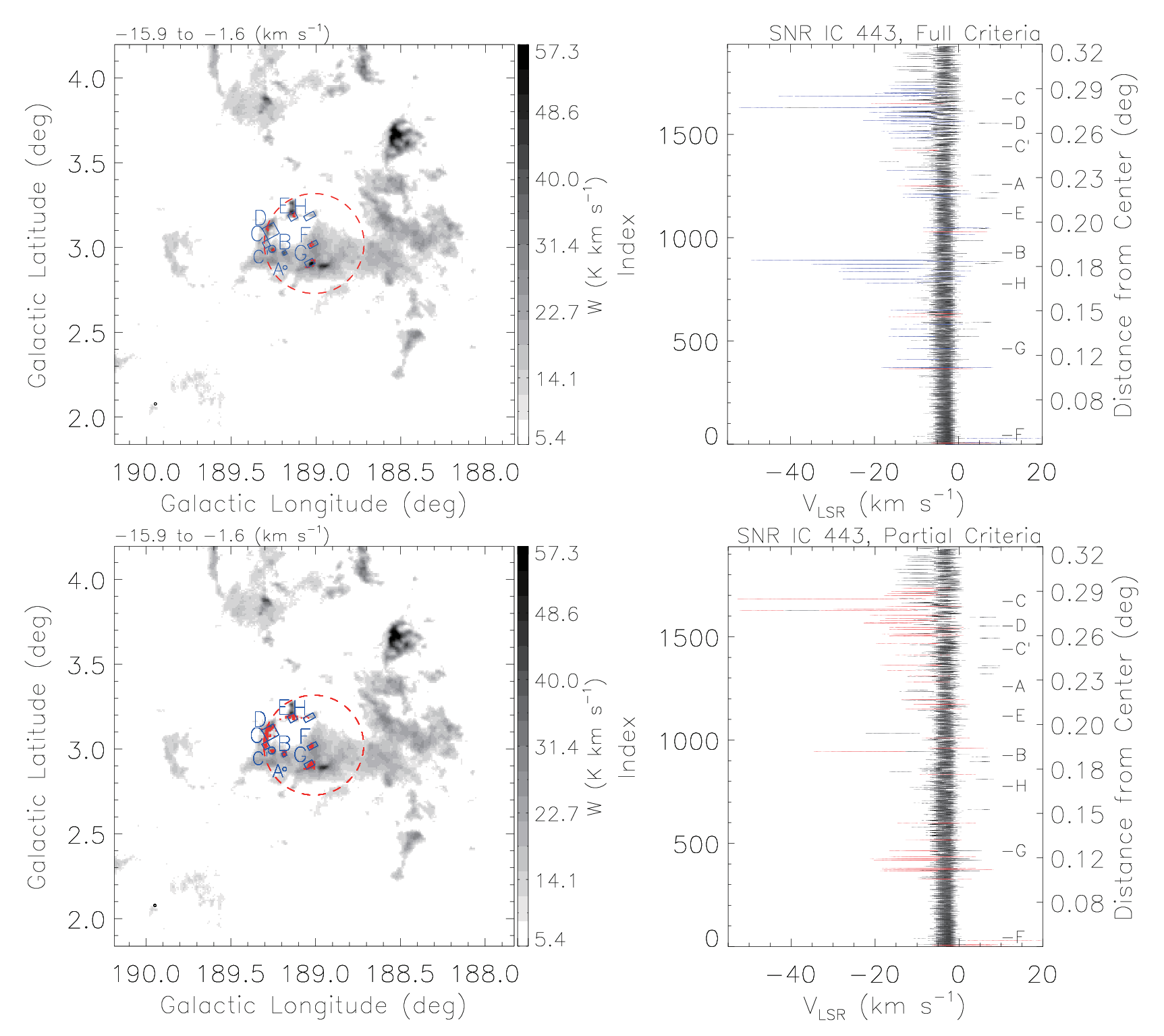, height=4.5in,angle=0, clip=}
\hfil\hfil}}
%\caption{{\fontsize{9pt}{10pt} \selectfont Integrated intensity and index-velocity maps of \twCO~(J=1--0) emission toward SNR IC~443. 
\caption{{\small Integrated intensity and index-velocity maps of \twCO~(J=1--0) emission toward SNR IC~443. 
Both intensity maps in the left column are integrated over the velocity range of $-$15.9 to $-$1.6~\km\ps, which covers most of the broad lines.
The red dashed circle denotes the remnant region, defined by visual inspection according to the bright radio continuum shell of the remnant. %[ref]
Regions A, B, C, D, E, F, G, and H are adopted from \cite{Denoyer1979b}, \cite{Dickman+1992}, where shocked molecular gases were detected by \twCO~(J=1--0) emission. Region C$'$ is an additional shocked molecular clump, which is adjacent to region C.
Locations of broad lines identified by full ({\sl upper}) and partial ({\sl lower}) criteria are denoted by red tiny stars.
The minimum value of \twCO\ emission intensity shown in the colorbar is at $5\sigma$ confidence level.
The beam is represented by a black circle in the lower left corner. This beam designation is used in all intensity maps in the paper.
Index-velocity maps in the right column contain all significant \twCO~(J=1--0) emission lines in the 1.1 times enlarged remnant region, whose intensity is normalized and logarithmically scaled for better visibility.
The index indicates the arrangement of the spectrum at each point, sorted by the distance of the point from the SNR center. Some distances are shown on the right axis. 
The center of each region adopted from previous works in intensity maps is also shown in index-velocity maps.
Broad lines identified by full ({\sl upper}) and partial ({\sl lower}) criteria are in red, and remaining lines with FWHM over 6~\km\ps\ are in blue.
}}
\label{f:ic443}
\end{figure*}

The association between SNR IC~443 and MCs has been well-established in previous works.
The variety of molecular lines studied for this remnant is the most complete, providing a good understanding of the physics of the interaction between the SNR shock and the molecular gas \citep[e.g.,][]{vanDishoeck+1993}.
%\citep[e.g.,][see Table~\ref{tab:snrpre} for more references]{vanDishoeck+1993}.
%The north of this SNR is adjacent to a large \HII\ region at $-$1~\km\ps. The northeast of the remnant overlaps with SNR candidate G189.6+3.3. 
There are nine shocked molecular clumps in IC~443, which exhibit broad \twCO~(J=1--0) line emission \citep{Denoyer1979b, Huang+1986, Dickman+1992, Lee+2012}. 
These clumps are indicated in Figure~\ref{f:ic443}. Eight of them are labeled as A through H, following the nomenclature in \cite{Dickman+1992}, and the remaining one that is adjacent to clump C is labeled as C$'$, which was named as SC~05 in the study of \cite{Lee+2012}.
Note that we apply a circular region to denote the remnant in our analysis, which is defined following the northern bright radio continuum shell of the remnant, and there is also radio continuum emission in the south outside the region.
As shown in Figure~\ref{f:ic443}, we identify broad lines in five clumps by the full criteria, and in eight clumps by the partial criteria. 
The broad \twCO~(J=1--0) line emission in clump A is too weak, with $T_A^*\sim0.3$~K \citep{Denoyer1979b}, which is below our detection limit, hence, we found no broad line there.
The broad line emission in clump H is also weak, and we only identify broad lines there by partial criteria.
In clumps B and C, broad lines are far blue-shifted, separated from the associated narrow line component, which are only identified by partial identification criteria. %eliminated by full identification criteria.
We evaluate the accuracy of the broad line identification by $(n_{p,SNR}-n_{p,outside})/(n_{p,SNR}+n_{p,outside})$, where $n_{p,SNR}$ and $n_{p,outside}$ are numbers of broad line points per unit area inside and outside the SNR region, respectively.
For IC~443, the broad line identifications using either full or partial criteria have an accuracy of 1.
It benefits from the clean MC background toward the SNR in the outer Galactic region. 
We also evaluate the completeness of the broad line identification by the number ratio between identified broad lines and all lines with FWHM over 6~\km\ps.
For IC~443, the completeness of broad line identifications using full and partial criteria is $\sim$0.20 and 1, respectively. 
Some broad lines separated from the associated narrow component or stronger than the narrow component are eliminated by full criteria.
Since there is no contamination from background line emission, the identification by partial criteria is the most effective.
%and enough significance of all broad lines, 

For SNR IC~443, the SCC has a maximum value of 0.24 at $-2.1$~\km\ps, which is over $3\sigma$ but below $5\sigma$ levels. %, indicating no significant spatial correlation between the remnant and CO gas distribution. 
The velocity of the maximum SCC is consistent with that of molecular gas associated with the remnant.
As seen in Figure~\ref{f:ic443}, associated molecular gas presents some partial shell structures surrounding the remnant. 
These partial shell structures result in large SCCs. However, central bright CO emission makes the enhancement of the SCC less significant.
These associated partial shells were studied by \cite{Su+2014b}, which are possibly swept up by the stellar wind of the remnant's progenitor.

Both the kinematic evidence and the spatial correlation result indicate that SNR IC~443 is associated with the $-$3.0~\km\ps\ velocity component. 
Based on a full distance probability density function\footnote{http://bessel.vlbi-astrometry.org/node/378} \citep{Reid+2016, Reid+2019}, the kinematic distance of the $-$3.0~\km\ps\ component is estimated as 2.10$\pm0.03$~kpc, which is located in the Perseus spiral arm. 
Note that \cite{Yu+2019} also measured the distance of the associated MC by dust extinction estimation as 1729$_{-94}^{+116}$ pc.

\subsubsection{HB~3}\label{sec:hb3}
\begin{figure*}[ptbh!]
\centerline{{\hfil\hfil
\psfig{figure=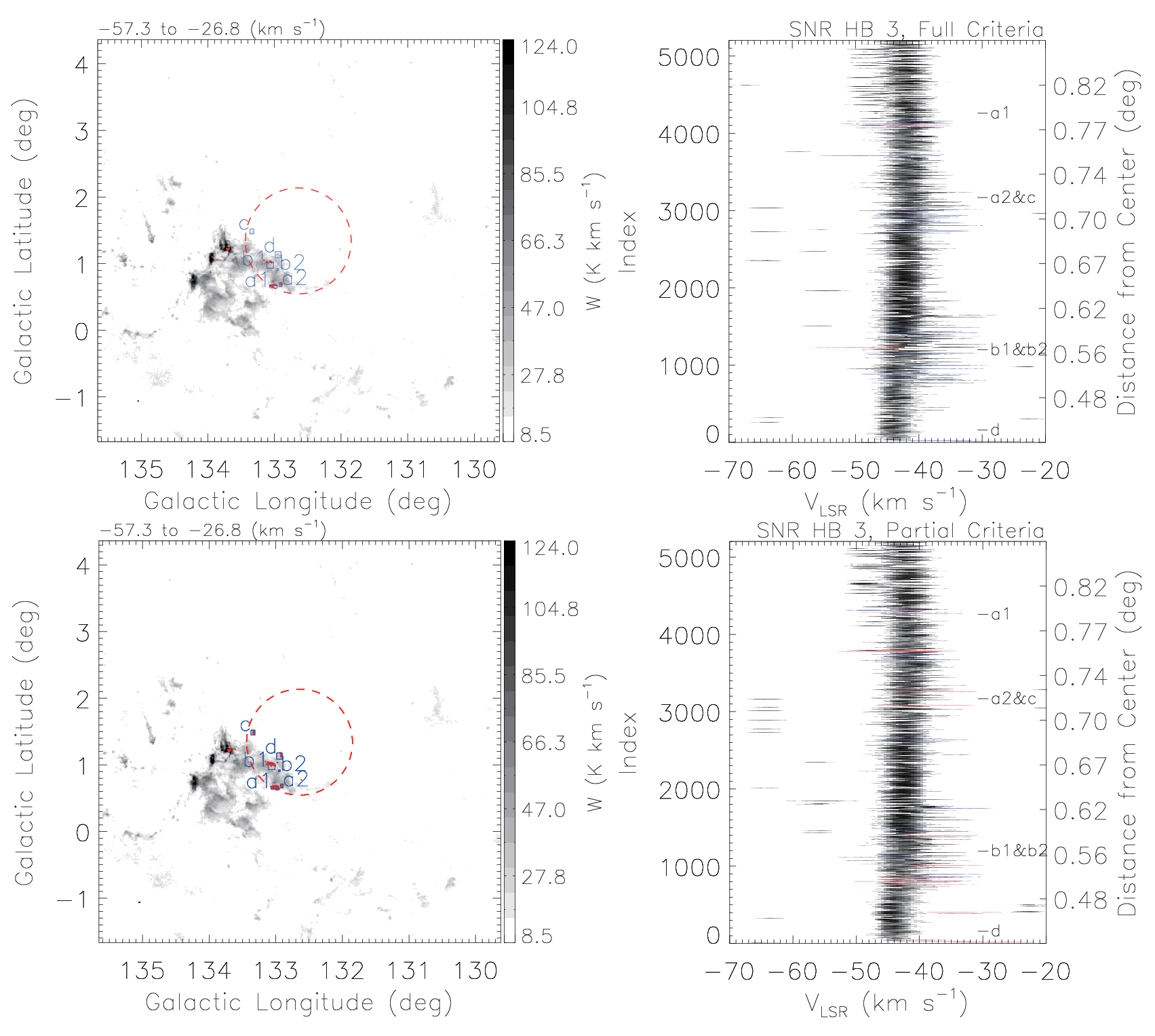, height=5.5in,angle=0, clip=}
\hfil\hfil}}
\caption{Same as Figure~\ref{f:ic443}, but for SNR HB~3.
Both integrated intensity maps in the left column are in the velocity range of $-57.3$ to $-26.8$~\km\ps.
Regions a1, a2, b1, b2, c, and d are adopted from \cite{Zhou+2016}, which indicate positions of shocked molecular clumps.
All broad lines outside the SNR region are located in the star-forming region W3.
}
\label{f:hb3}
\end{figure*}

\begin{figure*}[ptbh!]
\centerline{{\hfil\hfil
\psfig{figure=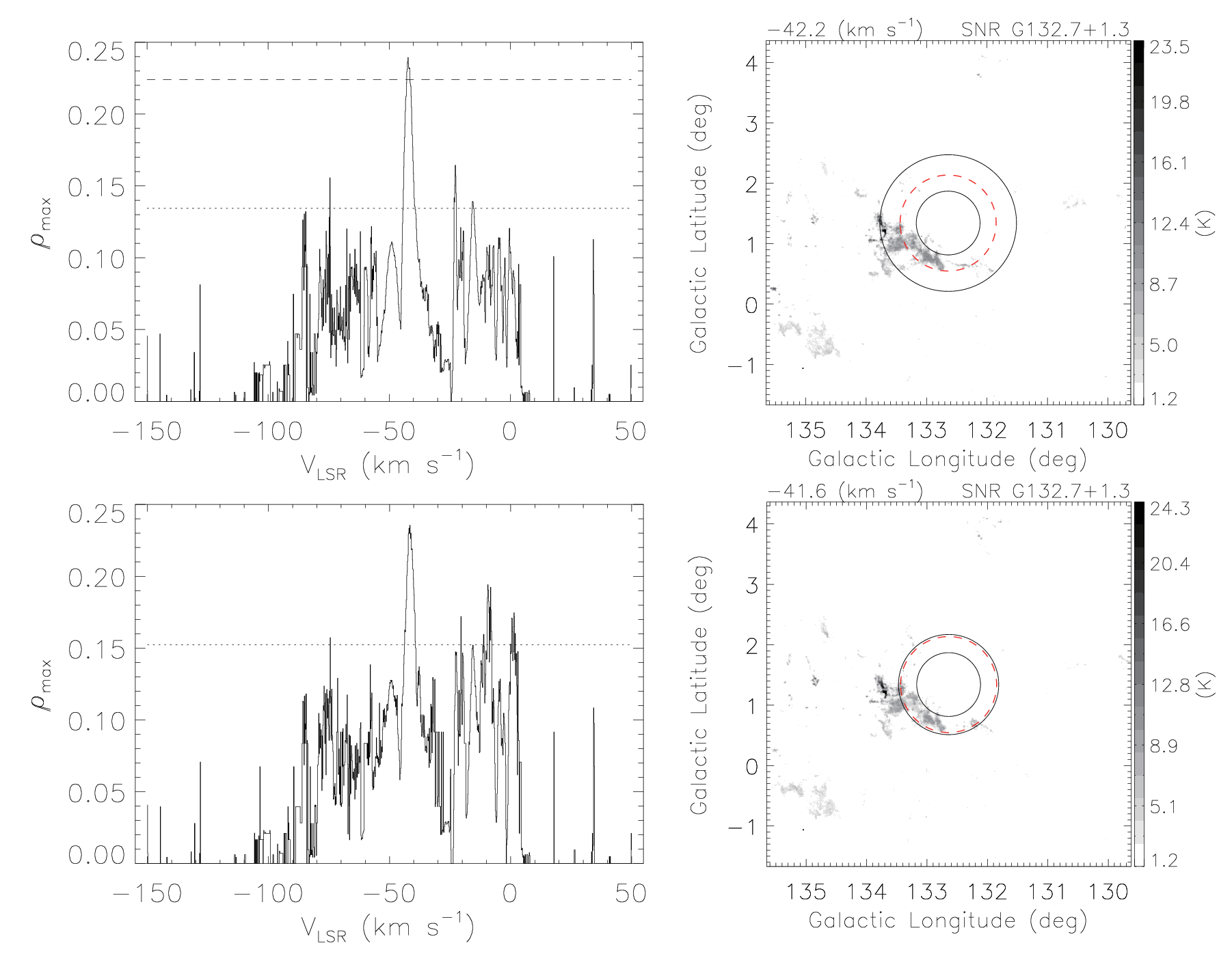, height=5.0in,angle=0, clip=}
\hfil\hfil}}
%\caption{{\sl Left:} maximum spatial correlation coefficient (SCC) at each velocity channel. The dashed and dotted lines denote $5\sigma$ and $3\sigma$ levels, respectively. {\sl Right:} \twCO~(J=1--0) intensity map at the velocity channel of the maximum SCC, i.e.\ $-$42.2\km\ps. Black circles indicate the annular template applied in the calculation to get the maximum SCC. The red dashed circle is the same as in Figure~\ref{f:hb3}.
\caption{Maximum spatial correlation coefficients (SCCs) at each velocity channel and \twCO~(J=1--0) intensity maps at the corresponding velocity channel at the SCC peak for SNR HB~3. 
Dashed and dotted lines in panels in the left column denote $5\sigma$ and $3\sigma$ levels, respectively. 
%{\sl Right:} \twCO~(J=1--0) intensity map at the velocity channel of the maximum SCC, i.e.\ $-$42.2\km\ps. 
Black circles in panels in the right column indicate the annular template applied in the calculation to obtain the maximum SCC. The red dashed circle is the same as in Figure~\ref{f:hb3}.
Upper panels show SCC results considering pixels inside the ring template or within twice the SNR radius, which has a significant peak at $\sim$$-$42.2~\km\ps.
For comparison, the peak of the SCC at $\sim$$-$41.6~\km\ps\ in lower panels, considering pixels inside the ring template or within the SNR radius, is not above $5\sigma$, however, it distinguishes the molecular shell structure around the SNR more precisely.
}
\label{f:hb3coeff}
\end{figure*}

%The east of this SNR is adjacent to several \HII\ regions at $\sim$$-$40~\km\ps.
%We identify one broad line component at $-$44.0~\km\ps\ by full criteria.
%The SCC peaks at $-$41.6~\km\ps, where molecular gas is surrounding the southeast of the remnant, spatially correlated with the bright radio continuum shell of the remnant.
%Based on the kinemtaic evidence and the spatial correlation result, we suggest that the remnant is associated with the $-$44.0~\km\ps\ MC, with the corresponding kinematic distance estimated as 1.96$\pm0.04$ kpc.\\
SNR HB~3 was initially suggested to be associated with MCs based on their spatial correlations \citep{Landecker+1987, Routledge+1991}, and this was further supported by broadened CO lines and H$_2$ 2.12 $\mu$m emission detected in later studies \citep{Zhou+2016, Rho+2021}.
Note that broadened \twCO (J=2--1) line emission toward this remnant was detected by \cite{Kilpatrick+2016}, but was suggested to be associated with the H~{\sc ii} region W3 (OH) rather than HB~3.
SNR HB~3 is associated with the nearby H~{\sc ii} region/MC complex W3 \citep[e.g.,][]{Zhou+2016}, hence, there is broad line emission in the background. Line overlapping effect in the direction of the outer galaxy toward this SNR is not significant.
All shocked molecular clumps found by detailed examination in \cite{Zhou+2016} are also identified here by partial criteria, i.e.\ in regions a1, a2, b1, b2, c, and d (see Figure~\ref{f:hb3}). 
Many broad lines outside the SNR region are also identified, and all of them are located in the star-forming region W3 (see Figure~\ref{f:hb3}), possibly associated with H~{\sc ii} regions within it.
The accuracy and completeness of the broad line identification using partial criteria are 0.98 and 0.61, respectively. 
Since the broad line identification is barely affected by line overlapping effect for this SNR, the accuracy of using partial criteria is high.
%There are fewer broad lines identified in the remnant using full criteria, and the corresponding accuracy is evaluated as 0.88. 
There are fewer broad lines identified in the remnant using full criteria, and the corresponding accuracy is evaluated as 0.94. 
The number ratio between broad lines identified inside and outside the remnant using full criteria is a bit less than that using partial criteria. %Broad lines identified using full criteria is still less affected by line overlapping effect.
%the accuracy of the identification using full criteria is not improved, which is 0.88. There are fewer 
%similar to that of the identification using partial criteria, which is $\sim$0.88.
%The completeness of the broad line identification using strict full criteria is as low as 0.13.
The completeness of the broad line identification using strict full criteria is as low as 0.11.

In the SCC plot in the upper left panel in Figure~\ref{f:hb3coeff}, when considering pixels inside ring templates or within twice the SNR radius, a component with a peak above the $5\sigma$ confidence level indicates a correlation between SNR HB~3 and molecular gas at $\sim$$-$42.2~\km\ps. The SCC result considering pixels inside ring templates or within the SNR radius also supports this correlation (see panels in the bottom row in Figure~\ref{f:hb3coeff}).
%As indicated in \cite{Zhou+2016}, the molecular shell corresponds to the bright radio continuum shell of the remnant well.
There is a molecular shell at $\sim$$-$42.2~\km\ps\ spatially correlated with the bright radio continuum shell of the remnant well \citep[also seen in][]{Zhou+2016}. %\citep[see][for reference]{Zhou+2016}.
Since the molecular shell occupies only about a quarter of the ring surrounding the remnant (see the right panel in Figure~\ref{f:hb3coeff}), the peak of the SCC is only 0.22, which is underestimated.

The kinematic evidence and the spatial correlation result indicate that SNR HB~3 is associated with the $-$44.0~\km\ps\ velocity component. 
Based on the full distance probability density function, the kinematic distance of the $-$44.0~\km\ps\ component is estimated as 1.96$\pm0.04$~kpc, which is located in the Perseus spiral arm. 

\subsubsection{G16.0-0.5}\label{sec:g16.0}
\begin{figure*}[ptbh!]
\centerline{{\hfil\hfil
\psfig{figure=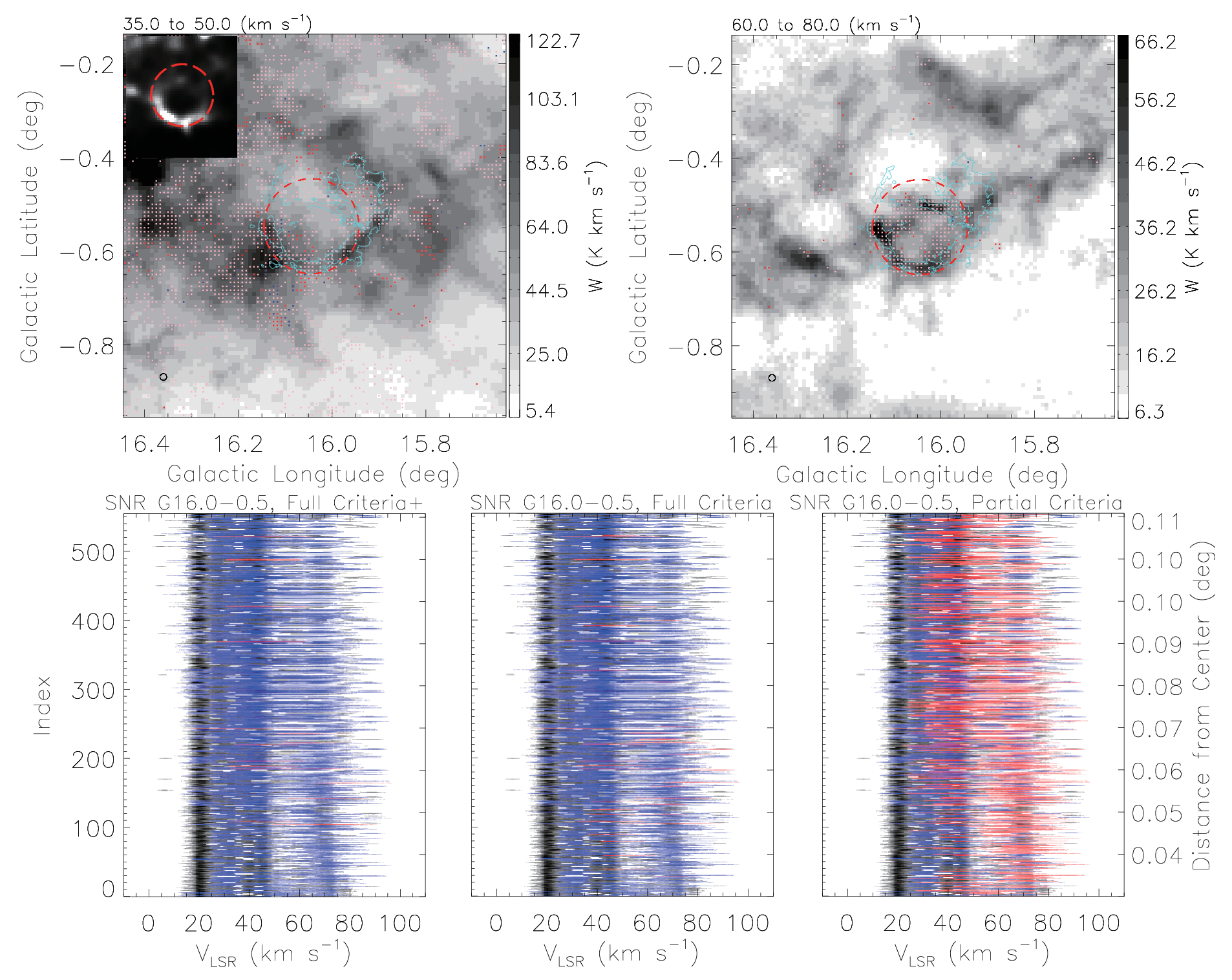, height=5.0in,angle=0, clip=}
\hfil\hfil}}
\caption{Same as Figure~\ref{f:ic443}, but for SNR G16.0$-$0.5.
Two integrate intensity maps of \twCO~(J=1--0) emission in the upper row are in the velocity ranges of +35 to +50~\km\ps\ ({\sl left}) and +60 to +80~\km\ps\ ({\sl right}).
Locations of broad lines at velocities of +43.7 and +67.9~\km\ps\ are denoted by tiny stars on the corresponding {\sl left} and {\sl right} integrated intensity maps, respectively. 
%Broad lines identified by partial criteria are in pink, and those by full criteria are in red.
Broad lines identified by partial criteria are in pink, those by full criteria are in red, and those by full criteria plus the clean subbackground region condition are in blue.
Cyan contours in both integrated intensity maps are at half the column density maximum of shocked molecular gas obtained by \cite{Beaumont+2011}.
The GLEAM 200~MHz radio continuum emission map in power-law scale is superimposed on the upper left corner of the {\sl left} integrated intensity map, overlaid with the same red dashed circle as in integrated intensity maps.
In the index-velocity maps in the bottom row, broad lines identified by full criteria plus the clean subbackground region condition ({\sl left}), full criteria ({\sl middle}), and partial criteria ({\sl right}) are in red.
}
\label{f:g16}
\end{figure*}

\begin{figure*}[ptbh!]
\centerline{{\hfil\hfil
\psfig{figure=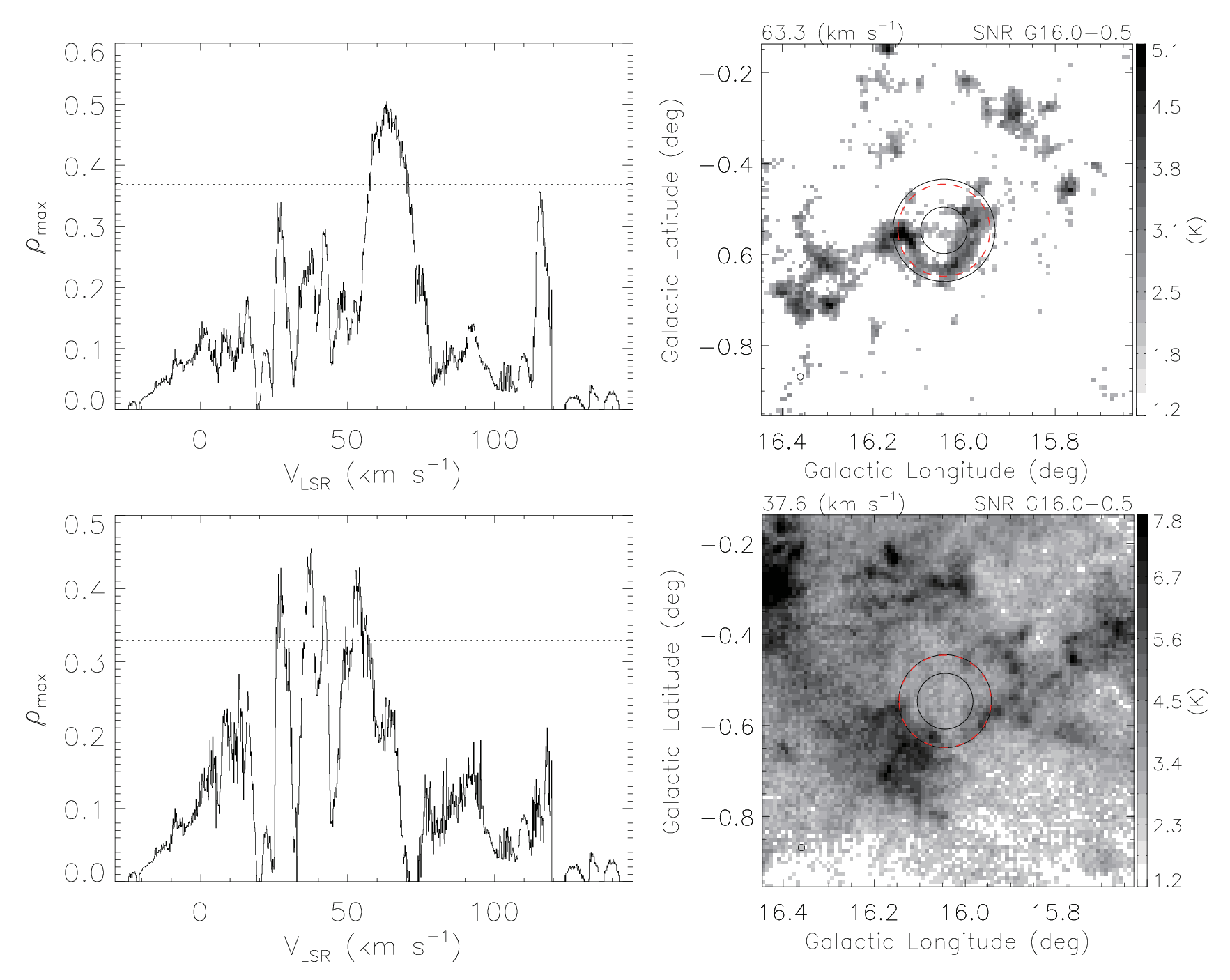, height=5.0in,angle=0, clip=}
\hfil\hfil}}
\caption{Same as Figure~\ref{f:hb3coeff}, but for SNR G16.0$-$0.5. The red dashed circle in the right panel is the same as in Figure~\ref{f:g16}.
The peak of the SCC at $\sim$+63.3~\km\ps\ in upper panels is above 0.5, nevertheless, the peak of the SCC in lower panels at $\sim$+37.6~\km\ps\ is not so significant.
}
\label{f:g16coeff}
\end{figure*}

\cite{Beaumont+2011} applied Support Vector Machines (SVMs) machine learning algorithm to classify broadened \twCO~(J=3--2) lines in SNR G16.0$-$0.5, and found shocked shell-like molecular gas in a wide velocity range from $-5$ to +90~\km\ps. Near-infrared H$_2$ lines at +51~\km\ps\ were also detected toward this SNR \citep{Lee+2020}.
%SNR G16.0-0.5: two vbl identified, all associated with the SNR and originate from one v of them, based on [Beaumont et al. 2001]
SNR G16.0$-$0.5 locates in the inner galaxy direction, where line overlapping effect is serious. It needs strict criteria to identify broad lines in such complicated background. %especially for \twCO~(J=1--0) line emission.
Here, applying full criteria, we identify three broad line components in the SNR at velocities of +43.7, +67.9, and +87.5~\km\ps. The identification accuracy for the +67.9~\km\ps\ component is the highest, which is 0.83, and accuracies of other two components are around 0.3. The total completeness is only $\sim$0.08. The high accuracy and low completeness indicate an efficient suppression of line overlapping effect.
%Applying partial criteria, broad lines are identified at many velocities. And, the broad line identification result is affected by line overlapping effect very much. %almost throughout the entire SNR region.
The +43.7 and +67.9~\km\ps\ broad line components are also identified when using full criteria plus the clean subbackground region condition. Accordingly, identification accuracies of two components are improved to 0.59 and 0.92, respectively, and the overall completeness drops to 0.03.
Overall, broad line identification using partial criteria seems greatly affected by line overlapping effect, with broad line candidates identified at many velocities. Nevertheless, many broad line candidates at +67.9~\km\ps\ are identified using partial criteria, and most of them are distributed in the remnant (see Figure~\ref{f:g16}). There are also many broad line candidates identified at +43.7~\km\ps\ by partial criteria, which are widely distributed inside and outside the remnant.
%\cite{Beaumont+2011} applied Support Vector Machines (SVMs) machine learning algorithm to classify broadened \twCO~(J=3--2) lines in this SNR, and found shocked shell-like molecular gas in a wide velocity range from $-5$ to +90~\km\ps. %(see Figure~\ref{f:g16}).
As shown in Figure~\ref{f:g16}, the shocked molecular shell detected by \twCO~(J=3--2) emission in the remnant \citep{Beaumont+2011} is mainly composed of gases of the +67.9~\km\ps\ velocity component, which also follows the remnant's radio continuum shell peak well. Some molecular gas of the +43.7~\km\ps\ velocity component contributes to the shocked molecular shell outside the remnant.

The peak of the SCC for SNR G16.0$-$0.5 is at $\sim$+63.3~\km\ps\ and above 0.5, which considers pixels inside the ring templates or within twice the radius of the SNR (see the upper left panel in Figure~\ref{f:g16coeff}).
In addition, the $\sim$+63.3~\km\ps\ component in the velocity versus SCC plot is wide.
%, and is the only component above $3\sigma$ confidence level. 
%It} indicates a spatial correlation between the remnant and molecular gas at $\sim$+63.3~\km\ps.
There is shell-like molecular gas at $\sim$+63.3~\km\ps\ spatially correlated with the southeastern radio continuum shell of the remnant.
%the correlation coefficient plot is wide, and 
It supports the association between the remnant and the $67.9$~\km\ps\ broad line component. %, which is with identified broad lines with good accuracy.
Nevertheless, when considering pixels inside the ring templates or within the radius of the SNR, the SCC has a nonsignificant peak at $\sim$+37.6~\km\ps. Molecular gas at $\sim$+37.6~\km\ps\ is distributed around the southeastern and southwestern boundaries, which belongs to the +43.7~\km\ps\ broad line component.

%As shown in \cite{Beaumont+2011}, there is not much \twCO~(J=3--2) emission around +67.9~\km\ps, 
There is not much \twCO~(J=3--2) emission detected around +67.9~\km\ps\ as shown by \cite{Beaumont+2011}. 
\twCO~(J=3--2) line emission is efficient to trace hot shocked molecular gas but not cold quiet molecular gas.
%Based on the spatial correlation between CO emission and the radio continuum emission of the remnant, as well as identified broad CO lines with higher accuracy, we suggest that the SNR is associated with the MC at the systemic velocity of +67.9~\km\ps. 
%Based on the spatial correlation result as well as identified broad CO lines with high accuracy, we suggest that the SNR is associated with the +67.9~\km\ps\ MC. %at the systemic velocity of +67.9~\km\ps. 
Based on the spatial correlation result as well as identified broad CO lines with high accuracy, the SNR is probably associated with the +67.9~\km\ps\ component. However, we cannot totally rule out the possibility of an association between the SNR and the +43.7~\km\ps\ component. 
Note that a further study of the expanding gas motion suggests that the +43.7~\km\ps\ component is associated with another object, which supports the association between SNR~G16.0$-$0.5 and the +67.9~\km\ps\ component (Zhou et al.\ 2023, in preparation). %at the systemic velocity of +67.9~\km\ps. 
%Incidentally, as in SNR IC~443, most of the broadened CO lines are blue-shifted.
Based on the full distance probability density function, the kinematic distance of the +67.9~\km\ps\ velocity component is estimated as 3.9$\pm0.3$~kpc, which is located in the Norma spiral arm. 
Kinematic distances of the +43.7~\km\ps\ component are estimated as 3.2$\pm0.3$ and 4.0$\pm0.3$ kpc.
The +43.7~\km\ps\ velocity component might be also located in the Norma spiral arm, as indicated by the full distance probability density function.

%SNR G11.2$-$0.3 is not consistent with Kilpatrick+2016 result: 0.8km/s here vs. 32km/s
\section{Results}\label{sec:result}
We universally search for evidence of kinematic and spatial correlation of SNR-MC association for 149 SNRs, 170 SNR candidates (SNRCs), and 18 pure PWNe in our coverage.
We apply full criteria to identify broad lines associated with SNRs at first, thereafter, apply the partial criteria to those with no broad lines found by full criteria. Broad lines identified by the partial criteria are considered as candidates. %since possible contamination of line overlapping effect.
%BL by full criteria may distribute good, mainly in SNR region, but may not good by partial criteria 
In some SNRs, more than one broad line components at different velocities are found.
Multiple broad line components found in one SNR might all belong to it, which originate from molecular gases pushed toward and away from us at the near side and far side of the remnant, respectively. 
%Furthermore, there may be more than one broad lines at different velocities all belonging to one SNR, which originate from molecular gases pushed toward and away from us at the near side and far side of the remnant.
However, such multiple components might also originate from contamination of the line overlapping effect or overlaid energetic objects.
In particular, in directions perpendicular to the direction of the circular motion of the solar system barycenter around the Galactic center, i.e.\ around l$\sim$$0^{\circ}$ and l$\sim$$180^{\circ}$, MCs in different spiral arms are with systemic velocities close to each other, where the line overlapping effect is severe.
Some lines are intrinsically indistinguishable in observations, which affects broad line identification results.
%Some lines are intrinsically indistinguishable in observations, especially in the inner Galactic region, where overlapping effects affect broad line identification.
For SNRs with many broad lines at different velocities identified even by full criteria, we use broad lines identified by full criteria plus the clean subbackground region condition as a reference.
We greatly eliminate the line overlapping effect in our searching method. Besides, further examination is performed in settling the systemic velocity of MC associated with SNRs. %based on automatically searching results.
%Broad lines, as kinematic evidence of association, are identified in / SNRs.
Spatial correlations are also examined to support kinematic evidence, or provided as independent evidence for SNR-MC association candidates.
For SNRs and SNRCs within 10$^{\circ}$ of l$\sim$0$^{\circ}$ or within 5$^{\circ}$ of l$\sim$180$^{\circ}$, where broad line identifications are affected by serious line overlapping effects, we search spatial correlation evidence first, and then choose those supported by identified broad line candidates.
For 18 pure PWNe, to identify candidates of associated MCs, we also search spatial correlation evidence over a large extent at first, and then further examine related broad lines identified by full criteria.

%table of all
%-snr
\begin{center}
%\begin{deluxetable}{cccccp{80pt}p{80pt}p{40pt}c}
%\begin{deluxetable}{ccccp{80pt}p{60pt}cp{40pt}c}
%\begin{deluxetable}{cccccp{80pt}p{60pt}cp{40pt}p{40pt}}
\begin{deluxetable}{cccccp{70pt}p{60pt}cp{40pt}p{40pt}}
\tabletypesize{\tiny} %\footnotesize \tiny
\tablecaption{SNR-MC Associations in Survey.\label{tab:snr}}
\tablewidth{0pt}
\tablehead{
%\colhead{SNR (Other Name)}&\colhead{L}&\colhead{B}&\colhead{Coverage}&\colhead{Type\tablenotemark{a}}&\colhead{BL \vlsr\ (Accuracy)}&\colhead{SC \vlsr\ (Coefficient)}&\colhead{$V_{\rm sys}$}&\colhead{Distance}\\
%&\colhead{$^{\circ}$}&\colhead{$^{\circ}$}&\colhead{$'\times '$}&&\colhead{\km\ps}&\colhead{\km\ps}&\colhead{\km\ps}&\colhead{kpc}
%\colhead{SNR (Other Name)}&\colhead{L}&\colhead{B}&\colhead{Coverage}&\colhead{Type\tablenotemark{a}}&\colhead{BL \vlsr\ (Accuracy)}&\colhead{SC \vlsr\ (Coefficient)}&\colhead{$V_{\rm sys}$}&\colhead{Distance}&\colhead{$V_{{\rm H{\sc II}}}$\tablenotemark{b}}\\
\colhead{SNR (Other Name)}&\colhead{GLon}&\colhead{GLat}&\colhead{Coverage}&\colhead{Type\tablenotemark{a}}&\colhead{BL \vlsr\tablenotemark{b}}&\colhead{SC \vlsr\ (Coefficient)\tablenotemark{c}}&\colhead{$V_{\rm sys}$\tablenotemark{d}}&\colhead{Distance\tablenotemark{e}}&\colhead{$V_{{\rm H{\sc II}}}$\tablenotemark{f}}\\
&\colhead{$^{\circ}$}&\colhead{$^{\circ}$}&\colhead{$'\times '$}&&\colhead{\km\ps}&\colhead{\km\ps}&\colhead{\km\ps}&\colhead{kpc}&\colhead{\km\ps}
}
\startdata
G1.4$-$0.1&1.460&$-$0.154&41$\times$41&S&+78.9, +127.7&[+57.0 (0.40)]&$-$4.8?&12.6 (0.23)?&$-$2.2\\
G1.9+0.3&1.871&0.324&7$\times$7&S&[$-$8.7, $-$2.1, +8.3, +10.0, +49.4]&-&-&-&-\\
G3.7$-$0.2&3.775&$-$0.288&57$\times$57&S&$-$30.6, +9.3, +17.3, +21.3, +83.5&+19.2 (0.48), +79.4 (0.55)&+9.3?, +83.5?&2.9 (0.47)?, 7.7 (0.66)?&$-$28.9, +4.1\\
G3.8+0.3&3.851&0.349&81$\times$81&S&$-$33.2, $-$0.6, +12.1, +162.9, +173.2, +185.1&+8.7 (0.38)&+12.1?, +162.9?&2.9 (0.60)?, 8.1 (1.00)?&+7.3\\
G4.5+6.8 (Kepler)&4.519&6.819&14$\times$14&S&-&[+11.7 (0.29)]&-&-&-\\
G4.8+6.2&4.776&6.231&65$\times$65&S&-&[+4.8 (0.46)]&+4.8?&1.6 (0.26) or 14.8 (0.26) ?&-\\
G5.2$-$2.6&5.173&$-$2.612&55$\times$55&S&-&+7.8 (0.30)&+7.8?&1.5 (0.75)?&-\\
G5.4$-$1.2 (Milne 56)&5.363&$-$1.220&148$\times$148&C&$-$24.9, $-$9.7, +0.3, +11.1, +22.1, +187.5&$-$14.8 (0.27), +8.6 (0.27), +29.7 (0.38)&$-$24.9&4.2 (0.56) or 12.1 (0.44)&-\\
G5.5+0.3&5.545&0.368&59$\times$59&S&+10.3&+7.0 (0.55)&+10.3&2.9 (0.54)&-\\
G6.1+0.5&6.104&0.479&73$\times$73&S&+7.0&+6.0 (0.33)&+7.0&3.0 (0.43)&-\\
...

\enddata
%\tablecomments{The full table is availible as online material.}
%\tablecomments{The full table is published in machine-readable format with this paper and is also availible online\footnote{http://?}.}
\tablecomments{The full table is published in machine-readable format with this paper.}
\tablenotetext{a}{Type of SNRs, S for shell, C for composite, F for filled center (also known as plerion), and "?" denotes uncertain ones.}
\tablenotetext{b}{\vlsr\ of detected broad lines. Those identified only by partial criteria are enclosed by square brackets.}
\tablenotetext{c}{\vlsr\ of spatially correlated molecular gas, with the corresponding spatial correlation coefficient enclosed in brackets. Those with coefficients below 5$\sigma$ confidence level and less than 0.5 are enclosed by square brackets.}
\tablenotetext{d}{Systemic velocity of associated MC. "?" denotes possible ones, see text for details.}
\tablenotetext{e}{Kinematic distance corresponding to $V_{\rm sys}$. 
All kinematic distances are estimated by a full distance probability density function \citep{Reid+2016, Reid+2019}, and estimated probabilities are enclosed in brackets.}%, which are denoted by "f".}
\tablenotetext{f}{$V_{\rm LSR}$ of overlapped \HII\ regions.}
\end{deluxetable}
\end{center}
%\clearpage

%snr candidate
\begin{center}
%\begin{deluxetable}{ccccp{80pt}p{80pt}cp{40pt}c}
%\begin{deluxetable}{cccccp{80pt}p{60pt}cp{40pt}p{40pt}}
\begin{deluxetable}{cccccp{70pt}p{60pt}cp{40pt}p{40pt}}
%\tabletypesize{\scriptsize} %\footnotesize \tiny
\tabletypesize{\tiny} %\footnotesize \tiny
\tablecaption{SNR candidate-MC Associations in Survey.\label{tab:snrcand}}
\tablewidth{0pt}
\tablehead{
%\colhead{SNR candidate}&\colhead{L}&\colhead{B}&\colhead{Coverage}&\colhead{Type\tablenotemark{a}}&\colhead{BL \vlsr\ (Accuracy)}&\colhead{SC \vlsr\ (Coefficient)}&\colhead{$V_{\rm sys}$}&\colhead{Distance}\\
%&\colhead{$^{\circ}$}&\colhead{$^{\circ}$}&\colhead{$'\times '$}&&\colhead{\km\ps}&\colhead{\km\ps}&\colhead{\km\ps}&\colhead{kpc}
%\colhead{SNR candidate}&\colhead{L}&\colhead{B}&\colhead{Coverage}&\colhead{Type\tablenotemark{a}}&\colhead{BL \vlsr\ (Accuracy)}&\colhead{SC \vlsr\ (Coefficient)}&\colhead{$V_{\rm sys}$}&\colhead{Distance}&\colhead{$V_{{\rm H{\sc II}}}$\tablenotemark{b}}\\
\colhead{SNR candidate}&\colhead{GLon}&\colhead{GLat}&\colhead{Coverage}&\colhead{Type}&\colhead{BL \vlsr\ (Accuracy)}&\colhead{SC \vlsr\ (Coefficient)}&\colhead{$V_{\rm sys}$}&\colhead{Distance}&\colhead{$V_{{\rm H{\sc II}}}$}\\
&\colhead{$^{\circ}$}&\colhead{$^{\circ}$}&\colhead{$'\times '$}&&\colhead{\km\ps}&\colhead{\km\ps}&\colhead{\km\ps}&\colhead{kpc}&\colhead{\km\ps}
}
\startdata
G1.95$-$0.10&1.929&$-$0.091&25$\times$25&F&$-$42.5, +36.7, +44.7, +88.4, +124.9&[+59.5 (0.40)]&+44.7?&8.1 (0.69)?&-\\
G1.98$-$0.46&1.975&$-$0.460&20$\times$20&S&+2.2, +12.6, +157.4&+5.1 (0.40), +8.6 (0.46)&+12.6?&2.8 (0.76)?&-\\
G2.23+0.06&2.228&0.058&9$\times$9&S&[$-$23.2, +11.3, +16.6, +23.7, +156.5]&$-$15.9 (0.54), $-$6.3 (0.67), $-$2.9 (0.53), $-$1.0 (0.55), +6.0 (0.65), +8.3 (0.51), +19.4 (0.52)&-&-&-\\
G2.28+0.40&2.248&0.385&27$\times$27&S&$-$66.8, $-$43.3, +4.8, +13.8, +88.4&+3.2 (0.40), +19.0 (0.40)&+4.8?&2.8 (0.54)?&+4.9\\
G2.91$-$0.18&2.910&$-$0.183&48$\times$48&F&$-$40.0, $-$13.3, +4.0, +13.3, +18.6, +52.7, +88.8, +102.6, +160.3&+14.3 (0.30), +43.3 (0.27)&$-$40.0&4.7 (0.92)&$-$3.5, $-$2.1, +2.5, +18.1\\
G3.10$-$0.09&3.101&$-$0.093&16$\times$16&C&+23.7&$-$18.1 (0.31)&+23.7?, +95.8?&2.9 (0.29) or 12.6 (0.35) ?, 8.2 (0.90)?&-\\
G3.10+0.11&3.103&0.110&9$\times$9&S&[+11.3, +22.7, +58.3, +104.9, +123.9, +135.0, +144.5, +153.6]&+1.7 (0.64), +11.3 (0.55), +17.6 (0.60), +81.9 (0.55), +84.3 (0.61), +87.6 (0.56)&-&-&-\\
G3.1$-$0.6&3.100&$-$0.600&83$\times$83&C&$-$41.0, $-$28.4, +6.8, +12.8, +17.4, +19.9, +52.7, +82.2, +102.7&[+21.4 (0.37)]&$-$41.0?&4.7 (0.71)?&+0.1, +5.1\\
G4.20$-$0.30&4.164&$-$0.333&32$\times$32&C?&$-$31.4, +7.1, +15.3&[+0.8 (0.40)]&$-$31.4?&4.7 (0.67)?&-\\
G4.49$-$0.39&4.482&$-$0.391&29$\times$29&F&+6.0, +14.1, +20.8, +43.2&+16.3 (0.37), +44.4 (0.41)&+14.1?&2.9 (0.61)?&-\\
...

\enddata
\tablecomments{Same as Table~\ref{tab:snr} but for SNR candidates. The full table is published in machine-readable format with this paper.}
%\tablenotetext{a}{S for shell, C for composite, F for filled center (also known as plerion), and ? for uncertain.}
%\tablenotetext{b}{$V_{\rm LSR}$ of overlapped \HII\ regions.}
\end{deluxetable}
\end{center}

\begin{center}
\onecolumn
\scriptsize
\begin{longtable}{p{80pt}p{80pt}p{50pt}p{60pt}p{80pt}p{50pt}}
\caption{SNR-MC Associations Studied in Previous Works in Our Coverage.}\label{tab:snrpre}\\
\hline\hline
\multicolumn{1}{c}{SNR (Other Name)}&\multicolumn{1}{c}{OH 1720 MHz Maser \vlsr\tablenotemark{a}}&\multicolumn{1}{c}{CO BL \vlsr\tablenotemark{b}}&\multicolumn{1}{c}{CO SC \vlsr\tablenotemark{c}}&\multicolumn{1}{c}{Other Evidences \vlsr\tablenotemark{d}}&\multicolumn{1}{c}{Reference\tablenotemark{e}}\\
\multicolumn{1}{c}{}&\multicolumn{1}{c}{\km\ps}&\multicolumn{1}{c}{\km\ps}&\multicolumn{1}{c}{\km\ps}&\multicolumn{1}{c}{\km\ps}&\multicolumn{1}{c}{}\\
\hline
\endfirsthead
 
\multicolumn{5}{c}{{\bfseries \tablename\ \thetable{} --Continued}} \\
\hline\hline
\multicolumn{1}{c}{SNR (Other Name)}&\multicolumn{1}{c}{OH 1720 MHz Maser \vlsr\tablenotemark{a}}&\multicolumn{1}{c}{CO BL \vlsr\tablenotemark{b}}&\multicolumn{1}{c}{CO SC \vlsr\tablenotemark{c}}&\multicolumn{1}{c}{Other Evidences \vlsr\tablenotemark{d}}&\multicolumn{1}{c}{Reference\tablenotemark{e}}\\
\multicolumn{1}{c}{}&\multicolumn{1}{c}{\km\ps}&\multicolumn{1}{c}{\km\ps}&\multicolumn{1}{c}{\km\ps}&\multicolumn{1}{c}{\km\ps}&\multicolumn{1}{c}{}\\
\hline
\endhead
 
\hline \multicolumn{5}{l}{{Continued on next page}}\\
\endfoot
 
\hline\hline
%\multicolumn{6}{l}{$^{\rm a}$\vlsr\ range of associated OH 1720 MHz maser emission, and 'D' for a detection of OH 1720 MHz line.}\\
%\multicolumn{6}{l}{$^{\rm b}$\vlsr\ of associated broad CO line.}\\
%\multicolumn{6}{l}{$^{\rm c}$\vlsr\ of spatially correlated CO emission, and 'Y' for a detection of spatially correlated CO emission without velocity information.}\\
%\multicolumn{6}{l}{$^{\rm d}$\vlsr\ by other evidence, where 'SC' represents spatial correlation.}\\
\endlastfoot
 
%\startdata
%\input{snrmc_pres.tex}
G1.4$-$0.1&$-$2.3 to $-$2.5&-&-&$-$32.6 to +29.9 (CH$_3$OH 36 GHz maser)&1, 2, 3\\
G1.9+0.3&D&-&-&-&1\\
G5.4$-$1.2 (Milne 56)&$-$21&-&$-$21&-&4, 5\\
G5.5+0.3&-&-&+12.5&-&5\\
G6.4$-$0.1 (W28)&+2.45 to +15.94&+5.8, +7&Y&$-$2.3 (OH 6035 MHz maser), +6.8 to +8.2 (CH$_3$OH 36 and 44 GHz maser), +6.1 to +13.1 (CS line), +6.3 to +7.4 (HCO+)&1, 10, 11, 12, 13, 14, 3, 6, 7, 8, 9\\
G8.7$-$0.1 (W30)&+36&-&-&-&4\\
G9.7$-$0.0&+43&-&-&-&4\\
G9.9$-$0.8&-&+31&-&+30 (H$_2$ line)&15, 16\\
G11.0$-$0.0&-&-&+21, +40&+30 (SiO and CO lines, CS and HC$_3$N SC)&17, 18, 19\\
G11.1+0.1&-&-&+56&-&18\\
G11.1$-$0.7&-&-&+33&-&18\\
G11.2$-$0.3&D&+32&+33&+45 (H~{\sc i} absorption), +48 (H$_2$ line)&1, 15, 16, 18, 20\\
G11.4$-$0.1&D&-&+30, +50&-&1, 18\\
G11.8$-$0.2&-&-&+49.8&-&18\\
G12.0$-$0.1&D&-&+37.4&-&1, 18\\
G12.2+0.3&-&+50&-&-&15\\
G13.3$-$1.3&D&-&-&-&1\\
G13.5+0.2&-&-&+24&+40 (H$_2$ line)&16, 18\\
G15.4+0.1&-&-&+34, +47.8&+60 (H~{\sc i} SC), +60 (H~{\sc i} absorption)&18, 21, 22\\
G15.9+0.2&-&-&+29&-&18\\
G16.0$-$0.5&-&+40&-&+51 (H$_2$ line)&16, 23\\
G16.7+0.1&+19.92 to +20.0&+25&+25.6, +47, +62&-&1, 15, 18, 24, 7\\
G17.0$-$0.0&-&-&+31, +93.4&-&18\\
G17.4$-$2.3&D&-&-&-&1\\
G17.8$-$2.6&D&-&-&-&1\\
G18.1$-$0.1&-&-&+53.1, +49&+100, +102, +103.74 (H~{\sc i} absorption), +73 to +85 (H$_2$ line)&16, 18, 25, 26, 27, 28\\
G18.6$-$0.2&-&+42&+62, +66&+62 (H~{\sc i} absorption), +62.84 (H~{\sc i} absorption)&15, 18, 25, 26, 29\\
G18.8+0.3 (Kes 67)&-&-&+19, +20, +21.35&+20 (CO (2--1)/(1--0) ratio of 1.25), +21.35 (H~{\sc i} absorption), +19 (H~{\sc i} SC)&18, 26, 30, 31, 32, 33, 34\\
G18.9$-$1.1&-&-&+25.6&+23 (H~{\sc i} absorption), +70 (H$_2$ line)&16, 35, 36, 37\\
G20.0$-$0.2&-&-&+65, +66, +66.4&+66.4 (H~{\sc i} absorption)&18, 26, 38\\
G21.5$-$0.9&D&-&-&$\sim$+68 (H~{\sc i} absorption)&1, 26, 39\\
G21.8$-$0.6 (Kes 69)&+69.3 to +69.76 (compact), +83.6 to +85.2 (extended)&-&+83, +85, +93.35&+85 (HCO+ line on radio peak), +86 to +93.35 (H~{\sc i} absorption), +61 (H$_2$ line)&1, 16, 18, 26, 39, 40, 7\\
G22.7$-$0.2&D&+77&+75, +76.63, +77&+76.63 (H~{\sc i} absorption)&1, 18, 26, 41, 42\\
G23.3$-$0.3 (W41)&D (at +71)&-&+63, +70, +77, +78.51&+63 (H~{\sc i} SC), +78.51 (H~{\sc i} absorption)&1, 18, 26, 42, 43, 44, 45\\
G24.7$-$0.6&D&-&+60, +60.67&-&18, 46, 47\\
G24.7+0.6&-&-&+112&-&18\\
G27.4+0.0 (Kes 73, 4C$-$04.71)&D&+100&+90, +101&+90 (CO (2--1)/(1--0) $\sim$ 1.1), +99.95 (H~{\sc i} absorption), +99 (H$_2$ line)&1, 15, 16, 18, 26, 48\\
G27.8+0.6&D&-&-&-&47\\
G28.6$-$0.1&-&-&+86&-&18, 46\\
G29.6+0.1&-&+94&+99.2&-&15, 18\\
G29.7$-$0.3 (Kes 75)&D&+53, +54&+52, +54, +95, +112&+95 to +102 (H~{\sc i} absorption)&1, 15, 18, 26, 49, 50\\
G31.5$-$0.6&-&-&+87.5, +97&-&18\\
G31.9+0.0 (3C 391)&+97.04 to +110.2&+103.9 to +112&+107, +100&+100 (H$_2$ line), +104.2 to +108.9 (broad CS line), +111.4 (broad HCO+ line)&1, 15, 16, 18, 47, 51, 52, 53, 54, 55, 7\\
G32.1$-$0.9&-&-&+95&+85 (H$_2$ line)&16, 18\\
G32.4+0.1&-&+43&+10.8, +42.6&-&15, 18\\
G32.8$-$0.1 (Kes 78)&+86.1&+81&+74, +81, +81.81, +103&+81 (CO (2--1)/(1--0) ratio enhanced), +81.81 (H~{\sc i} absorption), +90 (H$_2$ line)&16, 18, 26, 56, 57\\
G33.2$-$0.6&-&-&+54, +91&+82 (H$_2$ line)&16, 18\\
G33.6+0.1 (Kes 79)&D&$\sim$+105&+70, +80, +103, +105&+80 (CO (3--2)/(1--0) $>$ 0.8 and CO shell in P-V map), +105 (HCO+ line), +95 (H~{\sc i} SC), +57.9 (H~{\sc i} absorption)&1, 15, 18, 26, 58, 59, 60, 61, 62\\
G34.7$-$0.4 (W44)&+23.8 to +47.35&+40, +45, +46.6, +47.5, +48&+40, +48, +52&+40 and +45 (high-J CO lines; CO (2--1)/(1--0) $>$ 1), +40 and +47.5 (HCO+ line), +41 to +49 (H$_2$ line), +48 ($^{13}$CO (1--0)/$^{12}$CO (1--0) $\sim$ 0.03 and HCO+ (1--0)/$^{12}$CO (1--0) $\sim$ 0.3), +54.4 (OH 6035 MHz maser), +40.9 to +47.1 and +6.7 to +56.9 (CH$_3$OH 36 and 44 GHz maser), +50.48 (H~{\sc i} absorption)&1, 10, 11, 16, 18, 26, 6, 63, 64, 65, 66, 67, 68, 69, 7, 70, 71, 72, 8\\
G35.6$-$0.4&-&-&+55, +90&+63.67 (H~{\sc i} absorption)&18, 26, 73, 74\\
G36.6$-$0.7&-&-&+57, +79&-&18\\
G39.2$-$0.3 (3C 396)&D&+69, +77, +84&+51, +67, +69, +69.39, +84&+56 (H$_2$ line), +69.39 (H~{\sc i} absorption)&1, 15, 16, 18, 26, 75, 76, 77\\
G39.7$-$2.0 (W50)&-&+53, +77&+32&+53 (CO (2--1)/(1--0) $>$ 0.9, CN (3/1--1/2)), +77 (H~{\sc i} SC)&78, 79, 80\\
G40.5$-$0.5&-&-&+55, +58, +67&-&18, 81, 82\\
G41.1$-$0.3 (3C 397)&-&+31, +32&+32, +38, +40&-&15, 18, 83, 84\\
G41.5+0.4&-&-&+58&-&18\\
G42.0$-$0.1&-&-&+66&-&18\\
G43.3$-$0.2 (W49B)&-&+14&+10, +32, +40, +45, +62&+11 (OH 6035 MHz maser), +63 (HCO+ broad line, high HCO+ (1--0)/CO (1--0) ratio), +12.55 (H~{\sc i} absorption), +40 (H~{\sc i} absorption), +10 (CO high T$_{ex}$ and $\sim$+6 km~s$^{-1}$ expanding motion), +64 (H$_2$ line)&11, 15, 16, 18, 26, 85, 86, 87, 88\\
G43.9+1.6&-&+50&Y&-&89\\
G45.7$-$0.4&-&-&+26, +48.5&-&18\\
G46.8$-$0.3 (HC30)&-&+19&+19, +52&+0 to +40.4 (H~{\sc i} absorption)&18, 90\\
G49.2$-$0.7 (W51C)&+68.9 to +72.13&+60&+50, +60&+67.6 (OH 6035 MHz maser), +60 (HCO+ line), +57.6 to +76.2 (CH$_3$OH 36 GHz maser), $\sim$+70 (H~{\sc i} absorption)&1, 11, 18, 63, 7, 91, 92, 93\\
G53.6$-$2.2 (3C 400.2)&-&-&-&+27 (H~{\sc i} SC)&94\\
G54.1+0.3&-&+23&+23, +53, +53.66&+53.66 (H~{\sc i} absorption)&26, 95, 96, 97\\
G54.4$-$0.3 (HC40)&-&-&+36.66, +40&+44 (H$_2$ line)&16, 52, 98\\
G57.2+0.8 (4C$-$21.53)&+30&-&+12, +30&$-$46 (H~{\sc i} SC)&100, 99\\
G59.5+0.1&-&+28&Y&+28 (CO (3--2)/(2--1) $\sim$ 1.58)&101\\
G63.7+1.1&-&-&$\sim$+13, +21&$\sim$+13 (H~{\sc i} SC)&102, 103\\
G69.0+2.7 (CTB 80)&-&+11.5&+11.5&+13.5 (H~{\sc i} SC)&104\\
G73.9+0.9&-&-&+3&-&105\\
G74.9+1.2 (CTB 87)&-&$-$58&$-$58, $-$57.3, $-$57&$-$64 (H~{\sc i} SC)&106, 107, 108, 109\\
G78.2+2.1 (DR4)&-&-&$-$2 to +13.6&-&110, 111, 112\\
G84.2$-$0.8&-&-&$-$39, $-$17&-&105, 106, 113\\
G85.4+0.7&-&-&$-$41&-&105\\
G89.0+4.7 (HB 21)&-&$-$5, +3&$-$5, $-$3, $\sim$$-$1 ($-$10 to +8)&+3 (CO (2--1)/(1--0) ratio of 1.6 to 2.3), $\sim$$-$1 (H~{\sc i} SC)&114, 115, 116, 117\\
G93.7$-$0.2 (CTB 104A)&-&-&-&$-$6 (H~{\sc i} SC)&118\\
G94.0+1.0 (3C 434.1)&-&-&$-$13&$-$13 (CO (2--1)/(1--0) ratio $\sim$ 1.6)&105, 119\\
G106.3+2.7&-&-&$-$6.4&$-$6.4 (H~{\sc i} SC)&120\\
G109.1$-$1.0 (CTB 109)&-&$\sim$$-$55&$-$50 to $-$48&$-$50 (H~{\sc i} SC)&121, 122, 123, 124\\
G111.7$-$2.1 (Cas A)&-&$\sim$$-$40&Y&$-$38 (H$_2$CO distance matching), $-$40 (H~{\sc i} SC and multibands consistency)&125, 126, 127, 128, 129, 15\\
G116.9+0.2 (CTB 1)&-&-&-&$-$32 (H~{\sc i} SC)&130\\
G120.1+1.4 (Tycho)&-&$-$63.5&$-$63.5, $-$62, $-$61&$-$61 (high CO (2--1)/(1--0) ratio), $-$51.5 (H~{\sc i} SC)&131, 132, 133, 134, 135\\
G127.1+0.5 (R5)&-&$\sim$+5&Y&-&136\\
G130.7+3.1 (3C 58)&-&-&-&$-$39.1 to $-$34.1, $\sim$$-$38 (H~{\sc i} absorption), $\sim$$-$36 (H~{\sc i} absorption and SC)&137, 138, 139\\
G132.7+1.3 (HB 3)&-&$-$45, $-$42, $-$40&$-$45, $-$43, $-$42, $-$40.5&$-$30 (H~{\sc i} SC), $-$42 (high-J CO lines with co-spatial H$_2$ emission)&140, 141, 142, 143, 15\\
G141.2+5.0&-&-&-&$-$53 (H~{\sc i} SC)&144\\
G160.9+2.6 (HB 9)&-&-&-&$-$6 (H~{\sc i} SC)&145\\
G166.0+4.3 (VRO 42.05.01)&-&-&$-$22&$-$6 (H~{\sc i} SC and CO velocity gradient), $-$34 (H~{\sc i} SC)&106, 146, 147\\
G180.0$-$1.7 (S147)&-&-&-&$-$2.5 (optical extinction SC)&148\\
G182.4+4.3&-&-&+4&-&105\\
G189.1+3.0 (IC 443)&$-$32.0 to $-$3.5&$-$40.5 to $-$3.5, $-$5, $-$4.5, $-$4, $-$3, $-$2, +5&$-$4.5, $-$4, $-$3.6, +5&$-$5 (OH absorption line), $-$4.5 (high-J CO lines, CO line polarization), +9 (H$_\alpha$ line), $-$2 (CO (2--1)/(1--0) ratio $>$ +3, CO (3--2)/(2--1) $\sim$ 1.58), $-$4 (H~{\sc i} line with broad wing), $-$10 (H$_2$ line), $-$5 (HCO+ and HCN self-absorption, CN, SiO, CS, SO, and H$_2$CO lines), $-$44.2 to $-$4.2 (CS, HCN, HNC, DCN, HCO+, H$^{13}$CO+, H$_2$CO, H$_2$$^{13}$CO, CH$_3$OH, CH$_3$CN, HC$_3$N, CO+, C$_2$H, C$_3$H$_2$, HDO, SiO, SO, SO$_2$, H$_2$S, and H$_2$CS lines), $-$3.5 (H~{\sc i} emission, CO, HCO+, HCN, and H$_2$ lines)&1, 149, 15, 150, 151, 152, 153, 154, 155, 156, 157, 158, 159, 160, 161, 162, 163, 164, 165, 6, 68, 7\\
G205.5+0.5 (Monoceros Loop)&-&$\sim$+5, $\sim$+19&+5, +10, $\sim$+11, $\sim$+15, +19, +20&-&166, 167, 168, 18\\
G206.9+2.3 (PKS 0646+06)&-&-&+15&-&166\\
G213.0$-$0.6&-&$\sim$+9&+9, +21&-&166, 18\\
G5.7$-$0.1&+13&-&+12&$-$26.4 and $-$24.4 (CH$_3$OH 36 and 44 GHz maser)&4, 5, 63\\
G21.8$-$3.0&-&-&+5.9&-&169\\
G28.56+0.00&-&-&-&+96.6 (RRL of H~{\sc ii} region assumed to be associated by H~{\sc i} absorption)&37\\
G29.4+0.1&-&-&$\sim$+82&-&170\\
G44.5$-$0.2&-&+60&Y&+60 (H~{\sc i} SC)&171\\
G51.21+0.11 (G51.26+0.11)&-&-&-&+37 to +70 (H~{\sc i} absorption)&172\\

%\enddata
 
\end{longtable}
\end{center}
\scriptsize{$^{\rm a}$\vlsr\ range of associated OH 1720 MHz maser emission, and 'D' for a detection of OH 1720 MHz line.}\\
\scriptsize{$^{\rm b}$\vlsr\ of associated broad CO line.}\\
\scriptsize{$^{\rm c}$\vlsr\ of spatially correlated CO emission, and 'Y' for a detection of spatially correlated CO emission without velocity information.}\\
\scriptsize{$^{\rm d}$\vlsr\ by other evidence, where 'SC' represents spatial correlation.}\\
\scriptsize{$^{\rm e}$References.-- (1) \cite{Green+1997}; (2) \cite{Yusef-Zadeh+1999}; (3) \cite{Pihlstrom+2014}; (4) \cite{Hewitt+2009oh}; (5) \cite{Liszt2009}; (6) \cite{Claussen+1997}; (7) \cite{Hewitt+2008}; (8) \cite{GossRobinson1968}; (9) \cite{Arikawa+1999}; (10) \cite{Reach+2005}; (11) \cite{McDonnell+2008}; (12) \cite{Gusdorf+2012}; (13) \cite{Nicholas+2012}; (14) \cite{Wootten1981}; (15) \cite{Kilpatrick+2016}; (16) \cite{Lee+2020}; (17) \cite{Castelletti+2016}; (18) \cite{Sofue+2021}; (19) \cite{Voisin+2019}; (20) \cite{Green+1988}; (21) \cite{Castelletti+2013}; (22) \cite{Supan+2015}; (23) \cite{Beaumont+2011}; (24) \cite{ReynosoMangum2000}; (25) \cite{JohansonKerton2009}; (26) \cite{RanasingheLeahy2018a}; (27) \cite{Paron+2013}; (28) \cite{Leahy+2014}; (29) \cite{Voisin+2016}; (30) \cite{Tian+2007}; (31) \cite{Dubner+2004}; (32) \cite{Dubner+1999}; (33) \cite{Paron+2012}; (34) \cite{Paron+2015}; (35) \cite{Traverso+1999}; (36) \cite{Furst+1989}; (37) \cite{Ranasinghe+2020}; (38) \cite{Petriella+2013}; (39) \cite{TianLeahy2008b}; (40) \cite{Zhou+2009}; (41) \cite{Su+2014}; (42) \cite{Su+2015}; (43) \cite{Frail+2013}; (44) \cite{Tian+2007b}; (45) \cite{LeahyTian2008}; (46) \cite{RanasingheLeahy2018b}; (47) \cite{Frail+1996}; (48) \cite{Liu+2017}; (49) \cite{Su+2009}; (50) \cite{LeahyTian2008a}; (51) \cite{ReachRho1999}; (52) \cite{RanasingheLeahy2017}; (53) \cite{ReynoldsMoffett1993}; (54) \cite{Wilner+1998}; (55) \cite{Gusdorf+2014}; (56) \cite{Koralesky+1998}; (57) \cite{ZhoupChen2011}; (58) \cite{Zhoup+2016a}; (59) \cite{Kuriki+2018}; (60) \cite{GreenDewdney1992}; (61) \cite{Giacani+2009}; (62) \cite{Stanimirovic+2003}; (63) \cite{McEwen2016}; (64) \cite{Hoffman+2005}; (65) \cite{Anderl+2014}; (66) \cite{Sashida+2013}; (67) \cite{Seta+2004}; (68) \cite{Seta+1998}; (69) \cite{Wootten1977}; (70) \cite{Yamada+2017}; (71) \cite{Yoshiike+2013}; (72) \cite{Caswell+1975a}; (73) \cite{Zhu+2013}; (74) \cite{ParonGiacani2010}; (75) \cite{Su+2011}; (76) \cite{Lee+2009}; (77) \cite{deOnaWilhelmi+2020}; (78) \cite{Su+2018}; (79) \cite{Huang+1983}; (80) \cite{Liu+2020}; (81) \cite{Yang+2006}; (82) \cite{Duvidovich+2020}; (83) \cite{Jiang+2010}; (84) \cite{Safi-Harb+2005}; (85) \cite{Chen+2014}; (86) \cite{Zhu+2014}; (87) \cite{Sano+2021}; (88) \cite{Zhoup+2022}; (89) \cite{Zhou+2020}; (90) \cite{Supan+2022}; (91) \cite{Brogan+2000}; (92) \cite{KooMoon1997}; (93) \cite{TianLeahy2013}; (94) \cite{Giacani+1998}; (95) \cite{Lee+2012b}; (96) \cite{Koo+2008}; (97) \cite{Leahy+2008}; (98) \cite{Junkes+1992}; (99) \cite{Zhoup+2020}; (100) \cite{Kothes+2018}; (101) \cite{XuWang2012}; (102) \cite{Wallace+1997}; (103) \cite{Matheson+2016}; (104) \cite{Koo+1993}; (105) \cite{Jeong+2012}; (106) \cite{HuangThaddeus1986}; (107) \cite{Cho+1994}; (108) \cite{Kothes+2003}; (109) \cite{Liu+2018}; (110) \cite{Cong1977}; (111) \cite{Pollock1985}; (112) \cite{Higgs+1983}; (113) \cite{FeldtGreen1993}; (114) \cite{Koo+2001}; (115) \cite{Byun+2006}; (116) \cite{Dobashi+2019}; (117) \cite{Tatematsu+1990}; (118) \cite{Uyaniker+2002}; (119) \cite{Jeong+2013}; (120) \cite{Kothes+2001}; (121) \cite{Sasaki+2006}; (122) \cite{Tatematsu+1987}; (123) \cite{Tatematsu+1990a}; (124) \cite{Kothes+2002}; (125) \cite{Ma+2019}; (126) \cite{Kilpatrick+2014}; (127) \cite{Zhoup+2018}; (128) \cite{ReynosoGoss2002}; (129) \cite{Keohane+1996}; (130) \cite{WillisDickel1971}; (131) \cite{Cai+2009}; (132) \cite{Lee+2004}; (133) \cite{Zhoup+2016}; (134) \cite{Chenxp+2017}; (135) \cite{Reynoso+1999}; (136) \cite{Zhou+2014}; (137) \cite{GreenGull1982}; (138) \cite{Roberts+1993}; (139) \cite{Kothes2013}; (140) \cite{Zhou+2016}; (141) \cite{Routledge+1991}; (142) \cite{Landecker+1987}; (143) \cite{Rho+2021}; (144) \cite{Kothes+2014}; (145) \cite{LeahyTian2007}; (146) \cite{Arias+2019}; (147) \cite{Landecker+1989}; (148) \cite{Chenbq+2017}; (149) \cite{Denoyer1979b}; (150) \cite{Denoyer1979a}; (151) \cite{Hewitt+2006}; (152) \cite{Lee+2012}; (153) \cite{White+1987}; (154) \cite{Zhang+2010}; (155) \cite{Turner+1992}; (156) \cite{vanDishoeck+1993}; (157) \cite{Su+2014b}; (158) \cite{Cornett+1977}; (159) \cite{Ambrocio-Cruz+2017}; (160) \cite{Zhang+2013}; (161) \cite{Rosado+2007}; (162) \cite{Burton+1988}; (163) \cite{DellOva+2020}; (164) \cite{Hezareh+2013}; (165) \cite{Xu+2011}; (166) \cite{Su+2017a}; (167) \cite{Aharonian+2007}; (168) \cite{Oliver+1996}; (169) \cite{Gao+2020}; (170) \cite{Castelletti+2017}; (171) \cite{Su+2017b}; (172) \cite{RanasingheLeahy2022}.}\\
\normalsize
%\clearpage

%comparing with previous works
Results of SNRs and SNRCs are listed in Table~\ref{tab:snr} and \ref{tab:snrcand}, respectively. We only present short versions of Table~\ref{tab:snr} and \ref{tab:snrcand} here, and full versions are only available in electronic form.
Systemic velocities of MCs associated with SNRs and SNRCs are settled, and corresponding kinematic distances are estimated by the full distance probability density function \citep{Reid+2016, Reid+2019}.
Note that CO (J=1--0) emission is more advantageous in determining the systemic velocity of the associated MC, from which sufficient information about quiet molecular gases can be obtained.
The systemic velocity is a directly observed quantity, and the kinematic distance is model dependent. To avoid bias between different models, we uniformly apply the full distance probability density function to estimate kinematic distances here.
We also considered available \HI\ absorption results in previous works for some SNRs to discriminate between near and far kinematic distances.
%and compared with results in previous works. 
For comparison, we also list the results of SNR-MC associations studied in previous works in Table~\ref{tab:snrpre} \citep[see also Table 1 in][for a summary of known SNR distances]{RanasingheLeahy2022}.
For known SNRs, about 60\% of them were found to be associated with MCs in previous works, however, new SNR-MC associations have still been gradually discovered in recent years. For SNRCs and PWNe, the study of MCs around them is still very limited, of which only about 4 and 11\%, respectively, were found to be associated with MCs in previous works. 
It is worth noting that in some SNRs different evidence gives different velocity results.
%The overall knowledge of SNR-MC associations is unclear.
On the other hand, the interaction details of some SNR-MC associations were very well studied, with shocked molecular gases investigated through multiband observations, e.g., W28, W44, IC~443, etc.\ (see Table~\ref{tab:snrpre} for details).

We present details of searching results of SNRs and SNRCs in Sections~\ref{sec:snr} and \ref{sec:snrcand}, respectively.
In Section~\ref{sec:pwn}, we discuss searching results for pure PWNe. %, where 14 PWNe among 18 of them are found with candidates of associated MCs.
Corresponding figures of individual objects are available online \citep{Zhou+2023data}.

%SNR-MC associations suggested in previous works: consistent, non-detected, different velocity components.
%Some SNRs, e.g., those with more than one associated velocity components identified, are further examination (see Section~\ref{sec:individual}).

%\subsection{Individuals}\label{sec:individual}
\subsection{SNRs}\label{sec:snr}
%We list results of SNR-MC associations and SNRC-MC associations in Table~\ref{tab:snr} and \ref{tab:snrcand}, respectively.
SNR-MC association results are summarized in Table~\ref{tab:snr}.
In order to better confirm the association evidence, further investigations, e.g., on spatial correlations with partial radio continuum shells or locations of broad lines, are performed on some objects, especially those with more than one broad line components identified. 
We also make efforts to eliminate contaminations of other energetic sources overlapped, e.g., \HII\ regions. %, or line overlapping effects. 
All considered \HII\ regions are introduced from the WISE catalog of Galactic \HII\ regions \citep{Anderson+2014}, which is one of the most complete catalogs of \HII\ regions in the Galaxy.
Systemic velocities of \HII\ regions overlapping SNRs are listed in Table~\ref{tab:snr}.
\twCO\ or \thCO\ velocities of \HII\ regions are applied for better comparison with our data, and if they do not exist, velocities of other tracers are used.
As noted by \cite{Anderson+2014}, velocities of most \HII\ regions by different tracers are consistent, the mean of absolute velocity differences between molecular and ionized gas emission is $\sim$4~\km\ps.
%For comparison, we also list results of SNR-MC associations studied in previous works in Table~\ref{tab:snrpre}.

Among 149 SNRs studied in this paper, 57 of them are found to be associated with MCs, and 70 of them are considered to be possibly associated with MCs.
40 of the 57 SNR-MC associations and 43 of the 70 possible SNR-MC associations were studied in previous works (see Table~\ref{tab:snrpre}). There are also 7 SNRs with no associated MC found here, but suggested to be associated with surrounding clouds in previous works, mostly based on spatial correlations with \HI\ gases.
Few SNR-MC association results in this paper are not the same as those in previous works, including 7 of SNR-MC associations and 4 of possible SNR-MC associations, i.e.\ G9.9-0.8, G11.1+0.1, G15.4+0.1, G16.0-0.5, G18.9-1.1, G24.7+0.6, G41.5+0.4, G21.5-0.9, G29.7-0.3, G84.2-0.8, and G93.7-0.2. 
We discuss individual SNRs in Appendix~A.
%We discuss individual SNRs in Appendix~\ref{app:snr}.
%We discuss individual SNRs below.\\
%SNRs are suggested to be associated with MCs at new systematic velocities based on our results. %after full considerations.
%Among / SNR-MC associations detected in this paper, / of them are newly determined, / of them were studied in previous works, and results for / of them are consistent.
%For / possible SNRC-MC associations, / of them were studied in previous works, and results for / of them are consistent.
%In SNRs , we detect similar kinematic evidence as , 
%There are also 88 of SNRCs considered to be possibly associated with MCs. Kinematic distances of these SNRCs are also estimated based on systemic velocities of associated MCs.
%\\
%\begin{itemize}
%\item{G1.4$-$0.1.}
%\end{itemize}

\subsection{SNR Candidates}\label{sec:snrcand}
We list all results of SNRC-MC associations in Table~\ref{tab:snrcand}.
Most of SNRCs in our coverage have been recently discovered \citep{Kassim1988, Gorham1990, Gray1994, Trushkin2001, Brogan+2006, Helfand+2006, Gerbrandt+2014, Anderson+2017, Hurley-Walker+2019b, Gao+2020, Dokara+2021, Ranasinghe+2021}, which are not well studied by multiband observations.
A similar approach to study SNR-MC associations is performed to investigate associations between SNRCs and MCs. All types of SNRCs are examined in the same way, in particular, no enlargement of coverage is applied for plerionic SNRCs. 

170 SNRCs are examined, and 50 of them are suggested to be associated with MCs. 
There are also 91 of SNRCs considered to be possibly associated with MCs. 
For all SNRCs, only six of them were studied in previous works, of which five are considered as possible SNRC-MC associations and one as a fixed SNRC-MC association here. The result for one of them is not the same as that in previous works, i.e.\ G28.56+0.00.
%Kinematic distances of these SNRCs are also estimated based on systemic velocities of associated MCs.
%Among 48 SNRC-MC associations detected in this paper, / of them are newly determined, / of them were studied in previous works, and results for / of them are consistent. 
%For 88 possible SNRC-MC associations, / of them were studied in previous works, and results for / of them are consistent.
%SNRs / are suggested to be associated with MCs at new systematic velocities based on our results. %after full considerations.
We discuss individual SNRCs in Appendix~B.
%We discuss individual SNRCs in Appendix~\ref{app:snrc}.
%We discuss individual SNRCs below.\\
%\\

\begin{center}
%\begin{deluxetable}{cccccp{80pt}p{80pt}p{40pt}cc}
%\begin{deluxetable}{cp{40pt}cccp{70pt}ccc}
\begin{deluxetable}{cp{40pt}cccp{70pt}ccc}
\tabletypesize{\tiny} %\scriptsize\footnotesize \tiny
\tablecaption{PWN-MC Association Candidates in Survey.\label{tab:pwn}}
\tablewidth{0pt}
\tablehead{
%\colhead{PWN\tablenotemark{a}}&\colhead{PSR\tablenotemark{a}}&\colhead{L}&\colhead{B}&\colhead{Coverage}&\colhead{SC \vlsr\ (Coefficient)\tablenotemark{b}}&\colhead{BL Detection\tablenotemark{c}}&\colhead{$V_{\rm sys}\tablenotemark{d}$}&\colhead{Distance\tablenotemark{e}}&\colhead{Reference}\\
%&&\colhead{$^{\circ}$}&\colhead{$^{\circ}$}&\colhead{$'\times '$}&\colhead{\km\ps}&\colhead{}&\colhead{\km\ps}&\colhead{kpc}&\colhead{\km\ps}
\colhead{PWN\tablenotemark{a}}&\colhead{PSR\tablenotemark{a}}&\colhead{GLon}&\colhead{GLat}&\colhead{Coverage}&\colhead{SC \vlsr\ (Coefficient)\tablenotemark{b}}&\colhead{BL Detection\tablenotemark{c}}&\colhead{$V_{\rm sys}\tablenotemark{d}$}&\colhead{Distance\tablenotemark{e}}\\
&&\colhead{$^{\circ}$}&\colhead{$^{\circ}$}&\colhead{$'\times '$}&\colhead{\km\ps}&\colhead{}&\colhead{\km\ps}&\colhead{kpc}
}
\startdata
G18.00$-$0.69&J1826$-$1334&18.001&$-$0.691&28$\times$28&+5.4 (0.56)&N&+60.8&3.8 (1.00)\\
Eel&J1826$-$1256&18.500&$-$0.400&500$\times$500&[+51.9 (0.45)]&Y&+19.8&1.50 (0.60)\\
G21.88$-$0.10&J1831$-$0952&21.881&$-$0.102&261$\times$261&+40.2 (0.35)&Y&+55.0&3.4 (0.47)\\
G23.5+0.1?&J1833$-$0827&23.386&0.063&290$\times$290&[+96.2 (0.38)]&Y&+95.8&5.3 (0.84)\\
G25.24$-$0.19&J1838$-$0655&25.246&$-$0.196&65$\times$65&+3.5 (0.47), +9.0 (0.44), +107.1 (0.40)&N, N, Y&+102.1&5.9 (0.95)\\
G32.64+0.53?&J1849$-$0001&32.638&0.527&64$\times$64&+11.0 (0.44)&N&+98.0&5.2 (0.68)\\
G36.01+0.06?&J1856+0245&36.008&0.058&74$\times$74&[+34.9 (0.45)]&Y&+58.1?&9.5 (1.00)?\\
G47.38$-$3.88&J1932+1059&47.382&$-$3.884&375$\times$375&+6.7 (0.33)&N&+6.7?&0.361\\
G59.20$-$4.70&J1959+2048&59.197&$-$4.697&18$\times$18&[+13.0 (0.23)]&N&-&-\\
G63.7+1.1?&3XMM J194753.4+274357 (CXO J194753.3+274351)?&63.781&1.143&215$\times$215&[$-$14.1 (0.28)]&N&+5.6?&7.2 (0.65)?\\
CTB 87&CXOU J201609.2+371110?&74.944&1.114&236$\times$236&+5.4 (0.29)&Y&$-$51.6?&6.1?\\
G75.23+0.12&J2021+3651&75.222&0.111&52$\times$52&+5.6 (0.46)&[Y]&+6.0?&3.4 (0.99)?\\
G80.22+1.02&J2032+4127&80.224&1.028&161$\times$161&[$-$2.9 (0.32)]&N&$-$3.0?&2.6 (0.36)?\\
3C 58&J0205+6449&130.716&3.080&273$\times$273&$-$7.3 (0.35)&Y&-&-\\
G141.2+5.0&CXOU J033712.8+615302?&141.180&5.036&113$\times$113&[$-$9.5 (0.32)]&N&-&-\\
Mushroom&B0355+54&148.190&0.811&47$\times$47&[+5.2 (0.30)]&N&-&-\\
Crab&B0531+21&184.554&$-$5.786&211$\times$211&[+3.2 (0.36)]&N&+3.2?&1.6 (0.58)?\\
Geminga&J0633+1746&195.134&4.266&100$\times$100&$-$10.8 (0.31)&N&$-$10.8?&0.25\\

\enddata
%\tablecomments{}
\tablenotetext{a}{"?" denotes a PWN or pulsar (PSR) candidate.}
\tablenotetext{b}{\vlsr\ of spatially correlated (SC) molecular gas, with the corresponding spatial correlation coefficient enclosed in brackets. Those with coefficients below 5$\sigma$ confidence level and less than 0.5 are enclosed by square brackets.}
\tablenotetext{c}{Detection of broad line (BL) emission related to spatially correlated molecular gas. Those identified only by partial criteria are enclosed by squared brackets.} %accuracy condition is not applied
%\tablenotetext{d}{"f" denotes systemic velocities determined by further examination described in text, and "?" denotes possible ones.}
\tablenotetext{d}{Systemic velocity of candidate MC associated with the PWN, and "?" denotes possible ones, see text for details.}
\tablenotetext{e}{Kinematic distance corresponding to $V_{\rm sys}$, with the estimated probability enclosed in brackets. The same calculation method as in Table~\ref{tab:snr} is used, except the distance of CTB~87 estimated by the extinction-distance relation in previous work \citep{FosterRoutledge2002,Kothes+2003}, and the distances of G47.38$-$3.88 and Geminga estimated by parallax measurements \citep{Chatterjee+2004, Faherty+2007}.}%, which are denoted by "f".}
%\tablenotetext{e}{Kinematic distance corresponding to $V_{\rm sys}$, with the estimated probability enclosed in brackets. The same calculation method as in Table~\ref{tab:snr} is used, except for the distance of CTB~87 estimated by the extinction-distance relation in previous work \citep{FosterRoutledge2002,Kothes+2003} and the distance of Geminga estimated by parallax measurements \citep{Faherty+2007}.}%, which are denoted by "f".}
%\tablenotetext{e}{All kinematic distances are estimated by a full distance probability density function \citep{Reid+2016, Reid+2019}, except for the distance of CTB~87 estimated by the extinction-distance relation in previous work \citep{FosterRoutledge2002,Kothes+2003} and the distance of Geminga estimated by parallax measurements \citep{Faherty+2007}.}%, which are denoted by "f".}
\end{deluxetable}
\end{center}
\clearpage

\subsection{PWNe}\label{sec:pwn}
We examine MCs around 18 pure PWNe here, to search candidates of associated MCs.
Our results of PWN-MC association candidates are presented in Table~\ref{tab:pwn}.
%Kinematic distances of PWN-MC association candidates are estimated, which are compared with that in previous works.
%Kinematic distances of PWN-MC association candidates are estimated, which are comparable with that in previous works, except for PWN G21.88$-$0.10.
Kinematic distances of PWN-MC association candidates are estimated. Some of these PWNe have distance estimation results from previous works, which are comparable to the distance results here.
Note that most distance measurements in previous works are for associated pulsars. %, but not for PWN themselves.
Only two PWNe are with surrounding CO emissions studied in previous works, i.e.\ G63.7+1.1 and CTB~87, of which possible associations with MCs are found here.
There are another two PWNe with surrounding \HI\ emission studied in previous works, i.e.\ 3C 58 and G141.2+5.0, however, no associated MCs are found here. 

We discuss individual PWNe in Appendix~C.

\section{Discussion}\label{sec:discuss}
\subsection{Proportion of SNRs associated with MCs}
\begin{figure*}[ptbh!]
\centerline{{\hfil\hfil
\psfig{figure=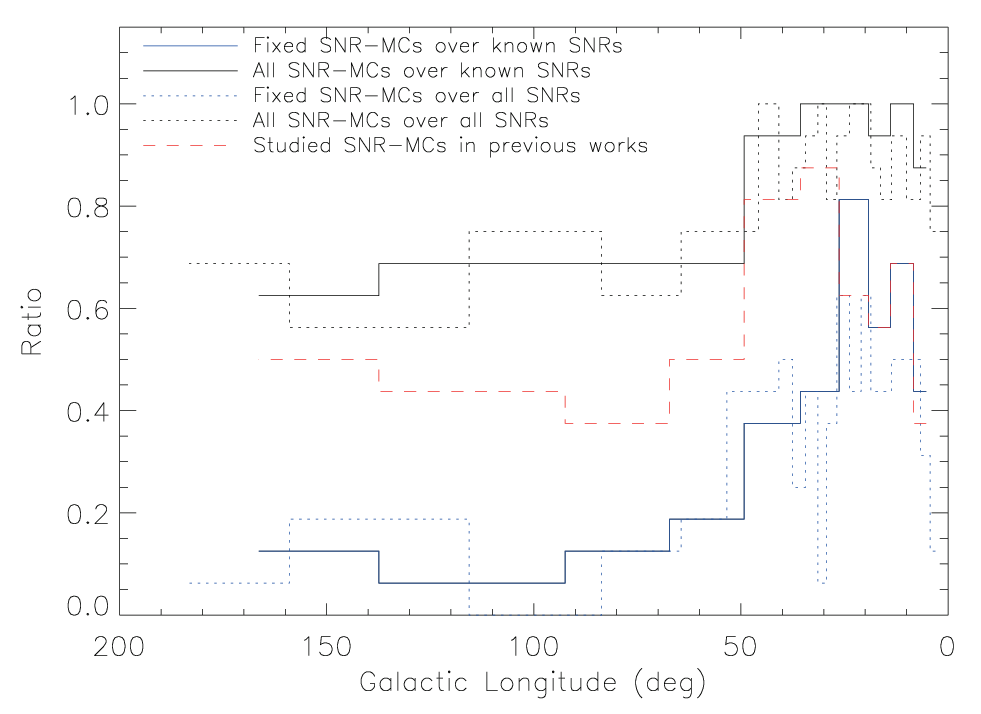,height=5in,angle=0, clip=}
\hfil\hfil}}
\caption{Ratios of SNRs associated with MCs along the Galactic longitude. Each histogram is adaptively binned with the number of corresponding SNRs to be sixteen. Black and blue histograms show ratios of all and fixed SNR-MC associations, respectively.
%and previously studied SNR-MC associations, respectively.
Solid lines are for known SNRs, and dotted lines for all known and candidate SNRs. Red dashed line is the ratio of previously detected SNR-MC associations in known SNRs.
}
\label{f:visl}
\end{figure*}
\begin{figure*}[ptbh!]
\centerline{{\hfil\hfil
\psfig{figure=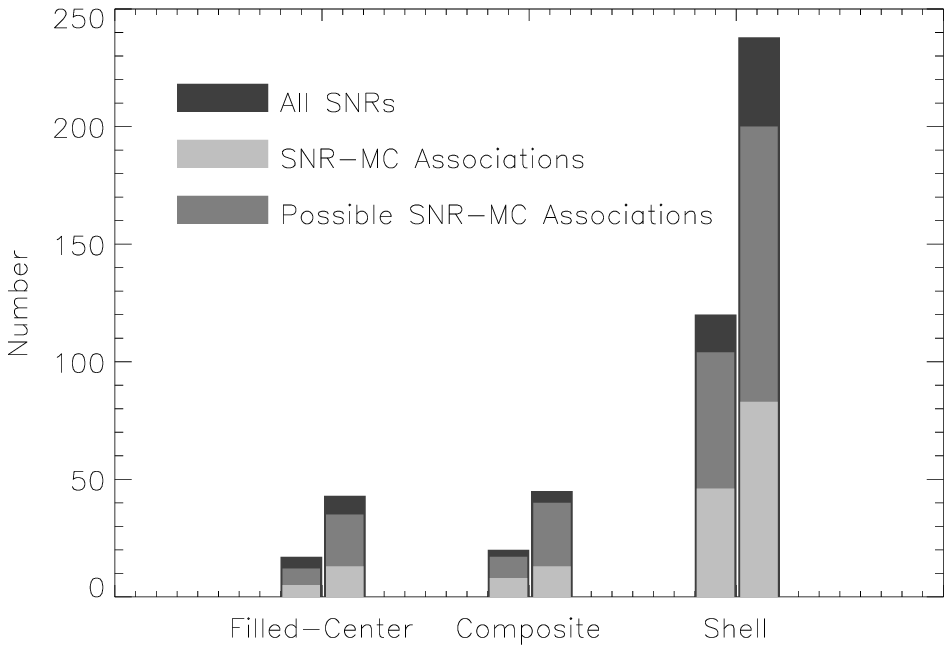,height=4in,angle=0, clip=}
\hfil\hfil}}
\caption{Type distributions of SNRs associated with MCs and all SNRs. For each pair of bars, the left one is for SNRs of certain type, and the right one for all SNRs of certain and possible type. The histogram of possible SNR-MC associations (dark gray) is stacked on top of that of fixed SNR-MC associations (light gray), and they are contained in that of all SNRs (black).
}
\label{f:type}
\end{figure*}
%accuracy and completeness based on ratio of previous detected sources
Ratios of SNRs associated with MCs along the Galactic longitude are shown in Figure~\ref{f:visl}, including fixed and all SNR-MC associations in known and all SNRs, respectively. All SNRs include known and candidate SNRs. The ratio of previously detected SNR-MC associations in known SNRs is also shown for reference.
All ratios have similar distributions, which are large within the Galactic longitude of $\sim$50$^\circ$. It is consistent with the distribution of the ratio of previously detected SNR-MC associations.
If all previously detected SNR-MC associations are real, the accuracy of fixed SNR-MC associations suggested in this paper would be at least 70\%, and at least 60\% for possible SNR-MC associations. 
%efficiency of \twCO~(J=1--0)
%Among previously detected SNR-MC associations, 91\% are identified here by \twCO~(J=1--0) line emission, and those missed are mostly based on spatial correlations with \HI\ gases. 
Among the previously detected SNR-MC associations, 91\% are identified here by \twCO~(J=1--0) line emission.
\twCO~(J=1--0) line emission of SNR-MC associations can be contaminated by background emission. However, it is still a good tracer for these associations.
Note that there are about 41\% of previously detected SNR-MC associations suggested as fixed SNR-MC associations here, and about 50\% considered as possible ones.

%SNR type dependent interaction rate
%in every respect, more composite type SNRs are associated with MCs.
Figure~\ref{f:type} shows the number of different types of SNRs and that associated with MCs.
Except for SNRs of certain composite type, all types of SNRs contain about 30\% fixed SNR-MC associations. 
Considering that only about half of previously detected SNR-MC associations are identified as fixed SNR-MC associations, there is probably more than 60\% of SNRs associated with MCs. As indicated by the number of all fixed and possible SNR-MC associations, the percentage of SNRs associated with MCs could be as high as about 80.
Among different types of SNRs, composite type SNRs have the highest proportion of being associated with MCs. This is reasonable, since composite SNRs are originated from core-collapse supernovae, and they are interacting with surrounding medium. Filled-center type SNRs are also originated from core-collapse supernovae; however, they have no radio continuum shell indicating effective interactions with surrounding medium.
As indicated by the number of all fixed and possible SNR-MC associations for composite type SNRs, there could be close to 90\% of core-collapse SNRs interacted with MCs during their lifetime.

%??associated Gamma-ray emission

%\subsection{Spatial Distribution in the Milky Way}
\subsection{Distribution of SNRs associated with MCs}
\begin{figure*}[ptbh!]
\centerline{{\hfil\hfil
\psfig{figure=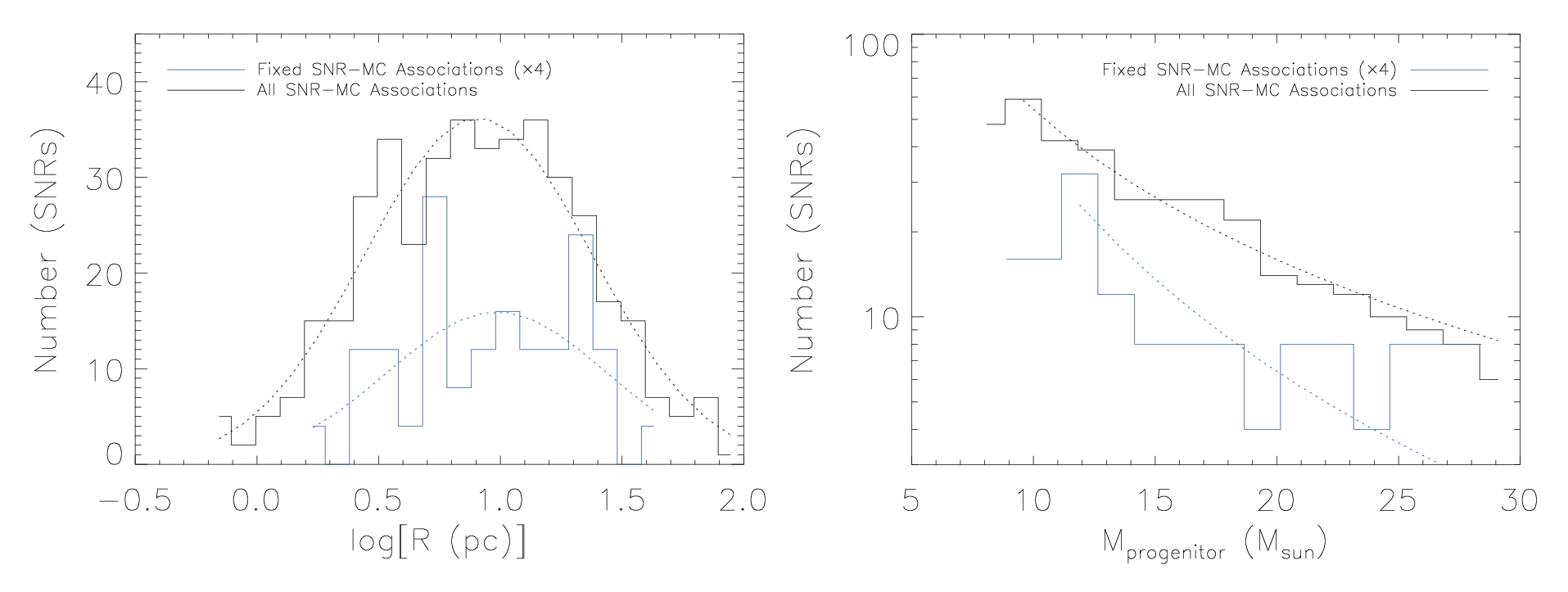,height=2.5in,angle=0, clip=}
\hfil\hfil}}
\caption{Radius ({\it left}) and progenitor initial mass ({\it right}) distributions for SNRs associated with MCs.
%Fixed SNR-MC associations in known SNRs are in blue, and all fixed and possible SNR-MC associations in all known and candidate SNRs are in black. 
Fixed SNR-MC associations in known SNRs are in blue except for those within 10$^{\circ}$ of l$\sim$0$^{\circ}$ and within 5$^{\circ}$ of l$\sim$180$^{\circ}$. All fixed and possible SNR-MC associations in all known and candidate SNRs are in black.
%Fixed SNR-MC associations are in blue, and all fixed and possible SNR-MC associations are in black.
The numbers of fixed SNR-MC associations are multiplied by 4 for better visibility.
Radii are mostly estimated based on SNRs' bright radio continuum shells. The progenitor initial mass is estimated based on its linear relationship with the size of the main-sequence interstellar bubble in a molecular environment reflected by the size of SNRs associated with MCs \citep[see details in][]{Chen+2013}.
%Lognormal fittings were performed, and fitting results are shown by dotted lines. %, which are peaks at $\sim$9 pc and $\sim$3400 yr.
Lognormal fitting results of radius distributions and power-law fitting results of progenitor initial mass distributions are shown as dotted lines.
SNR radius distributions peak at 9.7$_{-2.2}^{+2.8}$ and 8.1$\pm0.5$ pc, for the fixed and all SNR-MC associations, respectively. Progenitor initial mass distributions have indices of $-2.6\pm0.6$ and $-2.3\pm0.1$, for the fixed and all SNR-MC associations, respectively.
%, which are peaks at $\sim$9 pc and $\sim$3400 yr.
}
\label{f:snrprop}
\end{figure*}

We estimate radii of SNRs that are associated with MCs, based on their angular sizes and kinematic distances. Angular sizes are estimated from radio continuum emission of SNRs or associated molecular shell-like structures for those with no significant radio continuum emission detected. Kinematic distances of SNRs are estimated based on systemic velocities of their associated MCs by applying the full distance probability density function\footnote{http://bessel.vlbi-astrometry.org/node/378} \citep{Reid+2016, Reid+2019}.
%In addition, ages of SNRs are also estimated by assuming a constant supernova explosion energy as 10$^{51}$ ergs and a constant mean density of the evolutionary environment as 1 cm$^{-3}$.
%We also estimate ages of SNRs based on their radii, by assuming a constant supernova explosion energy as 10$^{51}$ ergs and a constant ambient particle density as 1 cm$^{-3}$.
%At first, we calculate the Sedov age of SNRs by $t=(2r)/(5v)=\sqrt{1.4 n_0 m_{\rm H} r^5/(\xi E_{\rm SN})}$, where $r$ is the radius, $v$ is the velocity of the remnant's forward shock, $n_0$ is the ambient particle density, $m_{\rm H}$ is the hydrogen atom mass, $E_{\rm SN}$ is the supernova's explosion energy, and $\xi=2.026$. If the Sedov age is larger than $2.7\times10^4 (E_{\rm SN}/10^{51}~{\rm erg})^{0.24} n_0^{-0.52}$ yr, the SNR is considered to have entered the radiative phase, and we calculate its age by $t=(2r)/(7v)=7.3\E{-11} n_0^{1.16} r^{4.16} \zeta_m^{0.161}/E_{SN}$, where $\zeta_m=Z/Z_\sun$ is set to 1 \citep{Cioffi+1988}. Since we use very simple assumptions, the estimated age is directly related to the radius of the SNR.
We also estimate progenitor initial masses of some SNRs based on their linear relationship with the size of the main-sequence interstellar bubble in a molecular environment, which is reflected by the size of SNRs associated with MCs \citep[see details in][]{Chen+2013}. Since this relationship is only valid for stars in the mass range of 8 to 25--30 $M_{\sun}$, those with estimated progenitor masses outside this range are excluded.
The number of SNRs with small-mass progenitors far exceeds the number of SNRs with large-mass progenitors. SNRs misidentified at large distances have a large impact on the mass distribution, even though their number is comparable to that of SNRs misidentified at small distances.
Therefore, when estimating progenitor masses, we use only the smallest distance of all possible distances for each possible SNR-MC association.
%SNRs associated with no molecular cavity or shell structures are also excluded from the mass distribution analysis.}

%Radius and age distributions of SNRs associated with MCs are shown in Figure~\ref{f:snrprop}, which are fitted by lognormal distribution functions. 
Distributions of the radii and the progenitor initial masses of the SNRs associated with MCs are shown in Figure~\ref{f:snrprop}, which are fitted by lognormal and power-law distribution functions, respectively.
Fitting of the mass distributions starts at the peak.
%For both radius and age distributions, fitting parameters for fixed SNR-MC associations except for those within 10$^{\circ}$ of l$\sim$0$^{\circ}$ and 5$^{\circ}$ of l$\sim$180$^{\circ}$ are consistent with that for all fixed and possible SNR-MC associations. %, except for their peak values. 
For both radius and progenitor initial mass distributions, fitting parameters for fixed SNR-MC associations except for those within 10$^{\circ}$ of l$\sim$0$^{\circ}$ and 5$^{\circ}$ of l$\sim$180$^{\circ}$ are consistent with that for all fixed and possible SNR-MC associations. %, except for their peak values. 
It indicates that possible SNR-MC associations do not introduce any significant systematic deviations.
%The radius distribution of SNRs associated with MCs peaks at $\sim$8.3 pc, and the age distribution peaks at $\sim$2900 yr.
The radius distribution of SNRs associated with MCs peaks at $\sim$8.1 pc. The progenitor initial mass distribution has an index of $\sim$$-2.3$, which is consistent with the Salpeter index of $-2.35$ \citep[][etc.]{Salpeter1955, Bastian+2010}.

\begin{figure*}[ptbh!]
\centerline{{\hfil\hfil
\psfig{figure=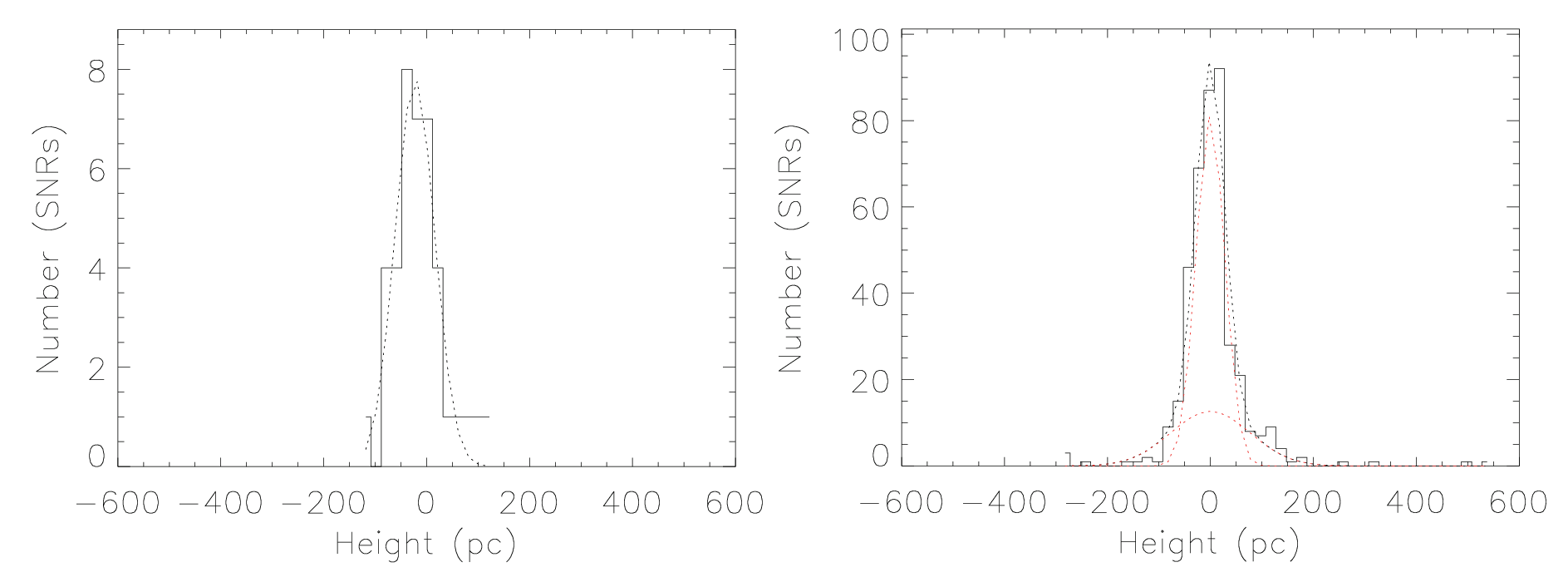,height=2.5in,angle=0, clip=}
\hfil\hfil}}
\caption{Distributions of heights from the Galactic plane for SNR-MC associations.
Fixed SNR-MC associations in known SNRs except for those within 10$^{\circ}$ of l$\sim$0$^{\circ}$ and 5$^{\circ}$ of l$\sim$180$^{\circ}$ are shown in the left panel. All fixed and possible SNR-MC associations in all known and candidate SNRs are shown in the right panel. 
%The position of the Galactic midplane is obtained from the b=$0^\circ$ plane corrected for the height of the Sun, which makes the overall vertical displacement of all SNR-MC associations from the Galactic plane the smallest. 
The position of the Galactic midplane is derived from the fixed SNR-MC associations by the least squares fitting.
%The height of the Sun above the Galactic midplane is estimated as $\sim$19.3 pc.
The height of the Sun above the Galactic midplane is estimated as $\sim$15.7 pc.
The same position of the Galactic midplane is also used to calculate heights of all SNR-MC associations.
Black dotted lines indicate fitting results by a Gaussian function in the left panel and by two Gaussian functions in the right panel (separated components shown by red dotted lines).
Note that all distances estimated for possible SNR-MC associations are used. However, if we use only the smallest distance for each possible SNR-MC association, the height distribution still has two components.
%the result is not significantly different than that using only small distances.
}
\label{f:hgal}
\end{figure*}

%fitting parameter
%fixed SNR-MC: FWHM 91$\pm15$ pc
%all SNR-MC: FWHM 69$\pm9$ and 217$\pm80$ pc
The height from the Galactic plane for SNR-MC associations is estimated according to their kinematic distances. 
Because of the Sun's vertical displacement from the Galactic midplane, there is a small angle between the Galactic plane and the $b=0^\circ$ plane.
%To obtain the height from the Galactic plane, we firstly estimate the Sun's vertical displacement from the Galactic plane. 
%The position of the Galactic midplane is derived from fixed SNR-MC associations in known SNRs by the least squares fitting. Thereby, the height of the Sun above the Galactic midplane is estimated as $\sim$19.3 pc. %, which is not well constrained becaused of the limited number of fixed SNR-MC associations in known SNRs.
%The position of the Galactic midplane is derived from fixed SNR-MC associations in known SNRs by the least squares fitting. Thereby, the height of the Sun above the Galactic midplane is estimated as $\sim$13.5 pc. %, which is not well constrained becaused of the limited number of fixed SNR-MC associations in known SNRs.
The position of the Galactic midplane is derived from fixed SNR-MC associations in known SNRs except for those within 10$^{\circ}$ of l$\sim$0$^{\circ}$ and 5$^{\circ}$ of l$\sim$180$^{\circ}$ by the least squares fitting. Thereby, the height of the Sun above the Galactic midplane is estimated as $\sim$15.7 pc. %, which is not well constrained becaused of the limited number of fixed SNR-MC associations in known SNRs.
We use the same Galactic plane position to calculate the height of all SNR-MC associations.
Distributions of the height from the Galactic plane for SNR-MC associations are shown in Figure~\ref{f:hgal}.
The height distribution of all SNR-MC associations has two components, a major narrow component and a minor broad component. Some possible SNR-MC associations with uncertain distances may contribute to the broad component; however, the broad component still exists when we use only the minimum distance among all possible distances for each possible SNR-MC association.
%The number of fixed SNR-MC associations in known SNRs is rather limited, and the height distribution of them is fitted by one Gaussian component, nevertheless, a weak broad component seems present but not significant.
%The number of fixed SNR-MC associations in known SNRs is limited, of which the height distribution may also have a weak broad component but not significant. It can be fitted well by one Gaussian function.
The number of fixed SNR-MC associations is limited, of which the height distribution may also have a weak broad component, not significant. It can be fitted well by one Gaussian function.
%For the height distribution of all SNR-MC associations, the thicknesses (i.e.\ FWHMs) of two components are estimated as $58\pm7$ and $161\pm54$ pc, respectively.
For the height distribution of all SNR-MC associations, the thicknesses (i.e.\ FWHMs) of two components are estimated as $65\pm6$ and $182\pm64$ pc, respectively.
The thickness (i.e.\ FWHM) is estimated as $90\pm9$ pc for the height distribution of the fixed SNR-MC associations, which is consistent with that of the thin CO disk revealed by \cite{Su+2019}. However, it may be just an intermediate value between the thicknesses of two components of the distribution of all SNR-MC associations. 
SNR-MC associations may trace MCs of active star-formation.
The thin and thick disks of MCs associated with SNRs found here may be inner layers of the thin and thick CO disks revealed by \cite{Su+2019}, respectively.
The ratio of the peak of the thin and thick disks of SNR-MC associations is about 6, which is larger than that of CO disks, i.e.\ about 2. It indicates that the star-formation may be more efficient in the thin disk.

\begin{figure*}[ptbh!]
\centerline{{\hfil\hfil
\psfig{figure=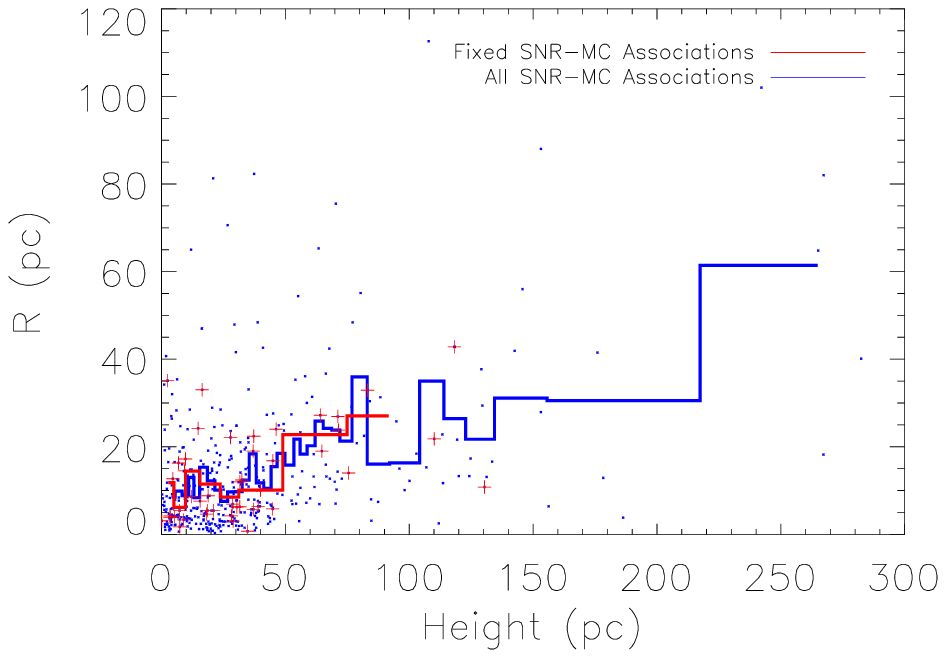,height=4in,angle=0, clip=}
\hfil\hfil}}
\caption{Radius vs. height from the Galactic plane for fixed (red pluses) and all (blue tiny stars) SNR-MC associations. Fixed SNR-MC associations are selected from known SNRs except for those within 10$^{\circ}$ of l$\sim$0$^{\circ}$ and 5$^{\circ}$ of l$\sim$180$^{\circ}$. All SNR-MC associations, including fixed and possible ones, are selected from all known and candidate SNRs. The position of the Galactic plane is the same as that applied in Figure~\ref{f:hgal}. 
%Histograms present the average of the radius and height for corresponding SNR-MC associations, binned with the number of sources to be at least five and the step of the height to be at least 5 pc.
%Two histograms respectively show the average radius of corresponding SNR-MC associations, which are binned with the number of sources to be at least sixteen and the step of the height to be at least 5 pc.
Two histograms respectively show the average radius of corresponding SNR-MC associations, which are binned with the number of sources to be at least ten and the step of the height to be at least 5 pc.
%both r and h are related to distance, so distance is critical; and error of r and h are related, distance related error are oblique.
}
\label{f:rh}
\end{figure*}
%plateau at $height<\sim$35 pc, and increasing above.
Figure~\ref{f:rh} shows correlations between the radius and the height from the Galactic plane for SNRs associated with MCs. Fixed and all SNR-MC associations selected from known SNRs except for those within 10$^{\circ}$ of l$\sim$0$^{\circ}$ and 5$^{\circ}$ of l$\sim$180$^{\circ}$ and all SNR-MC associations, respectively, are considered separately.
%For both of them, the distribution of the average radius in the height range of less than $\sim$40 pc are flat, where radii of individual SNRs are much scattered.
%When the height is greater than $\sim$40 pc, the average radii for both the fixed and all SNR-MC associations increase with the height. In the interval of the height less than $\sim$40 pc, the average radius distribution is roughly flat. Note that the average radius has a small peak at the height of $\sim$10 pc, where the radius dispersion is also very large.
In the interval where the height is less than $\sim$45 pc, the average radius distribution is roughly flat. When the height is greater than $\sim$45 pc, the average radius for both fixed and all SNR-MC associations increases with the height.
The turning point of the relation between the radius and the height, i.e.\ at $\sim$45 pc, is consistent with the thickness of the thin CO disk revealed by \cite{Su+2019}.
The radius of an SNR is related to its age, its ambient particle density, and the energy of its supernova explosion. 
The age and supernova explosion energy of SNRs are probably not correlated with their heights from the Galactic midplane. Different ages and explosion energies would make the radius scattered very much in Figure~\ref{f:rh}. 
The relationship between the average radius and the height may be caused by a height-dependent density distribution of molecular environments around SNRs. %massive stars.
Though, some SNRs associated with MCs are interacting with cavity or shell-like molecular structures that may be relics of wind-blown bubbles created by their progenitor stars, yet, the sizes of these molecular structures are still confined by ambient particle density. 
As the ambient particle density becomes larger, the radius of SNRs becomes smaller. 
Therefore, such relation between the average radius and the height probably indicates that the overall density of MCs with active star-formation does not vary much in the thin CO disk. 
%And, beyond the thin CO disk, it decreases with the increase of the height from the Galactic plane.
Beyond the thin CO disk, it decreases with increasing height from the Galactic plane.
Further investigation on MCs themselves is needed to draw a definite conclusion.

\section{Summary}\label{sec:sum}
The MWISP project is an unbiased \twCO/\thCO/\CeiO\ (J = 1--0) survey of the Galactic plane in the northern sky.
We universally search for evidence of kinematic and spatial correlation of SNR-MC associations for nearly all SNRs in the coverage of the MWISP CO survey, i.e.\ 149 SNRs, 170 SNRCs, and 18 pure PWNe in $1^\circ < l < 230^\circ$ and $-5.^\circ5 < b < 5.^\circ5$. 
Based on the high-quality CO data obtained from the MWISP survey, we apply automatic algorithms to identify broad lines and spatial correlations for molecular gas in each SNR region. The searching method is demonstrated to be efficient.
Among 149 SNRs studied in this paper, 57 of them are found to be associated with MCs, and 70 of them are considered to be possibly associated with MCs. We find 50 SNRCs to be associated with MCs, and 91 SNRCs to be possibly associated with MCs. We also find candidates of associated MCs for 14 pure PWNe.
Assuming all SNR-MC associations detected in previous works are real, the accuracy of fixed SNR-MC associations suggested in this paper would be at least 70\%, and at least 60\% for possible SNR-MC associations.
The 91\% previously detected SNR-MC associations are identified in this paper by CO line emission, which indicates that CO line emission is efficient for detecting SNR-MC associations.
The nine previously detected SNR-cloud associations (about 9\%) are not identified here, which are mostly based on spatial correlations with \HI\ gases.
We find that the proportion of SNRs associated with MCs is high within the Galactic longitude of $\sim$50$^\circ$, with the distribution consistent with that of previously detected SNR-MC associations. %which is consistent with the distribution of the ratio of previously detected SNR-MC associations in known SNRs.
Overall, there could be as high as 80\% of SNRs associated with MCs.

%statistics
Based on systemic velocities of associated MCs, kinematic distances of SNRs are estimated. Accordingly, distributions of their radii, progenitor initial masses, and heights from the Galactic plane are studied.
%We find that the radius and age distributions of SNRs associated with MCs follow lognormal distributions, which peak at $\sim$8.3 pc and $\sim$2900 yr, respectively.
We find that the radius and progenitor initial mass distributions of SNRs associated with MCs follow lognormal and power-law distributions, respectively. The radius distribution peaks at $\sim$8.1 pc, and the progenitor initial mass distribution has an index of $\sim$$-2.3$ that is consistent with the Salpeter index of $-2.35$.
SNR-MC associations are mainly distributed in a thin disk and also distributed in a faint thick disk along the Galactic plane. For only fixed SNR-MC associations except for those within 10$^{\circ}$ of l$\sim$0$^{\circ}$ and 5$^{\circ}$ of l$\sim$180$^{\circ}$, due to their limited number, the thick disk is hard to distinguish, and the overall thickness (FWHM) is estimated as 90$\pm9$ pc. For both fixed and possible SNR-MC associations, thicknesses (i.e.\ FWHMs) of the thin and thick disks are estimated as 65$\pm6$ and 182$\pm64$ pc, respectively. 
The thin and thick disks found here may be inner layers of the thin and thick CO disks revealed by \cite{Su+2019}, respectively.
The ratio of peaks of the thin and thick disks of SNR-MC associations is about 5, which is larger than that of CO disks, i.e.\ about 2. It indicates that star-formation may be more efficient in the thin disk.
%The Sun's vertical displacement from the Galactic midplane is estimated as $\sim$19.3 pc.
The Sun's vertical displacement from the Galactic midplane is estimated as $\sim$15.7 pc.
Radii of SNRs associated with MCs show some correlations with their heights from the Galactic plane, with a turning point at the height of $\sim$45 pc. When the height is below $\sim$45 pc, the average radius distribution is roughly flat, and the radii of individual SNRs are much scattered. When the height is above $\sim$45 pc, the average radius increases with the height. 
%It indicates that the overall density of MCs with active star-formation may not vary much in the thin CO disk, and it decreases with the increase of the height from the Galactic plane beyond the thin CO disk.
It indicates that the overall density of MCs with active star-formation may not vary much in the thin CO disk, and it decreases with increasing height from the Galactic plane beyond the thin CO disk.

\acknowledgments
We thank the anonymous referee for providing very helpful comments that improved the paper and its conclusions.
This research made use of the data from the Milky Way Imaging Scroll Painting (MWISP) project, which is a multiline survey in \twCO/\thCO/\CeiO\ along the northern Galactic plane with PMO 13.7m telescope. We are grateful to all the members of the MWISP working group, particularly the staff members at PMO 13.7m telescope, for their long-term support. 
MWISP was sponsored by National Key R\&D Program of China with grant 2017YFA0402701 and CAS Key Research Program of Frontier Sciences with grant QYZDJ-SSW-SLH047. 
%JY is supported by National Natural Science Foundation of China through grant 12041305. 
Y.Su and J.Y. are supported by National Natural Science Foundation of China through grants 12173090 and 12041305, respectively.
Y.Sun acknowledges support by the Youth Innovation Promotion Association, CAS (Y2022085), and the “Light of West China” Program (No. xbzg-zdsys-202212).

\paragraph{Data Availability}
Figures of individual objects are available online in https://www.scidb.cn/en, at https://doi.org/10.57760/sciencedb.08076 \citep{Zhou+2023data}. %\footnote{https://www.scidb.cn/s/bI7N7b}.
%Figures of individual objects are available online at https://www.scidb.cn/s/bI7N7b \citep{Zhou+2023data}.

\bibliographystyle{apj}
%\bibliography{/Users/zx/Documents/document/Literature/references/ref}
%\bibliography{/home/zx/Documents/document/Literature/references/ref}

%\input{ms.bbl}

\clearpage
\appendix
%\begin{appendix}
\section{Individual SNRs}\label{app:snr}
{\bf G1.4-0.1}:
This SNR is near the Galactic center, and presents a partial radio continuum shell in the west. %where the identification of broad line was affected by serious line overlapping effects.
%As indicated by the SCC, molecular gas at $\sim$+3.2~\km\ps\ is roughly surrounding the west of the remnant and also distributed around its eastern boundary. 
As indicated by the SCC, molecular gas at $\sim$+3.2~\km\ps\ is distributed around the remnant.
It belongs to the $-$4.8~\km\ps\ component with broad line candidates identified in the remnant only by partial criteria.
The association between the remnant and the $-$4.8~\km\ps\ component is also supported by OH 1720 MHz masers detected in the SNR at $\sim$$-$2.4~\km\ps\ (see Table~\ref{tab:snrpre}).
Molecular gas at the nonsignificant peak of the SCC at $\sim$+57.0~\km\ps\ is distributed around the northern boundary of the remnant, which is not well correlated with the remnant's radio continuum emission. %The identified broad line candidate at a nearby velocity of +78.9~\km\ps\ distributes widespread.
There is also molecular gas at $\sim$+99.1~\km\ps\ distributed around the western half of the remnant. It belongs to the +78.9~\km\ps\ component with broad line candidates identified in the remnant by full criteria, of which two alternative kinematic distances are estimated as 8.1$\pm2.1$ and 11.1$\pm0.3$ kpc. However, the association between the remnant and the +78.9~\km\ps\ component is not supported by OH 1720 MHz and CH$_3$OH 36 GHz maser emission detected in previous works (see Table ~\ref{tab:snrpre}).
If the remnant is associated with the $-$4.8~\km\ps\ component, its kinematic distance can be estimated as 12.6$\pm0.3$ kpc.\\
{\bf G1.9+0.3}:
The radius of this SNR is less than the beam size of our observation, which is too small to examine possible spatial correlations. No broad line candidate is identified in it by full criteria.\\
%Nevertheless, we get a peak of SCC at $\sim$+4.8~\km\ps. As shown in Table~\ref{tab:snrpre}, OH 1720 MHz line emission were detected in this remnant. The kinematic distance of $\sim$+4.8~\km\ps\ MC is estimated as 2.8$\pm0.3$ kpc.\\
{\bf G3.7-0.2}: 
%There is an \HII\ region overlapping this SNR, at +4 and $-28$~\km\ps.
This SNR presents radio continuum shells in the north and south.
Molecular gas at the peak of the SCC at $\sim$+19.2~\km\ps\ is roughly surrounding the SNR. It probably belongs to the +9.3~\km\ps\ component, since broad line candidates at +17.3 and +21.3~\km\ps\ can be attributed to the nearby prominent component at +9.3~\km\ps. Kinematic distances corresponding to these velocities are consistent.
In addition, molecular gas at $\sim$+2.5~\km\ps\ belonging to the +9.3~\km\ps\ component is distributed around the northern boundary of the remnant.
Molecular gas at another peak of the SCC at $\sim$+79.4~\km\ps\ is distributed around and within the northeastern and southeastern boundaries of the remnant, which belongs to the +83.5~\km\ps\ component with broad line candidates identified in the remnant.
There are also molecular gases at $\sim$$-$27.1~\km\ps\ distributed around the southern half of the remnant and also inside, which belongs to the $-$30.6~\km\ps\ component with broad line candidates identified in the remnant by full criteria.
Some broad line candidates at $-$30.6~\km\ps\ are associated with an overlapped \HII\ region in the northwest.
%This correlation is supported by associated broad line candidates identified by full criteria. 
%These evidences indicate a SNR-MC association.
%Based on kinematic evidence and the well spatial correlation result, we suggest that the SNR is associated with the +9.3~\km\ps\ component, and the corresponding kinematic distance is estimated as 2.9$\pm0.2$ kpc. 
Based on kinematic evidence and the spatial correlation result, the SNR is probably associated with the +9.3~\km\ps\ component, or may be associated with the +83.5~\km\ps\ component. Corresponding kinematic distances are estimated as 2.9$\pm0.2$ kpc for the +9.3~\km\ps\ component and 7.7$\pm2.1$ kpc for the +83.5~\km\ps\ component.
Note that the association between the SNR and the $-$30.6~\km\ps\ component cannot by totally ruled out, of which the kinematic distance is estimated as 4.7$\pm0.3$ kpc.
\\ 
{\bf G3.8+0.3}: 
%There is a small \HII\ region at +7~\km\ps, which locates at the eastern border of the remnant.
%For this SNR, the SCC peaks at $\sim$+7.9~\km\ps, but not significant. 
%As indicated by the SCC, molecular gases at $\sim$+7.9 and $\sim$+165.6~\km\ps\ are roughly surrounding the southeast and southwest of this SNR, respectively.
As indicated by the SCC, molecular gases at $\sim$+8.7 and $\sim$+165.6~\km\ps\ are roughly surrounding the southeast and southwest of this SNR, respectively.
These molecular gases belong to the +12.1~\km\ps\ and +162.9~\km\ps\ components, respectively, which are with broad line candidates identified in the remnant by full criteria. %plus the clean subbackground region condition
For the +12.1~\km\ps\ component, there is also molecular gas at $\sim$+17.5~\km\ps\ distributed around the northern boundary of the remnant, where radio continuum emission is bright.
Note that some broad line candidates at +12.1~\km\ps\ are probably originated from a small overlapped \HII\ region in the east, but not all. 
Based on these evidences, the SNR is probably associated with the +12.1~\km\ps\ component, or may be associated with the +162.9~\km\ps\ component. Kinematic distances are estimated as 2.9$\pm0.2$ kpc for the +12.1~\km\ps\ component and 8.1$\pm2.8$ kpc for the +162.9~\km\ps\ component.\\
{\bf G4.5+6.8 ({\it Kepler})}:
{\it Kepler}'s SNR (SN 1604) is believed to have originated from a Type Ia supernova \citep[][and references therein]{SunChen2019}. %, which locates at a high Galactic lattibue.
There is very little CO emission detected in the SNR region at velocities of $\sim$+4 and $\sim$+12~\km\ps. We find no broad line emission and no good spatial correlation result for the SNR.\\
{\bf G4.8+6.2}:
This SNR is located at a high Galactic latitude nearby {\it Kepler}'s SNR.
%, with CO emission detected at velocities of $\sim$+4 and $\sim$+12~\km\ps. 
Molecular gas at the peak of the SCC at $\sim$+4.8~\km\ps\ is roughly surrounding the eastern half of the remnant. 
No broad line candidate is identified in the remnant even by partial criteria. 
If the remnant is associated with molecular gas at $\sim$+4.8~\km\ps, its near and far kinematic distances can be estimated as 1.6$\pm0.3$ and 14.8$\pm0.3$ kpc, respectively. Considering that the remnant's height from the Galactic plane would be too high at the far kinematic distance, i.e.\ $\sim$1.6 kpc, it is probably located at the near kinematic distance at 1.6$\pm0.3$ kpc.\\
{\bf G5.2-2.6}:
A small amount of molecular gas is detected around this SNR.
At the peak of the SCC at $\sim$+7.8~\km\ps, there are three small molecular clumps distributed on the radio continuum shell of the remnant.
There is no broad line candidate identified in the remnant even by partial criteria. 
If the remnant is associated with molecular gas at $\sim$+7.8~\km\ps\, its kinematic distance can be estimated as 1.5$\pm0.2$ kpc.\\
{\bf G5.4-1.2 (Milne 56)}:
This SNR presents bright shell-like radio continuum emission in the northwest.
Molecular gas at the peak of the SCC at $\sim$+29.7~\km\ps\ is roughly surrounding the northeast of the remnant but not well correlated with its bright radio continuum shell.
%We have identified multiple broad line components in this SNR.
Among multiple components with broad line candidates identified in the remnant, only the $-$24.9 and +187.5~\km\ps\ components show some spatial correlations with the remnant's bright radio continuum shell.
Note that broad line candidates at $-$9.7~\km\ps\ are probably originated from the nearby prominent component at $-$24.9~\km\ps.
%The $-24.9$~\km\ps\ velocity component with broad line candidates identified contains molecular filaments correlated well with the partial radio shell, nevertheless, with also CO gas inside the remnant.
The association between the remnant and the $-24.9$~\km\ps\ component is also supported by the detection of the OH 1720 MHz maser at $-21$~\km\ps\ (see Table~\ref{tab:snrpre}).
%We also found the CO gas at the velocity of a peak of the spatial correlation coefficent, i.e.\ +198.1~\km\ps, circulates the right half of the SNR, which is with few related broad line candidates identified.
Therefore, the remnant is suggested to be associated with the $-24.9$~\km\ps\ component, of which the near and far kinematic distances are estimated as $4.2\pm1.1$ and $12.1\pm0.9$ kpc, respectively.\\
{\bf G5.5+0.3}:
As indicated by the SCC, molecular gas at $\sim$+7.0~\km\ps\ is roughly surrounding the southern half of this SNR, where its radio continuum emission is bright. There is also molecular gas at $\sim$+7.0~\km\ps\ distributed around the northern boundary of the remnant. These molecular gases belong to the +10.3~\km\ps\ component, which is the only component with broad line candidates identified in the remnant by full criteria.
A small part of the remnant in the west overlaps with SNRC G5.38+0.35, nevertheless, identified broad lines are not in this region.
%Both their spatial correlation and identified broad line candidates indicate that 
These evidences indicate that the remnant is associated with the +10.3~\km\ps\ velocity component. This result is consistent with that in a previous work (see Table~\ref{tab:snrpre}).
Accordingly, we estimate the kinematic distance of the remnant as $2.9\pm0.2$ kpc.\\
%{\bf G6.1+0.5, G6.12+0.39 (candidate), and G6.31+0.54 (candidate)}:
{\bf G6.1+0.5}:
Most of this SNR overlaps with SNRC G6.06+0.50 and SNRC G6.12+0.39, and it is also adjacent to SNRC G6.31+0.54. It is difficult to determine the morphology of radio continuum emission from the SNR.
Molecular gas at the peak of the SCC at $\sim$+6.0~\km\ps\ seems to be distributed around the northwestern half of the remnant, which belongs to the +7.0~\km\ps\ velocity component. The +7.0~\km\ps\ component is the only component with broad line candidates identified in the SNR by full criteria.
Most of the broad line points are located in the overlapping region with SNRC G6.06+0.50, but not all.
Based on these evidences, we suggest that the SNR is associated with the +7.0~\km\ps\ component, of which the kinematic distance is estimated as 3.0$\pm0.2$ kpc.\\
%The SCC indicates a correlation between SNRC G6.12+0.39 and the $\sim$+10.8~\km\ps\ molecular gas, which probably belongs to the +21.8~\km\ps\ velocity component with broad line candidate identified because of their consistent spatial distribution. 
%There is also broad line component candidate at +7.0~\km\ps\ identified in SNRC G6.12+0.39, which is probably associated with SNR G6.1+0.5.
%Therefore, we suggest that SNRC G6.12+0.39 is associated with the +21.8~\km\ps\ MC and locates at a kinematic distance of 3.6$\pm0.3$ kpc.
%For SNRC G6.31+0.54, we found no spatial correlation as well as broad line candidate by full criteria, hence, no associated MC settled.\\
{\bf G6.1+1.2}:
This SNR is circulated by molecular gas at $\sim$+23.2~\km\ps\ mostly to the east indicated by the SCC. 
There is also molecular gas at $\sim$+13.0~\km\ps\ distributed around the remnant but not well correlated.
These molecular gases may belong to the +12.2~\km\ps\ velocity component with broad line candidates identified in the remnant by partial criteria. 
No broad line candidate is identified in the remnant by full criteria.
If the remnant is associated with the +12.2~\km\ps\ component, its kinematic distance can be estimated as 1.5$\pm0.2$ kpc.\\
{\bf G6.4-0.1 (W28)}:
The association between this SNR and MCs was studied in multiple bands. \cite{Wootten1981} detected broad molecular lines in the remnant, and subsequent studies revealed more details of the interactions between the remnant and surrounding molecular gases (e.g., those in Table~\ref{tab:snrpre}).
We find that two molecular stripes at the minor peak of the SCC at $\sim$$-5.9$~\km\ps\ spatially correlate well with two inner radio continuum shells, where OH 1720 MHz masers were also detected in previous works (see Table~\ref{tab:snrpre}). Molecular gas at $\sim$$-5.9$~\km\ps\ belongs to the +6.1~\km\ps\ velocity component. Several groups of broad line candidates at the +6.1~\km\ps\ are identified in the remnant by full criteria, which are located on the remnant's radio continuum shells.
Molecular gases at other peaks of the SCC at $\sim$$-$21.1, $\sim$+22.1, and $\sim$+24.3~\km\ps\ are mostly distributed in the remnant, but not well correlated.
%We have identified multiple components of broad line candidates; however, only the +6.1~\km\ps\ component has such spatial correlation.
These evidences indicate that the SNR is associated with the +6.1~\km\ps\ MC, of which the near and far kinematic distances are estimated as 3.0$\pm0.2$ and 12.6$\pm0.3$ kpc, respectively.\\
{\bf G6.5-0.4}:
The northwestern half of this SNR overlaps with SNR W28, and it also overlaps with SNRC G6.45-0.56 and SNRC G6.54-0.60 in the south.
%The SCC peaks at $-$16.7~\km\ps\ but not significant.
%Nevertheless, this spatial correlation is supported by the presence of broad line candidates at $-13.4$~\km\ps\ identified by full criteria.
%Molecular gas at the peak of the SCC at $\sim$$-$16.7~\km\ps\ is roughly surrounding the northeastern half of the remnant, which belongs to the $-$13.4~\km\ps\ component with broad line candidates identified in the remnant by full criteria.
%Note that broad line candidates at +8.3~\km\ps\ identified in the remnant are probably originated from SNR W28, and that at +20.6~\km\ps\ may be originated from an overlapped \HII\ region.
Molecular gases at minor peaks of the SCC at $\sim$$-$9.7 and $\sim$$-$16.7~\km\ps\ are roughly surrounding the northwest and northeast of the remnant, respectively. They belong to the $-$13.4~\km\ps\ component with broad line candidates identified in the remnant by full criteria.
Molecular gas at the peak of the SCC at $\sim$+4.0~\km\ps\ is distributed around the boundary of the remnant but not well correlated, which belongs to the +6.5~\km\ps\ component.
Note that broad line candidates at +6.5~\km\ps\ identified in the remnant are probably originated from SNR W28, and that at +20.6~\km\ps\ may be originated from an overlapped \HII\ region.
There is no broad line candidate identified in SNRC G6.45-0.56 and SNRC G6.54-0.60 by full criteria.
Based on these evidences, we suggest that SNR G6.5-0.4 is associated with the $-13.4$~\km\ps\ component and locates at a kinematic distance of 4.7$\pm0.3$ kpc.\\
{\bf G7.0-0.1}:
%This SNR partly overlaps with a +13~\km\ps\ \HII\ region in its southern region.
%We find no significant spatial correlation result for this SNR by the SCC. 
%The SCC of this SNR has no significant peak.
Molecular gas at the peak of the SCC at $\sim$+47.3~\km\ps\ is mostly distributed in this SNR, not well correlated.
Molecular gas at the minor peak of the SCC at $\sim$+6.5~\km\ps\ is distributed around the remnant but not well correlated. Corresponding broad line candidates are identified at +18.3~\km\ps\ in the remnant by full criteria, and most of the broad line points are distributed around the eastern radio continuum shell of the remnant.
These evidences indicate that the remnant is associated with the +18.3~\km\ps\ component, accordingly, its kinematic distance can be estimated as 3.6$\pm0.4$ kpc.\\
{\bf G7.2+0.2}:
%There is a $-4$~\km\ps\ \HII region located within the SNR.
This SNR presents bright radio continuum emission in the southeast.
As indicated by the SCC, molecular gas at $\sim$$-$19.2~\km\ps\ is roughly surrounding the east of the remnant and distributed around its southwestern boundary, which belongs to the $-$12.1~\km\ps\ component. The $-$12.1~\km\ps\ component is with many broad line candidates identified by partial criteria, and all broad line points are located in the remnant. Some broad line points at $-$12.1~\km\ps\ are located on the radio continuum shell peak of the remnant.
Molecular gases at $\sim$+0.2 and $\sim$+13.3~\km\ps\ are roughly surrounding the southeastern half of the remnant. 
At the peak of the SCC at $\sim$+19.5~\km\ps, molecular gas is distributed around the remnant but not well correlated.
Broad line candidates corresponding to these molecular gases are identified at +7.0 and +20.4~\km\ps\ in the remnant by full criteria, respectively. However, all these broad line candidates may belong to the prominent component at +20.4~\km\ps.
%The SCC indicates a correlation between this SNR and the +19.5~\km\ps\ molecular gas, which is supported by the +20.4~\km\ps\ component of broad line candidate identified by full criteria.
%The remnant may be associated with either the $-$12.1 or +20.1~\km\ps\ components. Kinematic distances are estimated as 18.7$\pm0.5$ kpc for the $-$12.1~\km\ps\ component and 3.6$\pm0.3$ kpc for the +20.1~\km\ps\ component.\\
The remnant may be associated with either the $-$12.1 or +20.4~\km\ps\ components. Kinematic distances are estimated as 18.7$\pm0.5$ kpc for the $-$12.1~\km\ps\ component and 3.6$\pm0.3$ kpc for the +20.4~\km\ps\ component.\\
%{\bf G7.5-1.7}: This is a SNRC.
%{\bf G8.3-0.0}: not SNR by Dokara+2021. not covered.
{\bf G8.7-0.1 (W30)}: 
%There are several \HII\ regions overlapping this SNR, some of which are at velocities of +2, +20, +21, and +41~\km\ps.
%Our observation does not fully cover this SNR. 
%The partial shell of radio continuum emission is located from the south to west of the remnant.
This SNR presents bright radio continuum emission in the southwest.
%A small part of the southern side of the remnant is not covered by our observation.
%and some regions outside the remnant are not covered. %which does not affect our result.
Molecular gas at the nonsignificant peak of the SCC at $\sim$+13.0~\km\ps\ is roughly surrounding the west of the remnant, correlated with its radio continuum emission. It probably belongs to the +19.3~\km\ps\ component with broad line candidates identified in the remnant by full criteria.
%which belongs to the +15.1~\km\ps\ component of broad line candidate.
%Only one broad line candidate at +15.1~\km\ps\ is identified in the remnant, which can be attributed to the nearby prominent component at +20.4~\km\ps.
Molecular gas of the +33.9~\km\ps\ component is with enhanced CO emission distributed around the southwestern boundary of the remnant, which is best correlated with the remnant's radio continuum emission. 
Molecular gas at the peak of the SCC at $\sim$+30.0~\km\ps\ is distributed near and within the boundary of the remnant.
Molecular gas at the minor peak of the SCC at $\sim$+41.1~\km\ps\ belonging to the +33.9~\km\ps\ component is roughly surrounding the south of the remnant. An OH 1720 MHz maser was also detected at +36~\km\ps\ in previous work (see Table~\ref{tab:snrpre}), which also supports the association between the remnant and the +33.9~\km\ps\ component.
Molecular gas at the minor peak of the SCC at $\sim$+181.0~\km\ps\ is roughly surrounding the north of the remnant but not correlated with its radio continuum emission.
%Broad line points of these two components, i.e.\ +15.1 and +36.4~\km\ps, are not in the region of small overlapped SNRC G8.86-0.26.
Based on these evidences, we suggest that the remnant is associated with the +33.9~\km\ps\ component, and corresponding near and far kinematic distances are estimated as 2.9$\pm0.2$ and 12.5$\pm0.3$ kpc, respectively.\\
{\bf G8.9+0.4}:
As indicated by the SCC, molecular gas at $\sim$+23.5~\km\ps\ is roughly surrounding the southeastern half of the remnant, which belongs to the +21.1~\km\ps\ component. The +21.1~\km\ps\ component is with broad line candidates identified in the remnant by full criteria.
The spatial correlation result as well as the corresponding broad line candidates indicates the association between the remnant and the +21.1~\km\ps\ component, of which the kinematic distance is estimated as 3.7$\pm0.3$ kpc.
This SNR is close to SNR W30; however, they are probably irrelevant.\\
{\bf G9.7-0.0}:
%Almost the entire western half of this SNR is not covered by our observation.
The SCC of this SNR has no significant peak.
As indicated by the SCC, molecular gas at $\sim$+40.2~\km\ps\ is roughly surrounding the remnant, which belongs to the +43.0~\km\ps\ component with broad line candidates identified in the remnant by full criteria.
Molecular gas at $\sim$+77.5~\km\ps\ also shows a spatial correlation with the remnant, which is distributed around its southern boundary. It may belong to the +61.8~\km\ps\ component with only one broad line candidate identified in the remnant by full criteria.
%We identified broad line candidates at +43.2~\km\ps, %with the distribution of molecular gas surrounding the eastern part of the remnant.
The association between the remnant and the +43.0~\km\ps\ component is also supported by OH 1720 MHz maser emission detected at +43~\km\ps\ (see Table~\ref{tab:snrpre}).
Based on these evidences, we suggest that the remnant is associated with the +43.0~\km\ps\ MC, of which the kinematic distance is estimated as 3.8$\pm0.3$ kpc.
%Note that the association between the remnant and the +61.8~\km\ps\ component cannot be totally ruled out, of which the kinematic distance is estimated as 8.4$\pm0.5$ kpc.
\\
{\bf G9.8+0.6}:
%A small part of the eastern side of this SNR is not covered by our observation.
%As indicated by the SCC, molecular gases at $\sim$+15.1 and $\sim$+31.9~\km\ps\ are roughly surrounding the southwest of the remnant and distributed around its western boundary, respectively, which belong to the +29.1~\km\ps\ velocity component. 
%The +29.1~\km\ps\ component is the only component with broad line candidates identified in the remnant by full criteria.
The SCC indicates a spatial correlation between this SNR and molecular gas at $\sim$+31.3~\km\ps, which is roughly surrounding the remnant. It belongs to the +28.9~\km\ps\ component with broad line candidates identified in the remnant by full criteria.
Therefore, we suggest that the remnant is associated with the +28.9~\km\ps\ MC, with the kinematic distance estimated as 2.8$\pm0.2$ kpc.\\
{\bf G9.9-0.8}:
%A large part of the SNR overlaps with \HII\ regions at velocities of +30 and +32~\km\ps.
Broad \twCO (J=2--1) lines at +31~\km\ps\ and near-infrared H$_2$ lines at +30~\km\ps\ were detected toward this SNR \citep{Kilpatrick+2016, Lee+2020}.
Here, we find that molecular gas at a peak of the SCC at $\sim$+11.3~\km\ps\ is distributed around the northern boundary of the remnant, which belongs to the +4.6~\km\ps\ component with a broad line candidate identified in this SNR by full criteria.
Besides, molecular gas at $\sim$+16.2~\km\ps\ may belong to the +4.6~\km\ps\ component too, which is roughly surrounding the north of the remnant.
No good spatial correlation result is found for other velocity components.
Note that broad line candidates at +28.9~\km\ps\ identified in the remnant by full criteria seem to be originated from an overlapped \HII\ region.
%, or the broad line detected at +31~\km\ps\ in previous work (see Table~\ref{tab:snrpre}), 
These evidences indicate the association between the remnant and the +4.6~\km\ps\ MC, and the corresponding kinematic distance is estimated as 1.5$\pm0.1$ kpc.\\
%{\bf G10.5-0.0}: A large portion of the SNR overlaps with a +42.65~\km\ps\ \HII\ region.
{\bf G11.0-0.0}:
This SNR presents a partial radio continuum shell in the west.
Broad lines at +12.0, +29.7, and +38.4~\km\ps\ are identified in this SNR by full criteria.
Several broad line points are at +38.4~\km\ps, which may be originated from the nearby prominent component at +29.7~\km\ps.
Molecular gas around +38.4~\km\ps\ is distributed around the eastern and southern boundaries of the remnant, however, not well correlated with the remnant's radio continuum shell. %Note that the spatial correlation seen in previous work at +40~\km\ps\ (see \ref{tab:snrpre}). 
Molecular gas at the significant peak of the SCC at $\sim$+17.6~\km\ps\ is distributed around the remnant. %, which probably belongs to the +29.7~\km\ps\ component. %spatially correlated with the remnant's radio continuum emission.
%Since the distribution of the $\sim$+17.6~\km\ps\ molecular gas more resembles that around +29.7~\km\ps, it supports the association between the remnant and the +29.7~\km\ps\ MC.
Both the +12.0 and +29.7~\km\ps\ components may contribute to molecular gas at $\sim$+17.6~\km\ps; however, most of the gas seems to be from the +29.7~\km\ps\ component.
%Therefore, we suggest that the remnant is associated with the +28.9~\km\ps\ component, hence, it is probably associated with the nearby filamentary infrared dark cloud G11.11-0.12 \citep[e.g.,][]{Chenzw+2022}.
Therefore, we suggest that the remnant is associated with the +29.7~\km\ps\ component, hence, it is probably associated with the nearby filamentary infrared dark cloud G11.11-0.12 \citep[e.g.,][]{Chenzw+2023}.
The corresponding near and far kinematic distances are estimated as 2.9$\pm0.3$ and 12.4$\pm0.3$ kpc, respectively. The near kinematic distance is preferred, if it is associated with the G11.11-0.12.\\
{\bf G11.1+0.1}:
This SNR was previously called G11.18+0.11. \cite{Sofue+2021} found a partial shell structure around it, known as G11.2+0.12, at +56~\km\ps.
The SNR presents bright shell-like radio continuum emission in its southern half.
Here, several broad line components are identified in the remnant.
Only the +16.4~\km\ps\ component has many broad line points, and most of them are located on the southern radio continuum shell.
%The +16.4~\km\ps\ broad line component distributes around the bright radio continuum shell in the southern region of the SNR. The SCC gives no significant result, 
Molecular gas at $\sim$+24.8~\km\ps\ is distributed around the southern and eastern boundaries of the remnant. It supports the association between the remnant and the +16.4~\km\ps\ component. 
%Molecular gas at the peak of the SCC at $\sim$+104.6~\km\ps\ is distributed around the northern boundary of the remnant, not correlated with its radio continuum emission.
%Therefore, we suggest that the remnant is associated with the +22.1~\km\ps\ component, of which the near and far kinematic distances are estimated as 2.8$\pm0.3$ and 13.4$\pm0.3$ kpc, respectively.\\
Therefore, we suggest that the remnant is associated with the +16.4~\km\ps\ component, of which the near and far kinematic distances are estimated as 3.8$\pm0.3$ and 13.4$\pm0.3$ kpc, respectively.\\
{\bf G11.1-0.7}:
Broad lines at three velocities are identified in this SNR; however, only one broad line at +29.2~\km\ps\ and two broad lines at +41.0~\km\ps\ are probably originated from the nearby prominent component at +37.7~\km\ps.
%As indicated by the SCC, molecular gases at $\sim$+26.7 and $\sim$+34.0~\km\ps, belonging to the +37.7~\km\ps\ component, are spatially correlated with the remnant's radio continuum emission. %The spatial correlation indicated by the peak of the SCC 
As indicated by the SCC, molecular gas at $\sim$+33.7~\km\ps, belonging to the +37.7~\km\ps\ component, is spatially correlated with the remnant's radio continuum emission. %The spatial correlation indicated by the peak of the SCC 
%It is consistent with that found in previous work (see Table~\ref{tab:snrpre}).
These evidences imply an association between the SNR and the +37.7~\km\ps\ velocity component.
The corresponding kinematic distance is estimated as 3.0$\pm0.3$ kpc.\\
%{\bf G11.1-1.0}:
{\bf G11.2-0.3}:
Broad \twCO (J=2--1) lines at +32~\km\ps\ and near-infrared H$_2$ lines at +48~\km\ps\ were detected toward this SNR \citep[][see Table~\ref{tab:snrpre} for more information]{Kilpatrick+2016, Lee+2020}.
Here, broad lines at +0.5, +33.4, and +40.5~\km\ps\ are identified in the SNR by full criteria.
Only one broad line at +40.5~\km\ps\ may be originated from the nearby prominent component at +33.4~\km\ps.
The SNR is surrounded by molecular gas at the significant peak of the SCC at $\sim$+37.3~\km\ps, which belongs to the +33.4~\km\ps\ component. 
The +0.5~\km\ps\ broad line component is not well spatially correlated with the remnant.
Based on these evidences, we suggest that the remnant is associated with the +33.4~\km\ps\ MC, with the kinematic distance estimated as 2.9$\pm0.3$ kpc.
%It is consistent with previous results (see Table~\ref{tab:snrpre}).
The near kinematic distance obtained here is consistent with the \HI\ absorption result in previous work \citep{Green+1988}.
\\
{\bf G11.4-0.1}:
%This SNR presents bright shell-like radio continuum emission in the northeast.
Several broad line components are identified in this SNR.
As indicated by the SCC, molecular gases at $\sim$+16.5 and $\sim$+51.3~\km\ps\ are spatially correlated with radio continuum emission of the remnant, especially its radio continuum shell peak in the northeast. These molecular gases belong to the +12.6 and +45.5~\km\ps\ broad line components, respectively.
%show some spatial correlations with the shell peak of the remnant's radio continuum emission, which.
%, the distribution of the +45.5~\km\ps\ component is spatially correlated with the radio continuum shell of the SNR, as indicated by the SCC. 
%The +31.5~\km\ps\ MC also surrounds half of the remnant, but at the opposite side of the radio shell peak.
%The kinematic evidence as well as their spatial correlation indicate that the remnant is associated with the +45.5~\km\ps\ MC and locates at a kinematic distance of 3.8$\pm0.3$ kpc.\\
The remnant may be associated with either the +12.6 or +45.5~\km\ps\ components. Kinematic distances are estimated as 1.25$\pm0.05$ and 12.4$\pm0.3$ kpc for the +12.6~\km\ps\ component, and 3.8$\pm0.3$ kpc for the +45.5~\km\ps\ component.\\
{\bf G11.8-0.2}:
%A large part of this SNR overlaps with \HII\ regions at velocities of +4 and +46~\km\ps.
No broad line is identified in this SNR by full criteria.
Three components of broad line candidates are identified in this SNR by partial criteria.
Only the +41.3~\km\ps\ component shows some spatial correlation with the remnant.
%As indicated by the SCC, the +41.3~\km\ps\ component is spatially correlated with the remnant.
Molecular gas at $\sim$+47.9~\km\ps\ is roughly surrounding the southeast of the remnant, where radio continuum emission is bright.
%If the remnant is associated with the +41.5~\km\ps\ MC, its kinematic distance can be estimated as 3.8$\pm0.3$ kpc.\\
If the remnant is associated with the +41.3~\km\ps\ MC, its kinematic distance can be estimated as 3.8$\pm0.3$ kpc.\\
{\bf G12.0-0.1}:
There is only one broad line component at +43.5~\km\ps\ identified in this SNR by full criteria.
Broad line points are mainly distributed on the radio continuum shell of the remnant in the southeast. %, away from overlapped \HII\ regions at +39~\km\ps\ in the north.
The SCC gives no significant spatial correlation results. 
The +43.5~\km\ps\ MC is mainly distributed in the south, where radio continuum emission is bright.
Based on these evidences, we suggest that the remnant is associated with the +43.5~\km\ps\ MC, of which the kinematic distance is estimated as 3.8$\pm0.3$ kpc.\\
{\bf G12.2+0.3}:
This SNR presents a partial radio continuum shell in the southeast \citep{Brogan+2006}.
Broad \twCO (J=2--1) lines at +50~\km\ps\ were detected toward this SNR by \cite{Kilpatrick+2016}.
Here, two broad line components at +20.3 and +30.6~\km\ps\ are identified in the SNR by full criteria. %however, they show no spatial correlation with the SNR.
As indicated by the SCC, molecular gas at $\sim$+7.6~\km\ps\ is distributed around the western half of the remnant, not correlated with the southeastern radio continuum shell. It probably belongs to the +20.3~\km\ps\ component.
No good spatial correlation result is found for the +30.6~\km\ps\ component.
We find that molecular gas at the peak of SCC at $\sim$+49.2~\km\ps\ is surrounding the southeastern half of the remnant, correlated with the radio continuum shell well.
It belongs to the +54.1~\km\ps\ velocity component with broad line candidates identified in the remnant by partial criteria. This broad line component is also detected in previous work (see Table~\ref{tab:snrpre}).
Broad line points at +54.1~\km\ps\ are located on the remnant's radio continuum shell peak in the southeast.
%Considering that this remnant is still close to the Galactic center, line overlapping effect 
%It is possible that broad lines identified by full criteria are still contaminated by line overlapping effect in the direction still close to the Galactic center.
Based on these evidences, we suggest that the remnant is associated with the +54.1~\km\ps\ component, with the corresponding kinematic distance estimated as 3.9$\pm0.3$ kpc. Note that we cannot totally rule out the possibility of the association between the remnant and the +20.3~\km\ps\ component, of which two alternative kinematic distances are estimated as 1.2$\pm0.2$ and 3.9$\pm0.3$ kpc.\\
{\bf G12.5+0.2}:
In this SNR, broad lines at several velocities are identified by full criteria. 
Two broad lines at +53.0~\km\ps\ may be originated from the nearby prominent component at +44.6~\km\ps.
As indicated by the SCC, molecular gas of the +30.4~\km\ps\ component shows some spatial correlations with the remnant, which is distributed around the remnant, mostly in the northeast and southwest.
Moreover, molecular gas at $\sim$+46.5~\km\ps\ is surrounding the southeast of the remnant and also distributed around its northern and western boundaries, which belongs to the +44.6~\km\ps\ broad line component.
Therefore, the remnant may be associated with either the +30.4 or +44.6~\km\ps\ components. Kinematic distances are estimated as 2.9$\pm0.3$ kpc for the +30.4~\km\ps\ component and 3.8$\pm0.2$ kpc for the +44.6~\km\ps\ component.\\
{\bf G12.7-0.0}:
%The northeast part of the SNR overlaps with an \HII\ region at +34~\km\ps.
This SNR presents three sections of partial radio continuum shells in the northeast, southeast, and west, and the southeastern one is the longest.
There is only one broad line at +9.1~\km\ps\ identified in the remnant by full criteria. Molecular gas at a minor peak at $\sim$+0.2~\km\ps\ belonging to the +9.1~\km\ps\ component is distributed around the northern boundary of the remnant; however, it is not well correlated with the remnant's radio continuum emission.
%The SCC have multiple peaks, all with molecular gas more or less distributed around the remnant. We examined all these peaks, and find that 
Molecular gas at the significant peak of the SCC at $\sim$+57.8~\km\ps\ is roughly surrounding the southeastern half of the remnant, following the southeastern radio shell.
It belongs to the +53.0~\km\ps\ component that is with many broad line candidates identified in the remnant by partial criteria.
We find that most of the broad line candidates at +53.0~\km\ps\ are distributed within the remnant; furthermore, they are mainly distributed around the southeastern radio continuum shell.
For the +53.0~\km\ps\ component, both the distribution of molecular gas as well as locations of broad line candidates are correlated with radio continuum emission of the remnant. It strongly indicates that the remnant is associated with the +53.0~\km\ps\ MC, of which the kinematic distance is estimated as 3.8$\pm0.3$ kpc.\\
{\bf G12.8-0.0}:
We identify only one broad line at +11.1~\km\ps\ in this SNR by full criteria; however, corresponding molecular gas shows no good spatial correlation with the remnant.
As indicated by the SCC, molecular gas of the +21.3~\km\ps\ component is roughly  surrounding the remnant, which is with broad line candidates identified in the remnant by partial criteria.
Moreover, molecular gases at $\sim$+32.1 and $\sim$+49.5~\km\ps\ are roughly surrounding the north and east of the remnant but at a distance, respectively. They belong to the +38.0 and +47.2~\km\ps\ components, respectively, both of which are with broad line candidates identified in the remnant by partial criteria.
Based on these evidences, the remnant is probably associated with the +21.3~\km\ps\ component with the kinematic distance estimated as 13.4$\pm0.3$ kpc.
Nevertheless, we cannot exclude the possibility of the association between the remnant and either the +38.0 or +47.2~\km\ps\ components.
The near and far kinematic distances of the +38.0~\km\ps\ component are estimated as 3.9$\pm0.3$ and 12.4$\pm0.3$ kpc, respectively, and the kinematic distance of the +47.2~\km\ps\ component is estimated as 3.9$\pm0.3$ kpc.\\
{\bf G13.3-1.3}:
The northwest of this SNR overlaps with SNRC G13.10-0.50.
%This SNR overlaps with \HII\ regions, one of which is in the northeast at +24~\km\ps\ with an associated broad line identified.
Broad lines at multiple velocities are identified in the remnant by full criteria.
Broad lines at +20.6~\km\ps\ can be attributed to the nearby prominent component at +15.4~\km\ps, and broad lines at +33.7, +45.2, and +51.8~\km\ps\ may be originated from the nearby prominent component at +42.7~\km\ps.
As indicated by the SCC, molecular gases at $\sim$+14.0 and $\sim$+29.2~\km\ps, belonging to the +15.4~\km\ps\ broad line component, are surrounding the north of the remnant. Moreover, molecular gas at the peak of the SCC at $\sim$+41.3~\km\ps\ is roughly surrounding the northwest of the remnant, which belongs to the +42.7~\km\ps\ broad line component.
%Among multiple broad line components identified in this SNR by full criteria, broad lines of the +42.7~\km\ps\ velocity component are mostly distributed within the remnant region and outside the overlapping region with SNRC G13.10-0.50 in the northwest. It is also supported by the spatial correlation between the remnant and the molecular gas at the peak of SCC at $\sim$+41.3~\km\ps.
%Note that an broad line associated with an overlapped \HII\ region at +24~\km\ps\ is also identified.
Based on these evidences, the remnant may be associated with either the +15.4 or +42.7~\km\ps\ components. Kinematic distances of them are estimated as 1.25$\pm0.06$ and 3.6$\pm0.5$ kpc, respectively.\\
{\bf G13.5+0.2}:
Broad lines at +17.9, +24.9, and +38.9~\km\ps\ are identified in this SNR by full criteria. 
As indicated by the SCC, molecular gas of the +17.9~\km\ps\ component is roughly surrounding the eastern half of the remnant. Molecular gas of the +38.9~\km\ps\ component is also distributed around the remnant. No good spatial correlation result is found for the +24.9~\km\ps\ component.
The remnant may be associated with either the +17.9 or +38.9~\km\ps\ components, and corresponding kinematic distances are estimated as 1.8$\pm0.1$ kpc and 3.8$\pm0.3$ kpc, respectively.\\
{\bf G14.1-0.1}:
%This SNR overlaps with \HII\ regions at $\sim$+35~\km\ps.
This SNR presents bright radio continuum emission in the northwest.
Broad lines at +37.9~\km\ps\ are identified in the SNR.
Some broad lines may be originated from a large overlapped \HII\ region.
As indicated by the SCC, molecular gas at $\sim$+47.1~\km\ps\ belonging to the +37.9~\km\ps\ component is distributed around the remnant, mostly in the northwest.
%Molecular gas at the peak of the SCC at $\sim$+21.7~\km\ps\ is distributed around the remnant but mostly inside it, and no corresponding broad line is identified in the remnant even by partial criteria. Molecular gas at $\sim$+57.3~\km\ps\ is roughly surrounding the northeast of the remnant, and no corresponding broad line is identified too.
Molecular gas at the peak of the SCC at $\sim$+58.1~\km\ps\ is distributed around the remnant, however, no corresponding broad line is identified in the remnant even by partial criteria.
%The peak of the SCC indicates a possible association between the remnant and molecular gas at $\sim$+21.7~\km\ps. The corresponding kinematic distance is estimated as 12.3$\pm0.3$ kpc.\\
%Based on these evidences, we suggest that the remnant is associated with the +38.4~\km\ps\ component, of which the kinematic distance is estimated as 3.9$\pm0.3$ kpc.\\
Based on these evidences, we suggest that the remnant is associated with the +37.9~\km\ps\ component, of which the kinematic distance is estimated as 3.9$\pm0.3$ kpc.\\
%{\bf G14.3+0.1}:
{\bf G15.1-1.6}:
One broad line at +62.7~\km\ps\ is identified in this SNR. 
As indicated by the SCC, molecular gas of the +62.7~\km\ps\ component is roughly surrounding the north of the remnant. %We consider this as a weak evidence of spatial correlation between the remnant and the +62.7~\km\ps\ velocity component.
Molecular gas at the nonsignificant peak of the SCC at $\sim$+19.0~\km\ps\ is distributed around the northeastern and southwestern boundaries of the remnant; however, there is no corresponding broad line identified in the remnant even by partial criteria.
The remnant is probably associated with the +62.7~\km\ps\ component, of which the near and far kinematic distances are estimated as 3.8$\pm0.7$ and 10.4$\pm1.8$ kpc, respectively.\\
{\bf G15.4+0.1}: 
%There are two \HII region locating on the eastern and southwestern border of this SNR, of which the velocities are +27 and +41~\km\ps, respectively.
Molecular gases at +34 and +47.8~\km\ps\ were found to be spatially correlated with this SNR \citep{Castelletti+2013, Sofue+2021}, however, \HI\ spatial correlations and absorption results indicate an association with the +60~\km\ps\ cloud \citep{Castelletti+2013, Supan+2015}.
Here, two components at +30.9 and +46.8~\km\ps\ are with broad lines identified in the SNR. 
Some broad lines at +46.8~\km\ps\ may be originated from the nearby prominent +30.9~\km\ps\ component.
%The broad line component identified at +46.8~\km\ps\ is consistent with the broad CO line result in previous work (see Table~\ref{tab:snrpre}).
%As indicated by the SCC, molecular gas of the +30.9~\km\ps\ component is roughly surrounding the remnant.
As indicated by the SCC, molecular gas at $\sim$+34.3~\km\ps\ is roughly surrounding the southern half of the remnant, which belongs to the +30.9~\km\ps\ broad line component.
The +46.8~\km\ps\ component is not well spatially correlated with the remnant.
%the northern and southern radio continuum shells of the remnant is spatially correlated with molecular gas at $\sim$+23.3~\km\ps, which belongs to the +30.7~\km\ps\ component. Note that molecular gas at a minor peak of the SCC at $\sim$34.1 is also surrounding the remnant except for its northeast side, which was also found in previous work. It is uncertain whether molecular gas at $\sim$34.1 belong to the +30.7~\km\ps\ component.
%Therefore, we suggest that the remnant is associated with the +30.7~\km\ps\ MC, of which the near and far kinematic distances are estimated as 3.0$\pm0.4$ and 13.2$\pm0.3$ kpc, respectively.\\
%Therefore, we suggest that the remnant is associated with the +30.9~\km\ps\ MC, of which the near and far kinematic distances are estimated as 3.0$\pm0.4$ and 13.2$\pm0.3$ kpc, respectively.\\
Therefore, we suggest that the remnant is associated with the +30.9~\km\ps\ MC. No \HI\ absorption around the tangent point velocity shown by \cite{Castelletti+2013} indicates that the SNR is located at a near kinematic distance. The near kinematic distance of the +30.9~\km\ps\ MC is estimated as 3.0$\pm0.4$ kpc.\\
{\bf G15.9+0.2}:
Broad lines at two velocities are identified in this SNR, nevertheless, only one broad line at +29.1~\km\ps\ can be attributed to the nearby +26.4~\km\ps\ component.
%At a peak of the SCC at $\sim$+28.4~\km\ps, molecular gas is roughly surrounding the whole remnant except for its east side, where radio continuum emission is bright. 
As indicated by the SCC, molecular gas of the +26.4~\km\ps\ component is roughly surrounding the remnant, correlated with its radio continuum emission.
This spatial correlation was also found in previous work (see Table~\ref{tab:snrpre}).
Based on the kinematic evidence as well as the spatial correlation result, we suggest that the remnant is associated with the +26.4~\km\ps\ MC, of which the near and far kinematic distances are estimated as 3.1$\pm0.4$ and 13.3$\pm0.3$ kpc, respectively.\\
{\bf G16.0-0.5}: This SNR is discussed in Section~\ref{sec:g16.0}.\\
%{\bf G16.2-2.7}:
{\bf G16.4-0.5}: 
%The west of this SNR overlaps with an \HII\ region at +42~\km\ps.
There are three components with broad lines identified in this SNR by full criteria.
Some broad lines at +39.5~\km\ps\ may be originated from an overlapped \HII\ region. %For the +70.2~\km\ps\ component, there is only one broad line identified.
As indicated by the SCC, molecular gases at $\sim$+14.8 and $\sim$+20.2~\km\ps, belonging to the +23.6~\km\ps\ velocity component, are roughly surrounding the northeast of the remnant and distributed around the remnant, respectively. %spatially correlated with radio continuum emission of the remnant.
No good spatial correlation result is found for other velocity components.
Based on identified broad lines as well as the spatial correlation result, we suggest that the remnant is associated with the +23.6~\km\ps\ MC, of which the kinematic distance is estimated as 3.2$\pm0.4$ kpc.\\
{\bf G16.7+0.1}: 
%OH 1720 MHz maser emission at $\sim$+20~\km\ps\ was detected in this SNR (see Table~\ref{tab:snrpre}).
Both OH 1720 MHz maser emission at $\sim$+20~\km\ps\ and broad \twCO (J=2--1) lines at +25~\km\ps\ were detected in this SNR \citep[][see Table~\ref{tab:snrpre} for more information]{Green+1997, Hewitt+2008, Kilpatrick+2016}.
Here, three broad line components are identified in the SNR at +29.8, +43.8, and +63.3~\km\ps.
There are broad line points at +29.8~\km\ps\ located near the position of the OH 1720 MHz maser detected by \cite{Green+1997}.
As indicated by the SCC, molecular gas at $\sim$+19.5~\km\ps\ is distributed around the remnant, which belongs to the +29.8~\km\ps\ component.
For the +43.8~\km\ps\ component, molecular gas at $\sim$+47.3~\km\ps\ is roughly surrounding the whole remnant.
The +63.3~\km\ps\ component shows no good spatial correlation with the remnant.
We also find that molecular gas at $\sim$+84.3~\km\ps\ is surrounding the north of the remnant; however, no corresponding broad line is identified in the remnant even by partial criteria.
%This spatial correlation result is also found in previous work (see Table~\ref{tab:snrpre}).
Based on the kinematic and spatial correlation evidences, the remnant is probably associated with the +29.8~\km\ps\ component, or may be associated with the +43.8~\km\ps\ component.
Kinematic distances are estimated as 13.2$\pm0.3$ kpc for the +29.8~\km\ps\ component and 4.0$\pm0.3$ for the +43.8~\km\ps\ component.\\
%This SNR-MC association may be another example with a large difference between the velocity of the OH maser and the systemic velocity of the surrounding MC, such as the case of IC~443.
{\bf G17.0-0.0}: 
%The south of this SNR overlaps with three \HII\ regions at +92 and +93~\km\ps, which is with broad line candidates identified by partial criteria.
%Broad line candidates associated with overlapped \HII\ regions at $\sim$+92~\km\ps\ are identified by partial criteria.
Only one broad line component at +30.4~\km\ps\ is identified in this SNR by full criteria.
%The spatial correlation result from the SCC also supports the association between this component and the remnant.
As indicated by the SCC, molecular gas of the +30.4~\km\ps\ component is distributed around the remnant.
%The spatial correlation between the remnant and molecular gas at $\sim$+30.0~\km\ps\ was also found in previous work (see Table~\ref{tab:snrpre}).
Molecular gas at $\sim$+94.0~\km\ps\ is roughly surrounding the southwestern half of the remnant, which is probably a part of a molecular hub-filament structure.
Based on the kinematic distance as well as the spatial correlation result, we suggest that the remnant is associated with the +30.4~\km\ps\ component, of which the kinematic distance is estimated as 12.1$\pm0.3$ kpc.\\
{\bf G17.4-0.1}: 
%The northeast of this SNR overlaps with an \HII\ region at +63~\km\ps.
Two broad line components at +41.2 and +65.6~\km\ps\ are identified in this SNR.
Only one broad line at +65.6~\km\ps\ is probably originated from an overlapped \HII\ region. 
Molecular gas at a minor peak of the SCC at $\sim$+39.8~\km\ps\ is distributed around the northeastern and southwestern boundaries of the remnant, which belongs to the +41.2~\km\ps\ component.
If the remnant is associated with the +41.2~\km\ps\ component, its near and far kinematic distances can be estimated as 3.2$\pm0.3$ and 12.0$\pm0.3$ kpc, respectively.\\
{\bf G17.4-2.3}: 
This SNR presents a partial radio continuum shell in the southwest.
OH 1720 MHz line emission was detected in the SNR (see Table~\ref{tab:snrpre}).
Only one component at +22.6~\km\ps\ is with broad lines identified in the remnant by full criteria. Molecular gas at the SCC peak at $\sim$+18.4~\km\ps\ is distributed around the southwestern and northern boundaries of the remnant, which belongs to the +22.6~\km\ps\ component.
Therefore, we suggest that the remnant is associated with the +22.6~\km\ps\ MC, accordingly, its kinematic distance can be estimated as 1.5$\pm0.1$ kpc.\\
{\bf G17.8-2.6}: 
There is no broad line identified in this SNR even by partial criteria.
%Little molecular gas is detected in the remnant.
At the peak of the SCC at $\sim$+7.5~\km\ps, there is little molecular gas distributed around the remnant but not well correlated.
Considering that OH 1720 MHz line emission was detected in the remnant in a previous work (see Table~\ref{tab:snrpre}), the remnant may be associated with molecular gas at $\sim$+7.5~\km\ps\ with the kinematic distance estimated as 0.26$\pm0.14$ kpc.\\
{\bf G18.1-0.1}: 
%Two \HII\ regions overlap with the eastern and southern boundary of this SNR at +51 and +56~\km\ps, respectively.
Three broad line components at +50.0, +77.0, and +94.1~\km\ps\ are identified in this SNR by full criteria.
%The SCC has no significant peak.
Molecular gas at the peak of the SCC at $\sim$+79.4~\km\ps\ is distributed around the northeastern and western boundaries of the remnant, which belongs to the +77.0~\km\ps\ component.
There is only one broad line identified for the +77.0~\km\ps\ component.
%Molecular gas at $\sim$+64.3~\km\ps\ is roughly surrounding the southeastern half of the remnant, which probably belongs to the +50.0~\km\ps\ component.
As indicated by the SCC, molecular gas of the +50.0~\km\ps\ component is roughly surrounding the southeastern half of the remnant.
%The distribution of molecular gas at the peak of the SCC at $\sim$+51.9~\km\ps\ is roughly correlated with the remnant, which was also detected in previous work (see Table~\ref{tab:snrpre}).
%There is only one broad line identified for the +77.0~\km\ps\ component.
For the +94.1~\km\ps\ component, nine broad line points are identified in the remnant, and no broad line point is identified outside the remnant.
The overall spatial distribution of the +94.1~\km\ps\ component is not well correlated with the remnant.
However, there is a molecular clump belonging to the +94.1~\km\ps\ component that is located on the eastern boundary of the remnant, where radio continuum emission is bright.
The association between the remnant and the +94.1~\km\ps\ component is also supported by \HI\ absorption results in previous works (see Table~\ref{tab:snrpre}).
%There is only one broad line at +77.0~\km\ps, and no good spatial correlation result is found for this broad line component.
Molecular gas at $\sim$+14.0~\km\ps\ is roughly surrounding the southwest of the remnant; however, no corresponding broad line is identified in the remnant even by partial criteria.
%Therefore, the remnant is probably associated with the +94.1~\km\ps\ component, or may be associated with the +52.0~\km\ps\ component.
Therefore, the remnant is probably associated with the +94.1~\km\ps\ component, or may be associated with the +50.0 or +77.0~\km\ps\ components.
As shown in previous works, the lack of \HI\ absorption around the tangent point velocity toward the SNR indicates that the SNR is at a near kinematic distance \citep{Leahy+2014, JohansonKerton2009, RanasingheLeahy2018a}.
%The near and far kinematic distances of the +94.1~\km\ps\ component are estimated as 5.8$\pm1.4$ and 9.7$\pm1.4$ kpc, respectively.
%%The kinematic distance of the +52.0~\km\ps\ component is estimated as 4.0$\pm0.4$ kpc.\\
%The kinematic distance of the +50.0~\km\ps\ component is estimated as 4.0$\pm0.4$ kpc, and the near and far kinematec distances of the +77.0~\km\ps\ component are estimated as 4.2$\pm0.5$ and 9.3$\pm2.5$ kpc, respectively.\\
The near kinematic distance of the +94.1~\km\ps\ component is estimated as 5.9$\pm1.4$ kpc, and that of the +50.0 and +77.0~\km\ps\ components are estimated as 4.0$\pm0.4$ and 4.2$\pm0.4$ kpc, respectively.\\
{\bf G18.6-0.2}: 
%The southeast of this SNR overlaps with region of PWN Eel (SNR G18.5-0.4).
Broad \twCO (J=2--1) lines at +42~\km\ps\ were detected toward this SNR by \cite{Kilpatrick+2016}.
Here, two broad line components at +46.7 and +64.6~\km\ps\ are identified in the SNR.
The SCC peaks at $\sim$+49.5~\km\ps, where molecular gas is distributed around the remnant.
Molecular gas at a minor peak of the SCC at $\sim$+66.5~\km\ps\ is surrounding the remnant except for its northwest side, correlated well with its shell-like radio continuum emission, which belongs to the +64.6~\km\ps\ component.
Broad lines around +46.7~\km\ps\ and the spatial correlation between the remnant and molecular gas at $\sim$+66.5~\km\ps\ were also found in previous work (see Table~\ref{tab:snrpre}). 
The association between the remnant and the +64.6~\km\ps\ component is also supported by \HI\ absorption results in previous works (see Table~\ref{tab:snrpre}).
%Here we cannot determine which component is associated with the remnant.
Based on these evidences, the remnant is probably associated with the +64.6~\km\ps\ component, or may be associated with the +46.7~\km\ps\ component.
The kinematic distance of the +46.7~\km\ps\ component is estimated as 4.1$\pm0.4$ kpc, and that of the +64.6~\km\ps\ component is estimated as 3.9$\pm0.4$ kpc.
Note that the lack of \HI\ absorption around the tangent point velocity toward this SNR shown by \cite{JohansonKerton2009}, \cite{RanasingheLeahy2018a} supports the near kinematic distances obtained here.\\
{\bf G18.8+0.3 (Kes 67)}: 
%Multiple \HII\ regions at $\sim$+20~\km\ps\ overlap with a small part of the southern boundary of this SNR.
High-$J$ CO lines were detected toward this SNR, which, together with their spatial correlations, indicate that the remnant is in association with $\sim$+20~\km\ps\ MC \citep[e.g.,][see Table~\ref{tab:snrpre} for more information]{Dubner+2004, Paron+2015}.
Here, broad lines at +20.4 and +124.6~\km\ps\ are identified in the SNR.
Most of the broad line points at +20.4~\km\ps\ are distributed around the southern boundary of the remnant.
Molecular gases at peaks of the SCC at $\sim$+11.4 and $\sim$+17.6~\km\ps\ are distributed around the southwestern boundary of the remnant, which belongs to the +20.4~\km\ps\ component.
Only one broad line is identified for the +124.6~\km\ps\ component, and the +124.6~\km\ps\ component shows no good spatial correlation with the remnant.
At the peak of the SCC at $\sim$+4.6~\km\ps, there are two roughly parallel molecular filaments, which extend far beyond the remnant and are partly surrounding the north and south of the remnant. No corresponding broad line around +4.6~\km\ps\ is identified in the remnant even by partial criteria.
Based on identified broad lines and the spatial correlation result, we suggest that the remnant is associated with the +20.4~\km\ps\ MC.
This result is consistent with that in previous works, where \HI\ and CO morphologies, \HI\ absorption spectra, and high-$J$ CO lines were studied (see Table~\ref{tab:snrpre}).
Based on \HI\ absorption results, \cite{Tian+2007} indicated that the SNR is at a far kinematic distance.
%The kinematic distance of the +20.4~\km\ps\ MC is estimated as 1.5$\pm0.1$ kpc.\\
The far kinematic distance of the +20.4~\km\ps\ MC is estimated as 13.3$\pm0.3$ kpc.\\
{\bf G18.9-1.1}: 
A spatial correlation between this SNR and the MC at +25.6~\km\ps\ was found \citep{Traverso+1999}, and \HI\ absorption features were detected up to $\sim$+23~\km\ps\ \citep{Furst+1989, Ranasinghe+2020}. Nevertheless, near-infrared H$_2$ lines were detected at +70~\km\ps\ by \cite{Lee+2020}.
Here, broad lines at +44.7, +56.0, and +64.4~\km\ps\ are identified in the SNR by full criteria.
There is only one broad line at +56.0~\km\ps, which is probably originated from the nearby prominent component at +64.4~\km\ps.
As indicated by the SCC, molecular gas of the +64.4~\km\ps\ component is roughly surrounding the north of the remnant, where radio continuum emission is bright.
%For the +64.4~\km\ps\ component, broad lines are located in the northern radio bright region.
Molecular gases at $\sim$+6.2 and $\sim$+26.0~\km\ps\ are distributed around the northwestern and eastern boundaries of the remnant, respectively; however, there is no corresponding broad line identified in the remnant even by partial criteria.
We find no good spatial correlation between the remnant and the +44.7~\km\ps\ component.
Based on these evidences, we suggest that the remnant is associated with the +64.4~\km\ps\ MC, with the kinematic distance estimated as 3.8$\pm0.4$ kpc.
The near kinematic distance obtained here is consistent with the \HI\ absorption result in previous work \citep{Ranasinghe+2020}.\\
%Our result is different from that in previous work, which suggest an association with the $\sim$+25~\km\ps\ MC (see Table~\ref{tab:snrpre}).\\
{\bf G19.1+0.2}:
%This SNR overlaps with multiple small \HII\ regions around its boundary at +19, +25, +52, +68, +70, and +109~\km\ps.
%Many velocity components have broad lines identified in this SNR. %The SCC gives no significant result.
Many velocity components have broad lines identified in this SNR by full criteria.
%The SCC gives no significant result.
%Broad lines at +56.8 and +73.1~\km\ps\ may be originated from the nearby prominent component at +60.6~\km\ps.
%As indicated by the SCC, molecular gas of the +60.6~\km\ps\ broad line component is distributed around the southern boundary of the remnant.
As indicated by the SCC, molecular gas of the +56.8~\km\ps\ broad line component is distributed around the southwestern half of the remnant, where its radio continuum emission is bright.
Molecular gas of the +112.7~\km\ps\ broad line component is roughly surrounding the south of the remnant. 
However, broad lines are identified at +56.8~\km\ps\ but not at +112.7~\km\ps\ by full criteria plus the clean subbackground region condition.
At the peak of the SCC at $\sim$+5.4~\km\ps, many molecular gases are distributed within the remnant.
No good spatial correlation result is found for other broad line components.
%Among different components, +74.4~\km\ps\ and +86.0~\km\ps\ components are spatially correlated with the remnant better.
%The +74.4~\km\ps\ component mainly surrounds the south of the remnant, with broad lines distributed in the southeast of the remnant. 
%The +86.0~\km\ps\ component surrounds the south of the remnant, and there also some clumps located in the north, of which broad lines are from a clump in the north. 
%The remnant may be associated with either the +60.6 or +112.7~\km\ps\ components. Kinematic distances are estimated as 4.2$\pm0.5$ kpc for the +60.6~\km\ps\ component and 9.5$\pm0.4$ kpc for the +112.7~\km\ps\ component.\\
Based on these evidences, we suggest that the remnant is associated with the +56.8~\km\ps\ component, and its kinematic distance is estimated as 4.2$\pm0.5$ kpc.
Note that we cannot totally rule out the possibility of the association between the remnant and the +112.7~\km\ps\ component, of which the kinematic distance is estimated as 9.5$\pm0.4$ kpc.
\\
{\bf G20.0-0.2}:
%Multiple \HII\ regions overlap with the eastern and southeastern boundary of this SNR at +42 and +67~\km\ps, respectively.
The south of this SNR overlaps with SNRC G19.96-0.33.
Three broad line components are identified in the remnant, i.e.\ +40.8, +65.8, and +88.5~\km\ps.
There is only one broad line at +40.8~\km\ps, which may be associated with overlapped \HII\ regions.
There are many broad lines at +65.8~\km\ps\ outside the overlapping regions with \HII\ regions and the SNRC.
Molecular gas at a peak of the SCC at $\sim$+73.2~\km\ps, which belongs to the +65.8~\km\ps\ component, is roughly surrounding the northwestern half of the remnant. In addition, at the peak of the SCC at $\sim$+65.6~\km\ps, an elongated molecular clump is roughly distributed along the western boundary of the remnant.
Molecular gases of other broad line components show no good spatial correlation with the remnant.
The kinematic evidence as well as the spatial correlation result indicates an association between the remnant and the +65.8~\km\ps\ MC. %, of which the kinematic distance is estimated as 4.1$\pm0.6$~\km\ps.
The result here is consistent with CO morphology and \HI\ absorption results from previous studies (see Table~\ref{tab:snrpre}).
Considering the presence of \HI\ absorption at the tangent point velocity shown by \cite{RanasingheLeahy2018a}, the kinematic distance of the SNR is estimated as 10.6$\pm1.3$ kpc.\\
%{\bf G20.4+0.1}: not SNR by Anderson+2017
{\bf G21.0-0.4}:
%This SNR overlaps with a large \HII\ region at +38~\km\ps.
There is only one component at +41.2~\km\ps\ with broad lines identified in this SNR. Note that some broad lines may be originated from a large overlapped \HII\ region.
Broad line points are mainly distributed around the remnant.
Molecular gas at a minor peak of the SCC at $\sim$+49.8~\km\ps, which belongs to the +41.2~\km\ps\ component, is distributed around the southeastern and northern boundaries of the remnant.
%Based on these evidences, we suggest that the remnant is associated with the +41.3~\km\ps\ MC with the kinematic distance estimated as 3.4$\pm0.4$ kpc.\\
Based on these evidences, we suggest that the remnant is associated with the +41.2~\km\ps\ MC with the kinematic distance estimated as 3.4$\pm0.4$ kpc.\\
%{\bf G21.5-0.1}: not SNR now.
{\bf G21.5-0.9}:
\cite{RanasingheLeahy2018a}, \cite{TianLeahy2008b} suggest an association between this SNR and the $\sim$+68~\km\ps\ cloud, based on \HI\ absorption results (see Table~\ref{tab:snrpre} for more information).
Here, no broad line is identified in the SNR by full criteria. %, but many broad line candidates identified by partial criteria.
As indicated by the SCC, molecular gases at $\sim$+54.6 and $\sim$+58.6~\km\ps\ are roughly surrounding the remnant. % except for its northwest side. 
It supports the association between the remnant and the +56.0~\km\ps\ component that is with broad line candidates identified in the remnant by partial criteria.
If the remnant is associated with the +56.0~\km\ps\ MC, its kinematic distance can be estimated as 3.4$\pm0.3$ kpc.
The near kinematic distance obtained here is consistent with the \HI\ absorption result in previous works \citep[][and references therein]{TianLeahy2008b, RanasingheLeahy2018a}.\\
{\bf G21.6-0.8}:
Broad lines at three velocities are identified in this SNR.
As indicated by the SCC, molecular gases at $\sim$+49.5 and $\sim$+63.5~\km\ps\ are distributed around the remnant, which belong to the +52.5~\km\ps\ broad line component.
No good spatial correlation is found for other broad line components.
Therefore, we suggest that the remnant is associated with the +52.5~\km\ps\ MC and locates at a kinematic distance of 3.4$\pm0.3$ kpc.\\
{\bf G21.8-0.6 (Kes 69)}: 
%The northern boundary of this SNR overlaps with an \HII\ region at +79~\km\ps.
Several components are with broad lines identified in this SNR.
Broad lines at +71.7~\km\ps\ can be attributed to the nearby prominent component at +83.2~\km\ps.
Only the +83.2~\km\ps\ component is with broad line points distributed on radio continuum shells of the remnant in the northeast and southwest. 
Extended OH 1720 MHz maser emission at $\sim$+84~\km\ps\ was also detected around the southwestern radio continuum shell (see Table~\ref{tab:snrpre}).
Molecular gas at $\sim$+66.5~\km\ps\ is roughly surrounding the southwest of the remnant, which seems to belong to the +83.2~\km\ps\ component.
There is molecular gas of the +83.2~\km\ps\ component distributed along radio continuum shells and also inside the remnant.
No good spatial correlation result is found for other velocity components.
The SCC peaks at +44.8~\km\ps, where filamentary molecular gas is distributed inside the remnant.
These evidences indicate an association between the remnant and the +83.2~\km\ps\ MC, of which the kinematic distance is estimated as 4.4$\pm0.7$ kpc.
The result here is consistent with previous multiband studies (see Table~\ref{tab:snrpre}).\\
%{\bf G21.9-0.1}: PWN
{\bf G22.7-0.2}: 
%This SNR overlaps with multiple \HII\ regions at +2, +36, +69, +70, +71, +81, +87, +88, +106, and +112~\km\ps.
Broad CO lines were detected in this SNR, indicating its association with the +77~\km\ps\ MC \citep[][see Table~\ref{tab:snrpre} for more information]{Su+2014}.
Here, we identify numerous broad line components in the SNR by full criteria.
Some broad lines are probably originated from overlapped \HII\ regions.
When using full criteria plus the clean subbackground region condition, only two broad line components at +73.1 and +106.4~\km\ps\ are identified in the SNR.
As indicated by the SCC, molecular gas at $\sim$+77.1~\km\ps, belonging to the +73.1~\km\ps\ broad line component, is surrounding radio continuum emission of the remnant in the southeast very well.
Molecular gas at the peak of the SCC at $\sim$+6.3~\km\ps\ is distributed around the eastern boundary of the remnant but not well correlated. %, which is probably a chance overlap.
No good spatial correlation is found for other velocity components.
Therefore, we suggest that the remnant is associated with the +73.1~\km\ps\ MC with the kinematic distance estimated as 4.6$\pm0.5$ kpc.
This is consistent with results in previous works (see Table~\ref{tab:snrpre}).\\
{\bf G23.3-0.3 (W41)}:
%This SNR overlaps with many small \HII\ regions at velocities of $\sim$+60, $\sim$+70, $\sim$+75, $\sim$+80, and $\sim$+100~\km\ps.
Many broad line components are identified in this SNR even by full criteria plus the clean subbackground region condition.
Broad lines at +52.8 and +77.5~\km\ps\ can be attributed to nearby prominent components at +56.6 and +74.1~\km\ps, respectively.
%We examined all of them, and find that two of them are very likely to be associated with the remnant, i.e.\ the +56.6 and +74.1~\km\ps\ components.
More than two hundreds points in the remnant are with broad lines identified at +74.1~\km\ps, but very few outside the remnant. Most of these broad line points are located on radio continuum shells of the remnant, and some of them are located in overlapped \HII\ regions.
Furthermore, OH 1720 MHz line emission at +71~\km\ps\ was detected in this remnant (see Table~\ref{tab:snrpre}).
The +74.1~\km\ps\ component is not well spatially correlated with the remnant.
There is molecular gas of the +74.1~\km\ps\ component distributed around radio continuum shells of the remnant, nevertheless, there is also a lot of molecular gas of the component distributed in the remnant, which is associated with overlapped \HII\ regions.
%In the west of the remnant, two bright radio comtinuum filaments are nearly perpendicular to each other.
As indicated by the SCC, molecular gas at $\sim$+52.1~\km\ps, belonging to the +56.6~\km\ps\ component, is distributed along two bright radio continuum filaments of the remnant in the west.
Molecular gas at the nonsignificant peak of the SCC at $\sim$+8.7~\km\ps\ is roughly surrounding the southeast of the remnant; however, it is not well correlated with radio continuum emission of the remnant.
No good spatial correlation result is found for other broad line components.
Based on these evidences, the remnant may be associated with either the +56.6 or +74.1~\km\ps\ components.
Kinematic distances are estimated as 3.5$\pm0.4$ kpc for the +56.6~\km\ps\ component and 4.8$\pm0.6$ kpc for the +74.1~\km\ps\ component.
The near kinematic distances obtained here are consistent with the \HI\ absorption result in previous works \citep[][and references therein]{RanasingheLeahy2018a}.\\
%{\bf G23.5+0.1}:
%{\bf G23.6+0.3}: not SNR by Anderson+2017
{\bf G24.7-0.6}:
Broad lines at +57.9 and +97.3~\km\ps\ are identified in this SNR. The +97.3~\km\ps\ component is with only one broad line point identified, and shows no good spatial correlation with the remnant.
As indicated by the SCC, the remnant is spatially correlated with molecular gas at $\sim$+51.4~\km\ps, which is roughly surrounding the remnant. It supports the association between the remnant and the +57.9~\km\ps\ component.
Therefore, we suggest that the remnant is associated with the +57.9~\km\ps\ MC with the kinematic distance estimated as 3.7$\pm0.4$ kpc.
The near kinematic distance obtained here is consistent with the \HI\ absorption result in previous work \citep{RanasingheLeahy2018b}.\\
{\bf G24.7+0.6}: 
%Two \HII\ regions at $\sim$+100~\km\ps\ overlap with the southwestern boundary of this SNR. 
%This SNR presents radio continuum shells in the southeast and west.
\cite{Sofue+2021} found a spatial correlation between this SNR and molecular gas at +112~\km\ps.
Here, there are several broad line components identified in the SNR even by full criteria plus the clean subbackground region condition.
Some broad lines at +102.6~\km\ps\ are originated from overlapped \HII\ regions. 
At peaks of the SCC at $\sim$+7.0 and $\sim$+43.2~\km\ps, molecular gases are distributed around the remnant and also inside it, which are not well correlated with the remnant.
As indicated by the SCC, the +17.5 and +114.9~\km\ps\ broad line components also show some spatial correlations with the remnant.
For the +17.5~\km\ps\ component, broad line points are located on the remnant's partial radio shells in the east and west. Molecular gas at $\sim$+16.3~\km\ps\ belonging to the +17.5~\km\ps\ component is also distributed around the remnant's radio continuum shells.
Molecular gas at $\sim$+113.0~\km\ps\ is distributed around the southern boundary of the remnant but not well correlated with the remnant's radio continuum emission.
Broad lines at +114.9 and +106.8~\km\ps\ are probably originated from a molecular hub-filament structure overlapping the south of the remnant. 
Therefore, we suggest that the remnant is associated with the +17.5~\km\ps\ MC, and the corresponding kinematic distance is estimated as 1.6$\pm0.3$ kpc.
The near kinematic distance obtained here is consistent with the \HI\ absorption result in previous work \citep{Ranasinghe+2018}.\\
{\bf G25.1-2.3}:
Only one component at +53.1~\km\ps\ is with broad lines identified in this SNR.
At a peak of the SCC at $\sim$+53.8~\km\ps, molecular gas is distributed around the northern shell-like radio continuum emission of the remnant.
Molecular gas at the peak of the SCC at $\sim$+10.0~\km\ps\ is roughly surrounding the southwestern radio continuum shell of the remnant; however, no corresponding broad line is identified in the remnant even by partial criteria. 
Therefore, we suggest that the remnant is associated with the +53.1~\km\ps\ MC with the kinematic distance estimated as 3.9$\pm0.8$ kpc.\\
%{\bf G26.6-0.1}: Considered as a part of SNRC G26.53+0.07. Two small \HII\ regions overlap the north of this SNR, and there is also a small \HII\ region at $-35$~\km\ps\ adjacent to the south.
{\bf G27.4+0.0 (Kes 73, 4C-04.71)}:
Broad \twCO (J=2--1) lines at +100~\km\ps\ and near-infrared H$_2$ lines at +99~\km\ps\ were detected toward this SNR \citep[][see Table~\ref{tab:snrpre} for more information]{Kilpatrick+2016, Lee+2020}.
We find only one component at +99.3~\km\ps\ with broad lines identified in the SNR by full criteria. 
%Broad line points are distributed in the southeast of the remnant.
Molecular gas at a peak of the SCC at $\sim$+103.3~\km\ps\ is surrounding the northeast of the remnant, which belongs to the +99.3~\km\ps\ component. 
At the peak of the SCC at $\sim$+76.4~\km\ps, there are several molecular clumps distributed around the remnant but not well correlated.
Based on these evidences, the remnant is probably associated with the +99.3~\km\ps\ MC. 
%The corresponding near and far kinematic distances are estimated as 6.0$\pm0.6$ and 8.4$\pm0.5$ kpc, respectively.\\
\cite{RanasingheLeahy2018a} find no \HI\ absorption around the tangent point velocity, indicating that the SNR is at a near kinematic distance.
The near kinematic distance of the +99.3~\km\ps\ MC is estimated as 6.0$\pm0.6$ kpc.\\
{\bf G27.8+0.6}: 
%Two \HII\ regions at +35 and +98~\km\ps\ overlap with the southwest and southeast boundaries of this SNR, respectively.
Multiple broad line components are identified in this SNR even by full criteria plus the clean subbackground region condition.
Broad lines at +99.8 and +105.1~\km\ps\ can be attributed to the nearby prominent component at +97.0~\km\ps.
Some broad lines at +35.7 and few at +97.0~\km\ps\ are probably originated from two overlapped \HII\ regions.
As indicated by the SCC, the +97.0~\km\ps\ component is spatially correlated with the southern radio continuum emission of the remnant well.
%Molecular gas at $\sim$+41.0~\km\ps\ is distributed around the southern boundary of the remnant but not correlated with its southern radio continuum emission.
Molecular gases at $\sim$+10.0 and $\sim$+41.0~\km\ps\ are distributed around the northeastern and southern boundaries of the remnant, respectively. However, these molecular gases are not correlated with the southern radio continuum emission of the remnant.
There are also molecular clumps at $\sim$+68.9~\km\ps\ distributed around the eastern boundary of the remnant but mostly inside it.
No good spatial correlation is found for other velocity components.
Therefore, we suggest that the remnant is associated with the +97.0~\km\ps\ MC, and the near and far kinematic distances are estimated as 5.3$\pm1.0$ and 8.1$\pm0.5$ kpc, respectively.\\
{\bf G28.6-0.1}: 
%The north of this SNR is adjacent to \HII\ regions at $\sim$+43~\km\ps.
Broad lines at +74.9 and +94.3~\km\ps\ are identified in this SNR by full criteria.
Some broad lines at +74.9~\km\ps\ are probably originated from the nearby +94.3~\km\ps\ component.
As indicated by the SCC, molecular gas of the +94.3~\km\ps\ component is distributed around the remnant, mostly surrounding the southwestern half of it.
%Broad lines identified by full criteria as well as the spatial correlation indicated by the SCC imply that this SNR is associated with the +86.8~\km\ps. 
No good spatial correlation result is found for the +74.9~\km\ps\ component. 
%These evidences indicate an association between the remnant and the +86.8~\km\ps\ component, of which the kinematic distance is estimated as 4.5$\pm0.4$ kpc.\\
These evidences indicate an association between the remnant and the +94.3~\km\ps\ component, of which the kinematic distance is estimated as 8.1$\pm0.5$ kpc.
The far kinematic distance obtained here is consistent with the \HI\ absorption result in previous work \citep{RanasingheLeahy2018b}.\\
{\bf G28.8+1.5}: 
%This SNR is large, overlapping a small \HII\ region at $-$39~\km\ps.
There is only one component at +11.7~\km\ps\ with broad lines identified in this SNR by full criteria.
As indicated by the SCC, molecular gas at $\sim$+5.7~\km\ps\ is distributed around the remnant, which belongs to the +11.7~\km\ps\ component.
These evidences indicate that the remnant is associated with the +11.7~\km\ps\ MC, and the corresponding kinematic distance is estimated as 0.4$\pm2.2$ or 1.6$\pm0.3$ kpc.\\
%{\bf G29.4+0.1}:
{\bf G29.6+0.1}:
Broad \twCO (J=2--1) lines at +94~\km\ps\ were detected toward this SNR by \cite{Kilpatrick+2016} (see Table~\ref{tab:snrpre} for more information).
We identify three broad line components in the SNR by full criteria. 
There are more points with broad lines identified for the +80.4~\km\ps\ component than the other two components.
As indicated by the SCC, the +80.4 and +97.0~\km\ps\ components show some spatial correlations with the remnant.
Molecular gas at $\sim$+86.0~\km\ps\ belonging to the +80.4~\km\ps\ component is roughly surrounding the eastern half of the remnant. Molecular gas of the +97.0~\km\ps\ component is distributed around the remnant except for its northwest side.
Note that molecular gas at the peak of the SCC at $\sim$+65.1~\km\ps\ is roughly surrounding the whole remnant except for its east side; however, no corresponding broad line is identified even by partial criteria.
Molecular gas at another peak of the SCC at $\sim$+42.2~\km\ps\ is mainly distributed in the remnant but not well correlated.
%However, the +97.0~\km\ps\ component has a better spatial correlation with the remnant.
Therefore, the remnant may be associated with either the +80.4 or +97.0~\km\ps\ components.
%The kinematic distance of the +80.4~\km\ps\ component is estimated as 4.5$\pm0.5$ kpc. The near and far kinematic distances of the +97.0~\km\ps\ component are estimated as 4.8$\pm0.5$ and 7.7$\pm0.7$ kpc, respectively.\\
As shown by \cite{Ranasinghe+2018}, there is no \HI\ absorption at the tangent point velocity, indicating that the SNR is at a near kinematic distance. The kinematic distance of the +80.4~\km\ps\ component is estimated as 4.5$\pm0.5$ kpc, of which the near kinematic distance is consistent with the \HI\ absorption result. For the +97.0~\km\ps\ component, the near kinematic distance is estimated as 4.8$\pm0.5$ kpc, consistent with that of the +80.4~\km\ps\ component.\\
{\bf G29.7-0.3 (Kes 75)}:
\cite{Su+2009}, \cite{Kilpatrick+2016} studied CO line emission toward this SNR, and found broad line features around +54~\km\ps. On the other hand, \cite{LeahyTian2008a}, \cite{RanasingheLeahy2018a} analyzed \HI\ absorption spectra toward the SNR, and found no significant \HI\ absorption feature for a \thCO\ velocity component at $\sim$+102~\km\ps. Together with a significant \HI\ absorption feature at $\sim$+95~\km\ps, they suggest an LSR velocity range of $\sim$95--102~\km\ps. Nevertheless, the \HI\ absorption at $\sim$+102~\km\ps\ is not strong but still exists as suggested by \cite{Su+2009}, hence, \HI\ absorption features no longer provide strong constrains on the LSR velocity of the SNR.
These indicate complicated ISM distributions toward the remnant (see Table~\ref{tab:snrpre} for more information).
%The discrepency between CO and \HI\ observational results reveals complexity toward the inner Galactic region.
Here, there is only one broad line component at +87.3~\km\ps\ identified in the SNR by full criteria. As indicated by the SCC, the +87.3~\km\ps\ component also shows some spatial correlations with the remnant, of which molecular gas is roughly surrounding the remnant. Note that multiple components are with broad line candidates identified in the remnant by partial criteria, e.g., at +53.7~\km\ps.
Based on evidences we found here, the remnant is probably associated with the +87.3~\km\ps\ component, of which the kinematic distance is estimated as 4.6$\pm0.6$ kpc.\\
%The near kinematic distance obtained here is consistent with the \HI\ absorption result in previous works \citep[][and references therein]{RanasingheLeahy2018a}.\\
{\bf G30.7+1.0}:
Three velocity components at +29.9, +51.1, and +80.5~\km\ps\ are with broad lines identified in this SNR.
As indicated by the SCC, the +51.1 and +80.5~\km\ps\ components are spatially correlated with the remnant. Molecular gas at $\sim$+49.2~\km\ps\ is roughly surrounding the remnant. Molecular gas at $\sim$+80.8~\km\ps\ is surrounding the south of the remnant.
If the remnant is associated with the +51.1~\km\ps\ component, the near and far kinematic distances are estimated as 3.4$\pm1.0$ and 10.7$\pm0.9$ kpc, respectively. If associated with the +80.5~\km\ps\ component, the near and far kinematic distances are estimated as 3.7$\pm0.5$ and 7.6$\pm3.3$ kpc, respectively.\\
{\bf G30.7-2.0}:
There is only one broad line component at +80.3~\km\ps\ identified in this SNR by full criteria. Molecular gas at the peak of the SCC at $\sim$+8.7~\km\ps\ is roughly surrounding the remnant except for its east side, which belongs to the +7.5~\km\ps\ component with broad line candidates identified in the remnant by partial criteria. Molecular gas of the +80.3~\km\ps\ component is mainly distributed in the east of the remnant but not well correlated.
The remnant is probably associated with the +7.5~\km\ps\ component, or may be associated with the +80.3~\km\ps\ component. Corresponding kinematic distances are estimated as 0.24$\pm0.02$ and 7.8$\pm2.2$ kpc, respectively.\\
{\bf G31.5-0.6}:
Two broad line components at +81.7 and +95.8~\km\ps\ are identified in this SNR by full criteria. 
As indicated by the SCC, molecular gas at $\sim$+79.4~\km\ps\ is roughly surrounding the north of the remnant, which belongs to the +81.7~\km\ps\ component.
%However, more broad line points are identified for the +95.8~\km\ps\ component than the +81.7~\km\ps\ component.
Molecular gas at $\sim$+106.5~\km\ps\ component is distributed around the eastern boundary of the remnant, which belongs to the +95.8~\km\ps\ component.
The remnant may be associated with either the +81.7 or +95.8~\km\ps\ components, of which kinematic distances are estimated as 4.2$\pm0.7$ and 5.5$\pm1.5$ kpc, respectively. Both velocity components are probably in the near Scutum spiral arm.\\
{\bf G31.9+0.0 (3C 391)}: 
%This SNR overlaps with large \HII\ regions at velocities of +42, +95, and +111~\km\ps.
Both OH 1720 MHz maser emission and different kinds of broadened molecular lines were detected in this SNR at around +105~\km\ps\ \citep[][see Table~\ref{tab:snrpre} for more information]{Green+1997, Frail+1996, Wilner+1998, ReachRho1999, Hewitt+2008, Gusdorf+2014, Kilpatrick+2016}. \cite{Lee+2020} also detected near-infrared H$_2$ lines at +100~\km\ps toward the SNR.
Here, only one broad line component at +104.1~\km\ps\ is identified in the SNR by full criteria.
As indicated by the SCC, the +104.1~\km\ps\ component is also spatially correlated with the remnant. Molecular gas at $\sim$+107.1~\km\ps\ is roughly surrounding the remnant.
These evidences indicate that the remnant is associated with the +104.1~\km\ps\ MC, and the corresponding kinematic distance is estimated as 7.2$\pm0.8$ kpc.
Note that the result here is consistent with the \HI\ absorption result in previous work \citep{RanasingheLeahy2017}.\\
{\bf G32.0-4.9 (3C 396.1)}:
There is no broad line identified in this SNR even by partial criteria.
%Molecular gas at the peak of the SCC at $\sim$+3.7~\km\ps\ is roughly surrounding the northeast of the remnant.
Molecular gas at the peak of the SCC at $\sim$+0.6~\km\ps\ is distributed around the southwestern boundary of the remnant. 
In addition, molecular gas at an adjacent peak of the SCC at $\sim$+3.8~\km\ps\ is roughly surrounding the northeast of the remnant.
If the remnant is associated with molecular gas at $\sim$+0.6~\km\ps, its kinematic distance can be estimated as 0.24$\pm0.02$ kpc.\\
{\bf G32.1-0.9}: 
This SNR is adjacent to SNRC G32.37-0.51 to the northeast.
There is only one component at +91.3~\km\ps\ with broad lines identified in the northeast of the remnant.
Molecular gas at a peak of the SCC at $\sim$+100.8~\km\ps\ is distributed around the northeastern boundary of the remnant, which belongs to the +91.3~\km\ps\ component. %, but not correlated well.
The remnant is probably associated with the +91.3~\km\ps\ component with the corresponding kinematic distance estimated as 5.0$\pm0.9$ kpc.\\
{\bf G32.4+0.1}: 
%The south of this SNR overlaps with several \HII\ regions at $\sim$+43~\km\ps.
Broad \twCO (J=2--1) lines at +43~\km\ps\ were detected toward this SNR by \cite{Kilpatrick+2016} (see Table~\ref{tab:snrpre} for more information).
Here, only one component at +42.0~\km\ps\ is with broad lines identified in the SNR. Broad line points are located in the southeastern overlapping \HII\ regions and around the northeastern boundary of the remnant. Molecular gas at a peak of the SCC at $\sim$+48.3~\km\ps\ is surrounding the north and west of the remnant, which belongs to the +42.0~\km\ps\ broad line component.
Therefore, we suggest that the remnant is associated with the +42.0~\km\ps\ MC. The corresponding kinematic distance is estimated as 10.1$\pm0.4$ kpc.\\
{\bf G32.8-0.1 (Kes 78)}: 
%This SNR overlaps with many \HII\ regions at +15, +30, +37, +77, and $\sim$+100~\km\ps. 
This SNR presents shell-like radio continuum emission in the southeast \citep{Caswell+1975}.
The north of the SNR overlaps with the small SNRC G32.73+0.15.
OH 1720 MHz maser emission at +86.1~\km\ps, broad CO lines at +81~\km\ps, and near-infrared H$_2$ lines at +90~\km\ps\ were detected in the SNR \citep[][see Table~\ref{tab:snrpre} for more information]{Koralesky+1998, ZhoupChen2011, Lee+2020}.
Here, multiple broad line components are identified in the SNR by full criteria.
Some broad lines are probably originated from overlapped \HII\ regions. %, e.g., around +9.3~\km\ps.
As indicated by the SCC, the +64.1~\km\ps\ broad line component shows some spatial correlations with the remnant, of which a part of a long molecular filament is roughly surrounding its southeast.
The SCC peaks at $\sim$+92.7~\km\ps, where molecular gas is mainly distributed in the remnant.
There is also a molecular filament at $\sim$+91.4~\km\ps\ surrounding the south of the remnant but at a distance. %Some molecular gases at $\sim$+95.9~\km\ps\ are also distributed within the remnant. 
These molecular gases belong to the +94.5~\km\ps\ broad line component.
%, which belongs to the +94.5~\km\ps\ broad line component, but not well correlated.
Broad line points of two components, i.e.\ at +64.1 and +94.5~\km\ps, are outside the overlapping region with SNRC G32.73+0.15.
Nevertheless, only the +94.5~\km\ps\ component is with broad lines identified in the SNR by full criteria plus the clean subbackground region condition.
%Based on these evidences, the remnant may be associated with either the +64.1 or +94.5~\km\ps\ components.
%Note that an OH 1720 MHz maser at +86.1~\km\ps\ was detected in this remnant \citep{Koralesky+1998}.
%In addition, 
The highest velocity of \HI\ absorption features is $\sim$+90~\km\ps\ \citep[e.g.,][]{ZhoupChen2011}, indicating that the remnant is in front of the tangent point and behind the MC at $\sim$+90~\km\ps. It favors the +94.5~\km\ps\ MC being associated with the remnant.
%Based on these evidences, the remnant is probably associated with the +94.5~\km\ps\ component, or may be associated with the +64.1~\km\ps\ component.
Based on these evidences, we suggest that the remnant is associated with the +94.5~\km\ps\ component, of which the kinematic distance is estimated as 5.1$\pm0.5$ kpc. The near kinematic distance obtained is consistent with the \HI\ absorption result in previous works \citep[][and references therein]{RanasingheLeahy2018a}.
%The kinematic distance is estimated as 5.1$\pm0.5$ kpc for the +94.5~\km\ps\ component.
%And, the near and far kinematic distances of the +64.1~\km\ps\ component are estimated as as 4.9$\pm1.0$ and 9.9$\pm0.4$ kpc, respectively.
\\
{\bf G33.2-0.6}:
No broad line is identified in this SNR by full criteria.
Only one component at +78.9~\km\ps\ is with broad line candidates identified in the remnant by partial criteria.
%We find no significant spatial correlation result by the SCC.
Molecular gas of the +78.9~\km\ps\ component is distributed around the northern radio continuum shell of the remnant.
The remnant is probably associated with the +78.9~\km\ps\ component, of which the kinematic distance is estimated as 4.7$\pm0.8$ kpc.\\
{\bf G33.6+0.1 (Kes 79)}: 
%This SNR overlaps with two small \HII\ regions at +103~\km\ps.
Broad CO lines at $\sim$+105~\km\ps\ were detected toward this SNR \citep[][see Table~\ref{tab:snrpre} for more information]{Kilpatrick+2016, Zhoup+2016a}.
Here, several broad line components are identified in the SNR by full criteria.
No significant spatial correlation result is found through the SCC.
Some molecular gas of the +71.6~\km\ps\ broad line component is surrounding the western radio continuum shell of the remnant; however, there is also some molecular gas inside the remnant.
%also some molecular gas around +105.3~\km\ps\ around the western radio shell of the remnant and some inside. 
There is molecular gas of the +105.3~\km\ps\ component distributed around the radio continuum shell of the remnant in the southeast, especially, matching a radio continuum protrusion. Some broad line points at +105.3~\km\ps\ are also distributed on the top and in the wings of the radio continuum protrusion.
No good spatial correlation result is found for other broad line components.
The remnant is probably associated with the +105.3~\km\ps\ component, or may be associated with the +71.6~\km\ps\ component.
Kinematic distances are estimated as 5.1$\pm0.6$ kpc for the +71.6~\km\ps\ component and 6.7$\pm0.6$ kpc for the +105.3~\km\ps\ component.\\
{\bf G34.7-0.4 (W44)}: 
%This SNR overlaps with several \HII\ regions at velocities of +46, +52, +73, and +82~\km\ps. 
The association between this SNR and MCs was confirmed by multiband observations of shocked molecular gases, e.g., OH masers, CH$_3$OH masers, CO lines, HCO+ lines, near-infrared H$_2$ lines , etc.\ \citep[][see Table~\ref{tab:snrpre} for more information]{Caswell+1975a, Wootten1977, Green+1997, Claussen+1997, Seta+2004, Hewitt+2008, McDonnell+2008, Sashida+2013, Yoshiike+2013, Anderl+2014, McEwen2016, Yamada+2017, Lee+2020}.
As a well-established case of SNR-MC association, the interaction details were further studied as well, which also explains the hadronic $\gamma$-ray production in the remnant \citep{Yoshiike+2013}.
Note that a small part of the SNR in the northeast overlaps with SNRC G34.93-0.24.
Here, broad lines at several velocities are identified in the remnant even by full criteria plus the clean subbackground region condition.
Broad lines at +42.7 and +46.8~\km\ps\ can be attributed to the nearby prominent component at +46.0~\km\ps.
Most of the broad line points at +46.0~\km\ps\ are distributed around the southeastern radio continuum shell of the remnant.
Molecular gas at the peak of the SCC at $\sim$+48.9~\km\ps\ is also roughly surrounding the southeastern radio shell, which belongs to the +46.0~\km\ps broad line component.
We find no good spatial correlation between the remnant and other broad line components.
Note that a large portion of broad lines at +81.9~\km\ps\ is probably originated from an overlapped \HII\ region in the northwest.
Based on the kinematic evidence and the spatial correlation result, we suggest that the remnant is associated with the +46.0~\km\ps\ MC with the kinematic distance estimated as 2.5$\pm0.2$ kpc.
The near kinematic distance obtained here is consistent with the \HI\ absorption result in previous works \citep[][and references therein]{RanasingheLeahy2018a}.\\
{\bf G35.6-0.4}:
We identified several broad line components in this SNR by full criteria.
As indicated by the SCC, molecular gas at $\sim$+11.9~\km\ps\ is spatially correlated with the radio continuum emission of the remnant well; however, no corresponding broad line is identified in the remnant even by partial criteria.
Molecular gases at minor peaks of the SCC at $\sim$+26.0 and $\sim$+37.5~\km\ps, which probably belong to the +27.4~\km\ps\ broad line component, are roughly surrounding the northwest and east of the remnant, respectively.
Molecular gas of the +54.6~\km\ps\ component is distributed around the northern boundary of the remnant.
In addition, broad line points of the +54.6~\km\ps\ component are distributed around the radio continuum emission of the remnant.
No spatial correlation is found for other broad line components.
The remnant may be associated with either the +27.4 or +54.6~\km\ps\ components.
\HI\ absorption results in previous works favor the SNR at a near kinematic distance \citep{Zhu+2013, RanasingheLeahy2018a}.
%The kinematic distance is estimated as 2.1$\pm0.2$ kpc for the +27.4~\km\ps\ component. And, the near and far kinematic distances of the +54.6~\km\ps\ component are estimated as 2.9$\pm0.4$ and 9.4$\pm0.5$ kpc, respectively.
The kinematic distance is estimated as 2.1$\pm0.2$ kpc for the +27.4~\km\ps\ component. The near kinematic distance of the +54.6~\km\ps\ component is estimated as 2.9$\pm0.4$ kpc.
\\
{\bf G36.6-0.7}: 
%The northeast of this SNR overlaps with an \HII\ region at +54~\km\ps\ and SNRC G36.66-0.50.
The northeast of this SNR overlaps with SNRC G36.66-0.50, and there is an \HII\ region at +54.1~\km\ps\ located in the overlapping region.
There are two broad line components identified in the remnant by full criteria, i.e.\ at +56.6 and +78.7~\km\ps.
Broad line points at +56.6~\km\ps\ are located in the overlapping region with the SNRC and the \HII\ region.
As indicated by the SCC, molecular gas at $\sim$+54.4~\km\ps\ belonging to the +56.6~\km\ps\ component is distributed around the northeastern half of the remnant.
One broad line point is at +78.7~\km\ps, which is outside the overlapping region with the SNRC and the \HII\ region.
Molecular gas of the +78.7~\km\ps\ component is distributed around the northeastern boundary of the remnant.
The remnant may be associated with either the +56.6 or +78.7~\km\ps\ components. Kinematic distances are estimated as 3.0$\pm0.5$ kpc for the +56.6~\km\ps\ component and 4.2$\pm0.5$ kpc for the +78.7~\km\ps\ component.\\
{\bf G36.6+2.6}:
This SNR presents radio continuum shells in the northwest and south.
No broad line is identified in this SNR even by partial criteria.
%The SCC peaks at $\sim$+18.3~\km\ps, where molecular gas is roughly surrounding the southern radio continuum bright region of the remnant.
The SCC peaks at $\sim$+19.5~\km\ps, where molecular gas is distributed around the southern and western boundaries of the remnant.
%If the remnant is associated with molecular gas at $\sim$+18.3~\km\ps, its kinematic distance can be estimated as 0.4$\pm0.4$ kpc.\\
If the remnant is associated with molecular gas at $\sim$+19.5~\km\ps, its kinematic distance can be estimated as 0.5$\pm0.4$ kpc.\\
{\bf G38.7-1.3}: 
The northeast of this SNR overlaps with SNRC G38.72-0.87.
Only one broad line component at +75.1~\km\ps\ is identified in the remnant by full criteria. 
Note that many broad line points are in the overlapping region with the SNRC.
The SCC also supports the association between the remnant and the +75.1~\km\ps\ component, of which molecular gas is roughly surrounding the northeastern radio continuum shell.
We suggest that the remnant is associated with the +75.1~\km\ps\ component, and the corresponding kinematic distance is estimated as 4.2$\pm0.6$ kpc.\\
{\bf G39.2-0.3 (3C 396)}: 
%This SNR overlaps with two \HII\ regions at +53 and +55~\km\ps.
Broad CO lines at +69, +77, and +84~\km\ps\ and near-infrared H$_2$ lines at +56~\km\ps\ were detected toward this SNR in previous works \citep[][see Table~\ref{tab:snrpre} for more information]{Su+2011, Kilpatrick+2016, Lee+2020}.
Here, two broad lines at +44.1 and +61.4~\km\ps\ are identified in this SNR by full criteria.
The SCC peaks at $\sim$+56.4~\km\ps, where molecular gas is roughly surrounding the remnant. It supports the association between the remnant and the +61.4~\km\ps\ component.
No spatial correlation is found between the remnant and the +44.1~\km\ps\ component.
Molecular gas at a minor peak of the SCC at $\sim$+12.7~\km\ps\ is roughly surrounding the remnant except for its southwest side, but no corresponding broad line is identified in the remnant even by partial criteria. %may absorption
At another minor peak of the SCC at $\sim$+86.2~\km\ps, molecular gas is also roughly surrounding the whole remnant except for its southwest side. It belongs to the +84.2~\km\ps\ component with broad line candidates identified in the remnant by partial criteria.
The remnant may be associated with either the +61.4 or +84.2~\km\ps\ components.
\cite{Su+2011}, \cite{RanasingheLeahy2018a} showed \HI\ absorption up to the tangent point velocity, indicating that the SNR is at a far kinematic distance.
%Kinematic distances are estimated as 8.7$\pm0.5$ kpc for the +61.4~\km\ps\ component and 4.5$\pm0.6$ kpc for the +84.2~\km\ps\ component.
The kinematic distance of the +61.4~\km\ps\ component is estimated as 8.7$\pm0.5$ kpc, and the far kinematic distance of the +84.2~\km\ps\ component is estimated as 7.4$\pm0.9$ kpc.\\
{\bf G39.7-2.0 (W50)}: %no evidence
The radio continuum emission of this SNR is bipolar spindle shape, hosting the X-ray binary system SS~433 with relativistic jets precessing around the major axis of W50 \citep[e.g.,][]{Downes+1986, Brinkmann+1996, Dubner+1998, Gao+2011}.
Broad CO lines at +77 and +53~\km\ps\ were detected toward this SNR by \cite{Su+2018}, \cite{Liu+2020}, respectively (see Table~\ref{tab:snrpre} for more information). However, broad CO lines at +77~\km\ps\ are outside the SNR region we examined, probably due to jet-ISM interactions \citep{Su+2018}, and the width of asymmetric broad CO lines at +53~\km\ps\ is $\lesssim$6~\km\ps\ \citep{Liu+2020}.
%%The north of this SNR is adjacent to a small \HII\ region at +44~\km\ps.
Here, we identify no broad line in the SNR by full criteria.
We also find no good spatial correlation between the remnant and molecular gas at different velocities.\\
{\bf G40.5-0.5}:
We identify several broad line components in this SNR by full criteria.
Only one broad line is at +65.3~\km\ps, which can be attributed to the nearby component at +67.7~\km\ps.
As indicated by the SCC, only the +67.7~\km\ps\ broad line component is spatially correlated with the remnant. Molecular gas at the peak of the SCC at $\sim$+53.0~\km\ps\ is roughly surrounding the northeastern half of the remnant, where radio continuum emission is bright.
Based on these evidences, we suggest that the remnant is associated with the +65.3~\km\ps\ component.
The corresponding kinematic distance is estimated as 8.4$\pm0.4$ kpc.\\
{\bf G41.1-0.3 (3C 397)}: 
%The north of this SNR is adjacent to a large \HII\ region at +59~\km\ps.
Broad CO lines at $\sim$+32~\km\ps\ were detected toward this SNR \citep[][see Table~\ref{tab:snrpre} for more information]{Jiang+2010, Kilpatrick+2016}.
Here, two broad line components at +35.7 and +60.4~\km\ps\ are identified in the SNR by full criteria.
Broad lines at +60.4~\km\ps\ may be originated from an overlapped \HII\ region.
The SCC peaks at $\sim$+43.0~\km\ps, where molecular gas is roughly surrounding the remnant. It belongs to the +35.7~\km\ps\ component.
Both the kinematic evidence and the spatial correlation result indicate that the remnant is associated with the +35.7~\km\ps\ component.
As shown by \cite{LeahyRanasinghe2016}, the presence of \HI\ absorption up to the tangent point velocity indicates that the SNR is at a far kinematic distance.
%The corresponding kinematic distance is estimated as 2.3$\pm0.4$ kpc.\\
The far kinematic distance of the +35.7~\km\ps\ component is estimated as 8.8$\pm1.7$ kpc.\\
{\bf G41.5+0.4}: 
%The west of this SNR is adjacent to an \HII\ region at +71~\km\ps.
\cite{Sofue+2021} found a spatial correlation between this SNR and molecular gas at +58~\km\ps.
Here, there are two broad line components at +41.0 and +57.1~\km\ps\ identified in the SNR by full criteria.
%The +41.0~\km\ps\ component is supported to be associated with the remnant by a spatial correlation.
As indicated by the SCC, molecular gas at $\sim$+35.4~\km\ps\ is roughly surrounding the remnant, which belongs to the +41.0~\km\ps\ component.
No good spatial correlation result is found for the +57.1~\km\ps\ component.
Therefore, we suggest that the remnant is associated with the +41.0~\km\ps\ component, of which the kinematic distance is estimated as 2.6$\pm0.5$ kpc.
The near kinematic distance obtained here is consistent with the \HI\ absorption result in previous work \citep{Ranasinghe+2018}.\\
{\bf G42.0-0.1}: 
%This SNR overlaps with two \HII\ regions at +18 and +20~\km\ps, in the northwest and southwest. 
The south of this SNR overlaps with SNRC G41.95-0.18.
Three broad line components are identified in the remnant by full criteria, i.e.\ at +19.8, +56.6, and +65.6~\km\ps.
There are also many broad lines at +56.6~\km\ps\ identified outside the remnant, which are unrelated to the remnant.
We also find no good spatial correlation between the remnant and the +56.6~\km\ps\ component.
Broad line points at +19.8~\km\ps\ are located in the overlapping region with an \HII\ region in the northwest. However, there are also many broad line candidates at +19.8~\km\ps\ identified in the remnant but outside the overlapping region with the \HII\ region by partial criteria.
Molecular gas at a minor peak of the SCC at $\sim$+16.2~\km\ps\ is surrounding the east of the remnant and distributed in the northwest of the remnant, which belongs to the +19.8~\km\ps\ component.
There is only one broad line at +65.6~\km\ps, which locates in the overlapping region with SNRC G41.95-0.18.
Three molecular clumps at the nonsignificant peak of the SCC at $\sim$+73.3~\km\ps\ are distributed around the remnant, which belongs to the +65.6~\km\ps\ component.
The remnant may be associated with either the +19.8 or +65.6~\km\ps\ components.
Kinematic distances are estimated as 11.0$\pm0.4$ kpc for the +19.8~\km\ps\ component and as 7.9$\pm0.5$ kpc for the +65.6~\km\ps\ component.\\
{\bf G42.8+0.6}: 
A small part of this SNR in the east overlaps with SNRC G43.02+0.73.
No broad line is identified in this SNR by full criteria.
%Broad line candidates are identified only by partial criteria.
Broad line candidates at two velocities are identified in the remnant by partial criteria. There is only one broad line candidate at +25.6~\km\ps, which can be attributed to the nearby +23.7~\km\ps\ component.
Most of the broad line candidate points are outside the overlapping region with SNRC G43.02+0.73.
Molecular gases at minor peaks of the SCC at $\sim$+26.5 and $\sim$+33.3~\km\ps\ are roughly surrounding the northwest and southeast of the remnant, respectively, which belong to the +23.7~\km\ps\ component.
The remnant may be associated with the +23.7~\km\ps\ component, and the corresponding kinematic distance is estimated as 10.8$\pm0.3$ kpc.\\
{\bf G43.3-0.2 (W49B)}: 
%An \HII\ region at +58~\km\ps\ overlaps with the southeast of this SNR.
More or less spatial correlations were found for all three major velocity components in the direction of this SNR, i.e.\ at about +10, +45, and +60~\km\ps\ \citep{Chen+2014, Zhu+2014, Sano+2021, Sofue+2021}.
Broad \twCO (J=2--1) lines at +14~\km\ps\ were detected toward the SNR by \cite{Kilpatrick+2016}, and warm molecular gases at $\sim$+10~\km\ps\ and the expanding gas motion at $\sim$+10 and $\sim$+40~\km\ps\ were detected by \cite{Sano+2021}. 
Nevertheless, \cite{Zhoup+2022} detected very broad HCO+ lines in the SNR, providing strong evidence for the association between the remnant and the $\sim$+63~\km\ps\ velocity component. \cite{Lee+2020} also detected near-infrared H$_2$ lines at +64~\km\ps\ toward the remnant. The SNR was also studied by other spectral lines supporting one or the other of the three components (see Table~\ref{tab:snrpre} for more information).
Here, two broad line components at +39.8 and +60.1~\km\ps\ are identified in the SNR by full criteria.
There is only one broad line identified for the +39.8~\km\ps\ component, which locates in the northwestern radio bright region of the remnant.
Some broad lines of the +60.1~\km\ps\ component are in an overlapped \HII\ region in the southeast.
The SCC has two significant peaks at $\sim$+15.6 and $\sim$+43.3~\km\ps.
Molecular gas at $\sim$+43.3~\km\ps\ that belongs to the +39.8~\km\ps\ component is surrounding the whole remnant except for its northwest side.
A minor peak at $\sim$+56.5~\km\ps\ also indicates a roughly spatial correlation between the remnant the +60.1~\km\ps\ component, where molecular gas is distributed around the remnant's boundary.
Based on these evidences, either the +39.8 or +60.1~\km\ps\ components may be associated with the remnant.
%We cannot finally determine which component is associated with the remnant. 
As shown by \cite{RanasingheLeahy2018a}, the presence of \HI\ absorption up to the tangent point velocity indicates that the SNR is at a far kinematic distance.
%Kinematic distances are estimated as 2.7$\pm0.4$ kpc for the +39.8~\km\ps\ component and 7.7$\pm0.6$ kpc for the +60.1~\km\ps\ component.\\
The far kinematic distance of the +39.8~\km\ps\ component is estimated as 9.3$\pm0.8$ kpc, and the kinematic distance of the +60.1~\km\ps\ component is estimated as 7.7$\pm0.6$ kpc.\\
{\bf G43.9+1.6}:
Broad CO lines at +50~\km\ps\ were detected toward this SNR by \cite{Zhou+2020}.
Here, no broad line is identified in the SNR by full criteria. One broad line candidate at +47.9~\km\ps\ is identified in the SNR by partial criteria.
As indicated by the SCC, there are several molecular clumps belonging to the +47.9~\km\ps\ component distributed around the eastern half of the remnant.
These evidences indicate that the remnant may be associated with the +47.9~\km\ps\ component.
This association is consistent with the result in previous work (see Table~\ref{tab:snrpre}).
Corresponding near and far kinematic distances are estimated as 2.8$\pm0.6$ and 8.9$\pm0.6$ kpc, respectively.\\
%{\bf G44.5-0.2}: It's an SNRC, of which the northwest overlaps with small SNRC G44.08+0.13.
{\bf G45.7-0.4}: 
%\HII\ regions at +19, +51, and +60~\km\ps\ overlap with the northeastern boundary of this SNR. 
This SNR is adjacent to SNRC G45.35-0.37 to the west.
Broad lines at several velocities are identified in the remnant by full criteria, nevertheless, only the +59.6~\km\ps\ component is with more than one broad line identified.
Besides, we also find spatial correlation between the remnant and the +59.6~\km\ps\ component. Molecular gas at the peak of the SCC at $\sim$+71.4~\km\ps\ is surrounding the southeast and northwest of the remnant, where partial radio continuum shells are present.
No good spatial correlation is found for other velocity components.
These evidences indicate that the remnant is associated with the +59.6~\km\ps\ MC with the kinematic distance estimated as 7.0$\pm0.6$ kpc.\\
{\bf G46.8-0.3 (HC30)}: 
Three broad line components are identified in this SNR by full criteria.
Two of them at +19.4 and +59.3~\km\ps\ are found to be spatially associated with the remnant, as indicated by the SCC.
For the +19.4~\km\ps\ component, molecular gas at $\sim$+21.4~\km\ps\ is roughly surrounding the north of the remnant. For the +59.3~\km\ps\ component, molecular gas at $\sim$+56.4~\km\ps\ presents as a cavity-like structure enclosing the remnant.
Therefore, the remnant may be associated with either the +19.4 or +59.3~\km\ps\ components.
%Near and far kinematic distances of the +19.4~\km\ps\ component are estimated as 1.1$\pm0.6$ and 10.1$\pm0.6$ kpc, respectively.
%The kinematic distance of the +59.3~\km\ps\ component is estimated as 6.9$\pm0.5$ kpc.
Note that there is \HI\ absorption at the tangent point velocity toward this SNR detected by \cite{Sato1979}, \cite{RanasingheLeahy2018a}, and \cite{Supan+2022}, indicating that the SNR is at a far kinematic distance.
%Kinematic distances of the +19.4 and +59.3~\km\ps\ components are estimated as 10.1$\pm0.6$ and 6.7$\pm0.6$ kpc, respectively.\\
The far kinematic distance of the +19.4~\km\ps\ component is estimated as 10.1$\pm0.6$ kpc, and the kinematic distance of the +59.3~\km\ps\ component is estimated as 6.9$\pm0.5$ kpc.\\
{\bf G49.2-0.7 (W51C)}: 
%The north of this SNR overlaps with two \HII\ regions at +67 and +75~\km\ps.
Both OH 1720 MHz maser emission with velocities from +68.9 to +72.13~\km\ps\ and broad \twCO (J=2--1) lines at +60~\km\ps\ were detected in this SNR \citep[][see Table~\ref{tab:snrpre} for more information]{Green+1997, KooMoon1997, Brogan+2000, Hewitt+2008}.
Here, broad lines at three velocities are identified in the SNR by full criteria.
%These broad lines are with velocity ranges overlapping each other, all of which probably originate from the most prominent component at +61.1~\km\ps.
All of these broad lines are probably originated from the same component at +61.1~\km\ps.
Note that part of broad lines may be associated with overlapped \HII\ regions.
As indicated by the SCC, the +61.1~\km\ps\ component is spatially correlated with the remnant, and most of its gas is surrounding the north of the remnant.
As indicated by the kinematic evidence and the spatial correlation result, the remnant is associated with the +61.1~\km\ps\ MC with the corresponding kinematic distance estimated as 5.5$\pm0.3$ kpc. This is consistent with multiband results in previous works (see Table~\ref{tab:snrpre}).\\
%{\bf G53.4+0.0}: %no evidence
%The southeast of this SNR is adjacent to an \HII\ region at +23~\km\ps.
%No broad line is identified even by partial criteria.
%We also find no good spatial correlation between the remnant and MCs.\\
{\bf G53.6-2.2 (3C 400.2)}: %no evidence
We identify no broad line and find no good spatial correlation result for this SNR.
The SCC peaks at $\sim$+18.3~\km\ps, where a molecular clump is located around the eastern boundary of the remnant.
It may support the spatial correlation between the remnant and \HI\ gas revealed in previous work (see Table~\ref{tab:snrpre}).\\ %Here, we cannot determine which MC is associated with the remnant.\\
{\bf G54.1+0.3}: 
%This SNR overlaps with a large \HII\ region at +38~\km\ps.
Toward this SNR, broad \twCO (J=2--1) lines were detected at +23~\km\ps\ with widths (FWHM) $\lesssim$7~\km\ps\ \citep[][see Table~\ref{tab:snrpre} for more information]{Lee+2012b}.
Here, no broad line is identified in the SNR even by partial criteria.
The SCC peaks at $\sim$+51.3~\km\ps, where molecular gas is roughly surrounding the northwest of the remnant and also distributed around the southeastern boundary. If the remnant is associated with molecular gas at $\sim$+51.3~\km\ps, its kinematic distance can be estimated as 4.4$\pm0.6$ kpc.\\
{\bf G54.4-0.3 (HC40)}: 
This SNR presents a bright radio continuum shell in the north.
The northwest of this SNR is adjacent to two \HII\ regions at about +38 and +42~\km\ps.
Broad lines associated with these \HII\ regions are identified, which are outside the remnant.
Broad lines at velocities of +35.5, +40.2, and +51.5~\km\ps\ in the remnant are identified by full criteria.
Two broad lines at +40.2~\km\ps\ component can be attributed to the nearby +35.5~\km\ps\ component.
At the minor peak of the SCC at $\sim$+39.8~\km\ps, molecular gas is distributed around the northern boundary of the remnant but not well correlated, which belongs to the +35.5~\km\ps\ component. We find no good spatial correlation between the remnant and the +51.5~\km\ps\ component.
%Therefore, we suggest that the remnant and the +35.5~\km\ps\ component are associated, of which the kinematic distance is estimated as 4.4$\pm0.7$ kpc.\\
Therefore, we suggest that the remnant and the +35.5~\km\ps\ component are associated.
Considering that \HI\ absorption is present up to the tangent point velocity shown by \cite{RanasingheLeahy2017}, the kinematic distance of the +35.5~\km\ps\ component is estimated as 5.1$\pm0.3$ kpc.\\
{\bf G55.0+0.3}: 
The south of this SNR is adjacent to an \HII\ region at about +31~\km\ps.
Two broad line components at +10.5 and +28.4~\km\ps\ are identified in the remnant.
No good spatial correlation is found between the remnant and the +10.5~\km\ps\ component.
Molecular gas at the peak of the SCC at $\sim$+38.9~\km\ps\ is roughly surrounding the east and south of the remnant, which belongs to the +28.4~\km\ps\ broad line component.
Therefore, we suggest that the remnant is associated with the +28.4~\km\ps\ component with the kinematic distance estimated as 4.3$\pm0.7$ kpc.\\
{\bf G55.7+3.4}:
This SNR presents shell-like radio continuum emission around its southwestern half.
No broad line is identified in this SNR even by partial criteria.
Molecular gas at the peak of the SCC at $\sim$+6.8~\km\ps\ is roughly surrounding the south of the remnant.
The remnant may be associated with molecular gas at $\sim$+6.8~\km\ps, of which the near and far kinematic distances are estimated as 0.1$\pm0.6$ and 9.0$\pm0.6$ kpc, respectively. If the remnant locates at the far distance of 9.0 kpc, its height from the Galactic plane would be as high as $\sim$540 pc.\\
{\bf G57.2+0.8 (4C-21.53)}:
No broad line is identified in this SNR even by partial criteria.
As indicated by the SCC, molecular gases at $\sim$+13.0 and $\sim$+30.0~\km\ps\ show some spatial correlations with the remnant.
Molecular gas at $\sim$+13.0~\km\ps\ is roughly surrounding the south of the remnant. Molecular gas at $\sim$+30.0~\km\ps\ presents as a short filament connected to a shell-like structure located in the south of the remnant.
The remnant may be associated with molecular gases at $\sim$+13.0 or $\sim$+30.0~\km\ps. 
%For molecular gas at $\sim$+13.0~\km\ps, the kinematic distance is estimated as 8.3$\pm0.7$ kpc. For molecular gas at $\sim$+30.0~\km\ps, the near and far kinematic distances are estimated as 3.0$\pm0.8$ and 6.3$\pm0.2$ kpc, respectively.\\
The presence of \HI\ absorption up to the tangent point velocity shown by \cite{Ranasinghe+2018} indicates that the SNR is at the far side of the tangent point.
For molecular gas at $\sim$+13.0~\km\ps, the kinematic distance is estimated as 8.3$\pm0.7$ kpc. 
For molecular gas at $\sim$+30.0~\km\ps, the far kinematic distance is estimated as 6.3$\pm1.4$ kpc.\\
{\bf G59.5+0.1}: 
Broad CO lines at +28~\km\ps\ were detected toward this SNR by \cite[][see Table~\ref{tab:snrpre} for more information]{XuWang2012}.
The SNR is close to \HII\ regions at about $-$3 and +29~\km\ps.
%An \HII\ region at +29~\km\ps\ is adjacent to the south of this SNR.
Here, three broad line components at +0.7, +21.8, and +28.9~\km\ps\ are identified in the SNR by full criteria.
There are two broad lines identified at +28.9~\km\ps, which can be attributed to the nearby +21.8~\km\ps\ component.
Broad line points of the +0.7~\km\ps\ component are beside the nearby \HII\ region at about $-$3~\km\ps.
Molecular gas at the peak of the SCC at $\sim$+1.9~\km\ps\ is distributed around the northern boundary of the remnant, however, not well correlated.
Molecular gas of the +21.8~\km\ps\ component is distributed around the eastern half of the remnant, which is also not well correlated.
The remnant may be associated with either the +0.7 or +21.8~\km\ps\ components.
Kinematic distances are estimated as 8.3$\pm0.4$ kpc for the +0.7~\km\ps\ component and 2.2$\pm0.1$ kpc for the +21.8~\km\ps\ component.\\
%{\bf G59.8+1.2}: Not SNR now. This SNR overlaps with an \HII\ region at $-$50~\km\ps. The west of the remnant partly overlaps with SNRC G59.68+1.25.
%{\bf G63.7+1.1}:
%{\bf G64.5+0.9}: %no evidence
%We identify no broad line and find no good spatial correlation result in this SNR.\\
{\bf G65.1+0.6}:
This SNR presents a bright radio continuum shell in the southwest and some radio continuum emission in the northeast.
Broad lines at several velocities are identified in the SNR by full criteria.
Broad lines at $-$3.5 and +11.3~\km\ps\ are probably originated from nearby components at $-$16.4 and +16.0~\km\ps, respectively. 
As indicated by the SCC, molecular gas at $\sim$$-$15.2~\km\ps\ shows some spatial correlations with the remnant, which is distributed around the southwestern radio continuum shell of the remnant. It belongs to the $-$16.4~\km\ps\ broad line component.
%is distributed around the southwestern radio continuum shell of the remnant, which belongs to the $-$16.4~\km\ps\ broad line component.
Molecular gas at $\sim$+18.6~\km\ps\ belonging to the +16.0~\km\ps\ broad line component is distributed around the southwestern and northeastern boundaries of the remnant.
The remnant may be associated with either the $-$16.4 or +16.0~\km\ps\ components. The kinematic distance of the $-$16.4~\km\ps\ component is estimated as 7.9$\pm0.5$ kpc. The near and far kinematic distances of the +16.0~\km\ps\ component are estimated as 1.6$\pm1.1$ and 5.7$\pm0.9$ kpc, respectively.\\
%{\bf G65.7+1.2 (DA 495)}: %no evidence
%We identify no broad line and find no good spatial correlation result in this SNR too.\\
%{\bf G65.8-0.5}:
{\bf G66.0-0.0}:
No broad line is identified in this SNR by full criteria; however, broad line candidates at velocities of +1.7, +9.1, and +15.6~\km\ps\ are identified in this SNR by partial criteria.
Broad lines at +9.1~\km\ps\ can be attributed to the nearby prominent component at +15.6~\km\ps.
Molecular gases of both the +1.7 and +15.6~\km\ps\ components are roughly surrounding the remnant.
The remnant may be associated with either of these two components.
Kinematic distances are estimated as 7.1$\pm0.4$ kpc for the +1.7~\km\ps\ component and 4.4$\pm0.5$ kpc for the +15.6~\km\ps\ component.\\
{\bf G67.6+0.9}: 
%The east and southeast of this SNR are adjacent to \HII\ regions at +0 and $-3$~\km\ps, respectively.
Only one broad line component at $-$21.3~\km\ps\ is identified in this SNR by full criteria.
However, the $-$21.3~\km\ps\ component shows no good spatial correlation with the remnant.
At the peak of the SCC at $\sim$+1.7~\km\ps, there are molecular clumps roughly surrounding the remnant, mainly in the southeast, which belongs to the +9.8~\km\ps\ component with a broad line candidate identified in the remnant by partial criteria.
The remnant is probably associated with the +9.8~\km\ps\ component, or may be associated with the $-$21.3~\km\ps\ component.
Kinematic distances are estimated as 4.3$\pm0.7$ kpc for the +9.8~\km\ps\ component and 7.7$\pm0.6$ for the $-$21.3~\km\ps\ component.\\
%{\bf G67.7+1.8}: %no evidence
%No associated MC is found based on our result in this SNR.\\
{\bf G67.8+0.5}: 
%This SNR overlaps with a large \HII\ region at $-3$~\km\ps.
No broad line is identified in this SNR even by partial criteria.
%Molecular gas at the peak of the SCC at $\sim$$-$1.7~\km\ps\ is surrounding the northwest of the remnant.
Molecular gas at the peak of the SCC at $\sim$$-$1.7~\km\ps\ is distributed around the northern and western boundaries of the remnant.
If the remnant is associated with molecular gas at $\sim$$-$1.7~\km\ps, its kinematic distance can be estimated as 6.9$\pm0.5$ kpc.\\
{\bf G68.6-1.2}:
Only one broad line component at +10.0~\km\ps\ is identified in this SNR, with broad line points distributed in the north.
As indicated by the SCC, molecular gas at $\sim$+11.9~\km\ps\ is roughly surrounding the remnant, which belongs to the +10.0~\km\ps\ component.
Based on these results, we suggest that the remnant is associated with the +10.0~\km\ps\ component with the corresponding kinematic distance estimated as 4.2$\pm0.6$ kpc.\\
%{\bf G69.0+2.7 (CTB~80)}: %no evidence
%The radio continuum morphology of this SNR is irregular. We identify no broad line and find no good spatial correlation result in this remnant.\\
{\bf G69.7+1.0}: 
No broad line is identified in this SNR even by partial criteria. We also find no good spatial correlation result. However, there are several peaks of the SCC. Molecular gas at $\sim$$-$64.3~\km\ps\ is distributed around the southeastern boundary of the remnant, and molecular gas at $\sim$$-$48.9~\km\ps\ is distributed around the northeastern boundary. There is more molecular gas at $\sim$+13.7~\km\ps, which is widely distributed around the east of the remnant.\\
{\bf G73.9+0.9}:
%This SNR overlaps with a small \HII\ region at $-53$~\km\ps\ in the north and an \HII\ region at $-$2~\km\ps\ in the east.
We only identify one broad line candidate at +0.3~\km\ps\ in this SNR by partial criteria. 
Molecular gas at a peak of the SCC at $\sim$+6.0~\km\ps\ is distributed around the southern boundary of the remnant, which belongs to the +0.3~\km\ps\ component. The remnant may be associated with the +0.3~\km\ps\ component, of which the kinematic distance is estimated as 3.5$\pm0.6$ kpc.\\
%{\bf G76.9+1.0}: %no evidence
%This SNR overlaps with a large \HII\ region at +0~\km\ps.
%No broad line is identified in the remnant, and we also find no good spatial correlation result.\\
{\bf G78.2+2.1 (DR4)}: 
%More than half of this SNR overlaps with \HII\ regions at $-5$, $-3$, +0, and +3~\km\ps\ on the east side.
One broad line component at $-$0.8~\km\ps\ is identified in this SNR by full criteria.
Molecular gas at the peak of the SCC at $\sim$+1.3~\km\ps\ is roughly surrounding the remnant, which belongs to the $-$0.8~\km\ps\ component.
These evidences indicate that the remnant is associated with the $-$0.8~\km\ps\ component. The corresponding kinematic distance is estimated as 2.5$\pm0.8$ kpc.\\
{\bf G82.2+5.3 (W63)}:
A small part of this SNR in the north is not covered in our observation.
No broad line is identified in the SNR even by partial criteria.
Molecular gas at the nonsignificant peak of the SCC at $\sim$$-$18.7~\km\ps\ is distributed around the southwestern boundary of the remnant, where the remnant's radio continuum emission is bright.
If the remnant is associated with molecular gas at $\sim$$-$18.7~\km\ps, its kinematic distance can be estimated as 5.1$\pm0.8$ kpc.\\
{\bf G83.0-0.3}:
Only one broad line candidate at +12.2~\km\ps\ is identified in this SNR by partial criteria.
No significant spatial correlation result is found through the SCC.
Nevertheless, molecular gas at a peak of the SCC at $\sim$+12.4~\km\ps\ is distributed around the northeastern and western boundary of the remnant, which belongs to the +12.2~\km\ps\ component.
The remnant may be associated with the +12.2~\km\ps\ component, and the corresponding kinematic distance is estimated as 1.9$\pm0.5$ kpc.\\
{\bf G84.2-0.8}: 
%The east of this SNR overlaps with a large \HII\ region at +5~\km\ps.
Spatial correlations between this SNR and MCs at $-39$ and $-17$~\km\ps\ were found \citep{HuangThaddeus1986, FeldtGreen1993, Jeong+2012}.
Here, no broad line is identified in the SNR even by partial criteria.
The SCC peaks at $\sim$$-$17.5~\km\ps\, where molecular gas is mainly distributed in the remnant, not well correlated.
Molecular gas at a peak of the SCC at $\sim$+6.0~\km\ps\ is roughly cavity-like, surrounding the remnant.
If the remnant is associated with molecular gas at $\sim$+6.0~\km\ps, its kinematic distance can be estimated as 1.4$\pm0.2$ kpc.\\
{\bf G85.4+0.7}: 
%The south and southwest of this SNR is adjacent to \HII\ regions at +5 and $-41$~\km\ps, respectively.
Broad line candidates are identified in this SNR by partial criteria. No broad line is identified in the SNR by full criteria.
%Broad line points of the $-$78.2~\km\ps\ component are located in a molecular clump in the east of the remnant.
As indicated by the SCC, the remnant is spatially correlated with the $-$41.7 and $-$0.6~\km\ps\ components.
The $-$41.7~\km\ps\ component is well surrounding the east and southwest of the remnant, and the $-$0.6~\km\ps\ component is roughly surrounding its south and northwest.
The remnant may be associated with the $-$41.7 or $-$0.6~\km\ps\ components.
Kinematic distances are estimated as 5.1$\pm0.6$ kpc for the $-$41.7~\km\ps\ component and 1.7$\pm0.4$ kpc for the $-$0.6~\km\ps\ component.\\
%{\bf G85.9-0.6}: %no evidence%may use a better region
%This SNR overlaps with a large \HII\ region at +5~\km\ps.
%No broad line is identified in the remnant even by partial criteria.
%We also find no good spatial correlation result.\\
{\bf G89.0+4.7 (HB~21)}:
%This SNR is not fully covered by our observation, though, more than half of the remnant in the south is covered.
Broad CO lines at $-$5 and +3~\km\ps\ were detected toward this SNR \citep[][see Table~\ref{tab:snrpre} for more information]{Koo+2001, Byun+2006}.
Here, only one broad line component at $-$16.3~\km\ps\ is identified in the remnant by full criteria; however, it shows no good spatial correlation with the remnant.
As indicated by the SCC, molecular gas at $\sim$$-$8.3~\km\ps\ is spatially correlated with the remnant, which is distributed around the remnant. It belongs to the $-$4.9~\km\ps\ component with broad lines identified in the remnant by partial criteria.
Based on these evidences, the remnant is probably associated with the $-$4.9~\km\ps\ component, or may be associated with the $-$16.3~\km\ps\ component.
We estimate kinematic distances to be 3.7$\pm0.9$ kpc for the $-$16.3~\km\ps\ component and 1.6$\pm0.2$ kpc for the $-$4.9~\km\ps\ component. The association with the $-$4.9~\km\ps\ component is consistent with multiband results in previous works (see Table~\ref{tab:snrpre}).\\
{\bf G93.3+6.9 (DA 530)}: 
Part of this SNR in the east is not covered in our observation. 
No broad line is identified in the remnant even by partial criteria.
We also find no good spatial correlation result for this remnant.\\
{\bf G93.7-0.2 (CTB~104A)}: 
%The east of this SNR overlaps with two \HII\ regions at $-45$ and $-38$~\km\ps.
\cite{Uyaniker+2002} found a spatial correlation between this SNR and \HI\ gas at $-$6~\km\ps.
Here, only one broad line candidate at +19.4~\km\ps\ is identified in the SNR by partial criteria.
%The SCC peaks at +12.9~\km\ps, where molecular gas is surrounding the northwest of the remnant. It belongs to the +19.4~\km\ps\ component.
The SCC peaks at $\sim$+11.3~\km\ps, where molecular gas is surrounding the northwest of the remnant. It belongs to the +19.4~\km\ps\ component.
The remnant may be associated with the +19.4~\km\ps\ component.
In the direction of this remnant, the velocity of +19.4~\km\ps\ is beyond the Galactic rotation curve; hence, no corresponding kinematic distance is obtained.\\
{\bf G94.0+1.0 (3C 434.1)}:
No broad line is identified in this SNR even by partial criteria.
%There is no significant peak in the SCC, 
%Molecular gas at a minor peak of the SCC at $\sim$$-$11.3~\km\ps\ is surrounding the northwest of the remnant.
Molecular gas at the peak of the SCC at $\sim$$-$13.3~\km\ps\ is roughly surrounding the northwest of the remnant.
No good spatial correlation result is found for molecular gases at other velocities.
The spatial correlation result indicates a possible association between the remnant and molecular gas at $\sim$$-$13.3~\km\ps. The corresponding kinematic distance is estimated as 2.9$\pm0.8$ kpc.\\
{\bf G96.0+2.0}:
This SNR presents shell-like radio continuum emission in the southwest.
Only one component at $-$18.6~\km\ps\ is with broad line candidates identified in the SNR by partial criteria, while no broad line is identified in it by full criteria.
Molecular gas at the peak of the SCC at $\sim$$-$15.7~\km\ps\ is distributed around the northwestern boundary but not correlated well, which belongs to the $-$18.6~\km\ps\ component.
The SNR may be associated with the $-$18.6~\km\ps\ component, of which the kinematic distance is estimated as 3.0$\pm0.8$ kpc.\\
{\bf G106.3+2.7}:
We identify no broad line in this SNR even by partial criteria.
The SCC peaks at $\sim$$-$6.7~\km\ps, where molecular gas is roughly surrounding the north of the remnant.
If the remnant is associated with molecular gas at $\sim$$-$6.7~\km\ps, its kinematic distance can be estimated as 0.8$\pm0.1$ kpc.\\
%{\bf G107.5-1.5}: The northwest of this SNR overlaps with an \HII\ region at $-$36~\km\ps.
{\bf G108.2-0.6}: 
%The south of this SNR overlaps with two \HII\ region at $\sim$$-$53~\km\ps.
No broad line is identified in this SNR even by partial criteria.
There are many molecular clumps at the peak of the SCC at $\sim$$-$49.5~\km\ps\ roughly surrounding the east of the remnant.
The remnant may be associated with molecular gas at $\sim$$-$49.5~\km\ps, of which the kinematic distance is estimated as 2.9$\pm0.3$ kpc.\\
{\bf G109.1-1.0 (CTB~109)}:
%The association between this SNR and MCs was studied in multiple bands. \cite{Wootten1981} detected broad molecular lines in the remnant, and later studies revealed more interaction details between the remnant and surrounding molecular gases (e.g., those in Table~\ref{tab:snrpre}).
An anticorrelation between the CO and X-ray distributions was found and studied in detail by \cite{Tatematsu+1987, Tatematsu+1990a}, and an association between this SNR and MCs was suggested (see Table~\ref{tab:snrpre} for more information). \cite{Sasaki+2006} detected an asymmetric CO line profile at $\sim$$-$55~\km\ps\ but only broadened by $\lesssim3$~\km\ps. No broad CO emission above 0.4 K was detected by \cite{Tatematsu+1990a}.
We identify no broad line in the SNR even by partial criteria.
Molecular gas at the peak of the SCC at $\sim$$-$51.1~\km\ps\ shows some spatial correlations with the remnant, which is surrounding the west of it.
If the remnant is associated with molecular gas at $\sim$$-$51.1~\km\ps, we can estimate its kinematic distance as 2.8$\pm0.3$ kpc.\\
{\bf G111.7-2.1 (Cas A)}:
Broad CO lines at $\sim$$-$40~\km\ps\ were detected toward this SNR \citep[][see Table~\ref{tab:snrpre} for more information]{Kilpatrick+2014, Kilpatrick+2016, Ma+2019}. Nevertheless, no broad CO lines with widths greater than 7~\km\ps\ were detected by \cite{Zhoup+2018}.
We identify broad line candidates in only one component at $-$39.0~\km\ps\ in the SNR by partial criteria.
Molecular gas of the $-$39.0~\km\ps\ component is mainly distributed in the east and south of the remnant.
If the remnant is associated with the $-$39.0~\km\ps\ component, its kinematic distance can be estimated as 3.5$\pm0.3$ kpc.\\
%{\bf G113.0+0.2}: %no evidence
%We identify no broad line and find no good spatial correlation result in this SNR.\\
%{\bf G114.3+0.3}: %no evidence
%This SNR overlaps with small \HII\ regions at $-100$, $-47$, and $-13$~\km\ps.
%We identify no broad line and find no good spatial correlation result in this SNR.\\
%{\bf G116.5+1.1}: %no evidence
%We identify no broad line and find no good spatial correlation result in this SNR.\\
%{\bf G116.9+0.2 (CTB~1)}: %no evidence
%We identify no broad line and find no good spatial correlation result in this SNR.\\
{\bf G120.1+1.4 (Tycho)}:
Broad \twCO (J=1--0) lines ($>$ 5~\km\ps) at $-$63.5~\km\ps\ were detected toward this SNR by \cite{Cai+2009} (see Table~\ref{tab:snrpre} for more information).
Here, no broad line is identified in the SNR even by partial criteria.
The SCC peaks at $\sim$$-$63.0~\km\ps, where molecular gas is surrounding the northeast of the remnant and also distributed around its southwestern boundary.
The remnant may be associated with molecular gas at $\sim$$-$63.0~\km\ps, of which near and far kinematic distances are estimated as 2.7$\pm0.7$ and 5.5$\pm0.8$ kpc, respectively. Note that the absence of \HI\ absorption feature within the velocity range of $-$46 to $-$41~\km\ps\ favors the near kinematic distance \citep{TianLeahy2011}.\\
{\bf G126.2+1.6}:
One broad line component at +3.7~\km\ps\ is identified in this SNR by full criteria. As indicated by the SCC, molecular gas at $\sim$$-$3.8~\km\ps\ is surrounding the northeast and northwest of the remnant, spatially correlated with the partial radio continuum shell in the northwest. In addition, molecular gas at $\sim$+6.7~\km\ps\ is distributed around the southeastern and western boundary of the remnant. These molecular gases belong to the +3.7~\km\ps\ component. The kinematic evidence as well as the spatial correlation result indicates an association between the remnant and the +3.7~\km\ps\ component, of which two alternative kinematic distances are estimated as 0.1$\pm0.2$ and 1.3$\pm0.1$ kpc.\\
{\bf G127.1+0.5 (R5)}:
Broad CO lines at $\sim$+5~\km\ps\ were detected toward this SNR by \cite{Zhou+2014}.
Here, only one component at +2.7~\km\ps\ is with broad line candidates identified in the SNR by partial criteria.
Molecular gas at a peak of the SCC at $\sim$+6.0~\km\ps, as part of a long filament near the remnant, is surrounding the northeast of the remnant, which belongs to the +2.7~\km\ps\ component.
It is spatially correlated with the radio continuum shell of the remnant.
Based on these evidences, the remnant is probably associated with the +2.7~\km\ps\ component, and the corresponding kinematic distance is estimated as 0.23$\pm0.02$ kpc.\\
%{\bf G128.5+2.6}:
{\bf G132.7+1.3 (HB~3)}: This SNR is discussed in Section~\ref{sec:hb3}.\\
%The east of this SNR is adjacent to several \HII\ regions at $\sim$$-$40~\km\ps.
%We identify one broad line component at $-$44.0~\km\ps\ by full criteria.
%The SCC peaks at $-$41.6~\km\ps, where molecular gas is surrounding the southeast of the remnant, spatially correlated with the bright radio continuum shell of the remnant.
%Based on the kinemtaic evidence and the spatial correlation result, we suggest that the remnant is associated with the $-$44.0~\km\ps\ MC, with the corresponding kinematic distance estimated as 1.96$\pm0.04$ kpc.\\
%{\bf G149.5+3.2}:
{\bf G150.3+4.5}:
The middle and southwest of this SNR overlap with two SNRCs, i.e.\ G150.8+3.8 and G149.5+3.2.
No broad line is identified in the remnant even by partial criteria.
The SCC indicates a roughly spatial correlation between the remnant and molecular gas at $\sim$+1.6~\km\ps, which distributes around eastern and southern boundaries of the remnant.
If the remnant is associated with molecular gas at $\sim$+1.6~\km\ps, its kinematic distance can be estimated as 0.27$\pm0.01$ kpc.\\
%{\bf G150.8+3.8}: SNRC.
%{\bf G151.2+2.9}: SNRC. This SNR is adjacent to an \HII\ region at $-$32~\km\ps\ in the south.
{\bf G152.4-2.1}:
Only one broad line candidate at $-$16.3~\km\ps\ is identified in this SNR by partial criteria.
As indicated by minor peaks of the SCC, molecular gas of the $-$16.3~\km\ps\ component is distributed around eastern and northern boundaries of the remnant. However, they are not spatially correlated very well.
If the remnant is associated with the $-$16.3~\km\ps\ component, its kinematic distance can be estimated as 1.9$\pm0.4$ kpc.\\
%{\bf G160.1-1.1}:
{\bf G160.9+2.6 (HB~9)}: 
%The west of this SNR overlaps with a small \HII\ region at $-$25~\km\ps.
We identify only one broad line component in this SNR at $-$21.6~\km\ps; however, it shows no good spatial correlation with the remnant.
%However, broad line points are located at the conjunction of two molecular strips in the remnant, which may be not disturbed by the remnant.
The SCC peaks at $\sim$$-$2.5~\km\ps, where molecular gas is surrounding the northeast of the remnant.
The remnant may be associated with either the $-$21.6~\km\ps\ component or molecular gas at $\sim$$-$2.5~\km\ps. Kinematic distances are estimated as 2.1$\pm0.4$ kpc for the $-$21.6~\km\ps\ component and 0.5$\pm0.2$ kpc for molecular gas at $\sim$$-$2.5~\km\ps.\\
%{\bf G166.0+4.3 (VRO 42.05.01)}: %no evidence
%We identify no broad line and find no good spatial correlation result in this SNR.\\
%{\bf G172.8+1.5}: Many small \HII\ regions overlap with this SNR, which are at velocities of $-25$, $\sim$$-$21, $\sim$$-$17, $-$12, $-$9, and +0~\km\ps.
%{\bf G178.2-4.2}: %no evidence
%We identify no broad line and find no good spatial correlation result in this SNR.\\
%{\bf G179.0+2.6}: %no evidence
%We identify no broad line and find no good spatial correlation result in this SNR.\\
{\bf G180.0-1.7 (S147)}:
The peak of the SCC for this SNR is not significant, where molecular gas at $\sim$+6.5~\km\ps\ is distributed around the remnant. Part of the molecular gas belongs to the +1.7~\km\ps\ component that is the only component with a broad line candidate identified in the remnant by partial criteria. We identify no broad line in the remnant by full criteria.
%We find no significant spatial correlation result for this SNR.
If the remnant is associated with the +1.7~\km\ps\ component, its kinematic distance can be estimated as 1.59$\pm0.03$ or 4.3$\pm0.5$ kpc.\\
{\bf G182.4+4.3}:
Molecular gas at the peak of the SCC at $\sim$+5.6~\km\ps\ is surrounding the north and northwest of this SNR. %; however, it is not correlated well with radio continuum emission of the remnant.
However, we identify no broad line in the remnant even by partial criteria. 
If the remnant is associated with molecular gas at $\sim$+5.6~\km\ps, its kinematic distance can be estimated as 0.23$\pm0.02$ kpc.\\
{\bf G189.1+3.0 (IC~443)}: This SNR is discussed in Section~\ref{sec:ic443}.\\
%The north of this SNR is adjacent to a large \HII\ region at $-$1~\km\ps. The northeast of the remnant overlaps with SNRC G189.6+3.3. Note that we use a circular region following the northern bright radio continuum shell of the remnant in our analysis, and there is also radio continuum emission outside the region in the south.
%{\bf G189.6+3.3}: SNRC. The northwest of this SNR overlaps with a large \HII\ region at $-$1~\km\ps.
{\bf G190.9-2.2}:
We identify one broad line component at +1.0~\km\ps\ in this SNR by full criteria.
%This component is not well spatially correlated with the remnant. Nevertheless,
%Molecular gas at the peak of the SCC at $\sim$+5.1~\km\ps\ is distributed around the southeastern and western boundaries of the remnant. which belongs to the +1.0~\km\ps\ component.
At the peak of the SCC at $\sim$$-$4.1~\km\ps, there is a filamentary shape MC located in the northern part of the remnant but not correlated well.
Additionally, molecular gas at another peak of the SCC at $\sim$+5.1~\km\ps\ is distributed around the southeastern and western boundaries of the remnant. These molecular gases belong to the +1.0~\km\ps\ component.
Therefore, the remnant is probably associated with the +1.0~\km\ps\ component, with the corresponding kinematic distance estimated as 1.6$\pm0.2$ kpc.\\
%{\bf G192.8-1.1 (PKS 0607+17)}: This SNR overlaps with \HII\ regions at $\sim$+2, +11, and +21~\km\ps.
{\bf G205.5+0.5 (Monoceros Loop)}: 
%The south of this SNR overlaps with an \HII\ region at +18~\km\ps. %It also overlaps with SNR G201.1+8.3, which is very large, with radius of $\sim$13$^\circ$.
Broad CO lines at $\sim$+5 and $\sim$+19~\km\ps\ were detected toward this SNR with widths of several kilometers per second \citep[][see Table~\ref{tab:snrpre} for more information]{Su+2017a}.
Here, we identify one broad line component at +21.3~\km\ps\ in the SNR only by partial criteria. Broad line points are located near an overlapped \HII\ region at +18.1~\km\ps\ in the south.
Molecular gas at the peak of the SCC at $\sim$+13.3~\km\ps\ is around northern and southern boundaries of the remnant, which belongs to the +21.3~\km\ps\ component.
The remnant may be associated with the +21.3~\km\ps\ component, of which the kinematic distance is estimated as 2.3$\pm0.4$ kpc.\\
{\bf G206.9+2.3 (PKS 0646+06)}: 
%This SNR also overlaps with very large SNR G201.1+8.3.
No broad line is identified in this SNR.
The SCC indicates a spatial correlation between the remnant and molecular gas at $\sim$+13.7~\km\ps, which surrounds the south of the remnant.
If the remnant is associated with molecular gas at $\sim$+13.7~\km\ps, its kinematic distance can be estimated as 2.3$\pm0.4$ kpc.\\
{\bf G213.0-0.6}: %no evidence
%The south and west of this SNR overlap with \HII\ regions at +45.2 and +8.1~\km\ps, respectively. %The northwest of the remnant is adjacent to very large SNR 201.1+8.3.
Broad CO lines at $\sim$+9~\km\ps\ were detected toward this SNR with widths of several kilometers per second \citep[][see Table~\ref{tab:snrpre} for more information]{Su+2017a}.
Here, we identify no broad line in the SNR and find no good spatial correlation result. Molecular gas at the peak of the SCC at $\sim$+22.9~\km\ps\ is distributed within the remnant.%\\

\section{Individual SNR Candidates}\label{app:snrc}
{\bf G1.95-0.10}:
The SCC peaks at $\sim$+59.5~\km\ps, where molecular gas is roughly surrounding the west of this SNRC. It belongs to the +44.7~\km\ps\ component with a broad line candidate identified in the SNRC by full criteria. These evidences indicate a possible association between the SNRC and the +44.7~\km\ps\ component. The corresponding kinematic distance is estimated as 8.1$\pm2.9$ kpc.\\
{\bf G1.98-0.46}:
Molecular gas at the peak of the SCC at $\sim$+8.6~\km\ps\ is distributed around this SNRC, which belongs to the +12.6~\km\ps\ velocity component.
%The association between the SNRC and +12.6~\km\ps\ component is also supported by identified broad line candidates in the SNRC.
Identified broad line candidates in the SNRC also support the association between the SNRC and the +12.6~\km\ps\ component.
If the SNRC is associated with the +12.6~\km\ps\ component, its kinematic distance can be estimated as 2.8$\pm0.3$ kpc.\\
{\bf G2.23+0.06}: %r=1.3beamsiz, F
This SNRC is classified as the plerion type, and it is small. We find no good spatial correlation result and identify no broad line candidate in it by full criteria.\\
{\bf G2.28+0.40}:
%The south of this SNRC is adjacent to an \HII\ region at +4~\km\ps.
%No significant spatial correlation result is found for this SNRC.
Molecular gas at the peak of the SCC at $\sim$+3.2~\km\ps\ is roughly surrounding this SNRC, which belongs to the +4.8~\km\ps\ velocity component with many broad line candidates identified in the SNRC's west by full criteria.
The SNRC may be associated with the +4.8~\km\ps\ component, of which the kinematic distance is estimated as 2.8$\pm0.3$ kpc.\\
{\bf G2.91-0.18 and G3.1-0.6}:
The southeast of SNRC G2.91-0.18 is adjacent to SNRC G3.1-0.6.
%The north of SNRC G2.91-0.18 overlaps with several \HII\ regions at $\sim$$-$2, +2, and +18~\km\ps, 
%There are also two \HII\ regions at +0 and +5~\km\ps\ that overlap the eastern and northeastern boundaries of SNRC G3.1-0.6, respectively.
For SNRC G2.91-0.18, molecular gas at a peak of the SCC at $\sim$$-$42.7~\km\ps\ is surrounding the south and southwest of the SNRC, where shell-like radio continuum emission is present. It belongs to the $-$40.0~\km\ps\ velocity component that is with broad line candidates identified in the SNRC by full criteria.
These evidences indicate that SNRC G2.91-0.18 is associated with the $-$40.0~\km\ps\ component, of which the kinematic distance is estimated as 4.7$\pm0.3$ kpc.
For SNRC G3.1-0.6, molecular gas at a minor peak of the SCC at $\sim$$-$32.2~\km\ps\ is roughly surrounding the north of the SNRC, which belongs to the $-$41.0~\km\ps\ component with broad line candidates identified in the SNRC by full criteria.
SNRC G3.1-0.6 may be associated with the $-$41.0~\km\ps\ component with the corresponding kinematic distance estimated as 4.7$\pm0.3$ kpc.
Therefore, these two SNRCs may be related to each other.\\
{\bf G3.10-0.09}:
This SNRC overlaps with SNRC G2.91-0.18.
%The SCC gives no significant spatial correlation result, nevertheless, 
Molecular gases at minor peaks of the SCC at $\sim$+33.3 and $\sim$+92.9~\km\ps\ are roughly surrounding the shell-like radio continuum emission of the SNRC in the west. 
Molecular gas at $\sim$+33.3~\km\ps\ probably belongs to the +23.7~\km\ps\ component, which is with a broad line candidate identified in the SNRC by full criteria. Molecular gas at $\sim$+92.9~\km\ps\ belongs to the +95.8~\km\ps\ component with broad line candidates identified in the SNRC only by partial criteria.
Based on these evidences, either the +23.7 or +95.8~\km\ps\ components may be associated with the SNRC.
Near and far kinematic distances are estimated as 2.9$\pm0.3$ and 12.6$\pm0.3$ kpc for the +23.7~\km\ps\ component, respectively. For the +95.8~\km\ps\ component, the kinematic distance is estimated as 8.2$\pm2.9$ kpc.
This SNRC is not likely to be associated with SNRC G2.91-0.18.\\
{\bf G3.10+0.11}: %F
This SNRC is classified as the plerion type. No good spatial correlation result is found for the SNRC, because of its small size. We also identify no broad line candidate in it by full criteria.\\
{\bf G4.20-0.30}:
Molecular gas at a peak of the SCC at $\sim$$-$27.8~\km\ps\ is roughly surrounding the northern radio continuum shell of this SNRC, which belongs to the $-$31.4~\km\ps\ component with broad line candidates identified in the SNRC by full criteria.
As supported by the spatial correlation result and identified broad line candidates, the SNRC is probably associated with the $-$31.4~\km\ps\ component, of which the kinematic distance is estimated as 4.7$\pm0.3$ kpc.\\
{\bf G4.49-0.39 and G4.57-0.24}:
The northeast of SNRC G4.49-0.39 is adjacent to SNRC G4.57-0.24.
For SNRC G4.49-0.39, molecular gas at a peak of the SCC at $\sim$+16.3~\km\ps\ shows some spatial correlations with radio continuum emission of the SNRC, which is roughly surrounding the east and south of the SNRC. It belongs to the +14.1~\km\ps\ velocity component that is with broad line candidates identified in the SNRC by full criteria. If SNRC G4.49-0.39 and the +14.1~\km\ps\ component are associated, the kinematic distance of the SNRC can be estimated as 2.9$\pm0.2$ kpc.
For SNRC G4.57-0.24, molecular gas at $\sim$+17.0~\km\ps\ is roughly surrounding its east and northwest, as indicated by the SCC, which belongs to the +12.3~\km\ps\ velocity component.
There is only one broad line candidate identified in the SNRC by full criteria, which is at +12.3~\km\ps.
%These evidences indicate that SNRC G4.57-0.24 is associated with the +12.1~\km\ps\ component, and the corresponding kinematic distance is estimated as 2.9$\pm0.2$ kpc.
These evidences indicate that SNRC G4.57-0.24 is associated with the +12.3~\km\ps\ component, and the corresponding kinematic distance is estimated as 2.9$\pm0.2$ kpc.
These two SNRCs may be related to each other.\\
{\bf G5.11+0.33}:
%The southwest of this SNRC overlaps with an \HII\ region at +12~\km\ps.
This SNRC presents a radio continuum shell in the west.
Molecular gas at the peak of the SCC at $\sim$$-$28.6~\km\ps\ is roughly surrounding the north and southeast of the SNRC; however, it is with no broad line candidate identified in the SNRC even by partial criteria.
The distribution of molecular gas at $\sim$+11.7~\km\ps\ roughly follows the western radio continuum shell. It belongs to the only velocity component with broad line candidates identified in the SNRC by full criteria, which is at +13.8~\km\ps.
Note that part of broad line candidates may be originated from an overlapped \HII\ region at +12.0~\km\ps. %, and some are outside of it.
The SNRC is probably associated with the +13.8~\km\ps\ component with the kinematic distance estimated as 2.9$\pm0.2$ kpc. Accordingly, the SNRC may be related to the overlapped \HII\ region.\\
{\bf G5.16-0.32}:
%The northeast of this SNRC overlaps with an \HII\ region at +1~\km\ps.
Molecular gas at the peak of the SCC at $\sim$+5.4~\km\ps\ is distributed around this SNRC but not correlated well, which belongs to the +7.6~\km\ps\ component with a broad line candidate identified in the SNRC by full criteria.
Moreover, molecular gas at a minor peak of the SCC at $\sim$+59.5~\km\ps\ shows good spatial correlation with the SNRC, which is surrounding its southern half.
It belongs to the +58.9~\km\ps\ component that is with broad line candidates identified only by partial criteria.
Based on these evidences, either the +7.6 or +58.9~\km\ps\ components may be associated with the SNRC.
Corresponding kinematic distances are estimated as 2.9$\pm0.2$ and 10.9$\pm0.3$ kpc for the +7.6 and +58.9~\km\ps\ components, respectively.\\
{\bf G5.36-0.71}:
%The SCC peaks at $\sim$+21.0~\km\ps, where molecular gas is roughly surrounding this SNRC. 
Molecular gas at a peak of the SCC at $\sim$+21.0~\km\ps\ is roughly surrounding this SNRC.
It belongs to the +13.0~\km\ps\ component that is with broad line candidates identified in the SNRC by full criteria.
The SNRC is probably associated with the +13.0~\km\ps\ component, and the corresponding kinematic distance is estimated as 2.9$\pm0.2$ kpc.\\
{\bf G5.38-0.28}: %F
This SNRC is classified as the plerion type. Since the SNRC is small, no good spatial correlation result is found. No broad line candidate is identified in it by full criteria too.\\
{\bf G5.38+0.35}:
The east of this SNRC overlaps with SNR G5.5+0.3, and it presents a partial radio continuum shell in the northwest.
%A coarse spatial correlation between the SNRC and the +8.5~\km\ps\ component is indicated by minor peaks of the SCC, of which molecular gas is distributed around the SNRC's radio continuum shell.
As indicated by minor peaks of the SCC, the +8.5~\km\ps\ component shows a coarse spatial correlation with the SNRC, of which a portion of molecular gas is distributed around the SNRC's radio continuum shell.
The +8.5~\km\ps\ component is the only component with broad line candidates identified in the SNRC by full criteria. 
If the SNRC is associated with the +8.5~\km\ps\ component, its kinematic distance can be estimated as 2.9$\pm0.2$ kpc.\\
{\bf G5.7-0.1 (also known as G5.67-0.12)}:
%SNRC G5.7-0.1 seems related to SNRC G005.673-0.118 \citep{Dokara+2021}, and we consider them as one SNRC. %described by a circular region following the shell-like radio continuum emission.
We consider SNRC G5.7-0.1 and SNRC G005.673-0.118 \citep[see][]{Dokara+2021} as one SNRC.
For this SNRC, we find no prominent spatial correlation result. 
Two components at velocities of $-26.1$ and +8.3~\km\ps\ are with broad line candidates identified in the SNRC by full criteria.
Molecular gases at minor peaks of the SCC belonging to these two components are roughly surrounding the southwest and northeast of the SNRC, respectively.
%indicate roughly spatial correlations between the SNRC and them.
We cannot determine which of the velocity components at $-26.1$ or +8.3~\km\ps\ is associated with the SNRC. In previous works, OH 1720 MHz maser and broad CO emission were detected at $\sim$+12~\km\ps, and the CH$_3$OH 36 and 44 GHz masers were detected at $\sim$$-$25~\km\ps\ (see Table~\ref{tab:snrpre}). 
We estimate the kinematic distance of the $-26.1$~\km\ps\ component to be 4.7$\pm0.3$ kpc, and the near and far kinematic distances of the +8.3~\km\ps\ component to be 2.9$\pm0.2$ and 12.6$\pm0.3$ kpc, respectively.\\
{\bf G5.76+0.52}:
Molecular gas at a peak of the SCC at $\sim$+11.7~\km\ps\ is roughly surrounding this SNRC.
It belongs to the +6.6~\km\ps\ velocity component with broad line candidates identified in the SNRC.
The SNRC may be associated with the +6.6~\km\ps\ component with the kinematic distance estimated as 2.9$\pm0.2$ kpc.\\
{\bf G5.99+0.02}:
The east of this SNRC is adjacent to SNR W28.
As indicated by minor peaks of the SCC, the spatial distribution of the +8.0~\km\ps\ velocity component is roughly correlated with the SNRC, which is distributed around the SNRC. It is also with broad line candidates identified in the SNRC.
If the SNRC is associated with the +8.0~\km\ps\ component, its near and far kinematic distances can be estimated as 3.0$\pm0.2$ and 13.4$\pm0.3$ kpc, respectively.\\
{\bf G6.06+0.50, G6.12+0.39 and G6.31+0.54}:
All of these three SNRCs overlap with SNR G6.1+0.5.
For SNRC G6.06+0.50, molecular gas at the peak of the SCC at $\sim$+8.7~\km\ps\ is roughly surrounding it, which belongs to the +6.8~\km\ps\ velocity component with broad line candidates identified in it.
%It is also with broad line candidates identified.
These evidences indicate that SNRC G6.06+0.50 is associated with the +6.8~\km\ps\ component with the corresponding kinematic distance estimated as 3.0$\pm0.2$ kpc.
This SNRC seems to be at the same distance as SNR G6.1+0.5.
For SNRC G6.12+0.39, the SCC indicates a spatial correlation with molecular gas at $\sim$+10.8~\km\ps, which is surrounding the whole SNRC except for its west side.
Its spatial distribution is consistent with that of the +21.8~\km\ps\ component that is with a broad line candidate identified in the SNRC.
The +7.0~\km\ps\ component is also with broad line candidates identified in the SNRC, which are probably originated from SNR G6.1+0.5 or SNRC G6.06+0.50.
Therefore, we suggest that SNRC G6.12+0.39 is associated with the +21.8~\km\ps\ MC and locates at a kinematic distance of 3.6$\pm0.3$ kpc.
SNRC G6.31+0.54 presents a flat radio continuum shell in the northeast.
For SNRC G6.31+0.54, we find no good spatial correlation result and identify no broad line candidate in it by full criteria, hence, no associated MC settled.\\
{\bf G6.45-0.56 and G6.54-0.60}: %r_G6.45-0.56=1.9beamsiz
%These two SNRCs overlap with a large \HII\ region at +19~\km\ps, and they also overlap with the south and southwest of SNR G6.5-0.4.
%These two SNRCs are a bit small, and 
These two SNRCs overlap with the south and southwest of SNR G6.5-0.4, respectively.
%For SNRC G6.54-0.60, there is a molecular clump distributed around the its northern boundary at the peak of the SCC at $\sim$+4.8~\km\ps, where radio continuum emission is bright. It belongs to the +12.6~\km\ps\ component with broad line candidates identified only by partial criteria.
We find no good spatial correlation result and identify no broad line candidate in these two SNRCs by full criteria.\\
{\bf G7.4+0.3}:
%A small \HII\ region at $-$12~\km\ps\ overlaps with this SNRC around its central region.
%No good spatial correlation result is found for this SNRC.
%We find no significant spatial correlation result for this SNRC by the SCC. 
Molecular gas at a peak of the SCC at $\sim$+13.7~\km\ps\ is roughly surrounding this SNRC.
It belongs to the +16.4~\km\ps\ velocity component, which is with broad line candidates identified in the SNRC by full criteria.
The SNRC may be associated with the +16.4~\km\ps\ component, of which the kinematic distance is estimated as 3.6$\pm0.4$ kpc.\\
{\bf G7.5-1.7}:
%The southwest of this SNRC overlaps with three \HII\ regions around +5~\km\ps.
This SNRC exhibits shell-like radio continuum emission in the north.
%The surrounding region of the SNRC is not fully covered by our observation. Regions close to the SNRC to the east and farther to the south are not covered.
We find that velocity components at +10.5 and +136.2~\km\ps\ show some spatial correlations with the SNRC, of which broad line candidates are also identified in the SNRC.
For the +10.5~\km\ps\ component, there are shell-like molecular filaments roughly surrounding the southwest and north of the SNRC. Molecular gas of the +136.2~\km\ps\ component is distributed around the northern radio continuum shell of the SNRC, correlated not so well.
Based on these evidences, the SNRC is probably associated with the +10.5~\km\ps\ component, or may be associated with the +136.2~\km\ps\ component.
Kinematic distances of the +10.5 and +136.2~\km\ps\ components are estimated as 1.5$\pm0.1$ and 8.1$\pm1.9$ kpc, respectively.\\
{\bf G8.04+0.57}:
Molecular gas at the peak of the SCC at $\sim$+20.3~\km\ps\ distributes around the southeast of this SNRC, where its radio continuum emission is bright.
It supports the association between the SNRC and the +12.3~\km\ps\ component that is the only component with broad line candidates identified in the SNRC by full criteria.
If the SNRC is associated with the +12.3~\km\ps\ component, its kinematic distance can be estimated as 1.6$\pm0.1$ kpc.\\
{\bf G8.86-0.26}: 
The south side of this SNRC is adjacent to an \HII\ region at +31.0~\km\ps, and it also overlaps with large SNR W30.
%Since this SNRC is small, with the radius less than two times beam size of our observation, no good spatial correlation result is obtained.
The SCC peaks at $\sim$+20.0~\km\ps, where molecular gas is distributed around the SNRC. %but not well correlated.
It belongs to the +18.4~\km\ps\ component, which is with broad line candidates identified by full criteria.
If the SNRC is associated with the +18.4~\km\ps\ component, its kinematic distance can be estimated as 3.7$\pm0.3$ kpc.
%Broad line candidates are identified at velocities of +1.3, +18.4, and +35.7~\km\ps. We cannot determine if the SNRC is associated with any of these velocity components.
\\
{\bf G11.55+0.33}:
Broad lines at three velocities, i.e.\ $-$3.0, +11.8, and +29.9~\km\ps, are identified in this SNRC.
As indicated by the SCC, molecular gas at $\sim$+18.6~\km\ps\ belonging to the +11.8~\km\ps\ broad line component shows some spatial correlations with the SNRC, which is roughly surrounding the southwest of it and also distributed around its northern boundary.
There is also molecular gas at $\sim$+31.9~\km\ps\ distributed around the western half of the SNRC, which belongs to the +29.9~\km\ps\ broad line component.
No good spatial correlation is found between the SNRC and the $-$3.0~\km\ps\ component.
Based on these evidences, the SNRC is probably associated with the +11.8~\km\ps\ component, or may be associated with the +29.9~\km\ps\ component.
Kinematic distances are estimated as 1.25$\pm0.05$ kpc for the +11.8~\km\ps\ component and 2.9$\pm0.3$ kpc for the +29.9~\km\ps\ component.\\
{\bf G13.10-0.50}:
The southeast of this SNRC overlaps with SNR G13.3-1.3. %, and its northwest overlaps with a small \HII\ region at +31~\km\ps.
There are several broad line components identified in the SNRC.
Nevertheless, there is only one broad line at both +53.5 and +56.5~\km\ps, which are probably originated from the nearby +50.6~\km\ps\ component. %And, most of broad line points of the +50.6~\km\ps\ component are distributed in the radio continuum bright region of the SNR.
%When using full criteria plus the clean subbackground region condition, only one broad line component at +38.4~\km\ps\ is identified in the SNR.}
Molecular gas at a peak of the SCC at $\sim$+53.3~\km\ps\ is surrounding the bright radio continuum shell of the SNRC in the north, which belongs to the +50.6~\km\ps\ broad line component.
Both the kinematic evidence and the spatial correlation result indicate that the SNRC is associated with the +50.6~\km\ps\ component. The corresponding kinematic distance is estimated as 3.9$\pm0.3$ kpc.
At the peak of the SCC at $\sim$+18.4~\km\ps, there are molecular clumps distributed around the eastern and southern boundaries of the SNRC. %, which is not spatially correlated with the radio continuum emission of the SNRC. 
It belongs to the +19.9~\km\ps\ velocity component, with only one broad line identified in the south. The +19.9~\km\ps\ component and the SNRC do not seem to be associated.
We find no spatial correlation between the SNRC and other broad line components.\\
{\bf G13.50+0.07}: %F
This SNRC is a plerion type. We find no good spatial correlation result for the SNRC, because of its small size. We also identify no broad line in it by full criteria. Based on our observation, we cannot determine if the SNRC is associated with an MC.\\
{\bf G13.55+0.35, G13.63+0.30, and G13.65+0.26}:
SNRCs G13.55+0.35 and G13.65+0.26 overlap with SNRC G13.63+0.30.
Several broad line components are identified in SNRC G13.55+0.35 by full criteria. As indicated by the SCC, molecular gas at $\sim$+9.4~\km\ps\ that belongs to the +16.9~\km\ps\ broad line component is spatially correlated with the SNRC, which is roughly surrounding the southwestern radio continuum shell of the SNRC. No good spatial correlation is found between the SNRC and other broad line components. Therefore, SNRC G13.55+0.35 is probably associated with the +16.9~\km\ps\ component, with the corresponding kinematic distance estimated as 1.8$\pm0.1$ kpc.
For SNRC G13.63+0.30, only one broad line component at +22.2~\km\ps\ is identified in it. The +22.2~\km\ps\ component is also spatially correlated with the SNRC, as indicated by the SCC, of which molecular gas is roughly surrounding the SNRC. We suggest that SNRC G13.63+0.30 is associated with the +22.2~\km\ps\ component, and its kinematic distance can be estimated as 3.0$\pm0.4$ kpc.
For SNRC G13.65+0.26, no broad line is identified in it by full criteria. As indicated by the SCC, molecular gas at $\sim$+26.2~\km\ps\ is distributed around the southeastern boundary of the SNRC, which belongs to the +25.1~\km\ps\ component with broad line candidates identified by partial criteria. SNRC G13.65+0.26 may be associated with the +25.1~\km\ps\ component, of which the kinematic distance is estimated as 3.0$\pm0.4$ kpc. SNRCs G13.63+0.30 and G13.65+0.26 may be in the same neighborhood, but SNRC G13.55+0.35 is not likely.\\
{\bf G13.66-0.24}:
%The east of this SNRC overlaps with a small \HII\ region at velocities of +8, +36, and +97~\km\ps.
There are multiple broad line components identified in this SNRC. 
Broad lines at +43.5 and +50.1~\km\ps\ can be attributed to the nearby prominent component at +43.2~\km\ps.
Some broad lines at other velocities are probably originated from an overlapped \HII\ region in the east.
As indicated by the SCC, only the +43.2~\km\ps\ broad line component shows some spatial correlations with the SNRC, of which molecular gas is roughly surrounding the SNRC.
The kinematic evidence and the spatial correlation result indicate that the SNRC is associated with the +43.2~\km\ps\ component. The corresponding kinematic distance is estimated as 3.9$\pm0.3$ kpc.\\
{\bf G14.52+0.14}:
%This SNRC overlaps with several \HII\ regions at +26, +30, and $\sim$+38~\km\ps, and its west is also adjacent to SNR G14.3+0.1----not SNR now.
%This SNRC presents shell-like radio continuum emission in its northeast and southern half.
This SNRC presents bright radio continuum emission in the southeast and shell-like radio continuum emission around its southern half.
Several broad line components are identified in this SNRC.
Three of them at velocities of +26.2, +38.2, and +136.0~\km\ps\ show some spatial correlations with the SNRC.
The +26.2~\km\ps\ component is with some molecular clumps distributed around the northeastern boundary of the SNRC. Molecular gas at $\sim$+53.0~\km\ps\ that probably belongs to the +38.2~\km\ps\ component is roughly surrounding the south of the SNRC. Molecular gas of the +136.0~\km\ps\ component is roughly surrounding the southeast of the SNRC.
Any of these three components may be associated with the SNRC. 
The near and far kinematic distances of the +26.2~\km\ps\ component are estimated as 3.0$\pm0.4$ and 13.3$\pm0.3$ kpc, respectively.
Kinematic distances of the +38.2 and +136.0~\km\ps\ components are estimated as 3.9$\pm0.3$ and 7.9$\pm1.9$ kpc, respectively.\\
{\bf G15.51-0.15}:
Only one broad line component at +50.6~\km\ps\ is identified in this SNRC.
%The SCC peaks at $\sim$+58.3~\km\ps, where molecular gas is roughly surrounding the SNRC, which supports the association between the SNRC and the +50.6~\km\ps\ broad line component. %, and it belongs to the broad line component.
At a peak of the SCC at $\sim$+58.3~\km\ps, molecular gas is roughly surrounding the SNRC, which supports the association between the SNRC and the +50.6~\km\ps\ broad line component. %, and it belongs to the broad line component.
Therefore, we suggest that the SNRC is associated with the +50.6~\km\ps\ component with the corresponding kinematic distance estimated as 3.9$\pm0.3$ kpc.\\
{\bf G15.86+0.52, G16.02+0.75, and G16.13+0.69}:
Both SNRCs G15.86+0.52 and G16.02+0.75 overlap with SNRC G16.13+0.69. % that is large. 
%There is also an \HII\ region at +18~\km\ps\ overlaps with the northeast of SNRC G15.86+0.52.
No broad line is identified in SNRC G15.86+0.52 even by partial criteria. Molecular gas at a nonsignificant peak of the SCC at $\sim$+33.2~\km\ps\ is distributed around the eastern boundary of it but not well correlated.
For SNRC G16.02+0.75, two broad line components are identified in it. As indicated by the SCC, molecular gases at $\sim$+15.7 and $\sim$+27.5~\km\ps\ show spatial correlation with it. Molecular gas at $\sim$+15.7~\km\ps\ is surrounding its southern half, which belongs to the +14.6~\km\ps\ component with a broad line candidate identified in the SNRC by partial criteria. Molecular gas at $\sim$+27.5~\km\ps\ is surrounding its east and south, which belongs to the +30.1~\km\ps\ broad line component. The other broad line component at +42.7~\km\ps\ shows no spatial correlation with the SNRC. Therefore, SNRC G16.02+0.75 may be associated with the +14.6 or +30.1~\km\ps\ components, of which corresponding kinematic distances are estimated as 1.4$\pm0.3$ and 1.50$\pm0.04$ kpc. Kinematic distances of both velocity components are consistent, both of which are probably located at the near side of the Sagittarius spiral arm.
SNRC G16.13+0.69 is large, and its radio continuum emission is weak.
Several broad line components are identified in it.
As indicated by the SCC, the +19.9 and +52.6~\km\ps\ broad line components show some spatial correlations with the SNRC.
%As indicated by the SCC, spatial distributions of molecular gases at $\sim$+23.5 and $\sim$48.7 are weakly correlated with the SNRC.
Molecular gas at $\sim$+23.5~\km\ps\ is roughly surrounding the north of the SNRC, and some molecular clumps at $\sim$+50.2~\km\ps\ are distributed around the boundary of the SNRC's southern half.
When using full criteria plus the clean subbackground region condition, two broad line components at +27.6 and +36.4~\km\ps\ are identified in the SNR.
The +27.6~\km\ps\ component has substantial molecular gas distributed in the remnant, which seems not correlated with the remnant. The +36.4~\km\ps\ component has a shell-like structure surrounding the southeastern boundary of the remnant, nevertheless, it also has molecular gas distributed in the SNRC.
Note that some molecular gases of the +52.6~\km\ps\ component seems to originate from the +36.4~\km\ps\ component.
These spatial correlation results are undetermined, since the weak radio continuum emission of the SNRC has no clear spatial structure.
The +19.9, +36.4, or +52.6~\km\ps\ components may be associated with SNRC G16.13+0.69, and their kinematic distances are estimated as 1.8$\pm0.1$, 3.3$\pm0.9$, and 3.6$\pm0.4$ kpc, respectively.\\
{\bf G16.36-0.18}: %r=1.6beamsiz
%The south of this SNRC overlaps with two \HII\ regions at +49~\km\ps.
We identify no broad line in this SNRC by full criteria.
We also find no good spatial correlation result for the SNRC, because of its small size.\\
{\bf G16.96-0.93}:
%The southwest of this SNRC overlaps with a large \HII\ region at +20~\km\ps.
Broad lines are identified in this SNRC at +40.3 and +45.0~\km\ps. Broad lines at +40.3~\km\ps\ are probably originated from the nearby +45.0~\km\ps\ component.
Molecular gas at a peak of the SCC at $\sim$+28.6~\km\ps\ is distributed along the southern radio continuum shell of the SNRC, which probably originated from the +45.0~\km\ps\ component.
Based on these evidences, we suggest that the SNRC is associated with the +45.0~\km\ps\ component, and its kinematic distance can be estimated as 3.3$\pm0.3$ kpc.\\
{\bf G17.34-0.14}: %r=1.06beamsiz
%This SNRC overlaps with an \HII\ region at $-$6~\km\ps. 
No broad line is identified in this SNRC by full criteria. Only one broad line candidate at $-$8.7~\km\ps\ is identified in it by partial criteria, which is probably originated from an overlapped \HII\ region.
We also find no good spatial correlation result for the SNRC because of its small size.\\
{\bf G17.43+0.27}:
No broad line is identified in this SNRC by full criteria. The SCC peaks at $\sim$+21.7~\km\ps, where molecular gas is roughly surrounding the northeast of the SNRC; however, no corresponding broad line candidate is identified even by partial criteria.
%However, it is not a certain spatial correlation, since the SNRC is small.
\\
{\bf G17.59+0.24}: %r=1.8beamsiz
No broad line is identified in this SNRC by full criteria. The SNRC is a bit small, nevertheless, possible spatial correlations are found between it and molecular gas at $\sim$+21.3 and $\sim$+46.2~\km\ps, both of which are roughly surrounding the south and east of the SNRC. They belong to the +24.4 and +46.5~\km\ps\ components, respectively, which are with broad line candidates identified in the SNRC by partial criteria. The SNRC may be associated with the +24.4 or +46.5~\km\ps\ components, of which kinematic distances are estimated as 1.50$\pm0.05$ and 3.2$\pm0.3$ kpc, respectively.\\
{\bf G17.62+0.09}:
%The south and west of this SNRC overlap with several \HII\ regions at +16 and +18~\km\ps.
This SNRC presents shell-like radio continuum emission in the west, and it overlaps multiple \HII\ regions around +17~\km\ps.
Several broad line components are identified in the SNRC. As indicated by the SCC, only the +16.9~\km\ps\ broad line component is spatially correlated with the radio continuum shell of the SNRC, of which molecular gas at $\sim$+20.6~\km\ps\ is roughly surrounding the west half of it.
These evidences indicate that the SNRC is associated with the +16.9~\km\ps\ component, hence, associated with overlapped \HII\ regions.
The corresponding near and far kinematic distances are estimated as 1.5$\pm0.1$ and 13.3$\pm0.4$ kpc, respectively.\\
{\bf G17.80-0.02}:
Several broad line components are identified in this SNRC. As indicated by the SCC, molecular gas at $\sim$+35.2~\km\ps, belonging to the +47.0~\km\ps\ broad line component, presents a cavity-like structure around the SNRC. No spatial correlation is found for other broad line components.
These evidences indicate that the SNRC is associated with the +47.0~\km\ps\ component, and the corresponding near and far kinematic distances are estimated as 4.0$\pm0.4$ and 12.0$\pm0.3$ kpc, respectively.\\
{\bf G18.39-0.82 and G18.53-0.86}:
These two SNRC overlap each other.
%Two broad line components at nearby velocities are identified in SNRC G18.39-0.82, both of which probably belong to the prominent one at +48.8~\km\ps. 
Broad lines are identified in SNRC G18.39-0.82 at +48.8 and +51.6~\km\ps. 
Broad lines at +51.6~\km\ps\ can be attributed to the nearby component at +48.8~\km\ps. 
Molecular gas at a minor peak of the SCC at $\sim$+61.7~\km\ps\ is surrounding the northern half of the SNRC, which belongs to the +48.8~\km\ps\ broad line component.
Based on these evidences, SNRC G18.39-0.82 is probably associated with the +48.8~\km\ps\ component, with the corresponding kinematic distance estimated as 3.3$\pm0.3$ kpc.
SNRC G18.53-0.86 presents bright shell-like radio continuum emission in the southeast.
Three broad line components are identified in this SNRC. However, broad line points at +48.8 and +51.6~\km\ps\ are in the overlapping region with SNRC G18.39-0.82, all of which are probably originated from the +48.8~\km\ps\ component.
As indicated by the SCC, molecular gas at $\sim$+39.7~\km\ps\ is distributed around the SNRC, but not well correlated with its radio continuum shell. It belongs to the +48.8~\km\ps\ broad line component.
Moreover, molecular gas at $\sim$+71.7~\km\ps, belonging to the +65.6~\km\ps\ broad line component, is surrounding the northeast of the SNRC.
SNRC G18.53-0.86 may be associated with either the +48.8 or +65.6~\km\ps\ components. Kinematic distances are estimated as 3.3$\pm0.3$ kpc for the +48.8~\km\ps\ component and 3.9$\pm0.4$ kpc for the +65.6~\km\ps\ component.\\
{\bf G19.1-3.1}:
Only one broad line component at +6.0~\km\ps\ is identified in this SNRC. Molecular gas at the peak of the SCC at $\sim$+8.3~\km\ps\ is surrounding the eastern half of the SNRC, which belongs to the +6.0~\km\ps\ component.
These evidences indicate that the SNRC is associated with the +6.0~\km\ps\ component, and the corresponding kinematic distance is estimated as 0.25$\pm0.14$ kpc.\\
{\bf G19.48-0.11}:
%The east and southeast of this SNRC overlap with several \HII\ regions at +32, +39, +54, and +58~\km\ps.
We identify several broad line components in this SNRC even by full criteria plus the clean subbackground region condition.
%The SCC peaks at $\sim$+72.9~\km\ps, which is not significant. 
Molecular gas at a nonsignificant peak of the SCC at $\sim$+72.9~\km\ps\ is roughly surrounding the SNRC, which belongs to the +66.6~\km\ps\ broad line component. Moreover, molecular gas at $\sim$+32.1~\km\ps\ is also distributed around the boundary of the SNRC, which belongs to the +26.1~\km\ps\ broad line component.
%The SNRC may be associated with the +66.6~\km\ps\ component with the corresponding kinematic distance estimated as 4.1$\pm0.5$ kpc.\\
The SNRC may be associated with either the +26.1 or +66.6~\km\ps\ components. The near and far kinematic distances of the +26.1~\km\ps\ component are estimated as 1.5$\pm0.05$ and 11.8$\pm0.3$ kpc, respectively. The kinematic distance of the +66.6~\km\ps\ component is estimated as 4.1$\pm0.5$ kpc.\\
{\bf G19.7-0.7}:
Two broad line components at +44.0 and +64.6~\km\ps\ are identified in this SNRC.
As indicated by the SCC, molecular gas at $\sim$+67.1~\km\ps\ is distributed around the SNRC but not correlated well. No good spatial correlation is found between the SNRC and the +44.0~\km\ps\ broad line component.
These evidences indicate a possible association between the SNRC and the +64.6~\km\ps\ component, of which the kinematic distance is estimated as 4.1$\pm0.4$ kpc.\\
{\bf G19.75+0.20}:
%The north of this SNRC overlaps with two \HII\ regions at $-$9 and +19~\km\ps.
Several broad line components are identified in this SNRC.
A broad line at +108.9~\km\ps\ can be attributed to the nearby prominent component at +123.5~\km\ps.
%Molecular gas at the peak of the SCC at $\sim$+125.4~\km\ps\ is roughly surrounding the southwest of the SNRC, which belongs to the +123.5~\km\ps\ broad line component. 
Molecular gas at the peak of the SCC at $\sim$+109.8~\km\ps\ is roughly surrounding the southeast of the SNRC. In addition, molecular gas at $\sim$+125.4~\km\ps\ is roughly surrounding the southwest of the SNRC. These molecular gases belong to the +123.5~\km\ps\ broad line component.
The SNRC may be associated with the +123.5~\km\ps\ component with the corresponding kinematic distance estimated as 9.4$\pm0.4$ kpc.\\
{\bf G19.96-0.33}:
The north of this SNRC overlaps with SNR G20.0-0.2, and the east of it overlaps with an \HII\ region at +67.3~\km\ps.
Three broad line components are identified in the SNRC. Only one broad line is identified at +57.5~\km\ps, which is probably originated from the nearby prominent component at +65.8~\km\ps. The SCC indicates a spatial correlation between the SNRC and the +65.8~\km\ps\ broad line component, of which molecular gas at $\sim$+58.3~\km\ps\ is roughly surrounding the SNRC.
No spatial correlation is found between the SNRC and the +28.9~\km\ps\ broad line component.
Therefore, we suggest that the SNRC is associated with the +65.8~\km\ps\ MC, and the kinematic distance can be estimated as 4.0$\pm0.6$ kpc.
SNRC G19.96-0.33, SNR G20.0-0.2, and the overlapped \HII\ region are probably in the same neighborhood.\\
{\bf G20.20+0.03}:
Three broad line components are identified in this SNRC. However, only one broad line is identified at +71.1~\km\ps, which can be attributed to the nearby prominent component at +77.5~\km\ps. As indicated by the SCC, the +77.5~\km\ps\ broad line component is spatially correlated with the SNRC, of which molecular gas at $\sim$+64.8~\km\ps\ is roughly surrounding the whole SNRC except for its southeast side.
The +25.6~\km\ps\ broad line component shows no spatial correlation with the SNRC.
These evidences indicate that the SNRC is associated with the +77.5~\km\ps\ component, and its kinematic distance can be estimated as 4.4$\pm0.5$ kpc.\\
{\bf G20.26-0.86}:
Among three broad line components identified in this SNRC, only that at +40.0~\km\ps\ shows some spatial correlations with the SNRC. Molecular gas at the peak of the SCC at $\sim$+48.7~\km\ps\ is distributed around the SNRC.
Therefore, we suggest that the SNRC is associated with the +40.0~\km\ps\ component with the corresponding kinematic distance estimated as 3.3$\pm0.3$ kpc.\\
{\bf G20.30-0.06}:
Two broad line components are identified in this SNRC.
No significant spatial correlation result is found. Molecular gas of the +50.5~\km\ps\ broad line component is distributed around the northeastern and southwestern boundaries of the SNRC, and molecular gas of the +73.2~\km\ps\ broad line component is mostly distributed in the eastern half of it.
The SNRC may be associated with either the +50.5 or +73.2~\km\ps\ components.
Kinematic distances are estimated as 4.2$\pm0.5$ kpc for the +50.5~\km\ps\ component and 4.4$\pm0.4$ kpc for the +73.2~\km\ps\ component. Both components are probably located at the near side of the Norma spiral arm.\\
{\bf G21.49-0.01 and G21.60-0.18}:
A small part of SNRC G21.49-0.01 overlaps with the northwest of SNRC G21.60-0.18.
%SNRC G21.60-0.18 overlaps with several \HII\ regions at $-$4, +20, +72, and +120~\km\ps, and also overlaps with SNR G21.5-0.1.  %This SNRC overlaps with SNR G21.5-0.1 (not SNR now).
Several broad line components are identified in SNRC G21.49-0.01. As indicated by the SCC, the +67.2 and +91.2~\km\ps\ broad line components show some spatial correlations with the SNRC. Molecular gas at $\sim$+72.1~\km\ps\ belonging to the +67.2~\km\ps\ component is surrounding the northwest and southeast of the SNRC, and molecular gases at $\sim$+91.6 and $\sim$+102.5~\km\ps\ belonging to the +91.2~\km\ps\ component are roughly surrounding the north and southeast of the SNRC, respectively. The +91.2~\km\ps\ component is also with broad lines identified by full criteria plus the clean subbackground region condition.
SNRC G21.49-0.01 may be associated with either the +67.2 or +91.2~\km\ps\ components. The kinematic distance of the +67.2~\km\ps\ component is estimated as 4.7$\pm0.4$ kpc, and the near and far kinematic distances of the +91.2~\km\ps\ component are estimated as 4.6$\pm0.6$ and 8.7$\pm2.3$ kpc, respectively.
For SNRC G21.60-0.18, multiple broad line components are identified in it by full criteria.
Broad lines at +61.3 and +93.3~\km\ps\ are probably originated from the nearby prominent components at +69.7 and +85.0~\km\ps, respectively.
When using full criteria plus the clean subbackground region condition, only one broad line at +69.7~\km\ps\ is identified in the SNRC.
Shell-like radio continuum emission is present in the southeastern half of the SNRC.
%The SCC gives no significant spatial correlation result. 
%Molecular gas of the +6.1~\km\pc\ broad line component is mostly distributed around the northwestern broundary.
Molecular gas of the +45.3~\km\ps\ broad line component is surrounding the southeastern half of the SNRC, which shows good spatial correlation with its radio continuum shell.
Molecular clumps of the +85.0~\km\ps\ broad line component are also distributed around the southeastern boundary of the SNRC.
SNRC G21.60-0.18 may be associated with either the +45.3 or +85.0~\km\ps\ components. Kinematic distances are estimated as 3.3$\pm0.4$ kpc for the +45.3~\km\ps\ component and 4.6$\pm0.5$ kpc for the +85.0~\km\ps\ component.
Note that molecular gases of the +6.1 and +69.7~\km\ps\ broad line components are not spatially correlated with the radio continuum shell of the SNRC; however, some molecular gases of them at minor peaks of the SCC are distributed around the SNRC. The association between SNRC G21.60-0.18 and either of these two components is not totally ruled out. Kinematic distances of the +6.1 and +69.7~\km\ps\ components are estimated as 15.0$\pm0.4$ and 4.6$\pm0.5$ kpc, respectively.\\
{\bf G21.68+0.13}:
%The south of this SNRC overlaps with an \HII\ region at +21 and +96~\km\ps.
Several broad line components are identified in this SNRC.
As indicated by the SCC, molecular gas at $\sim$+76.0~\km\ps\ is surrounding the south of the SNRC and is also distributed around its northern boundary, which belongs to the +78.2~\km\ps\ broad line component.
The +113.7~\km\ps\ broad line component also shows some spatial correlations with the SNRC, with molecular gas at $\sim$+108.6~\km\ps\ roughly surrounding its east and molecular gas at $\sim$+112.1~\km\ps\ distributed around its northern boundary.
The SNRC may be associated with either the +78.2 or +113.7~\km\ps\ components. Their kinematic distances are estimated as 4.7$\pm0.4$ and 8.0$\pm1.3$ kpc, respectively.\\
{\bf G21.8-3.0}:
No broad line is identified in this SNRC by full criteria.
The SCC indicates that molecular gas at $\sim$+6.7~\km\ps\ is distributed around the SNRC, mostly around its western half, which belongs to the +3.2~\km\ps\ component with broad lines identified in the SNRC by partial criteria.
The SNRC may be associated with the +3.2~\km\ps\ component with the corresponding kinematic distance estimated as 0.24$\pm0.13$ kpc.\\
{\bf G21.86+0.17}:
Several broad line components are identified in this SNRC.
As indicated by the SCC, molecular gas of the +47.8~\km\ps\ broad line component is roughly surrounding the east and south of the SNRC, where a partial radio continuum shell is present.
Therefore, we suggest that the SNRC is associated with the +47.8~\km\ps\ component, of which the kinematic distance can be estimated as 3.3$\pm0.4$ kpc.\\
{\bf G22.05-0.03}:
%This SNRC overlaps with a large \HII\ region at +83~\km\ps, and its southwest is adjacent to SNR G21.9-0.1. The southwest of this SNRC is adjacent to SNR G21.9-0.1 (PWN region).
We identify two broad components in this SNRC at +74.6 and +91.7~\km\ps. 
%However, no good spatial correlation result is found. 
Both broad line components have molecular gases roughly distributed around the SNRC.
The SNRC may be associated with the +74.6 or +91.7~\km\ps\ components, of which kinematic distances are estimated 4.7$\pm0.4$ and 4.7$\pm0.5$ kpc, respectively.
Both of the two components are probably located at the near side of the Norma spiral arm. \\
{\bf G22.18+0.31 and G22.32+0.11}:
These two SNRCs are adjacent to each other.
%The southwest of SNRC G22.18+0.31 overlaps with several \HII\ regions at $\sim$+50~\km\ps. %The southeast of SNRC G22.32+0.11 overlaps with two \HII\ regions at +82 and +87~\km\ps.
SNRC G22.18+0.31 is relatively large, and its radio continuum emission is weak.
We identify multiple broad line components in it by full criteria. Broad lines at +13.3~\km\ps\ are probably originated from the nearby prominent component at +7.0~\km\ps. 
%Based on the SCC, no siginificant spatial correlation result is found. 
Three broad line components, i.e.\ at +7.0, +54.6, and +85.2~\km\ps, are with molecular gases roughly surrounding parts of the SNRC.
The +85.2~\km\ps\ component is also with one broad line identified in the SNRC by full criteria plus the clean subbackground region condition.
Any of these three components may be associated with SNRC G22.18+0.31.
%We cannot finally determine which component is associated with the SNRC. 
Kinematic distances are estimated as 15.0$\pm0.4$, 3.3$\pm0.4$, and 4.7$\pm0.5$ kpc, respectively.
For SNRC G22.32+0.11, we identify multiple broad line components in it by full criteria too.
Only one broad line is identified at +55.1~\km\ps, which is probably originated from the nearby +50.3~\km\ps\ component.
The +50.3~\km\ps\ component also has a broad line identified in the SNRC by full criteria plus the clean subbackground region condition.
As indicated by the SCC, both the +15.8 and +50.3~\km\ps\ components are with molecular gases roughly surrounding parts of the SNRC.
SNRC G22.32+0.11 may be associated with the +15.8 or +50.3~\km\ps\ components. Their kinematic distances are estimated as 1.5$\pm0.04$ and 3.3$\pm0.4$ kpc, respectively.\\
{\bf G22.95-0.31}:
%The middle and southeast of this SNRC overlap with \HII\ regions at +75~\km\ps.
Two broad line components are identified in this SNRC at +31.5 and +75.1~\km\ps.
%Since this SNRC is small, the spatial correlation result is not significant. 
As indicated by the SCC, only the +75.1~\km\ps\ broad line component shows some spatial correlations with the SNRC, of which molecular gas is roughly surrounding the SNRC.
Therefore, we suggest that the SNRC is associated with the +75.1~\km\ps\ component with the kinematic distance estimated as 4.8$\pm0.5$ kpc.\\
{\bf G23.1+0.1}:
There are several \HII\ regions adjacent to this SNRC at velocities of about +78, +90, and +108~\km\ps.
We identify multiple broad line components in the SNRC by full criteria. However, only the +92.7~\km\ps\ component is with broad lines identified in the SNRC by full criteria plus the clean subbackground region condition.
%Velocity ranges of broad lines at +75.1 and +92.7~\km\ps\ are wide, and they can be attributed to prominent component at +92.7~\km\ps. 
Broad lines at +75.1~\km\ps\ have wide velocity ranges, which can be attributed to nearby prominent component at +92.7~\km\ps.
Kinematic distances estimated for the +75.1 and +92.7~\km\ps\ components are consistent.
Some broad lines at +105.6~\km\ps\ are probably originated from the +92.7~\km\ps\ component too.
The SCC indicates a spatial correlation between the SNRC and molecular gas at $\sim$+83.5~\km\ps, which is roughly surrounding radio continuum bright parts of the SNRC, i.e.\ the southeastern half and northwest. It supports the association between the SNRC and the +92.7~\km\ps\ component.
The SNRC is probably associated with the +92.7~\km\ps\ component.
However, the +4.8 and +65.1~\km\ps\ broad line components also show some spatial correlations with the SNRC, of which molecular gases are roughly surrounding the south of it.
It is also possible that the SNRC is associated with the +4.8 or +65.1~\km\ps\ components.
The kinematic distance estimated for the +92.7~\km\ps\ component is 5.1$\pm0.6$ kpc. For the +4.8 and +65.1~\km\ps\ components, kinematic distances are estimated as 15.0$\pm0.4$ and 4.9$\pm0.5$ kpc, respectively.\\
{\bf G23.85-0.18 and G24.0-0.3}:
These two SNRCs overlap each other. SNRC G24.0-0.3 is much larger than SNRC G23.85-0.18.
%There are also multiple \HII\ regions at $\sim$+56, +69, $\sim$+75, $\sim$+89, +98 and +101~\km\ps\ that overlap with SNRC G24.0-0.3.
Two broad line components are identified in SNRC G23.85-0.18 at +44.8 and +57.3~\km\ps. %, and their velocities are close to each other. 
As indicated by the SCC, both show some spatial correlations with the SNRC. Molecular gas of the +44.8~\km\ps\ component is roughly surrounding the north and south of the SNRC, and molecular gas of the +57.3~\km\ps\ component roughly surrounding the east of it.
SNRC G23.85-0.18 may be associated with the +44.8 or +57.3~\km\ps\ components.
Their kinematic distances are estimated as 3.7$\pm0.4$ and 3.5$\pm0.4$ kpc, respectively, which are consistent.
Both of the two components are probably located at the near side of the Scutum spiral arm.
SNRC G24.0-0.3 presents shell-like radio continuum emission in its northern half.
There are multiple broad line components identified in this SNRC by full criteria.
The +55.3~\km\ps\ broad line component presents a bubble-like structure with inward protrusions, which has the same center as the SNRC but has a smaller radius. In addition, the +55.3~\km\ps\ component is also with a broad line identified by full criteria plus the clean subbackground region condition.
Molecular gas at a minor peak of the SCC at $\sim$+110.6~\km\ps\ also shows a spatial correlation with the SNRC, which is roughly surrounding the bright radio continuum shell in the northeast. It belongs to the +111.6~\km\ps\ broad line component.
%We find no good spatial correlation between the radio continuum shell of the SNRC and other broad line components.
We find no good spatial correlation between the SNRC and other broad line components.
Therefore, SNRC G24.0-0.3 may be associated with the +55.3 or +111.6~\km\ps\ components.
Their kinematic distances are estimated as 3.6$\pm0.4$ and 7.1$\pm0.9$ kpc, respectively.\\
{\bf G23.97+0.51}:
Three broad line components are identified in this SNRC at +94.5, +110.1, and +110.3~\km\ps. Only one broad line is identified at +110.3~\km\ps, which can be attributed to the nearby +110.1~\km\ps\ component.
As indicated by the SCC, both the +94.5 and +110.1~\km\ps\ components show some spatial correlations with the SNRC. Molecular gas at $\sim$+99.4~\km\ps\ is distributed around the northern boundary of the SNRC, and molecular gas at $\sim$+106.2~\km\ps\ is roughly surrounding the south of it.
Therefore, the SNRC may be associated with the +94.5 or +110.1~\km\ps\ components.
The near and far kinematic distances of the +94.5~\km\ps\ component are estimated as 5.2$\pm0.7$ and 7.7$\pm2.2$ kpc, respectively. The kinematic distance of the +110.1~\km\ps\ component is estimated as 6.2$\pm1.1$ kpc.\\
{\bf G24.06-0.81}: %F
This SNRC is small, and classified as the plerion type.
We find no good spatial correlation result, and identify no broad line in it by full criteria.\\
{\bf G24.19+0.28}:
%The south of this SNR overlaps with several \HII\ regions at +82, +109 and +114~\km\ps.
We identify two broad line components in this SNRC at +50.8 and +114.1~\km\ps.
As indicated by the SCC, molecular gas at $\sim$+115.9~\km\ps\ belonging to the +114.1~\km\ps\ component is roughly surrounding the south of the SNRC.
No good spatial correlation is found between the SNRC and the +50.8~\km\ps\ component.
Therefore, we suggest that the SNRC is associated with the +114.1~\km\ps\ component, and its kinematic distance can be estimated as 6.9$\pm1.1$ kpc.\\
{\bf G25.49+0.01}:
%This SNRC overlaps with several \HII\ regions at $\sim$$-$13, $\sim$+59, +96, +107~\km\ps.
Multiple broad line components are identified in this SNRC.
As indicated by the SCC, molecular gases at $\sim$+55.1 and $\sim$+94.0~\km\ps\ are distributed around the SNRC, which belong to the +54.3 and +94.5~\km\ps\ broad line components, respectively. 
%These spatial correlations are not crucial.
These evidences indicate that the SNRC may be associated with the +54.3 or +94.5~\km\ps\ components.
The near and far kinematic distances of the +54.3~\km\ps\ component are estimated as 3.8$\pm0.5$ and 11.1$\pm0.3$ kpc, respectively. 
For the +94.5~\km\ps\ component, the near and far kinematic distances are estimated as 5.7$\pm0.8$ and 8.9$\pm0.4$ kpc, respectively.\\
{\bf G26.04-0.42}:
Several broad line components are identified in this SNRC.
%Broad line components at +93.5, +104.6, and +112.7~\km\ps\ have wide velocity ranges and are close to each other, which are probably originated from the same prominent component at +104.6~\km\ps. 
Broad lines at +93.5 and +112.7~\km\ps\ are probably originated from the nearby prominent component at +104.6~\km\ps.
Their kinematic distance estimation results are also consistent, all at the near side of the Norma spiral arm.
%As indicated by minor peaks of the SCC, 
Only the +104.6~\km\ps\ component shows some spatial correlations with the SNRC, with molecular gases distributed around the northern boundary of it.
Therefore, we suggest that the SNRC is associated with the +104.6~\km\ps\ component, accordingly, its kinematic distance can be estimated as 5.6$\pm0.6$ kpc.\\
{\bf G26.13+0.13}:
%Several \HII\ regions are located around this SNRC, which are at velocities of $\sim$+28, +67, +88, +98, +104, and +111~\km\ps.
This SNRC overlaps with multiple \HII\ regions, and many broad line components are identified in it by full criteria.
When using full criteria plus the clean subbackground region condition, two broad lines at +70.7 and +97.1~\km\ps\ are identified in the SNRC by full criteria plus the clean subbackground region condition. Molecular gas at $\sim$+102.2~\km\ps\ belonging to the +97.1~\km\ps\ component presents a shell-like structure surrounding the southeastern radio continuum shell of the SNRC, which also has some inward protrusions. Based on these evidences, we suggest that the SNRC is associated with the +97.1~\km\ps\ component, of which the near and far kinematic distances are estimated as 5.8$\pm0.6$ and 8.8$\pm0.4$ kpc, respectively.
\\
%However, we find no good spatial correlation between the SNRC and these broad line components. We cannot determine which component is associated with the SNRC.\\
{\bf G26.53+0.07}:
%This SNRC overlaps with several \HII\ regions at +22, +69, +98, +102, and +112~\km\ps, and it also overlaps with SNR G26.6-0.1.
We consider SNRC G26.6-0.1 as a southeastern part of SNRC G26.53+0.07. This SNRC presents bright shell-like radio continuum emission in its eastern half.
%34.5 +100.8~\km\ps\ MCs may be associated with the SNRC.
We identify many broad lines in the SNRC by full criteria.
Only one broad line is identified at +113.2~\km\ps, which can be attributed to the nearby +103.1~\km\ps\ component.
When using full criteria plus the clean subbackground region condition, the +34.5 and +103.1~\km\ps\ components each have a broad line identified in the SNRC.
The SCC indicates a spatial correlation between the SNRC and the +103.1~\km\ps\ component, of which molecular gas is distributed around the SNRC.
The +34.5~\km\ps\ broad line component also shows a spatial correlation with the SNRC, with molecular gas roughly surrounding the northeast of it and a molecular clump located in the southeast of the SNRC.
No spatial correlation is found for other broad line components.
%Based on these evidences, the SNRC is probably associated with the +100.8~\km\ps\ component, or may be associated with the +34.5~\km\ps\ component.
Based on these evidences, the SNRC is probably associated with the +103.1~\km\ps\ component, or may be associated with the +34.5~\km\ps\ component.
%Kinematic distances are estimated as 8.7$\pm0.4$ kpc for the +100.8~\km\ps\ component and 12.5$\pm0.3$ for the +34.5~\km\ps\ component.\\
Kinematic distances are estimated as 6.1$\pm0.6$ and 8.7$\pm0.4$ kpc for the +103.1~\km\ps\ component, and 12.5$\pm0.3$ for the +34.5~\km\ps\ component.\\
{\bf G26.75+0.73}:
Radio continuum emission of this SNRC is bright in the north.
We identify only one broad line component in the SNRC at +86.5~\km\ps.
As indicated by the SCC, molecular gas at $\sim$+89.2~\km\ps\ is distributed around the southern boundary and a small section of the northern boundary.
It supports the association between the SNRC and the +86.5~\km\ps\ broad line component.
Note that molecular gas at the peak of the SCC at $\sim$+46.7~\km\ps\ is roughly surrounding the north of the SNRC; however, no corresponding broad line is identified even by partial criteria.
Based on these evidences, we suggest that the SNRC is associated with the +86.5~\km\ps\ component with the near and far kinematic distances estimated as 4.5$\pm0.5$ and 7.9$\pm2.4$ kpc, respectively.\\
{\bf G27.06+0.04}:
%This SNRC overlaps with \HII\ regions at +28, +69, +81, +85, +90, and +93~\km\ps.
This SNRC presents a bright radio continuum shell in the east, and some weak radio continuum emission in the west.
%The radio continuum emission of this SNRC is bright and shell-like in the east. There seems also some weak radio continuum emission in the west.
Multiple broad line components are identified in the SNRC. 
As indicated by the SCC, the +63.3~\km\ps\ component is spatially correlated with the SNRC well.
Molecular gas at $\sim$+66.7~\km\ps\ is surrounding the eastern radio continuum shell and distributed around the southern and western boundaries.
Molecular gas at $\sim$+61.0~\km\ps\ is distributed along the southwestern boundary of the SNRC, and there is also a half cavity-like structure around the east of the SNRC in large scale.
No good spatial correlation result is found for other broad line components.
%Therefore, we suggest that the SNRC is associated with the +63.3~\km\ps\ component, hence, we can estimate its kinematic distance as 4.2$\pm0.3$ kpc.\\
Therefore, we suggest that the SNRC is associated with the +63.3~\km\ps\ component. 
\cite{JohansonKerton2009} showed \HI\ absorption up to the tangent point, indicating that the SNR is at the far side of the tangent point.
The far kinematic distance of the +63.3~\km\ps\ component is estimated as 10.9$\pm0.4$ kpc.\\
{\bf G27.18+0.30}: %r=1.06*beamsiz
This SNRC is small, and has diffuse and elongated radio continuum emission. 
Only one broad line is identified in the SNRC at +73.1~\km\ps. 
As indicated by the SCC, the +73.1~\km\ps\ component shows a minor spatial correlation with the SNRC, of which molecular gas is roughly surrounding the east of it.
Therefore, the SNRC is probably associated with the +73.1~\km\ps\ component, of which the kinematic distance is estimated as 4.3$\pm0.3$ kpc.\\
{\bf G27.78-0.33}:
%The north of this SNRC overlaps with an \HII\ region at +46~\km\ps.
This SNRC presents a bright radio continuum shell in the northeast.
Two broad line components are identified in this SNRC at +45.8 and +88.5~\km\ps.
The +45.8~\km\ps\ component shows good spatial correlation with the SNRC.
Molecular gas at $\sim$+42.7~\km\ps\ is surrounding the northeastern radio continuum shell of the SNRC, and molecular gas at $\sim$+45.1~\km\ps\ is roughly surrounding the whole SNRC.
These evidences strongly support the association between the SNRC and the +45.8~\km\ps\ component.
Nevertheless, the +88.5~\km\ps\ component also shows minor spatial correlation with the SNRC, of which molecular gas at $\sim$+93.8~\km\ps\ is roughly surrounding the northeastern radio continuum shell of it.
It is also possible that the SNRC is associated with the +88.5~\km\ps\ component.
The kinematic distance of the +45.8~\km\ps\ component is estimated as 4.1$\pm0.7$ kpc, and 4.4$\pm0.4$ kpc for the +88.5~\km\ps\ component. With consistent distances estimated, both velocity components are probably located at the near side of the Scutum spiral arm.\\
{\bf G28.21+0.02}: %F
%The east of this SNRC overlaps with an \HII\ region at +106~\km\ps.
This SNRC is classified as the plerion type. 
The SCC indicates a good spatial correlation between the SNRC and molecular gas at $\sim$+101.0~\km\ps, which is shell-like and surrounding the SNRC.
%and presents shell-like structure outside in the east. 
It belongs to the +98.8~\km\ps\ component with broad lines identified in the SNRC by full criteria.
These evidences indicate that the SNRC is associated with the +98.8~\km\ps\ component with the kinematic distance estimated as 8.2$\pm0.5$ kpc.\\
{\bf G28.22-0.09}:
%The northwest of this SNRC is adjacent to an \HII\ region at +96~\km\ps.
%Radio continuum morphology of this SNRC is center-filled, and it is considered as a possible plerion type SNRC.
This SNRC is a plerion type but uncertain.
Since the SNRC is small, we find no significant spatial correlation result, and no broad line is identified in it by full criteria.
Note that molecular gas at the peak of the SCC at $\sim$+82.5~\km\ps\ is roughly surrounding the southeast of the SNRC.\\
{\bf G28.3+0.2 and G28.33+0.06}: %G28.33+0.06 F
%The southwest of SNRC G28.3+0.2 overlaps with an \HII\ region at +7~\km\ps.
These two SNRCs are adjacent to each other. 
SNRC G28.3+0.2 presents shell-like radio continuum emission in the southeast.
Several broad line components are identified in this SNRC. We find that the SNRC is spatially correlated with the +47.9~\km\ps\ broad line component, of which molecular gas is distributed along the radio continuum shell.
The SCC also indicates that molecular gases at $\sim$+73.7 and $\sim$+106.0~\km\ps\ are roughly surrounding the south of the SNRC, which belong to the +80.0 and +97.8~\km\ps\ broad line components, respectively.
SNRC G28.3+0.2 is probably associated with the +47.9~\km\ps\ component; however, it may be associated with either the +80.0 or +97.8~\km\ps\ components too.
Kinematic distances are estimated as 10.8$\pm0.5$ kpc for the +47.9~\km\ps\ component, 4.4$\pm0.4$ for the +80.0~\km\ps\ component, and 5.6$\pm0.9$ or 8.2$\pm0.5$ kpc for the +97.8~\km\ps\ component.
SNRC G28.33+0.06 is classified as the plerion type. There are several \HII\ regions located around this SNRC at velocities of about +46, +76, and +81~\km\ps.
%No significant spatial correlation result is found for the SNRC.
Molecular gas at the peak of the SCC at $\sim$+97.3~\km\ps\ is distributed around the SNRC, which belongs to the +100.6~\km\ps\ component with broad lines identified in the SNRC by full criteria.
SNRC G28.33+0.06 may be associated with the +100.6~\km\ps\ component with the near and kinematic distances estimated as 5.8$\pm0.9$ and 8.2$\pm0.5$ kpc, respectively.\\
{\bf G28.56+0.00}:
This SNRC was considered to be related to a nearby \HII\ region at +96.6~\km\ps, based on their very similar \HI\ absorption spectra \citep{Ranasinghe+2020}.
The east of the SNRC is adjacent to an \HII\ region at +43~\km\ps.
Two broad line components are identified in this SNRC at +43.3 and +75.7~\km\ps. 
As indicated by the SCC, both show some spatial correlations with the SNRC. Molecular gas at $\sim$+34.8~\km\ps\ is distributed around the northern boundary of the SNRC, and molecular gas at $\sim$+70.6~\km\ps\ is roughly surrounding the south of the SNRC.
Since the SNRC is small, the spatial correlation result is indefinite.
The SNRC may be associated with the +43.3 or +75.7~\km\ps\ components.
%Kinematic distances are estimated as 10.8$\pm0.4$ and 4.3$\pm0.3$ kpc for these two components, respectively.\\
The presence of \HI\ absorption up to the tangent point shown by \cite{Ranasinghe+2020} indicates that the SNR is at a far kinematic distance.
The kinematic distance of the +43.3~\km\ps\ component is estimated as 10.8$\pm0.4$ kpc, and the far kinematic distance of the +75.7~\km\ps\ component is estimated as 8.2$\pm0.6$ kpc.\\
{\bf G28.52+0.27 and G28.64+0.20}:
These two SNRCs overlap each other.
Several broad line components are identified in SNRC G28.52+0.27. As indicated by the SCC, two broad line components at +82.2 and +100.3~\km\ps\ show some spatial correlations with the SNRC. Molecular gas at $\sim$+72.7~\km\ps\ is roughly distributed around the SNRC, and molecular gas at $\sim$+98.1~\km\ps\ is roughly surrounding the south of the SNRC.
SNRC G28.52+0.27 may be associated with the +82.2 or +100.3~\km\ps\ components. Kinematic distances of these two components are estimated as 4.4$\pm0.4$ and 5.5$\pm0.9$ kpc, respectively.
SNRC G28.64+0.20 is larger, and several broad line components are identified in it too. %We find no significant spatial correlation result for it.
When using full criteria plus the clean subbackground region condition, broad lines are identified at +81.2 and +104.4~\km\ps\ in the SNRC.
Three broad line components at +52.1, +81.2, and +104.4~\km\ps\ show roughly spatial correlations with the SNRC, of which molecular gases are distributed around the SNRC. 
SNRC G28.64+0.20 may be associated with any of these three components. 
%Kinematic distances are estimated as 4.2$\pm0.8$ and 10.8$\pm0.5$ kpc for the +52.1~\km\ps\ component, 4.4$\pm0.4$ kpc for the +81.2~\km\ps\ component, and 5.6$\pm1.0$ or 8.1$\pm0.5$ kpc for the +104.4~\km\ps\ component.\\
As shown by \cite{JohansonKerton2009}, there is \HI\ absorption up to the tangent point, indicating that SNRC G28.64+0.20 is at a far kinematic distance.
Far kinematic distances are estimated as 10.8$\pm0.5$ kpc for the +52.1~\km\ps\ component, 8.1$\pm0.6$ kpc for the +81.2~\km\ps\ component, and 8.1$\pm0.5$ kpc for the +104.4~\km\ps\ component.\\
{\bf G28.7-0.4}:
Several broad line components are identified in this SNRC. Broad lines at +81.9~\km\ps\ are probably originated from the nearby prominent component at +73.9~\km\ps.
As indicated by the SCC, molecular gas of the +73.9~\km\ps\ broad line component is distributed around the SNRC.
These evidences indicate that the SNRC is associated with the +73.9~\km\ps\ component with the corresponding kinematic distance estimated as 4.3$\pm0.3$ kpc.\\
{\bf G28.87+0.62}:
This SNRC is small. We find no good spatial correlation result for it, and identify no broad line in it by full criteria.\\
{\bf G28.88+0.24, G28.88+0.41, and G28.93+0.25}: %G28.88+0.41 classified as F but with shell-like morphology
SNRC G28.93+0.25 is adjacent to the other two SNRCs, and SNRC G28.88+0.41 is also adjacent to SNRC G28.64+0.20.
There are several \HII\ regions at $\sim$+108~\km\ps\ located near the junction between SNRCs G28.88+0.41 and G28.64+0.20.
%SNRC G28.64+0.20 also overlaps with multiple \HII\ regions at $-$13, +39, +87, and $\sim$+103~\km\ps\ elsewhere.
%SNRC G28.88+0.41 and G28.93+0.25 also overlap with \HII\ regions at +21 and +28~\km\ps, respectively.
Since SNRC G28.88+0.24 is small, we find no good spatial correlation result for it, and no broad line is identified in it by full criteria.
SNRC G28.88+0.41 presents shell-like radio continuum emission in the southwest.
Several broad line components are identified in it. Note that only one broad line is identified at +73.7~\km\ps, which can be attributed to the nearby prominent component at +84.3~\km\ps.
As indicated by the SCC, molecular gas at $\sim$+73.8~\km\ps\ is surrounding the southwestern radio continuum shell of the SNRC. It supports the associated between the SNRC and the +84.3~\km\ps\ broad line component. The +53.7 and +96.1~\km\ps\ broad line components also have some molecular gases distributed around the radio continuum shell; however, they are not well spatially correlated with it.
Therefore, we suggest that SNRC G28.88+0.41 is associated with the +84.3~\km\ps\ component with the corresponding kinematic distance estimated as 4.4$\pm0.5$ kpc.
For SNRC G28.93+0.25, we identify only one broad line at +83.2~\km\ps\ in it. 
Molecular gas of the +83.2~\km\ps\ component is mainly distributed in the northwestern half of the SNRC, where its radio continuum emission is bright; however, they are not spatially correlated well. 
If SNRC G28.93+0.25 is associated with the +83.2~\km\ps\ component, its kinematic distance can be estimated as 4.5$\pm0.4$ kpc. SNRCs G28.88+0.41 and G28.93+0.25 may be related to each other and located at the near side of the Scutum spiral arm.\\
{\bf G29.33+0.28}:
Only one broad line is identified in this SNRC at +91.5~\km\ps. The SCC indicates that the broad line component is spatially correlated with the SNRC, with molecular gas at $\sim$+95.7 and $\sim$+100.2~\km\ps\ surrounding the southern half of the SNRC. The southern half of the SNRC is radio continuum bright.
Based on these evidences, we suggest that the SNRC is associated with the +91.5~\km\ps\ component with the kinematic distance estimated as 4.7$\pm0.5$ kpc.\\
{\bf G29.4+0.1}: %may contain PWN
This SNRC presents bright filamentary radio continuum emission in its central region, and there is also a faint radio continuum shell in the southeast.
Several broad line components are identified in this SNRC. 
Only one broad line is identified at +97.8~\km\ps, which can be attributed to the nearby prominent component at +91.0~\km\ps. 
When applying the additional clean subbackground region condition, only the +80.4~\km\ps\ component is with a broad line identified in the SNRC.
%As indicated by the SCC, three broad line components at +36.4, +80.4, and +91.0~\km\ps\ show more or less spatial correlations with the SNRC.
As indicated by the SCC, molecular gas at $\sim$+28.4 and $\sim$+46.0~\km\ps, belonging to the +36.4~\km\ps\ broad line component, is distributed around the eastern and southern boundaries, respectively. 
Molecular gas of the +58.9~\km\ps\ broad line component has an elongated molecular clump distributed along the southeastern boundary of the SNRC, and a molecular filament distributed along the bright filamentary radio continuum emission in the northwest of the SNRC.
Molecular gas at $\sim$+75.2~\km\ps\ is surrounding the southeastern radio continuum shell of the SNRC, which belongs to the +80.4~\km\ps\ broad line component. 
%It is noteworthy that there is also filamentary molecular gas at $\sim$+65~\km\ps\ distributed along the northwestern half of the inner filamentary radio continuum emission, which also belongs to the +80.4~\km\ps\ component.
Molecular gas at $\sim$+96.5~\km\ps, belonging to the +91.0~\km\ps\ component, is roughly distributed around the southeastern and northwestern boundaries of the SNRC.
%Based on these evidences, the SNRC is most likely to be associated with the +80.4~\km\ps\ component; however, it may be associated with either the +36.4 or +91.0~\km\ps\ components too.
Based on these evidences, the SNRC is probably associated with the +58.9 or +80.4~\km\ps\ components; however, it may be associated with the +36.4 or +91.0~\km\ps\ components too.
The presence of \HI\ absorption up to the tangent point shown by \cite{Castelletti+2017, JohansonKerton2009} indicates that the SNR is at a far kinematic distance.
If the SNRC is associated with the +58.9~\km\ps\ component, its kinematic distance can be estimated as 10.6$\pm0.4$ kpc, and if it is associated with the +80.4~\km\ps component, its kinematic distance can be estimated as 7.9$\pm0.6$ kpc.
%If the SNRC is associated with the +58.9~\km\ps\ component, its near and far kinematic distances can be estimated as 4.5$\pm0.8$ and 10.6$\pm0.4$ kpc, respectively, and if it is associated with the +80.4~\km\ps component, its kinematic distance can be estimated as 4.5$\pm0.5$ kpc.
Kinematic distances of the +36.4 and +91.0~\km\ps\ components are estimated as 12.2$\pm0.3$ and 7.9$\pm0.6$ kpc, respectively.
Both the +80.4 and +91.0~\km\ps\ components are probably located at the near side of the Scutum spiral arm.\\
{\bf G29.41-0.18}:
%This SNRC overlaps an \HII\ region at +105~\km\ps.
Multiple broad line components are identified in this SNRC. Broad lines at +59.8 and +93.8~\km\ps\ may be originated from the nearby prominent components at +77.5 and +93.1~\km\ps, respectively.
When applying the additional clean subbackground region condition, only the +77.5~\km\ps\ component is with broad lines identified in the SNRC.
As indicated by the SCC, there are molecular gases of the +77.5 and +93.1~\km\ps\ components roughly distributed around the SNRC. These spatial correlations are not significant.
The SNRC may be associated with the +77.5 or +93.1~\km\ps\ components. Kinematic distances are estimated as 4.4$\pm0.4$ for the +77.5~\km\ps\ component, and 4.7$\pm0.5$ or 7.9$\pm0.6$ for the +93.1~\km\ps\ component.\\
{\bf G29.92+0.21}: %F
This SNRC is classified as the plerion type.
As indicated by the SCC, molecular gas at $\sim$+37.5~\km\ps\ is roughly surrounding the north and east of the SNRC, which belongs to the +38.0~\km\ps\ component with broad lines identified. Molecular gas at $\sim$+97.6~\km\ps\ is surrounding the eastern half of the SNRC, and molecular gas at $\sim$+106.0~\km\ps\ is distributed around the eastern and western boundaries. They belong to the +90.3~\km\ps\ component that is with broad lines identified.
The SNRC is most likely to be associated with the +90.3~\km\ps\ component; however, it may be associated with the +38.0~\km\ps\ component too.
Kinematic distances are estimated as 4.7$\pm0.3$ kpc for the +90.3~\km\ps\ component, and 2.1$\pm0.3$ or 12.0$\pm0.3$ kpc for the +38.0~\km\ps\ component.\\
{\bf G30.30+0.13}: %r=1.2beamsiz, F
This SNRC is classified as the plerion type.
Since the SNRC is small, no significant spatial correlation result is found.
Nevertheless, molecular gas at $\sim$+86.4~\km\ps\ is roughly distributed around the SNRC, which belongs to the +84.8~\km\ps\ component with a broad line identified in the SNRC.
If the SNRC is associated with the +84.8~\km\ps\ component, its kinematic distance can be estimated as 4.9$\pm0.8$ kpc.\\
{\bf G30.36+0.62 and G30.51+0.57}:
These two SNRCs are overlapping each other. %and SNRC G30.51+0.57 also overlaps with an \HII\ region at +15 and +46~\km\ps.
Three broad line components are identified in SNRC G30.36+0.62. Only the +47.0 and +88.7~\km\ps\ broad line components show some spatial correlations with the SNRC, of which molecular gases are mostly distributed around the southeastern boundary of the SNRC. %, not correlated with the northerwestern radio continuum shell of the SNRC.
SNRC G30.36+0.62 may be associated with the +47.0 or +88.7~\km\ps\ components. Kinematic distances of these two components are estimated as 10.6$\pm0.5$ and 5.0$\pm0.8$ kpc, respectively.
For SNRC G30.51+0.57, two broad line components at +72.9 and +88.3~\km\ps\ are identified in it by full criteria.
As indicated by the SCC, molecular gases at $\sim$+69.2 and $\sim$+82.7~\km\ps\ are surrounding the southwest and east of the SNRC, respectively.
%Note that molecular gas at the peak of the SCC at $\sim$+16.8~\km\ps\ is surrounding the southeastern half of the SNRC; however, no corresponding broad line is identified in the SNRC by full criteria.
Molecular gas at the peak of the SCC at $\sim$+16.8~\km\ps\ is surrounding the southeastern half of the SNRC, which belongs to the +16.6~\km\ps\ component with broad line candidates identified in the SNRC only by partial criteria.
%SNRC G30.51+0.57 may be associated with the +72.9 or +88.3~\km\ps\ components. Kinematic distances of them are estimated as 3.5$\pm0.3$ and 4.4$\pm0.5$ kpc, respectively.
SNRC G30.51+0.57 may be associated with the +16.6, +72.9, or +88.3~\km\ps\ components. Kinematic distances of them are estimated as 12.4$\pm0.3$, 3.5$\pm0.3$, and 4.4$\pm0.5$ kpc, respectively.
Both the +72.9 and +88.3~\km\ps\ components are probably located at the near side of the Scutum spiral arm.
\\
{\bf G30.38+0.42}:
%The east of this SNRC overlaps with the same \HII\ region that overlaps with SNRC G30.51+0.57.
This SNRC presents a partial radio continuum shell in the southwest.
Three broad line components are identified in this SNRC. As indicated by the SCC, only the +17.9 and +93.0~\km\ps\ components show roughly spatial correlations with the SNRC. 
Molecular gas at $\sim$+15.4~\km\ps\ is surrounding the eastern half of the SNRC, and molecular gas at $\sim$+89.2~\km\ps\ is distributed around the SNRC.
The SNRC may be associated with the +17.9 or +93.0~\km\ps\ components, and their corresponding kinematic distances are estimated as 12.5$\pm0.3$ and 4.9$\pm0.7$ kpc, respectively.\\
{\bf G31.22-0.02 and G31.26-0.04}:
These two SNRCs overlap each other.
%The southwest of this SNRC G31.22-0.02 is adjacent to an \HII\ region at +41~\km\ps.
Several broad line components are identified in both of the two SNRCs. Since many broad line points are located in the overlapping region between two SNRCs, the identification of broad lines for each SNRC is not very clear.
In SNRC G31.22-0.02, only one broad line is identified at +84.7~\km\ps, which can be attributed to the nearby prominent component at +78.2~\km\ps. Almost all of broad line points in the SNRC are located in the overlapping region between two SNRCs, except for one at +25.6~\km\ps. 
As indicated by the SCC, the +44.3 and +78.2~\km\ps\ broad line components show some spatial correlations with the SNRC. 
At $\sim$+48.4~\km\ps, a molecular bar and a molecular clump are distributed around the eastern and western of the SNRC, respectively.
In addition, molecular gas at $\sim$+34.3~\km\ps\ is distributed around the northeastern and western boundaries of the SNRC. 
Molecular gas at $\sim$+82.5~\km\ps\ is distributed around the SNRC, and molecular gas at $\sim$+91.9~\km\ps\ is surrounding the southeast of the SNRC.
These evidences indicate that SNRC G31.22-0.02 is probably associated with the +78.2~\km\ps\ component, or may be associated with the +44.3~\km\ps\ component.
Kinematic distances are estimated as 4.9$\pm0.8$ kpc for the +78.2~\km\ps\ component and 10.2$\pm0.4$ kpc for the +44.3~\km\ps\ component.
For SNRC G31.26-0.04, as indicated by the SCC, molecular gas at $\sim$+38.3~\km\ps\ belonging to the +44.3~\km\ps\ broad line component is surrounding the whole SNRC except for its southeast side, and molecular gas at $\sim$+95.2~\km\ps\ belonging to the +98.8~\km\ps\ broad line component is roughly surrounding the eastern half of the SNRC. 
%There is also a molecular hub-filament structure at $\sim$+84.1~\km\ps\ located in the southwest, with filaments distribute around the southwestern and southeastern boundaries of the SNRC, which belongs to the +78.7~\km\ps\ component.
There is also some molecular gas at $\sim$+84.1~\km\ps\ distributed around the southwestern and southeastern boundaries of the SNRC, which belongs to the +78.7~\km\ps\ component; however, it seems to be a part of a molecular hub-filament structure located in the southwest.
In addition, most of the broad line points at +78.7~\km\ps\ are located in the overlapping region with SNRC G31.22-0.02.
SNRC G31.26-0.04 is probably associated with the +44.3~\km\ps\ component, or may be associated with the +98.8~\km\ps\ component. We also cannot rule out the possibility of the association with +78.7~\km\ps\ component. Kinematic distances are estimated as 10.2$\pm0.4$ kpc for the +44.3~\km\ps\ component, 4.9$\pm0.8$ kpc for the +78.7~\km\ps\ component, and 7.1$\pm0.8$ kpc for the +98.8~\km\ps\ component.
Based on these results, these two SNRCs are probably unrelated.\\
%{\bf G31.30-0.49}: r=0.21'
{\bf G31.44+0.36}:
%The southwest of this SNRC partly overlaps with an \HII\ region at +100~\km\ps.
Broad lines are identified in this SNRC at four velocities. However, there is only one broad line at +97.5~\km\ps, and it can be attributed to the nearby prominent component at +96.0~\km\ps.
All three broad line components at +23.2, +74.6, and +96.0~\km\ps\ show more or less spatial correlations with the SNRC, as indicated by the SCC.
Molecular gas at $\sim$22.7 is roughly surrounding the whole SNRC except for its southwest side, and molecular gas at $\sim$+77.6~\km\ps\ is roughly surrounding the south and northwest of the SNRC. Molecular gas at $\sim$+109.5~\km\ps\ is surrounding the SNRC and also has a small amount inside.
The +23.2, +74.6, or +96.0~\km\ps\ components may be associated with the SNRC.
Their kinematic distances are estimated as 12.3$\pm0.4$, 5.0$\pm0.7$, and 5.1$\pm0.6$ kpc, respectively.\\
{\bf G31.93+0.16}:
%The northwest of this SNRC partly overlaps with an \HII\ region at +99~\km\ps.
Two broad line components are identified in this SNRC at +44.7 and +102.1~\km\ps.
As indicated by the SCC, molecular gas at $\sim$+110.2~\km\ps\ is roughly surrounding the northwest of the SNRC, which belongs to the +102.1~\km\ps\ broad line component. 
The +44.7~\km\ps\ component shows no good spatial correlation with the SNRC.
These evidences indicate that the SNRC is probably associated with the +102.1~\km\ps\ component, with the corresponding kinematic distance estimated as 6.8$\pm0.8$ kpc. Accordingly, this SNRC could be related to an overlapped \HII\ region in its northwest.\\
{\bf G32.22-0.21}: %F
%The east of this SNRC partly overlaps with an \HII\ region at +22~\km\ps.
This SNRC is a plerion type. 
Molecular gas at a minor peak of the SCC at $\sim$+91.9~\km\ps\ is distributed around the northern boundary of the SNRC and surrounding the southwest of the SNRC at a far distance, which belongs to the +80.9~\km\ps\ component with broad lines identified.
The SNRC may be associated with the +80.9~\km\ps\ component, of which the kinematic distance is estimated as 4.9$\pm0.8$ kpc.\\
{\bf G32.37-0.51}:
The southwest of this SNR is adjacent to SNR G32.1-0.9.
No broad line is identified in this SNRC by full criteria.
The SCC indicates that the SNRC is spatially correlated with molecular gas at $\sim$+45.9~\km\ps, which is roughly surrounding the whole SNRC except for its west side.
It belongs to the +42.2~\km\ps\ component that is with a broad line candidate identified in the SNRC by partial criteria.
Therefore, the SNRC may be associated with the +42.2~\km\ps\ component with the corresponding kinematic distance estimated as 2.4$\pm0.2$ kpc.\\
{\bf G32.46-0.11}:
%The east of this SNRC is adjacent to an \HII\ region at +21 and +97~\km\ps.
Two broad line components are identified in this SNRC at +81.5 and +100.3~\km\ps.
As indicated by the SCC, molecular gas at $\sim$+83.7~\km\ps, belonging to the +81.5~\km\ps\ broad line component, is surrounding the east of the SNRC and also distributed around the southwestern boundary of it. We find no good spatial correlation between the SNRC and the +100.3~\km\ps\ component. Since this SNRC is a bit small, the spatial correlation result is indefinite.
The kinematic evidence and the spatial correlation result indicate that the SNRC is associated with the +81.5~\km\ps\ component with the kinematic distance estimated as 5.0$\pm0.8$ kpc.\\
{\bf G32.73+0.15}:
This SNRC is located around the northern boundary of SNR Kes 78, and its north is adjacent to an \HII\ region at +19~\km\ps.
Three broad line components are identified in this SNRC. As indicated by the SCC, only the +74.1~\km\ps\ broad line component shows some spatial correlations with the SNRC, of which molecular gas is distributed around the eastern and southern boundaries of the SNRC.
These evidences indicate that the SNRC is associated with the +74.1~\km\ps\ component with the kinematic distance estimated as 5.1$\pm0.8$ kpc.\\
{\bf G33.62-0.23}:
%The east of this SNRC overlaps with an \HII\ region at +102~\km\ps.
We identify two broad line components at +60.3 and +74.4~\km\ps\ in this SNRC.
As indicated by the SCC, molecular gas at $\sim$+73.5~\km\ps\ is distributed around the eastern boundary of the SNRC, which belongs to the +74.4~\km\ps\ component.
No good spatial correlation result is found for the +60.3~\km\ps\ component.
Therefore, we suggest that the SNRC is associated with the +74.4~\km\ps\ component with the kinematic distance estimated as 5.2$\pm0.6$ kpc.\\
{\bf G34.52-0.76}:
The north of this SNRC is adjacent to SNR W44.
No broad line is identified in this SNRC by full criteria.
The SCC indicates a spatial correlation between the SNRC and molecular gas at $\sim$+11.4~\km\ps, which is roughly surrounding the SNRC. However, there is no corresponding broad line identified even by partial criteria.\\
{\bf G34.62+0.24}: %F
This SNRC is classified as the plerion type.
The SCC indicates that the SNRC is spatially correlated with molecular gas at $\sim$+43.3~\km\ps, which is distributed around the SNRC mostly on its eastern half.
It belongs to the +44.8~\km\ps\ component that is with broad line candidates identified in the SNRC by partial criteria.
No broad line is identified in the SNRC by full criteria.
The SNRC may be associated with the +44.8~\km\ps\ component with the kinematic distance estimated as 2.5$\pm0.2$ kpc.\\
{\bf G34.93-0.24}:
The west of this SNRC overlaps with SNR W44.
Several broad line components are identified in this SNRC.
Only one broad line identified at +42.8~\km\ps\ and some of broad lines at +51.1~\km\ps\  can be attributed to the nearby prominent component at +44.8~\km\ps.
When applying the additional clean subbackground region condition, only the +83.2~\km\ps\ component is with a broad line identified in the SNRC.
As indicated by the SCC, molecular gas at $\sim$+41.0~\km\ps\ is roughly surrounding the northeast of the SNRC and distributed around the southwestern boundary of it, which belongs to the +44.8~\km\ps\ broad line component.
Molecular gas at $\sim$+72.2~\km\ps, belonging to the +83.2~\km\ps\ broad line component, is distributed around the eastern and western boundaries of the SNRC.
No good spatial correlation is found between the SNRC and other broad line components.
Therefore, the SNRC may be associated with the +44.8 or +83.2~\km\ps\ components, and corresponding kinematic distances are estimated as 2.5$\pm0.2$ kpc and 5.0$\pm0.6$ kpc, respectively. If this SNRC is associated with the +44.8~\km\ps\ component, it is probably related to SNR W44 and a northern overlapped \HII\ region.\\
{\bf G35.13-0.34}:
Two broad line components are identified at +43.7 and +76.0~\km\ps\ in this SNRC.
As indicated by the SCC, molecular gas of the +43.7~\km\ps\ component is roughly surrounding the northeast of the SNRC. The +76.0~\km\ps\ component shows no good spatial correlation with the SNRC.
Since this SNRC is small, the spatial correlation result is indefinite.
The SNRC may be associated with the +43.7~\km\ps\ component with the kinematic distance estimated as 2.5$\pm0.2$ kpc.\\
{\bf G35.3-0.0}:
We identify several broad line components in this SNRC.
Some broad lines at +53.1~\km\ps\ are originated from overlapped \HII\ regions, especially those in the east of the SNRC. 
%The velocity range of the +53.1~\km\ps\ component is wide, and broad lines at +42.7~\km\ps\ may be originated from it.
Broad lines at +42.7~\km\ps\ may be originated from the +53.1~\km\ps\ component that has a wide velocity range.
When applying the additional clean subbackground region condition, only the +53.1~\km\ps\ component is with a broad line identified in the SNRC.
As indicated by the SCC, molecular gases at $\sim$+40.5 and $\sim$+66.4~\km\ps\ are surrounding the west and south of the SNRC, which support the associated between the SNRC and the +53.1~\km\ps\ component. No good spatial correlation is found for other broad line components.
Therefore, we suggest that the SNRC is associated with the +53.1~\km\ps\ component, of which the kinematic distance can be estimated as 9.6$\pm0.5$ kpc.\\
{\bf G36.66-0.50}:
The southwest of this SNRC overlaps with SNR G36.6-0.7.
Three broad line components are identified in this SNRC. Only the +56.6 and +78.9~\km\ps\ components show some spatial correlations with the SNRC, as indicated by the SCC. Note that part of broad lines at +56.6~\km\ps\ may be originated from an overlapped \HII\ region in the southwest of the SNRC.
Molecular gas at $\sim$+52.7~\km\ps\ is distributed around the northeastern and southwestern boundaries of the SNRC, and molecular gas at $\sim$+85.6~\km\ps\ is distributed around the northern and southwestern boundaries of the SNRC.
Therefore, the SNRC may be associated with the +56.6 or +78.9~\km\ps\ components, with corresponding kinematic distances estimated as 9.2$\pm0.4$ and 4.3$\pm0.5$ kpc, respectively.\\
{\bf G36.68-0.14}:
Three broad line components are identified in this SNRC. The +54.6 and +78.2~\km\ps\ components show some spatial correlations with the SNRC. Molecular gases at $\sim$+49.2 and $\sim$+89.5~\km\ps\ are both distributed around the SNRC and have a small amount inside.
The SNRC may be associated with the +54.6 or +78.2~\km\ps\ components, of which kinematic distances are estimated as 9.3$\pm0.5$ and 4.5$\pm0.6$ kpc, respectively.\\
{\bf G36.84-0.43}: %F
This SNRC is classified as the plerion type, and its southeast is adjacent to an \HII\ region at about +61~\km\ps.
The SCC indicates that the SNRC is spatially correlated with molecular gas at $\sim$+80.5~\km\ps, which is surrounding the whole SNRC except for its southwest side.
It belongs to the +80.7~\km\ps\ component with a broad line identified in the SNRC by full criteria.
These evidences indicate that the SNRC is associated with the +80.7~\km\ps\ component, and its kinematic distance can be estimated as 4.4$\pm0.6$ kpc.\\
{\bf G36.85-0.25}:
Two broad line components are identified at +55.1 and +79.7~\km\ps. As indicated by the SCC, molecular gas at $\sim$+57.1~\km\ps\ is roughly surrounding the SNRC,
and molecular gas at $\sim$+78.3~\km\ps\ is surrounding the northwest and southeast of the SNRC. These molecular gases belong to the +55.1 and +79.7~\km\ps\ broad line components, respectively. The spatial correlation result is indefinite, since the SNRC is a bit small.
The SNRC may be associated with the +55.1 or +79.7~\km\ps\ components with kinematic distances estimated as 9.3$\pm0.5$ and 4.5$\pm0.6$ kpc, respectively.\\
{\bf G36.90+0.49}: 
We identify no broad line in this SNRC by full criteria. The SCC peaks at $\sim$+12.4~\km\ps; however, no corresponding broad line is identified in the SNRC even by partial criteria.\\
{\bf G37.34+0.42}:
Only one broad line is identified at +90.8~\km\ps\ in this SNRC, of which molecular gas is distributed around the southern boundary of the SNRC.
Therefore, we suggest that the SNRC is associated with the +90.8~\km\ps\ component with the kinematic distance estimated as 4.7$\pm0.6$ kpc.\\
{\bf G37.51+0.78}: %r=1.3beamsiz, F
This SNRC is small and classified as the plerion type. We find no good spatial correlation result and identify no broad line in it even by partial criteria.\\
{\bf G37.62-0.22}:
Two broad line components are identified at +18.8 and +48.2~\km\ps\ in this SNRC. The SNRC is small, and we find no good spatial correlation result.
If the SNRC is associated with the +18.8~\km\ps\ component, its kinematic distance can be estimated as 2.2$\pm0.3$ kpc, and if it is associated with the +48.2~\km\ps\ component, its kinematic distance can be estimated as 9.1$\pm0.6$ kpc.\\
{\bf G37.67-0.50}:
Two broad lines at +51.7 and +74.6~\km\ps\ are identified in this SNRC. 
As indicated by the SCC, the +74.6~\km\ps\ component shows some spatial correlations with the SNRC, of which molecular gases at $\sim$+65.1 and $\sim$+77.8~\km\ps\ are roughly surrounding the SNRC. No good spatial correlation is found for the +51.7~\km\ps\ component.
These evidences indicate that the SNRC is associated with the +74.6~\km\ps\ component with the kinematic distance estimated as 4.4$\pm0.5$ kpc.
%Note that there is a molecular clump in the southern region of the SNRC, with some broad line candidates at +45.5~\km\ps\ identified by partial criteria are concentrated.\\
Additionally, there are broad line candidates at +45.5~\km\ps\ identified by partial criteria, concentrated in a molecular clump in the southern region of the SNRC, which are probably unrelated to the SNRC.\\
{\bf G37.88+0.32 and G38.17+0.09}:
These two SNRCs are adjacent to each other.
Several broad line components are identified in SNRC G37.88+0.32. 
%%No significant spatial correlation is found for it by the SCC. 
%The SCC of this SNRC has no significant peak. Nevertheless, molecular gas at $\sim$+9.4~\km\ps\ belonging to the +3.0~\km\ps\ broad line component is roughly surrounding the northwest of the SNRC, and molecular gas at $\sim$+81.7~\km\ps\ belonging to the +85.5~\km\ps\ broad line component is distributed around the SNRC.
As indicated by the SCC, molecular gas at $\sim$+1.1~\km\ps\ is distributed around the northeastern boundary of the SNRC, and molecular gas at $\sim$+9.4~\km\ps\ is roughly surrounding the northwest of the SNRC, which belong to the +3.0~\km\ps\ broad line component.
Molecular gas at $\sim$+81.7~\km\ps\ belonging to the +85.5~\km\ps\ broad line component is distributed around the SNRC.
SNRC G37.88+0.32 may be associated with the +3.0 or +85.5~\km\ps\ components, of which kinematic distances are estimated as 12.8$\pm0.6$ and 4.5$\pm0.6$ kpc, respectively.
For SNRC G38.17+0.09, multiple broad line components are identified in it. Some broad lines at +62.1~\km\ps\ are probably originated from an overlapped \HII\ region in the southwest of the SNRC, nevertheless, there are still more broad line points at this velocity than at other velocities.
When applying the additional clean subbackground region condition, only the +82.2~\km\ps\ component is with broad lines identified in the SNRC. However, it shows no good spatial correlation with the SNRC.
As indicated by minor peaks of the SCC, the +2.7, +17.8, and +62.1~\km\ps\ broad line components show some spatial correlations with the SNRC. Molecular gas at $\sim$+0.2~\km\ps\ is distributed around the western boundary of the SNRC. Molecular gas at $\sim$+16.0~\km\ps\ is distributed around the northern and southeastern boundaries of the SNRC, and molecular gas at $\sim$+21.9~\km\ps\ is surrounding its southwest. Molecular gas at $\sim$+64.3~\km\ps\ is roughly surrounding the south of the SNRC and also distributed around its northern boundary.
SNRC G38.17+0.09 is most likely to be associated with the +62.1~\km\ps\ component; however, it may be associated with either the +2.7 or +17.8~\km\ps\ components too.
Kinematic distances are estimated as 4.1$\pm0.7$ kpc for the +62.1~\km\ps\ component, 11.5$\pm0.5$ kpc for the +2.7~\km\ps\ component, and 11.5$\pm0.4$ kpc for the +17.8~\km\ps\ component.\\
{\bf G38.62-0.24}:
Two broad line components are identified in this SNRC at +44.0 and +67.9~\km\ps. 
%This SNRC is small, no clear spatial correlation result is found. 
Molecular gas at a minor peak of the SCC at $\sim$+47.0~\km\ps\ is roughly surrounding the east of the SNRC and also distributed around its western boundary, which belongs to the +44.0~\km\ps\ broad line component.
These evidences support the association between the SNRC and the +44.0~\km\ps\ component, of which the kinematic distance is estimated as 2.5$\pm0.5$ kpc.\\
{\bf G38.68-0.43}: %F
This SNRC is classified as the plerion type.
As indicated by the SCC, molecular gas at $\sim$+20.3~\km\ps\ is roughly surrounding the north the SNRC. It belongs to the +18.9~\km\ps\ component that is with broad lines identified.
These evidences indicate an association between the SNRC and the +18.9~\km\ps\ component, with the corresponding kinematic distance estimated as 2.2$\pm0.2$ kpc.\\
{\bf G38.72-0.87}: %F
This SNRC is classified as the plerion type and overlaps with SNR G38.7-1.3.
%As indicated by the SCC, molecular gases at $\sim$+42.7 and $\sim$+70.2~\km\ps\ are roughly surrounding the northeast of the SNRC. 
As indicated by the SCC, molecular gases at $\sim$+42.7 and $\sim$+80.0~\km\ps\ are roughly surrounding the northeast of the SNRC. 
However, molecular gas at $\sim$+42.7~\km\ps\ is with no corresponding broad line identified. 
Molecular gas at $\sim$+80.0~\km\ps\ is with broad lines identified in the SNRC at +74.4~\km\ps.
There is also a broad line identified at +79.8~\km\ps, which can be attributed to the prominent component at +74.4~\km\ps\ too.
Therefore, we suggest that the SNRC is associated with the +74.4~\km\ps\ component, and the corresponding kinematic distance is estimated as 4.3$\pm0.4$ kpc.\\
{\bf G38.83-0.01}: %r=1.3beamsiz
No broad line is identified in this SNRC even by partial criteria. We also find no significant spatial correlation result, since this SNRC is small.
Note that molecular gas at a peak of the SCC at $\sim$+41.1~\km\ps\ is distributed around the SNRC except for its southwest side.\\
{\bf G39.19+0.52}:
Two broad line components are identified in this SNRC at +20.6 and +28.2~\km\ps. The SCC indicates that the SNRC is spatially correlated with molecular gas at $\sim$+45.6~\km\ps, which is surrounding the northwestern half of the SNRC well. It supports the association between the SNRC and the +28.2~\km\ps\ component. Nevertheless, there is also molecular gas at $\sim$+12.5~\km\ps\ roughly surrounding the southwest of the SNRC, which may belong to the +20.6~\km\ps\ component.
Therefore, the SNRC is probably associated with the +28.2~\km\ps\ component with the kinematic distance estimated as 1.90$\pm0.08$ kpc. However, it is also possible that the SNRC is associated with the +20.6~\km\ps\ component, of which the kinematic distance is estimated as 11.5$\pm0.3$ kpc.\\
{\bf G39.20+0.81}:
No broad line is identified in this SNRC even by partial criteria. Since the SNRC is a bit small, we also find no good spatial correlation result.\\
{\bf G39.54+0.37}:
Only one broad line component is identified in this SNRC at +36.2~\km\ps, which is spatially correlated with the SNRC indicated by the SCC. Molecular gas at $\sim$+42.2~\km\ps\ is roughly surrounding the southwest and northwest of the SNRC.
We suggest that the SNRC is associated with the +36.2~\km\ps\ component with the kinematic distance estimated as 1.91$\pm0.07$ kpc.\\
{\bf G39.56-0.32}:
This SNRC overlaps a small \HII\ region at +55.2~\km\ps, and is also adjacent to several \HII\ regions at velocities of about +14, +27, +51, +53, and +58~\km\ps.
We identify only one broad line component in this SNRC at +57.9~\km\ps, which is also spatially correlated with the SNRC. 
One of many broad line points seems to be originated from the overlapped \HII\ region in the southern region of the SNRC.
Molecular gas at $\sim$+52.9~\km\ps\ is roughly surrounding the northern half of the SNRC. The SCC also indicates a spatial correlation between the SNRC and molecular gas at $\sim$+15.2~\km\ps, which is surrounding the eastern half of the SNRC; however, no corresponding broad line is identified by full criteria.
Therefore, we suggest that the SNRC is associated with the +57.9~\km\ps\ component and may be related to the overlapped and adjacent \HII\ regions. The corresponding kinematic distance is estimated as 8.7$\pm0.5$ kpc.\\
{\bf G41.51-0.53}:
Several broad line components are identified in this SNRC. Only the +37.4~\km\ps\ component shows some spatial correlations with the SNRC, as indicated by the SCC. Molecular gas at $\sim$+37.6~\km\ps\ is roughly surrounding the northwestern half of the SNRC.
Based on these evidences, we suggest that the SNRC is associated with the +37.4~\km\ps\ component with the kinematic distance estimated as 1.92$\pm0.08$ kpc.\\
{\bf G41.63+0.26}:
The northwest of this SNRC is adjacent to SNR G41.5+0.4.
Only one broad line component is identified in this SNRC at +36.5~\km\ps, and as indicated by the SCC, molecular gas at $\sim$+34.3~\km\ps\ belonging to it is roughly surrounding the SNRC.
These evidences indicate that the SNRC is associated with the +36.5~\km\ps\ component with the kinematic distance estimated as 2.5$\pm0.5$ kpc. Both SNRC G41.63+0.26 and SNR G41.5+0.4 are located at the near side of the Sagittarius spiral arm, and they may be related.\\
{\bf G41.95-0.18}:
The north of this SNRC overlaps with SNR G42.0-0.1.
We identify only one broad line component in this SNRC at +65.6~\km\ps, which is not well spatially correlated with the SNRC.
Molecular gas at a minor peak of the SCC at $\sim$+73.7~\km\ps\ is distributed around the southeastern boundary and also in the northwest of the SNRC.
In addition, as indicated by the SCC, molecular gas at $\sim$+17.5~\km\ps\ is surrounding the eastern half of the SNRC, and molecular gas at $\sim$+29.5~\km\ps\ is surrounding the southeast and northwest of the SNRC.
These molecular gases belong to the +15.8 and +27.1~\km\ps\ components, which are with broad line candidates identified in the SNRC by partial criteria.
%The SNRC may be associated with the +68.6~\km\ps\ component, or with either the +15.8 or +27.1~\km\ps\ components.
The SNRC may be associated with the +65.6~\km\ps\ component, or with either the +15.8 or +27.1~\km\ps\ components.
%Kinematic distances are estimated as 3.5$\pm0.3$ and 7.8$\pm0.5$ kpc for the +68.6~\km\ps\ component, 11.1$\pm0.4$ kpc for the +15.8~\km\ps\ component, and 3.0$\pm1.0$ kpc for the +27.1~\km\ps\ component.\\
Kinematic distances are estimated as 7.9$\pm0.5$ kpc for the +65.6~\km\ps\ component, 11.1$\pm0.4$ kpc for the +15.8~\km\ps\ component, and 3.0$\pm1.0$ kpc for the +27.1~\km\ps\ component.\\
{\bf G42.62+0.14}:
We identify only one broad line at +64.3~\km\ps\ in this SNRC. As indicated by the SCC, molecular gas at $\sim$+57.5~\km\ps, belonging to the +64.3~\km\ps\ component, is roughly surrounding the southeast of the SNRC and also distributed around its northern boundary. Therefore, we suggest that the SNRC is associated with the +64.3~\km\ps\ component with the near and far kinematic distances estimated as 3.6$\pm0.4$ and 7.2$\pm1.1$ kpc, respectively.\\
{\bf G42.71-0.27}:
Several broad line components are identified in this SNRC. Nevertheless, there is only one broad line identified at +60.1 and +69.4~\km\ps, which can be attributed to the nearby prominent component at +60.4~\km\ps.
The SCC indicates that the +60.4~\km\ps\ component is spatially correlated with the SNRC, with molecular gas at $\sim$+55.9~\km\ps\ surrounding the whole SNRC except for its southwest side.
Based on these evidences, the SNRC is associated with the +60.4~\km\ps\ component with the kinematic distance estimated as 7.8$\pm0.5$ kpc.\\
{\bf G43.02+0.73}:
This SNRC is adjacent to SNR G42.8+0.6 in the west.
No broad line is identified in this SNRC by full criteria.
The SCC indicates a roughly spatial correlation between the SNRC and molecular gas at $\sim$+59.4~\km\ps, which is surrounding the northeast of the SNRC and also distributed in the southeast of it. There is also molecular gas at $\sim$+67.1~\km\ps\ roughly surrounding the southeast of the SNRC. These molecular gases belong to the +59.4~\km\ps\ component with broad line candidates identified in the SNRC by partial criteria. The SNRC may be associated with the +59.4~\km\ps\ component with the near and far kinematic distances estimated as 3.8$\pm0.9$ and 8.1$\pm0.9$ kpc, respectively.\\
{\bf G43.07+0.56}:
This SNRC presents a bright partial radio continuum shell in the south and weak diffuse radio continuum emission elsewhere. No broad line is identified in this SNRC by full criteria. %, and no significant spatial correlation result is found through the SCC.
Molecular gas at the peak of the SCC at $\sim$+42.4~\km\ps\ is roughly surrounding the north of the SNRC, which belongs to the +40.3~\km\ps\ component with a broad line candidate identified in the SNRC by partial criteria. %Since this SNRC is small, these spatial correlation results are indeterminate.
Molecular gas at a peak of the SCC at $\sim$+1.3~\km\ps\ roughly is surrounding the south of the SNRC; however, no corresponding broad line candidate is identified in it.
If the SNRC is associated with the +40.3~\km\ps\ component, its kinematic distance can be estimated as 2.7$\pm0.4$ kpc.\\
{\bf G44.08+0.13 and G44.5-0.2}:
These two SNRCs are overlapping each other.
Among broad lines identified in SNRC G44.08+0.13, one broad line is at +12.0~\km\ps, and three broad lines at +59.1~\km\ps\ can be attributed to the nearby +55.1~\km\ps\ component. As indicated by the SCC, the +55.1~\km\ps\ component shows spatial correlations with the SNRC, with molecular gases at $\sim$+45.1 and $\sim$+67.6~\km\ps\ surrounding the southeast and southern half of the SNRC, respectively.
Additionally, molecular gases at $\sim$+8.4 and $\sim$+21.1~\km\ps, belonging to the +12.0~\km\ps\ component, are surrounding the northeast and southeast of the SNRC, respectively.
SNRC G44.08+0.13 is most likely to be associated with the +55.1~\km\ps\ component; however, it is also possible that it is associated with the +12.0~\km\ps\ component. Kinematic distances are estimated as 3.6$\pm0.6$ and 8.0$\pm0.5$ kpc for the +55.1~\km\ps\ component and 10.7$\pm0.4$ kpc for the +12.0~\km\ps\ component.
SNRC G44.5-0.2 was suggested to be a counterpart of the shell-type TeV source HESS J1912+101, with broad CO lines detected at +60~\km\ps\ \citep[][see Table~\ref{tab:snrpre} for more information]{Su+2017b, ReichSun2019}. It is large and overlaps multiple \HII\ regions. Here, many broad lines are identified in this SNRC, and some of them are probably originated from overlapped \HII\ regions. Most of the broad lines are at +59.8~\km\ps, and that at +44.5 and +68.1~\km\ps\ can be attributed to the +59.8~\km\ps\ component too. Few broad lines are identified at +24.4~\km\ps, and some of them are located in the overlapping region with SNRC G44.08+0.13. 
%No significant spatial correlation result is found for SNRC G44.5-0.2. Molecular gas at $\sim$+14.0 and $\sim$+50.2~\km\ps\ is roughly distributed around the SNRC.
No significant spatial correlation result is found for SNRC G44.5-0.2. Molecular gases at $\sim$+15.1 and $\sim$+51.3~\km\ps\ are roughly distributed around the SNRC, which belong to the +24.4 and +59.8~\km\ps\ components, respectively.
SNRC G44.5-0.2 is probably associated with the +59.8~\km\ps\ component; however, it may be associated with the +24.4~\km\ps\ component too. Kinematic distances are estimated as 7.6$\pm0.4$ kpc for the +59.8~\km\ps\ component and 4.0$\pm0.7$ kpc for the +24.4~\km\ps\ component. It is possible that SNRCs G44.08+0.13 and G44.5-0.2 are related to each other.\\
{\bf G45.35-0.37}:
The east of this SNRC is adjacent to SNR G45.7-0.4.
Broad lines are identified in this SNRC at +56.1 and +66.5~\km\ps. However, only one broad line is identified at +66.5~\km\ps, which can be attributed to the nearby prominent component at +56.1~\km\ps. As indicated by the SCC, molecular gas at $\sim$+46.0~\km\ps, belonging to the +56.1~\km\ps\ component, is distributed around the SNRC.
Based on these evidences, we suggest that the SNRC is associated with the +56.1~\km\ps\ component with the kinematic distance estimated as 7.3$\pm0.6$ kpc. Both SNRC G45.35-0.37 and SNR G45.7-0.4 seem located at the far side of the Sagittarius spiral arm, and may be related to each other.\\
{\bf G45.51-0.03}:
Two broad line components are identified in this SNRC at +11.3 and +59.6~\km\ps.
As indicated by the SCC, molecular gas at $\sim$+51.6~\km\ps\ is surrounding the western half of the SNRC and also distributed around its eastern boundary. In addition, molecular gas at $\sim$+67.0~\km\ps\ is distributed around the SNRC. These spatial correlations support the association between the SNRC and the +59.6~\km\ps\ component. No spatial correlation is found for the +11.3~\km\ps\ component.
Associated with the +59.6~\km\ps\ component, we can estimate the kinematic distance of the SNRC as 7.3$\pm0.5$ kpc.\\
{\bf G46.18-0.02}: %mayCinA17 FinD21
%This SNRC is classified as the plerion type; however, its radio continuum emission presents shell-like structures.
This SNRC presents shell-like structures in radio continuum emission.
It overlaps with \HII\ regions at +15~\km\ps\ in the south, and is also adjacent to an \HII\ region at +63~\km\ps\ in the southwest.
Molecular gases at $\sim$+50.5 and $\sim$+54.8~\km\ps\ are roughly surrounding the southern half of the SNRC, indicated by the SCC. These molecular gases belong to the +56.0~\km\ps\ component that is the only broad line component identified in the SNRC.
Therefore, we suggest that the SNRC is associated with the +56.0~\km\ps\ component with the kinematic distance estimated as 7.0$\pm0.7$ kpc.\\
{\bf G46.54-0.03}:
%Only one broad line component is identified in this SNRC at +60.3~\km\ps. 
%No significant spatial correlation result is found through the SCC. For the +60.3~\km\ps\ component, little molecular gas at $\sim$+69.7~\km\ps\ is distributed around the western boundary of the SNRC.
%The SNRC may be associated with the +60.3~\km\ps\ component, and the corresponding kinematic distance is estimated as 6.8$\pm0.6$ kpc.\\
Two broad line components are identified in this SNRC at +15.8 and +57.5~\km\ps. Molecular gas at the peak of the SCC at $\sim$+16.0~\km\ps\ is roughly surrounding the SNRC, which belongs to the +15.8~\km\ps\ broad line component. 
%We find no good spatial correlation result for the +57.5~\km\ps\ component.
Molecular gas of the +57.5~\km\ps\ component shows no significant spatial correlation with the SNRC.
Based on these evidences, we suggest that the SNRC is associated with the +15.8~\km\ps\ component, and the corresponding kinematic distance is estimated as 10.3$\pm0.4$ kpc.\\
{\bf G47.15+0.73}:
The radius of this SNRC is less than the beam size of our observation, which is too small to examine possible spatial correlations. There is no broad line identified in this SNRC even by partial criteria.\\
{\bf G47.36-0.09}:
Three broad line components are identified in this SNRC. However, broad lines at +64.3~\km\ps\ can be attributed to the nearby prominent component at +55.1~\km\ps. %and some broad lines at +41.5~\km\ps\ may be attributed to it too.
Molecular gases at $\sim$+34.9 and $\sim$+53.5~\km\ps\ are distributed around the SNRC, which probably belong to the +41.5 and +55.1~\km\ps\ broad line components.
The SNRC may be associated with the +41.5 or +55.1~\km\ps\ components. Kinematic distances are estimated as 4.0$\pm0.8$ kpc for the +41.5~\km\ps\ component and 4.1$\pm0.5$ kpc for the +55.1~\km\ps\ component. Kinematic distances of these two velocity components are consistent, and both are probably located at the near side of the Sagittarius spiral arm.\\
{\bf G47.74-0.97}:
Only one broad line component is identified in this SNRC at +50.3~\km\ps. As indicated by the SCC, molecular gas at $\sim$+50.0~\km\ps\ is distributed around the northern boundary of the SNRC, which belongs to the +50.3~\km\ps\ component.
These evidences indicate that the SNRC is associated with the +50.3~\km\ps\ component, of which the near and far kinematic distances are estimated as 3.2$\pm0.7$ and 7.7$\pm0.7$ kpc, respectively.\\
{\bf G48.88+0.17}:
Broad lines at three velocities are identified in this SNRC. There are two broad lines at +55.0~\km\ps, which can be attributed to the nearby component at +49.6~\km\ps.
This SNRC presents shell-like radio continuum emission in its eastern half.
As indicated by the SCC, molecular gas at $\sim$+16.8~\km\ps, belonging to the +8.5~\km\ps\ broad line component, is surrounding the northwestern half of the SNRC. Molecular gas at $\sim$+58.4~\km\ps, belonging to the +49.6~\km\ps\ broad line component, is roughly surrounding the southeastern half of the SNRC.
Based on these evidences, the SNRC is probably associated with the +49.6~\km\ps\ component, and the corresponding kinematic distance is estimated as 5.3$\pm0.7$ kpc. It is also possible that the SNRC is associated with the +8.5~\km\ps\ component, of which the kinematic distance is estimated as 10.1$\pm0.4$ kpc.\\
{\bf G51.06+0.56}:
No broad line is identified in this SNRC even by partial criteria.
We also find no significant spatial correlation result by the SCC.
Nevertheless, molecular gas at a peak of the SCC at $\sim$+37.9~\km\ps\ is roughly surrounding the west of the SNRC, where shell-like radio continuum emission is present. If the SNRC is associated with molecular gas at $\sim$+37.9~\km\ps, its kinematic distance can be estimated as 8.0$\pm0.8$ kpc.\\
{\bf G51.21+0.11 (G51.26+0.11)}:
Two broad line components are identified in this SNRC.
Nevertheless, only one broad line is identified at +51.3~\km\ps, and it can be attributed to the nearby prominent broad line component at +54.6~\km\ps. As indicated by the SCC, molecular gas at $\sim$+62.9~\km\ps\ is distributed around the southeastern boundary of the SNRC, which belongs to the +54.6~\km\ps\ component.
Therefore, we suggest that the SNRC is associated with the +54.6~\km\ps\ component. Considering the presence of \HI\ absorption up to the tangent point velocity shown by \cite{RanasingheLeahy2022}, its kinematic distance is estimated as 5.8$\pm0.5$ kpc.\\
%and its kinematic distance can be estimated as 4.6$\pm0.2$ kpc.\\ %It may be related to an overlapped \HII\ region at +42~\km\ps.\\
{\bf G53.07+0.49}: %r=1.2beamsiz 
No broad line is identified in this SNRC even by partial criteria. Because of its small size, no good spatial correlation result is found for it too.\\
{\bf G53.84-0.75}: %maySinA17 FinD21
Two broad lines are identified in this SNRC at +25.1 and +37.9~\km\ps.
As indicated by the SCC, molecular gas at $\sim$+23.7~\km\ps\ is spatially correlated with radio continuum bright regions of the SNRC, which is roughly surrounding the west of the SNRC and also distributed around its eastern boundary. It belongs to the +25.1~\km\ps\ broad line component.
No spatial correlation is found between the SNRC and the +37.9~\km\ps\ component.
Based on these evidences, the SNRC is associated with the +25.1~\km\ps\ component, of which the kinematic distance is estimated as 1.6$\pm0.7$ kpc.\\
{\bf G56.56-0.75}: %F
This SNRC is classified as the plerion type.
Molecular gas at the peak of the SCC at $\sim$+31.9~\km\ps\ is surrounding the northwest of the SNRC. It belongs to the only component with broad line candidates identified in the SNRC by partial criteria, at +32.0~\km\ps. No broad line is identified in the SNRC by full criteria.
The SNRC is probably associated with the +32.0~\km\ps\ component with the kinematic distance estimated as 2.8$\pm0.7$ kpc.\\
{\bf G57.12+0.35}:
Two broad line components are identified in this SNRC at $-$1.1 and +10.3~\km\ps.
We find no significant spatial correlation result by the SCC, nevertheless, molecular gas at $\sim$+7.6~\km\ps\ is roughly surrounding the east of the SNRC and distributed around its northwestern boundary, which belongs to the +10.3~\km\ps\ component.
There is also molecular gas at $\sim$$-$8.7~\km\ps\ roughly surrounding the southwest of the SNRC and also distributed around its northern boundary, which belongs to the $-$1.1~\km\ps\ component, however, it is not well correlated with radio continuum emission in the northeastern half.
The SNRC may be associated with the $-$1.1 or +10.3~\km\ps\ components, with kinematic distances estimated as 8.9$\pm0.4$ and 8.5$\pm0.5$ kpc, respectively. Kinematic distances of two components are consistent, and both are located in the Perseus spiral arm.\\
{\bf G58.70-0.31}: %F
This SNRC is classified as the plerion type.
Molecular gas at the peak of the SCC at $\sim$$-$5.6~\km\ps\ is surrounding the east of the SNRC. However, no broad line is identified in the SNRC even by partial criteria.
The SNRC might be associated with molecular gas at $\sim$$-$5.6~\km\ps, of which the kinematic distance is estimated as 8.6$\pm0.5$ kpc.\\
{\bf G59.46+0.83}: %F
This SNRC is classified as the plerion type. 
%Molecular gas at the peak of the SCC at $\sim$+22.9~\km\ps\ is distributed around the SNRC, but not correlated well. 
Molecular gas at the peak of the SCC at $\sim$+24.3~\km\ps is distributed around the SNRC, but not correlated well. 
Moreover, no broad line is identified in the SNRC even by partial criteria.\\
{\bf G59.68+1.25}:
The east of this SNRC overlaps with SNR G59.8+1.2.
We identify no broad line in this SNRC even by partial criteria. The SCC indicates that molecular gas at $\sim$+9.5~\km\ps\ is spatially correlated with the SNRC, which is roughly surrounding the southwestern half of the SNRC.
The SNRC might be associated with molecular gas at $\sim$+9.5~\km\ps\ with the near and far kinematic distances estimated as 0.5$\pm0.7$ and 7.8$\pm0.7$ kpc, respectively.\\
{\bf G59.83-0.41}:
Only one broad line candidate at +55.1~\km\ps\ is identified in this SNRC by partial criteria. However, we find no good spatial correlation result for the +55.1~\km\ps\ component or other velocity components.\\
%{\bf G67.25-0.36}: We identify no broad line in this SNRC even by partial criteria, and find no significant spatial correlation result by the SCC.\\
%{\bf G107.5-1.5}: No broad line is identified in this SNRC even by partial criteria, and no significant spatial correlation result is found through the SCC.\\
{\bf G128.5+2.6}:
This SNRC presents shell-like radio continuum emission in the south.
Only one broad line is identified in the SNRC at $-$7.6~\km\ps.
The SCC indicates a spatial correlation between the SNRC and the $-$7.6~\km\ps\ component, of which molecular gas at $\sim$$-$2.9~\km\ps\ is roughly surrounding the south of the SNRC and also distributed around its northeastern boundary.
Based on these evidences, we suggest that the SNRC is associated with the $-$7.6~\km\ps\ component, and its kinematic distance can be estimated as 0.6$\pm0.2$ kpc.\\
{\bf G149.5+3.2, G150.8+3.8, and G151.2+2.9}:
SNRCs G149.5+3.2 and G151.2+2.9 are adjacent to SNRC G150.8+3.8, and all of them overlap with SNR G150.3+4.5.
No broad line is identified in these three SNRCs even by partial criteria. 
%We also find no significant spatial correlation result by the SCC. 
For SNRC G149.5+3.2, molecular gas at the peak of the SCC at $\sim$+1.3~\km\ps\ is roughly surrounding its southeast, however, not correlated with bright radio continuum emission of the SNRC. %in its northeast and southwest. 
Molecular gas at $\sim$+6.2~\km\ps\ is roughly surrounding the northwest of the SNRC, where there is bright radio continuum emission. If the SNRC G149.5+3.2 is associated with molecular gas at $\sim$+6.2~\km\ps, its near and far kinematic distances can be estimated as 0.25$\pm0.05$ and 2.05$\pm0.07$ kpc, respectively.
SNRC G150.8+3.8 presents shell-like radio continuum emission in its southeast. As indicated by the SCC, there is a molecular shell at $\sim$+1.7~\km\ps\ roughly surrounding its radio continuum shell, which also has some protrusions pointing toward its center.
Therefore, SNRC G150.8+3.8 may be associated with molecular gas at $\sim$+1.7~\km\ps\ with kinematic distance estimated as 0.4$\pm0.2$ kpc, and it may be related to SNR G150.3+4.5.
For SNRC G151.2+2.9, molecular gases at nonsignificant peaks of the SCC at $\sim$$-$9.4 and $\sim$+4.0~\km\ps\ are mainly distributed in the northwest, not correlated with shell-like radio continuum emission in the southeast of the SNRC.\\
%SNRC G151.2+2.9 presents shell-like radio continuum emission in its southeast too. 
%{\bf G160.1-1.1}: We identify no broad line in this SNRC even by partial criteria, and find no good spatial correlation result.\\
{\bf G172.8+1.5}:
Three broad line components at $-$16.4, $-$9.0, and $-$2.3~\km\ps\ are identified in this SNRC. Only one broad line is identified at $-$9.0~\km\ps, which can be attributed to the nearby prominent component at $-$16.4~\km\ps. No significant spatial correlation is found through the SCC. There is molecular gas at $\sim$$-$22.7~\km\ps\ distributed around the northern boundary of the SNRC, which belongs to the $-$16.4~\km\ps\ broad line component.
If the SNRC is associated with the $-$16.4~\km\ps\ component, its kinematic distance can be estimated as 1.7$\pm0.04$ kpc.\\
{\bf G189.6+3.3}:
The southwest of this SNRC overlaps with SNR IC~443.
All broad lines identified in this SNRC are in the overlapping region with SNR IC~443, which are probably originated from SNR IC~443.
Molecular gas at the peak of the SCC at $\sim$$-$3.0~\km\ps\ is distributed around the northwestern and southwestern boundaries of the SNRC, but not correlated well.
We cannot determine which MC is associated with the SNRC.\\
{\bf G190.2+1.1}:
This SNRC is revealed as a rapidly expanding \HI\ shell observed by \HI\ 21 cm line emission \citep{Koo+2006}.
%The south of this SNRC is adjacent to a small \HII\ region at +8~\km\ps.
No broad line is identified in the SNRC by full criteria.
Molecular gas at the peak of the SCC at $\sim$+5.1~\km\ps\ shows some spatial correlations with the SNRC, which is roughly surrounding its southern half. It belongs to the +7.5~\km\ps\ component with broad line candidates identified by partial criteria.
The SNRC may be associated with the +7.5~\km\ps\ component, of which the kinematic distance is estimated as 2.1$\pm0.03$ kpc.%\\

\section{Individual PWNe}\label{app:pwn}
\begin{figure*}[ptbh!]
\centerline{{\hfil\hfil
\psfig{figure=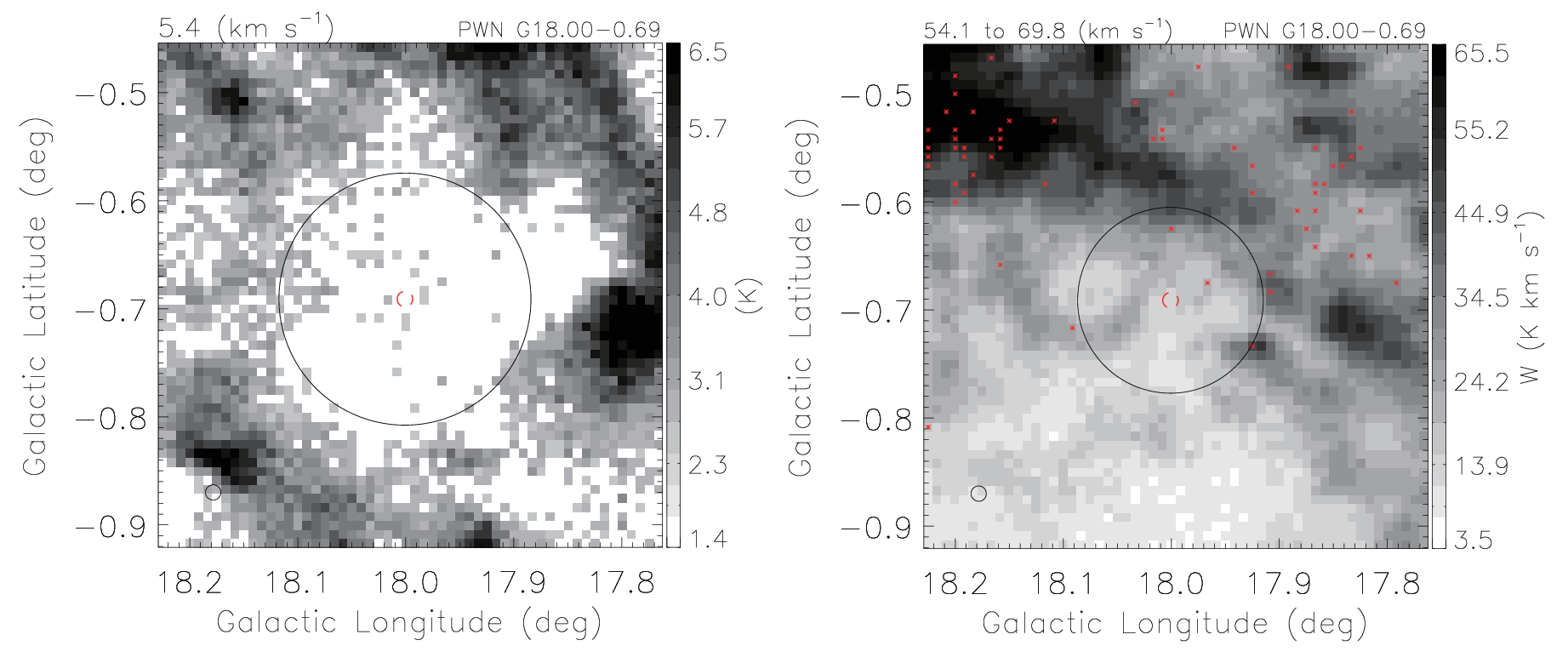,height=2.5in,angle=0, clip=}
\hfil\hfil}}
\caption{\twCO\ (J=1--0) intensity maps of PWN G18.00-0.69, at the +5.4~\km\ps\ velocity channel (left) and integrated over the velocity range of +54.1 to +69.8~\km\ps\ (right).
Minimum values of \twCO\ emission intensity shown in colorbars are at $3\sigma$ confidence level.
%The red dashed circle denotes the location of the PWN. 
%Locations of broad lines identified by full criteria are denoted by red tiny stars.
%The extent of the PWN is denoted by red dashed circles.
Red tiny stars denote locations of broad lines identified by full criteria.
Red dashed circles denote the extent of the PWN.
Black solid circles delineate the cavity edge (left) and the surrounding shell (right).
}
\label{f:pwn18}
\end{figure*}
{\bf G18.00-0.69}: As inferred from the radio dispersion measurement, the distance of the associated pulsar was estimated as $3.9\pm0.4$ kpc \citep{CordesLazio2002, Gaensler+2003}, while, the extinction-distance relation of red clump stars toward this PWN derived a distance of $3.1\pm0.2$ kpc \citep{Shan+2018}. 
The SCC peaks at $\sim$+5.4~\km\ps, where molecular gas presents a cavity-like structure around the PWN, as shown in the left panel in Figure~\ref{f:pwn18}. The corresponding kinematic distance is estimated as 1.26$\pm0.05$ kpc. However, no corresponding broad line is identified even by partial criteria.
Both a minor peak of the SCC and the detection of associated broad line emission indicate that the PWN is associated with the +60.8~\km\ps\ velocity component. As shown in Figure~\ref{f:pwn18}, molecular gas of the +60.8~\km\ps\ component is surrounding the PWN in the west. In addition, there is also a shell-like molecular structure besides the PWN in the east, which seems unrelated.
Therefore, we suggest that the PWN is associated with the +60.8~\km\ps\ component, with the corresponding kinematic distance estimated as 3.8$\pm0.4$ kpc. The radius of the western shell-like structure is $\sim$5.7 pc. %\\ %(r=0.086deg)
%These results are consistent with our measurements, with the distance of $\sim$3.3 kpc.\\
\\
\begin{figure*}[ptbh!]
\centerline{{\hfil\hfil
\psfig{figure=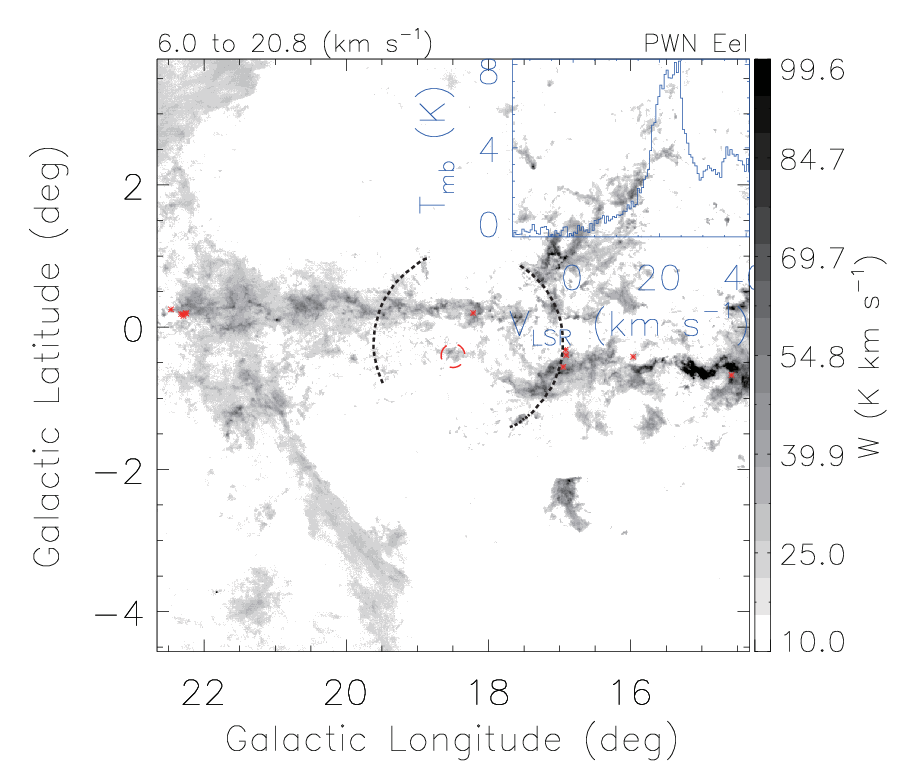,height=5.5in,angle=0, clip=}
\hfil\hfil}}
\caption{\twCO\ (J=1--0) intensity map of PWN Eel, integrated over the velocity range of +6.0 to +20.8~\km\ps. 
The minimum value of \twCO\ emission intensity shown in the colorbar is at $5\sigma$ level.
%The red dashed circle denotes the location of the PWN. 
%Locations of broad lines identified by full criteria are denoted by red tiny stars.
Red tiny stars denote locations of broad lines identified by full criteria.
The extent of the PWN is denoted by a red dashed circle.
Two shell-like structures are sketched by two black dotted lines.
The average \twCO\ (J=1--0) spectrum of three broad line points on the western shell-like structure is shown in the upper right corner.
}
\label{f:eel}
\end{figure*}
{\bf Eel}: This PWN might be associated with the TeV source HESS J1825-137 \citep{Voisin+2016}. The dispersion measure of the accompany pulsar is unknown; however, distance estimations were performed based on pulsar's properties, which give possible distances of $\sim$1.3 kpc \citep{Wang2011} and $\sim$3.9 kpc \citep{Kargaltsev+2017}.
%No significant spatial correlation is found through the SCC.
The SCC for the PWN has no significant peak.
In this direction, there presents a large bubble in local MCs at $\sim$+6~\km\ps, and the PWN locates near the southeastern boundary of it. 
Nevertheless, two shell-like CO structures of the +19.8~\km\ps\ component seem correlated, which circulate the PWN, and broad lines are also identified on them by full criteria (see Figure~\ref{f:eel}).
Note that there are some background MCs near these shell-like structures, which are distributed near the Galactic plane and probably at a far kinematic distance. 
Therefore, we suggest that the PWN is associated with the +19.8~\km\ps\ MC, hence, locates at a kinematic distance of 1.50$\pm0.04$~kpc. The radius of these shell-like structures is $\sim$35 pc. %\\%r=1.34deg
\\
{\bf G21.88-0.10}: The dispersion measure distance of PSR J1831-0952 was estimated as 4.3~kpc \citep{Manchester+2005, Sheidaei2011}. Here, the SCC indicates an association between the PWN and molecular gas at $\sim$+40.2~\km\ps, which is distributed around the PWN. 
%It belongs to the +36.4~\km\ps\ component with broad lines identified.
It belongs to the +55.0~\km\ps\ component with broad lines identified.
%There is no clear shell or cavity structure assciated.
%The corresponding kinematic distance is estimated as 11.6$\pm0.3$~kpc. %\\
The corresponding kinematic distance is estimated as 3.4$\pm0.4$~kpc. %\\
\\
{\bf G23.5+0.1}: Molecular gas at the peak of the SCC at $\sim$+96.2~\km\ps\ is roughly surrounding this PWN candidate, which belongs to the +95.8~\km\ps\ component with broad lines identified.
We suggest that the PWN candidate is associated with the +95.8~\km\ps\ component, with the corresponding kinematic distance estimated as 5.3$\pm0.7$~kpc. 
It is consistent with the \HI\ absorption result toward PSR J1833-0827 \citep{Weisberg+1995}.
The radius of the surrounding shell-like structure is $\sim$15 pc. %\\ %r=0.16deg
\\
{\bf G25.24-0.19}: A distance of 6.6~kpc was adopted for this PWN by assuming an association with an adjacent cluster of red supergiant stars \citep{GotthelfHalpern2008}. %, which seems consistent with the distance estimated by \cite{Kargaltsev+2013}. 
The SCC peaks at $\sim$+3.5, $\sim$+9.0, and $\sim$+107.1~\km\ps. There is no clear shell-like or cavity-like structure found for molecular gases at $\sim$+3.5 and $\sim$+9.0~\km\ps. Molecular gas at $\sim$+107.1~\km\ps\ is roughly surrounding the PWN in the northeast, which belongs to the +102.1~\km\ps\ component with corresponding broad lines identified.
%Both the +3.5 and +9.5~\km\ps\ MCs probably belong to the $\sim$+6~\km\ps\ velocity component.
Note that there is also a small shell-like structure at $\sim$+62.5~\km\ps\ roughly surrounding the PWN; however, it is probably related to a large-scale hub-filament structure but not the PWN.
%There are broad line emission in vincity, but not from the adjacent structures.
%Kinematic distances of +6 and +107~\km\ps\ MCs are $\sim$14.7 and $\sim$6.1 kpc, respectively. 
Therefore, we suggest that the PWN is associated with the +102.1~\km\ps\ component, with the kinematic distance estimated as 5.9$\pm0.7$ kpc.
Correspondingly, the radius of the surrounding shell-like structure is $\sim$24 pc. %(r=0.235deg)
Note that, the derived distance is consistent with that adopted by \cite{GotthelfHalpern2008}.
%The association with the +101.1~\km\ps\ MC seems consistent with the assumption in previous work. 
%However, we cannot exclude the possibility of the association between the PWN and the +6~\km\ps\ velocity component.\\
%We cannot determine which MC is associated with the PWN.
%However, these MCs present no good cavity or shell structure. %\\
\\
\begin{figure*}[ptbh!]
\centerline{{\hfil\hfil
\psfig{figure=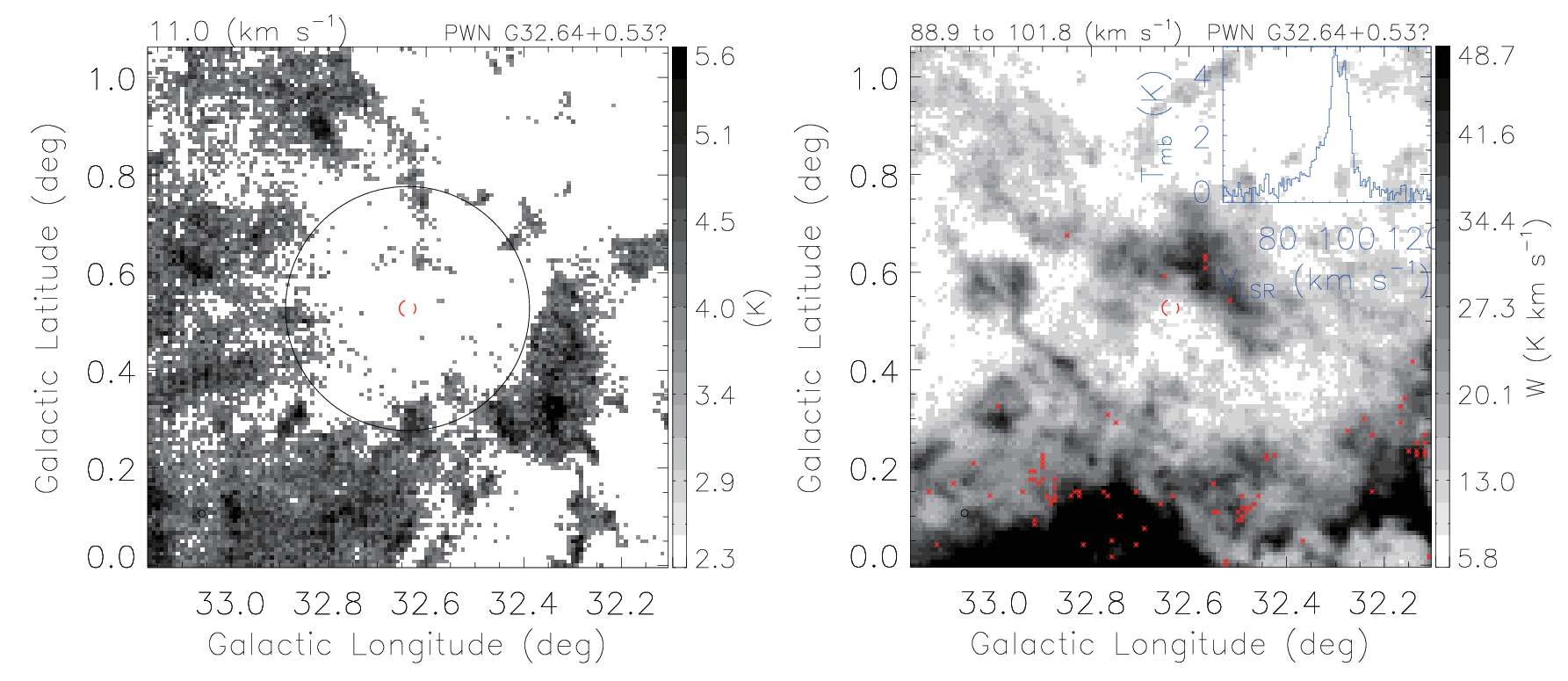,height=3.0in,angle=0, clip=}
\hfil\hfil}}
\caption{\twCO\ (J=1--0) intensity maps of PWN candidate G32.64+0.53, at the +11.0~\km\ps\ velocity channel (left) and integrated over the velocity range of +88.9 to +101.8~\km\ps\ (right).
Minimum values of \twCO\ emission intensity shown in colorbars are at $5\sigma$ confidence level.
%The red dashed circle denotes the location of the PWN. 
%Locations of broad lines identified by full criteria are denoted by red tiny stars.
%The extent of the PWN is denoted by red dashed circles.
Red tiny stars denote locations of broad lines identified by full criteria.
Red dashed circles denote the extent of the PWN.
Black solid circles delineate the cavity edge (left) and the surrounding shell (right).
The average \twCO~(J=1--0) spectrum of broad line emission from the surrounding shell is shown in the upper right corner of the right panel.
}
\label{f:pwn32}
\end{figure*}
{\bf G32.64+0.53}: 
Considering the large X-ray absorbing column density toward this source, a distance of 7~kpc, located at the tangent point of the Scutum arm, was assumed \citep{Gotthelf+2011}.
As indicated by the SCC, there is a cavity-like structure at $\sim$+11.0~\km\ps\ distributed around this PWN candidate (see Figure~\ref{f:pwn32}); however, no corresponding broad line emission is detected.
Molecular gas at a minor peak of the SCC at $\sim$+97.8~\km\ps\ also shows possible spatial correlation with the PWN candidate, which is roughly surrounding it in the north. It belongs to the +98.0~\km\ps\ component, which is the only broad line component identified in the nearby region by full criteria (see Figure~\ref{f:pwn32}).
Based on the spatial correlation result as well as the kinematic evidence, we suggest that the PWN candidate is associated with the +98.0~\km\ps\ MC, hence, locates at a kinematic distance of 5.2$\pm0.5$ kpc. The radius of the surrounding shell-like structure is $\sim$9 pc. %\\%r=0.1deg
\\
\begin{figure*}[ptbh!]
\centerline{{\hfil\hfil
\psfig{figure=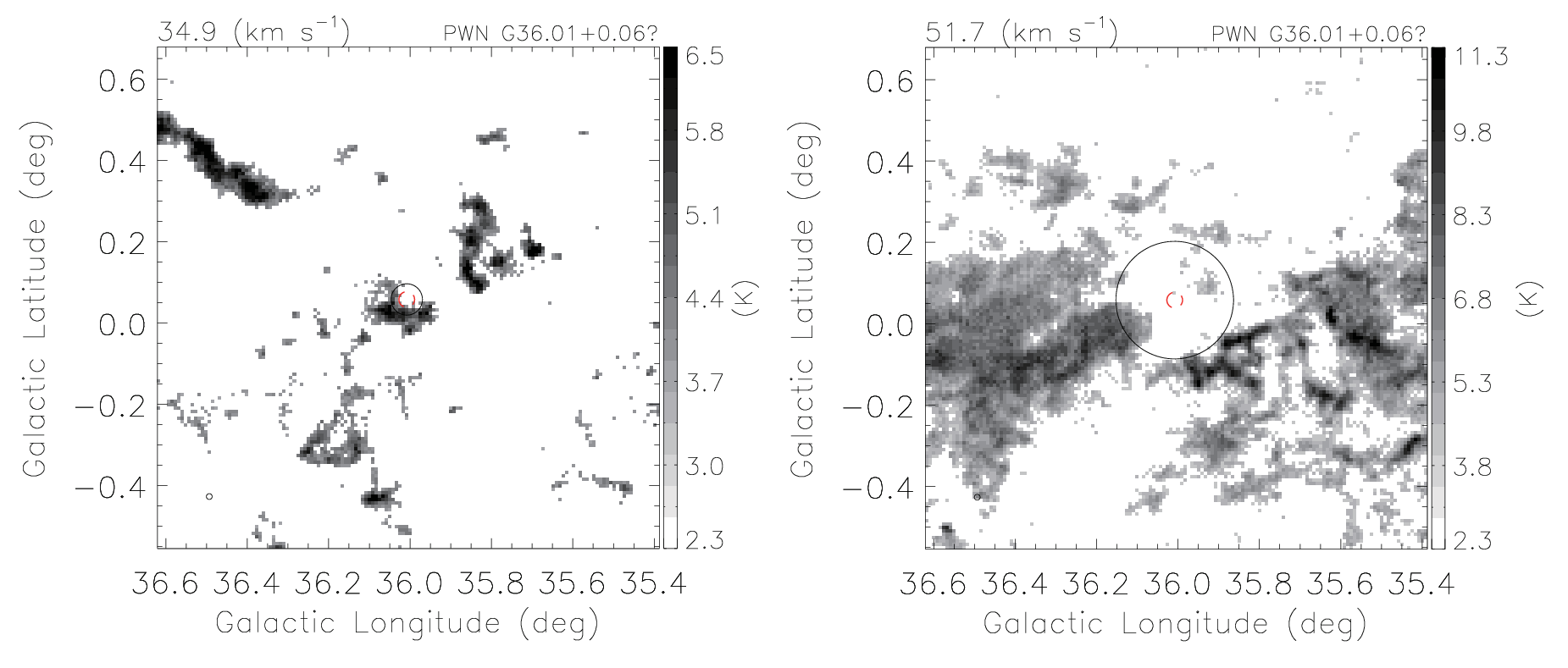,height=2.5in,angle=0, clip=}
\hfil\hfil}}
\caption{\twCO\ (J=1--0) intensity maps of PWN candidate G36.01+0.06, at +34.9~\km\ps\ (left) and +51.7~\km\ps\ (right) velocity channels.
Minimum values of \twCO\ emission intensity shown in colorbars are at $5\sigma$ confidence level.
%Locations of broad lines identified by full criteria are denoted by tiny stars.
%The extent of the PWN is denoted by red dashed circles.
Red dashed circles denote the extent of the PWN.
Black solid circles delineate edges of surrounding MCs.
%Average \twCO (J=1--0) spectrum of broad line emission on the surrounding shell is shown in the upper right corner of the right panel.
}
\label{f:pwn36}
\end{figure*}
{\bf G36.01+0.06}: 
This PWN candidate was detected as spatially extended $\gamma$-ray source, i.e.\ HESS J1857+026 \citep{Aharonian+2008}, powered by PSR J1856+0245 \citep{Hessels+2008}.
The distance of PSR J1856+0245 was estimated as $\sim$9~kpc based on its radio dispersion measure \citep{CordesLazio2002,Rousseau+2012}.
%There is no significant spatial correlation between the PWN candidate and MCs found.
Molecular gas at the peak of the SCC at $\sim$+34.9~\km\ps\ is roughly surrounding the southeastern half of the PWN candidate (see left panel in Figure~\ref{f:pwn36}), which belongs to the +37.7~\km\ps\ component with corresponding broad lines detected. Nevertheless, with the kinematic distance estimated as 2.4$\pm0.2$ kpc for the +37.7~\km\ps\ component, the radius of the surrounding molecular shell is just $\sim$1.6 pc.
Molecular gas at a peak of the SCC at $\sim$+51.7~\km\ps\ is also distributed around the PWN candidate (see right panel in Figure~\ref{f:pwn36}), which seems to be spatially correlated with very-high-energy $\gamma$-ray emission \citep[see Figure~7 in][]{Aharonian+2008}. 
It belongs to the +58.1~\km\ps\ component with corresponding broad lines identified.
%Molecular gases at both velocities are with corresponding broad line emission detected.
%Moreover, molecular gas at $\sim$+51.7~\km\ps\ 
If this source is associated with the +58.1~\km\ps\ component, its kinematic distance can be estimated as 9.5$\pm0.5$ kpc, and the distance to the edge of surrounding MCs is $\sim$24 pc. %r=0.145deg
It is consistent with the dispersion measure distance of the associated pulsar. %\\
%together with broad line emission indicate an association between the PWN and the $\sim$+37.7~\km\ps\ velocity component. As shown in Figure~\ref{f:pwn36} consistent with TeV emission. 
\\
{\bf G47.38-3.88}: The distance of PSR J1932+1059 (i.e.\ PSR B1929+10) was determined as 0.361$^{+0.010}_{-0.008}$~kpc based on parallax measurements \citep{Chatterjee+2004}.
Molecular gas at the peak of the SCC at $\sim$+6.7~\km\ps\ is roughly surrounding the PWN in the north at a radius of about $2.^{\circ}3$.
We identify no broad line emission around this PWN even by partial criteria.
If the PWN is associated with molecular gas at $\sim$+6.7~\km\ps, its near and far kinematic distances can be estimated as 0.16$\pm0.54$ and 10.8$\pm0.6$ kpc, respectively. The near kinematic distance is consistent with the parallax distance of the associated pulsar. The radius of the $\sim$+6.7~\km\ps\ diffuse shell-like structure at 0.361 kpc is $\sim$14 pc. %\\
\\
{\bf G59.2-4.70}: %The pulsar distance was derived as 1.5 kpc based on its dispersion measure.
The distance of the associated pulsar was derived as 2.5$\pm1.0$ kpc based on its radio dispersion measure \citep{CordesLazio2002, Huang+2012}.	%or 1.5kpc by Stappers+2003
\twCO\ (J=1--0) emission is very weak toward this PWN, and we found no clear shell or cavity-like structure in its vicinity. %\\
\\
\begin{figure*}[ptbh!]
\centerline{{\hfil\hfil
\psfig{figure=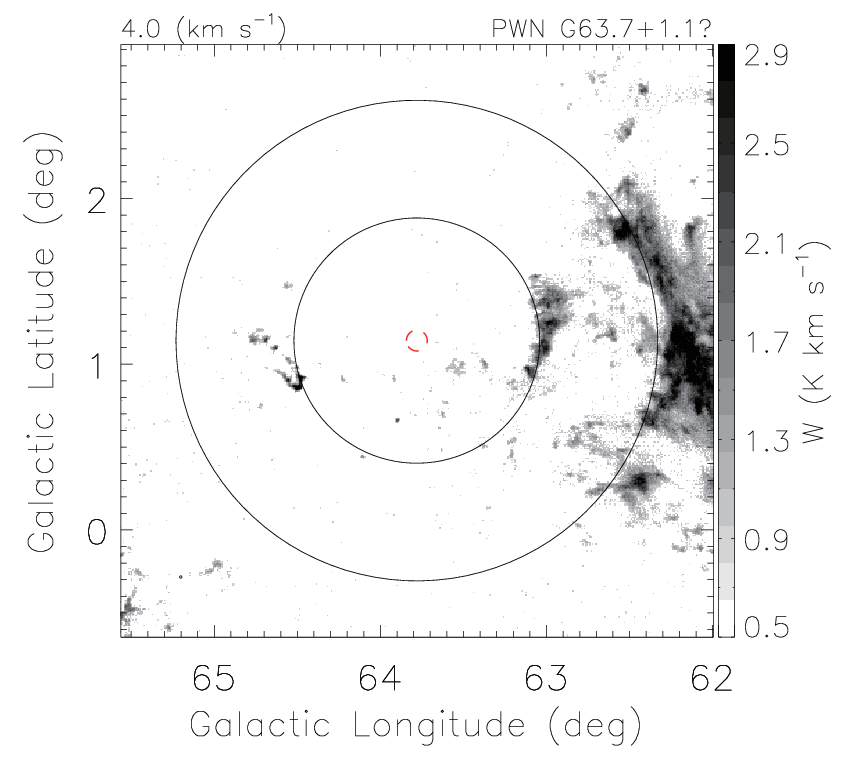,height=4.0in,angle=0, clip=}
\hfil\hfil}}
\caption{\twCO\ (J=1--0) intensity map of PWN candidate G63.7+1.1, at the +4.0~\km\ps\ velocity channel.
The minimum value of \twCO\ emission intensity shown in the colorbar is at $5\sigma$ level.
%Locations of broad lines identified by full criteria are denoted by tiny stars.
The extent of the PWN is denoted by a red dashed circle.
Black solid circles delineate surrounding shell-like structures, with radii of 0.$^\circ$74 and 1.$^\circ$45.
%Average \twCO (J=1--0) spectrum of broad line emission on the surrounding shell is shown in the upper right corner of the right panel.
}
\label{f:pwn63}
\end{figure*}
{\bf G63.7+1.1}:
This PWN candidate may be powered by the neutron star candidate 3XMM J194753.4+274357 \citep[CXO J194753.3+274351;][]{Matheson+2016}.
%It was suggested to be associated with MC at a systemic velocity of 
\cite{Wallace+1997} suggested that the PWN candidate is associated with an MC at the systemic velocity of +21~\km\ps. 
%, and locates at the tangent point at a distance of 3.8 kpc. 
However, as noted by \cite{Matheson+2016}, R.\ Kothes et al.\ (2016, in preparation) suggested an association with the +13~\km\ps\ MC.
Here, indicated by the SCC, we detect a double shell structure at $\sim$+4.0~\km\ps\ (see Figure~\ref{f:pwn63}), which seems to be associated with the PWN candidate.
It belongs to the +5.6~\km\ps\ component that is the only component with broad line candidates identified by partial criteria.
If the PWN candidate is associated with the +5.6~\km\ps\ component, its kinematic distance can be estimated as 7.2$\pm0.7$~kpc. Radii of the double shell structure are $\sim$88 and $\sim$176 pc. %\\%r=0.7 and 1.4 deg
\\
\begin{figure*}[ptbh!]
\centerline{{\hfil\hfil
\psfig{figure=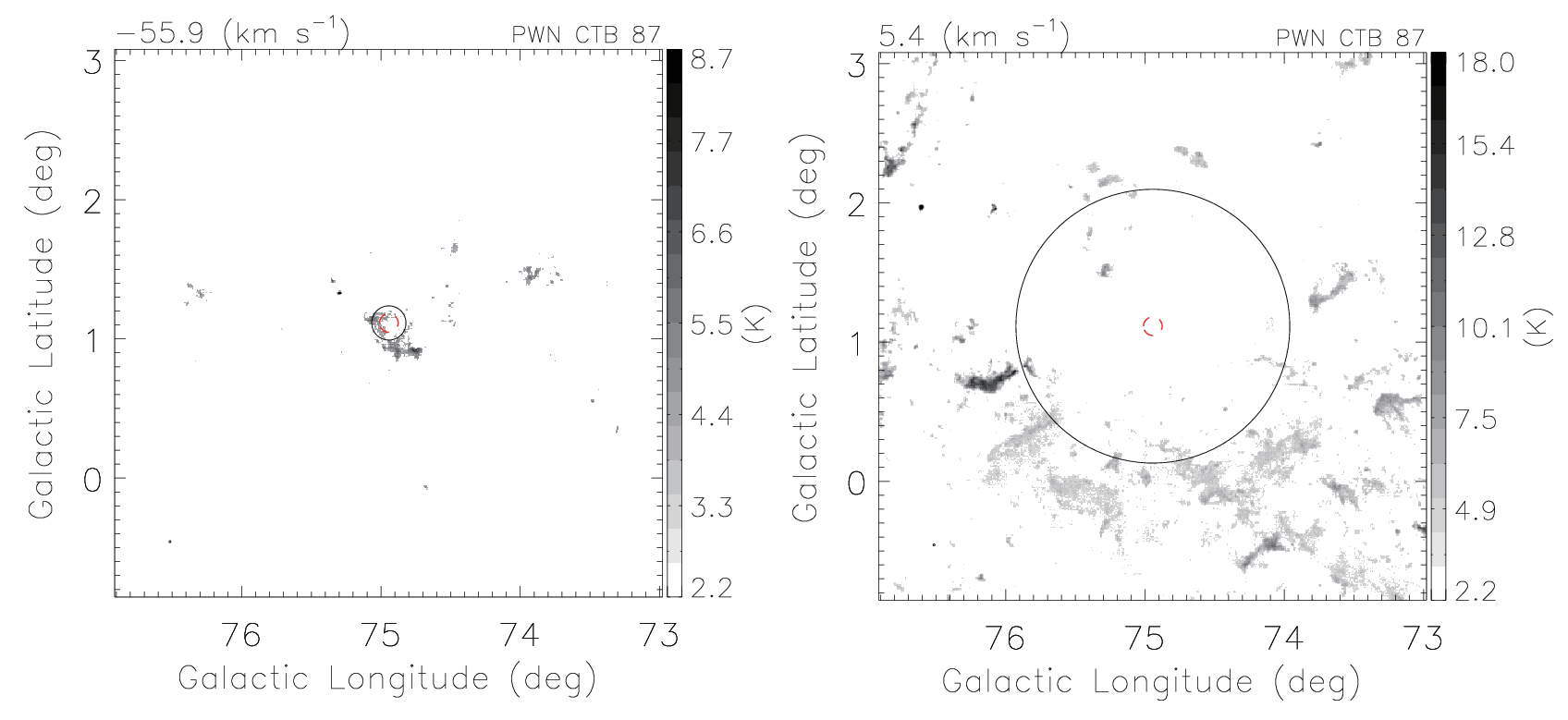,height=3.0in,angle=0, clip=}
\hfil\hfil}}
\caption{\twCO\ (J=1--0) intensity maps of PWN CTB~87, at $-$55.9~\km\ps\ (left) and +5.4~\km\ps\ (right) velocity channels.
Minimum values of \twCO\ emission intensity shown in colorbars are at $5\sigma$ confidence level.
%Locations of broad lines identified by full criteria are denoted by tiny stars.
%The extent of the PWN is denoted by red dashed circles.
Red dashed circles denote the extent of the PWN.
Black solid circles delineate the surrounding shell-like (left) and cavity-like (right) structures.
%Average \twCO (J=1--0) spectrum of broad line emission on the surrounding shell is shown in the upper right corner of the right panel.
}
\label{f:pwn74}
\end{figure*}
{\bf CTB~87}: Based on \HI\ absorption measurements and CO observations of nearby MCs, the PWN was suggested to be associated with $\sim$$-$58~\km\ps\ MC, and its distance was estimated to be 6.1 kpc \citep[][see Table~\ref{tab:snrpre} for more information]{KazesCaswell1977, HuangThaddeus1986, Cho+1994, Kothes+2003, Liu+2018}.
%\cite{Liu+2018} also found a $\sim$37$'$ \HI\ superbubble at $\sim$$-$64~\km\ps\ surrounding the PWN, which may be associated with the PWN.
Here, the SCC peaks at $\sim$+5.4~\km\ps; however, surrounding molecular gas shows no clear cavity or shell-like structure (see right panel in Figure~\ref{f:pwn74}). 
It belongs to the +0.5~\km\ps\ component with corresponding broad line points identified in the southwest.
Nevertheless, a shell-like structure surrounding the PWN is found at a peak of the SCC at $\sim$$-$55.9~\km\ps, with the center shifted a little to the south (see left panel in Figure~\ref{f:pwn74}). Molecular gas at $\sim$$-$55.9~\km\ps\ belongs to the $-$51.6~\km\ps\ component, of which broad line candidates are identified by partial criteria to be located a little farther to the northwest.
The northern part of the shell-like structure was detected by \cite{Liu+2018}.
Based on these evidences, the PWN is probably associated with the $-$51.6~\km\ps\ component. This is consistent with results in previous works, and the corresponding kinematic distance estimated in previous work is 6.1 kpc \citep{FosterRoutledge2002, Kothes+2003}.
%The PWN is probably associated with the surrounding shell-like structure at $\sim$$-$55.9~\km\ps, with the systemic velocity consistent with the previous results, hence, locates at the distance of 6.1 kpc as that estimated in previous work \citep{FosterRoutledge2002, Kothes+2003}.
The radius of the entire shell-like structure around the PWN is $\sim$13 pc at the distance of 6.1 kpc. %\\ %r_in=0.123deg
\\
{\bf G75.23+0.12}: The distance to PSR J2021+3651 is controversial, different distance estimations gave different results, e.g., $1.8^{+1.7}_{-1.4}$ kpc by hydrogen absorbing column density measurements on X-ray observations and an adopted $N_{\rm H}$-D relation \citep{Kirichenko+2015}, $\sim$12 kpc by the pulsar's dispersion measure \citep[e.g.,][]{Roberts+2002}, 3--4 kpc by comparing X-ray absorption column density with the total Galactic \HI\ column density \citep{VanEtten+2008}, and $\sim$1 kpc by using the empirical $\gamma$-ray "pseudo-distance" relation \citep[e.g.,][]{SazParkinson+2010}.
We find that molecular gas at the peak of the SCC at $\sim$+5.6~\km\ps\ is distributed around the PWN; however, it shows no clear shell or cavity-like structure. It belongs to the +6.0~\km\ps\ component with corresponding broad line candidates identified only by partial criteria, which are at a distance from the PWN.
If the PWN is associated with the +6.0~\km\ps\ component, its kinematic distance can be estimated as 3.4$\pm0.4$ kpc. %\\
\\
{\bf G80.22+1.02}: \cite{Camilo+2009} derived the dispersion measure distance of PSR 2032+4127 as $\sim$3.6 kpc using the electron density model of \cite{CordesLazio2002}; however, they suggested that the actual distance to the pulsar is $\sim$1.8 kpc based on a detailed $\gamma$-ray flux estimate. %hence, located within the Cyg OB2 stellar association.
We find that molecular gas at the peak of the SCC at $\sim$$-$2.9~\km\ps\ is distributed around the PWN at a radius of $\sim$0.$^\circ$29. It belongs to the $-$3.0~\km\ps\ component with corresponding broad lines identified. Note that some broad lines may be originated from overlapped \HII\ regions.
%however, we cannot distinguish the origion of these broad lines from overlapped \HII\ regions. %, which is not significant.
%There is no significant spatial correlation found by the SCC.
If the PWN is associated with the $-$3.0~\km\ps\ component, its kinematic distance can be estimated as 2.6$\pm0.3$ kpc. The distance to the edge of surrounding molecular gas is $\sim$13 pc. %\\
\\
\begin{figure*}[ptbh!]
\centerline{{\hfil\hfil
\psfig{figure=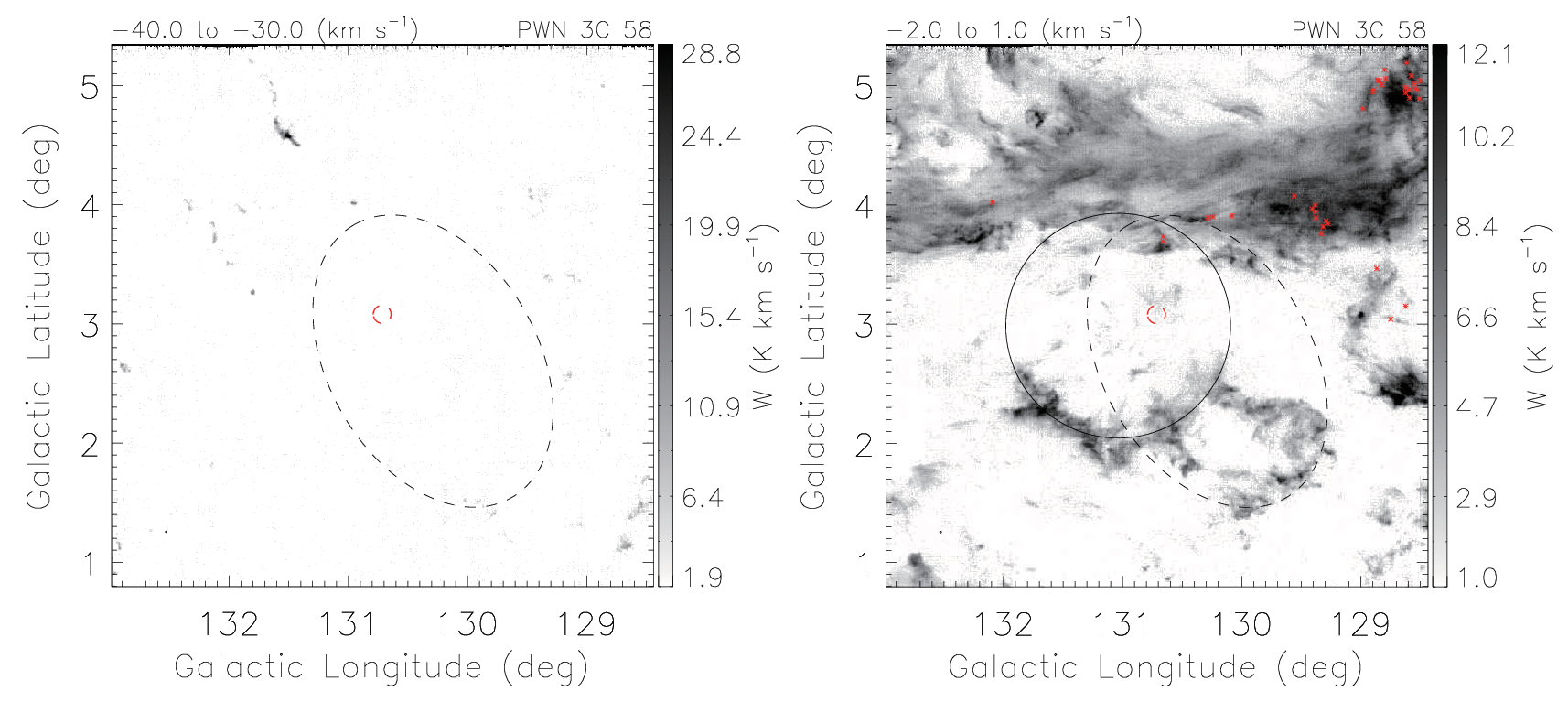,height=2.5in,angle=0, clip=}
\hfil\hfil}}
\caption{\twCO\ (J=1--0) intensity maps of 3C 58, integrated over velocity ranges of $-$40 to $-$30~\km\ps\ (left) and $-$2.0 to +1.0~\km\ps\ (right).
Minimum values of \twCO\ emission intensity shown in colorbars are at $3\sigma$ confidence level.
%Locations of broad lines at $-$2.5~\km\ps\ identified by full criteria are denoted by red tiny stars. % in the right panel.
%The extent of the PWN is denoted by red dashed circles.
Red tiny stars denote locations of broad lines at $-$2.5~\km\ps\ identified by full criteria.
Red dashed circles denote the extent of the PWN.
The black solid circle in the right panel delineates the CO shell, and black dashed ellipses delineate the \HI\ bipolar shell found by \citep{Wallace+1994}.
%cavity edge (left) and the surrounding shell (middle and right).
%Average \twCO (J=1--0) spectrum of broad line emission on the surrounding shell is shown in the upper right corner of the right panel.
}
\label{f:pwn130}
\end{figure*}
{\bf 3C 58}: The minimum velocity of \HI\ absorption features toward 3C 58 inferred a systemic velocity of $\sim$$-$36~\km\ps\ \citep{GreenGull1982, Roberts+1993, Kothes2013}, correspondingly, the kinematic distance was estimated as $\sim$2 kpc \citep{Kothes2013}.
A large bipolar shell structure was detected in \HI\ observations around $\sim$$-$36~\km\ps, which was suggested to be associated with the PWN \citep{Wallace+1994, Kothes2013}.
%approximately $3^\circ.3\times1^\circ.5$ in angular size, 
As shown in Figure~\ref{f:pwn130}, we detect very little CO emission associated with the \HI\ bipolar shell around $-$36~\km\ps. However, more CO emission was detected along the \HI\ bipolar shell around $-$2.5~\km\ps, and there are broad lines at $-$2.5~\km\ps\ identified on its northern part. %which may be a random matching.
The SCC peaks at $\sim$$-$7.3~\km\ps\ that belongs to the $-$2.5~\km\ps\ velocity component. The $-$2.5~\km\ps\ component also presents a clear CO shell structure accompanied with some inward filaments and outward protrusions. The kinematic distance of the $-$2.5~\km\ps\ component is estimated as 0.49$\pm0.18$ kpc, hence, located in the Local Arm.
The radius of the $-$2.5~\km\ps\ CO shell is about 56$'$, i.e.\ $\sim$8 pc at the distance of 0.49 kpc, and the offset of its center from 3C 58 is about 20$'$ (i.e.\ $\sim$2.9 pc).
Considering the \HI\ absorption results, the CO shell is probably not associated with the PWN.
The CO shell of such size may originate from a wind-blown bubble of an O- or B-type star. Further analysis is needed to better understand the distribution of MCs in this region. %\\
\\
{\bf G141.2+5.0}: The kinematic distance of this PWN was determined as 4.0$\pm0.5$ kpc, based on the association with a broken \HI\ shell at a systemic velocity of $-53$~\km\ps, which expands at +6~\km\ps\ and with a radius of about 6$'$ \citep{Kothes+2014}. %of which the radius is about 6$'$, %expanding at $\sim$6~\km\ps
We detect no significant \twCO (J=1--0) emission associated with the \HI\ shell here.
We also find no significant spatial correlation result and identify no broad line near the PWN even by partial criteria.
%{\bf Mushroom}: The parallax distance of PSR B0355+54 was estimated as $1.04^{+0.21}_{-0.16}$ \citep{Chatterjee+2004}. We find no spatial correlation or kinematic evidence supporting the association between the PWN and MCs.\\
\\
\begin{figure*}[ptbh!]
\centerline{{\hfil\hfil
\psfig{figure=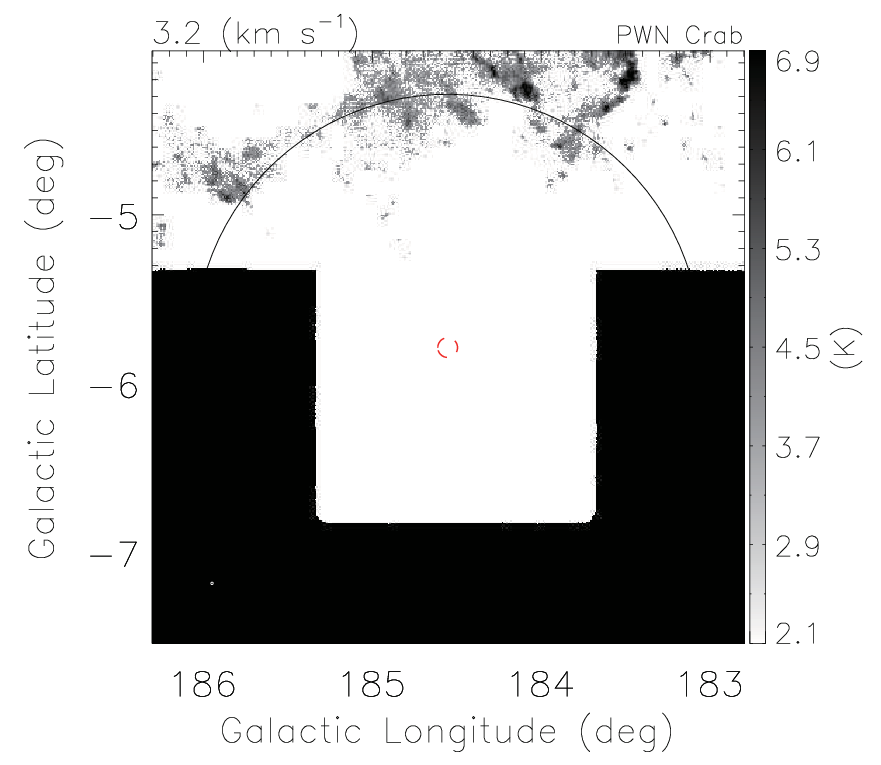,height=3.5in,angle=0, clip=}
\hfil\hfil}}
\caption{\twCO\ (J=1--0) intensity map of the Crab, at the +3.2~\km\ps\ velocity channel.
The minimum value of \twCO\ emission intensity shown in the colorbar is at $5\sigma$ level.
%Locations of broad lines identified by full criteria are denoted by tiny stars.
The extent of the PWN is denoted by a red dashed circle.
The black solid circle delineates the diffuse shell-like structure.
%Average \twCO (J=1--0) spectrum of broad line emission on the surrounding shell is shown in the upper right corner of the right panel.
}
\label{f:crab}
\end{figure*}
{\bf Crab}: As the distance to the Crab is not well constrained, a nominal distance of 2 kpc was commonly adopted \citep{Trimble1968}, but recent GAIA studies of the central pulsar suggested that this distance might be greater \citep{Bailer-Jones+2018, FraserBoubert2019}.
In our observation, molecular gas as the peak of the SCC at $\sim$+3.2~\km\ps\ is roughly surrounding this PWN in the north, of which the radius is $\sim$1.$^\circ$5 (see Figure~\ref{f:crab}). No corresponding broad line is identified even by partial criteria.
Note that the south of the region near the PWN, where the opposite part of the northern molecular shell might be, is not fully covered here.
The PWN may be associated with molecular gas at $\sim$+3.2~\km\ps.
Since the direction of the PWN is toward the anti-Galactic center, its kinematic distance cannot be determined well. Nevertheless, the kinematic distance of molecular gas at $\sim$+3.2~\km\ps\ is estimated as 1.59$\pm0.03$ kpc, correspondingly, the radius of the surrounding shell-like structure is $\sim$42 pc. %\\
\\
\begin{figure*}[ptbh!]
\centerline{{\hfil\hfil
\psfig{figure=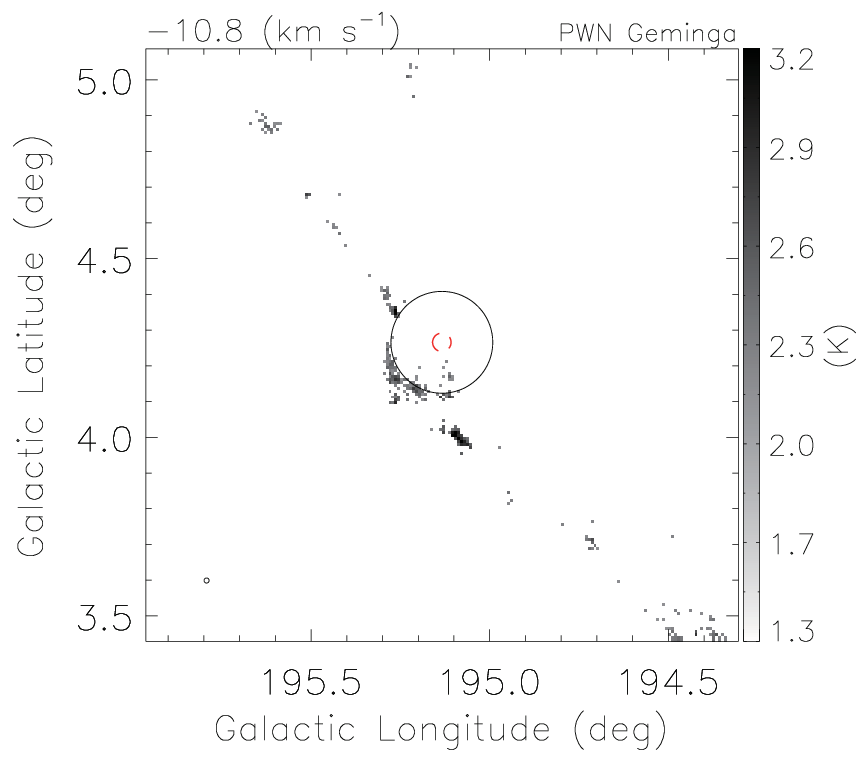,height=3.5in,angle=0, clip=}
\hfil\hfil}}
\caption{\twCO\ (J=1--0) intensity map of the Geminga, at the $-$10.8~\km\ps\ velocity channel.
The minimum value of \twCO\ emission intensity as shown in the colorbar is at $3\sigma$ level.
%Locations of broad lines identified by full criteria are denoted by tiny stars.
The extent of the PWN is denoted by a red dashed circle.
The black solid circle delineates the shell-like structure.
%Average \twCO (J=1--0) spectrum of broad line emission on the surrounding shell is shown in the upper right corner of the right panel.
}
\label{f:pwn195}
\end{figure*}
{\bf Geminga}: The distance to the Geminga pulsar by a latest parallax measurement was given as $250^{+120}_{-62}$ pc \citep{Faherty+2007}.
Here, the SCC indicates a possible association between the PWN and molecular gas at $\sim$$-$10.8~\km\ps, which presents a shell-like structure surrounding the PWN as shown in Figure~\ref{f:pwn195}. 
This shell-like structure seems to be a curved section of a long molecular filament that spreads from the northeast to the southwest. 
If the PWN and molecular gas at $\sim$$-$10.8~\km\ps\ are associated and both locate at the parallax distance of 250 pc, the radius of the shell-like structure would be only $\sim$0.7 pc. %\\	%r=0.15deg

%\end{appendix}

%\end{CJK*}

\begin{thebibliography}{276}
\expandafter\ifx\csname natexlab\endcsname\relax\def\natexlab#1{#1}\fi

\bibitem[{{Aharonian} {et~al.}(2008{\natexlab{a}}){Aharonian}, {Akhperjanian},
  {Bazer-Bachi}, {Behera}, {Beilicke}, {Benbow}, {Berge}, {Bernl{\"o}hr},
  {Boisson}, {Bolz}, {Borrel}, {Braun}, {Brion}, {Brown}, {B{\"u}hler},
  {Bulik}, {B{\"u}sching}, {Boutelier}, {Carrigan}, {Chadwick}, {Chounet},
  {Clapson}, {Coignet}, {Cornils}, {Costamante}, {Degrange}, {Dickinson},
  {Djannati-Ata{\"\i}}, {Domainko}, {O'C. Drury}, {Dubus}, {Dyks}, {Egberts},
  {Emmanoulopoulos}, {Espigat}, {Farnier}, {Feinstein}, {Fiasson},
  {F{\"o}rster}, {Fontaine}, {Fukui}, {Funk}, {Funk}, {F{\"u}{\ss}ling},
  {Gallant}, {Giebels}, {Glicenstein}, {Gl{\"u}ck}, {Goret}, {Hadjichristidis},
  {Hauser}, {Hauser}, {Heinzelmann}, {Henri}, {Hermann}, {Hinton}, {Hoffmann},
  {Hofmann}, {Holleran}, {Hoppe}, {Horns}, {Jacholkowska}, {de Jager},
  {Kendziorra}, {Kerschhaggl}, {Kh{\'e}lifi}, {Komin}, {Kosack}, {Lamanna},
  {Latham}, {Le Gallou}, {Lemi{\`e}re}, {Lemoine-Goumard}, {Lenain}, {Lohse},
  {Martin}, {Martineau-Huynh}, {Marcowith}, {Masterson}, {Maurin}, {McComb},
  {Moderski}, {Moriguchi}, {Moulin}, {de Naurois}, {Nedbal}, {Nolan}, {Olive},
  {Orford}, {Osborne}, {Ostrowski}, {Panter}, {Pedaletti}, {Pelletier},
  {Petrucci}, {Pita}, {P{\"u}hlhofer}, {Punch}, {Ranchon}, {Raubenheimer},
  {Raue}, {Rayner}, {Reimer}, {Renaud}, {Ripken}, {Rob}, {Rolland},
  {Rosier-Lees}, {Rowell}, {Rudak}, {Ruppel}, {Sahakian}, {Santangelo},
  {Saug{\'e}}, {Schlenker}, {Schlickeiser}, {Schr{\"o}der}, {Schwanke},
  {Schwarzburg}, {Schwemmer}, {Shalchi}, {Sol}, {Spangler}, {Stawarz},
  {Steenkamp}, {Stegmann}, {Superina}, {Takeuchi}, {Tam}, {Tavernet},
  {Terrier}, {van Eldik}, {Vasileiadis}, {Venter}, {Vialle}, {Vincent},
  {Vivier}, {V{\"o}lk}, {Volpe}, {Wagner}, \& {Ward}}]{Aharonian+2008a}
{Aharonian}, F., {Akhperjanian}, A.~G., {Bazer-Bachi}, A.~R., {et~al.}
  2008{\natexlab{a}}, \aap, 481, 401

\bibitem[{{Aharonian} {et~al.}(2008{\natexlab{b}}){Aharonian}, {Akhperjanian},
  {Barres de Almeida}, {Bazer-Bachi}, {Behera}, {Beilicke}, {Benbow},
  {Bernl{\"o}hr}, {Boisson}, {Bolz}, {Borrel}, {Braun}, {Brion}, {Brown},
  {B{\"u}hler}, {Bulik}, {B{\"u}sching}, {Boutelier}, {Carrigan}, {Chadwick},
  {Chounet}, {Clapson}, {Coignet}, {Cornils}, {Costamante}, {Dalton},
  {Degrange}, {Dickinson}, {Djannati-Ata{\"\i}}, {Domainko}, {Drury}, {Dubois},
  {Dubus}, {Dyks}, {Egberts}, {Emmanoulopoulos}, {Espigat}, {Farnier},
  {Feinstein}, {Fiasson}, {F{\"o}rster}, {Fontaine}, {Funk}, {F{\"u}{\ss}ling},
  {Gallant}, {Giebels}, {Glicenstein}, {Gl{\"u}ck}, {Goret}, {Hadjichristidis},
  {Hauser}, {Hauser}, {Heinzelmann}, {Henri}, {Hermann}, {Hinton}, {Hoffmann},
  {Hofmann}, {Holleran}, {Hoppe}, {Horns}, {Jacholkowska}, {de Jager}, {Jung},
  {Katarzy{\'n}ski}, {Kendziorra}, {Kerschhaggl}, {Kh{\'e}lifi}, {Keogh},
  {Komin}, {Kosack}, {Lamanna}, {Latham}, {Lemi{\`e}re}, {Lemoine-Goumard},
  {Lenain}, {Lohse}, {Martin}, {Martineau-Huynh}, {Marcowith}, {Masterson},
  {Maurin}, {Maurin}, {McComb}, {Moderski}, {Moulin}, {de Naurois}, {Nedbal},
  {Nolan}, {Ohm}, {Olive}, {de O{\~n}a Wilhelmi}, {Orford}, {Osborne},
  {Ostrowski}, {Panter}, {Pedaletti}, {Pelletier}, {Petrucci}, {Pita},
  {P{\"u}hlhofer}, {Punch}, {Ranchon}, {Raubenheimer}, {Raue}, {Rayner},
  {Renaud}, {Ripken}, {Rob}, {Rolland}, {Rosier-Lees}, {Rowell}, {Rudak},
  {Ruppel}, {Sahakian}, {Santangelo}, {Schlickeiser}, {Sch{\"o}ck},
  {Schr{\"o}der}, {Schwanke}, {Schwarzburg}, {Schwemmer}, {Shalchi}, {Sol},
  {Spangler}, {Stawarz}, {Steenkamp}, {Stegmann}, {Superina}, {Tam},
  {Tavernet}, {Terrier}, {van Eldik}, {Vasileiadis}, {Venter}, {Vialle},
  {Vincent}, {Vivier}, {V{\"o}lk}, {Volpe}, {Wagner}, {Ward}, {Zdziarski}, \&
  {Zech}}]{Aharonian+2008}
{Aharonian}, F., {Akhperjanian}, A.~G., {Barres de Almeida}, U., {et~al.}
  2008{\natexlab{b}}, \aap, 477, 353

\bibitem[{{Aharonian} {et~al.}(2007){Aharonian}, {Akhperjanian}, {Bazer-Bachi},
  {Behera}, {Beilicke}, {Benbow}, {Berge}, {Bernl{\"o}hr}, {Boisson}, {Bolz},
  {Borrel}, {Braun}, {Brion}, {Brown}, {B{\"u}hler}, {B{\"u}sching},
  {Boutelier}, {Carrigan}, {Chadwick}, {Chounet}, {Coignet}, {Cornils},
  {Costamante}, {Degrange}, {Dickinson}, {Djannati-Ata{\"\i}}, {Domainko},
  {O'C. Drury}, {Dubus}, {Egberts}, {Emmanoulopoulos}, {Espigat}, {Farnier},
  {Feinstein}, {Fiasson}, {F{\"o}rster}, {Fontaine}, {Funk}, {Funk},
  {F{\"u}{\ss}ling}, {Gallant}, {Giebels}, {Glicenstein}, {Gl{\"u}ck}, {Goret},
  {Hadjichristidis}, {Hauser}, {Hauser}, {Heinzelmann}, {Henri}, {Hermann},
  {Hinton}, {Hoffmann}, {Hofmann}, {Holleran}, {Hoppe}, {Horns},
  {Jacholkowska}, {de Jager}, {Kendziorra}, {Kerschhaggl}, {Kh{\'e}lifi},
  {Komin}, {Kosack}, {Lamanna}, {Latham}, {Le Gallou}, {Lemi{\`e}re},
  {Lemoine-Goumard}, {Lohse}, {Martin}, {Martineau-Huynh}, {Marcowith},
  {Masterson}, {Maurin}, {McComb}, {Moulin}, {de Naurois}, {Nedbal}, {Nolan},
  {Noutsos}, {Olive}, {Orford}, {Osborne}, {Panter}, {Pedaletti}, {Pelletier},
  {Petrucci}, {Pita}, {P{\"u}hlhofer}, {Punch}, {Ranchon}, {Raubenheimer},
  {Raue}, {Rayner}, {Reimer}, {Ripken}, {Rob}, {Rolland}, {Rosier-Lees},
  {Rowell}, {Ruppel}, {Sahakian}, {Santangelo}, {Saug{\'e}}, {Schlenker},
  {Schlickeiser}, {Schr{\"o}der}, {Schwanke}, {Schwarzburg}, {Schwemmer},
  {Shalchi}, {Sol}, {Spangler}, {Steenkamp}, {Stegmann}, {Superina}, {Tam},
  {Tavernet}, {Terrier}, {Tluczykont}, {van Eldik}, {Vasileiadis}, {Venter},
  {Vialle}, {Vincent}, {V{\"o}lk}, {Wagner}, {Ward}, {Moriguchi}, \&
  {Fukui}}]{Aharonian+2007}
{Aharonian}, F.~A., {Akhperjanian}, A.~G., {Bazer-Bachi}, A.~R., {et~al.} 2007,
  \aap, 469, L1

\bibitem[{{Ambrocio-Cruz} {et~al.}(2017){Ambrocio-Cruz}, {Rosado}, {de la
  Fuente}, {Silva}, \& {Blanco-Pi{\~n}on}}]{Ambrocio-Cruz+2017}
{Ambrocio-Cruz}, P., {Rosado}, M., {de la Fuente}, E., {Silva}, R., \&
  {Blanco-Pi{\~n}on}, A. 2017, \mnras, 472, 51

\bibitem[{{Anderl} {et~al.}(2014){Anderl}, {Gusdorf}, \&
  {G{\"u}sten}}]{Anderl+2014}
{Anderl}, S., {Gusdorf}, A., \& {G{\"u}sten}, R. 2014, \aap, 569, A81

\bibitem[{{Anderson} {et~al.}(2014){Anderson}, {Bania}, {Balser}, {Cunningham},
  {Wenger}, {Johnstone}, \& {Armentrout}}]{Anderson+2014}
{Anderson}, L.~D., {Bania}, T.~M., {Balser}, D.~S., {et~al.} 2014, \apjs, 212,
  1

\bibitem[{{Anderson} {et~al.}(2017){Anderson}, {Wang}, {Bihr}, {Rugel},
  {Beuther}, {Bigiel}, {Churchwell}, {Glover}, {Goodman}, {Henning}, {Heyer},
  {Klessen}, {Linz}, {Longmore}, {Menten}, {Ott}, {Roy}, {Soler}, {Stil}, \&
  {Urquhart}}]{Anderson+2017}
{Anderson}, L.~D., {Wang}, Y., {Bihr}, S., {et~al.} 2017, \aap, 605, A58

\bibitem[{{Arendt}(1989)}]{ArendtRichard1989}
{Arendt}, R.~G. 1989, \apjs, 70, 181

\bibitem[{{Arias} {et~al.}(2019){Arias}, {Dom{\v{c}}ek}, {Zhou}, \&
  {Vink}}]{Arias+2019}
{Arias}, M., {Dom{\v{c}}ek}, V., {Zhou}, P., \& {Vink}, J. 2019, \aap, 627, A75

\bibitem[{{Arikawa} {et~al.}(1999){Arikawa}, {Tatematsu}, {Sekimoto}, \&
  {Takahashi}}]{Arikawa+1999}
{Arikawa}, Y., {Tatematsu}, K., {Sekimoto}, Y., \& {Takahashi}, T. 1999, \pasj,
  51, L7

\bibitem[{{Bailer-Jones} {et~al.}(2018){Bailer-Jones}, {Rybizki}, {Fouesneau},
  {Mantelet}, \& {Andrae}}]{Bailer-Jones+2018}
{Bailer-Jones}, C.~A.~L., {Rybizki}, J., {Fouesneau}, M., {Mantelet}, G., \&
  {Andrae}, R. 2018, \aj, 156, 58

\bibitem[{{Bastian} {et~al.}(2010){Bastian}, {Covey}, \&
  {Meyer}}]{Bastian+2010}
{Bastian}, N., {Covey}, K.~R., \& {Meyer}, M.~R. 2010, \araa, 48, 339

\bibitem[{{Beaumont} {et~al.}(2011){Beaumont}, {Williams}, \&
  {Goodman}}]{Beaumont+2011}
{Beaumont}, C.~N., {Williams}, J.~P., \& {Goodman}, A.~A. 2011, \apj, 741, 14

\bibitem[{{Beuther} {et~al.}(2016){Beuther}, {Bihr}, {Rugel}, {Johnston},
  {Wang}, {Walter}, {Brunthaler}, {Walsh}, {Ott}, {Stil}, {Henning},
  {Schierhuber}, {Kainulainen}, {Heyer}, {Goldsmith}, {Anderson}, {Longmore},
  {Klessen}, {Glover}, {Urquhart}, {Plume}, {Ragan}, {Schneider},
  {McClure-Griffiths}, {Menten}, {Smith}, {Roy}, {Shanahan}, {Nguyen-Luong}, \&
  {Bigiel}}]{Beuther+2016}
{Beuther}, H., {Bihr}, S., {Rugel}, M., {et~al.} 2016, \aap, 595, A32

\bibitem[{{Brinkmann} {et~al.}(1996){Brinkmann}, {Aschenbach}, \&
  {Kawai}}]{Brinkmann+1996}
{Brinkmann}, W., {Aschenbach}, B., \& {Kawai}, N. 1996, \aap, 312, 306

\bibitem[{{Brogan} {et~al.}(2000){Brogan}, {Frail}, {Goss}, \&
  {Troland}}]{Brogan+2000}
{Brogan}, C.~L., {Frail}, D.~A., {Goss}, W.~M., \& {Troland}, T.~H. 2000, \apj,
  537, 875

\bibitem[{{Brogan} {et~al.}(2006){Brogan}, {Gelfand}, {Gaensler}, {Kassim}, \&
  {Lazio}}]{Brogan+2006}
{Brogan}, C.~L., {Gelfand}, J.~D., {Gaensler}, B.~M., {Kassim}, N.~E., \&
  {Lazio}, T.~J.~W. 2006, \apjl, 639, L25

\bibitem[{{Burton} {et~al.}(1988){Burton}, {Geballe}, {Brand}, \&
  {Webster}}]{Burton+1988}
{Burton}, M.~G., {Geballe}, T.~R., {Brand}, P.~W.~J.~L., \& {Webster}, A.~S.
  1988, \mnras, 231, 617

\bibitem[{{Byun} {et~al.}(2006){Byun}, {Koo}, {Tatematsu}, \&
  {Sunada}}]{Byun+2006}
{Byun}, D.-Y., {Koo}, B.-C., {Tatematsu}, K., \& {Sunada}, K. 2006, \apj, 637,
  283

\bibitem[{{Cai} {et~al.}(2009){Cai}, {Yang}, \& {Lu}}]{Cai+2009}
{Cai}, Z.-Y., {Yang}, J., \& {Lu}, D.-R. 2009, \caa, 33, 393

\bibitem[{{Camilo} {et~al.}(2009){Camilo}, {Ray}, {Ransom}, {Burgay},
  {Johnson}, {Kerr}, {Gotthelf}, {Halpern}, {Reynolds}, {Romani}, {Demorest},
  {Johnston}, {van Straten}, {Saz Parkinson}, {Ziegler}, {Dormody}, {Thompson},
  {Smith}, {Harding}, {Abdo}, {Crawford}, {Freire}, {Keith}, {Kramer},
  {Roberts}, {Weltevrede}, \& {Wood}}]{Camilo+2009}
{Camilo}, F., {Ray}, P.~S., {Ransom}, S.~M., {et~al.} 2009, \apj, 705, 1

\bibitem[{{Castelletti} {et~al.}(2016){Castelletti}, {Giacani}, \&
  {Petriella}}]{Castelletti+2016}
{Castelletti}, G., {Giacani}, E., \& {Petriella}, A. 2016, \aap, 587, A71

\bibitem[{{Castelletti} {et~al.}(2013){Castelletti}, {Supan}, {Dubner},
  {Joshi}, \& {Surnis}}]{Castelletti+2013}
{Castelletti}, G., {Supan}, L., {Dubner}, G., {Joshi}, B.~C., \& {Surnis},
  M.~P. 2013, \aap, 557, L15

\bibitem[{{Castelletti} {et~al.}(2017){Castelletti}, {Supan}, {Petriella},
  {Giacani}, \& {Joshi}}]{Castelletti+2017}
{Castelletti}, G., {Supan}, L., {Petriella}, A., {Giacani}, E., \& {Joshi},
  B.~C. 2017, \aap, 602, A31

\bibitem[{{Caswell} {et~al.}(1975{\natexlab{a}}){Caswell}, {Clark}, \&
  {Crawford}}]{Caswell+1975}
{Caswell}, J.~L., {Clark}, D.~H., \& {Crawford}, D.~F. 1975{\natexlab{a}},
  Australian Journal of Physics Astrophysical Supplement, 37, 39

\bibitem[{{Caswell} {et~al.}(1975{\natexlab{b}}){Caswell}, {Murray}, {Roger},
  {Cole}, \& {Cooke}}]{Caswell+1975a}
{Caswell}, J.~L., {Murray}, J.~D., {Roger}, R.~S., {Cole}, D.~J., \& {Cooke},
  D.~J. 1975{\natexlab{b}}, \aap, 45, 239

\bibitem[{{Celli} {et~al.}(2019){Celli}, {Morlino}, {Gabici}, \&
  {Aharonian}}]{Celli+2019}
{Celli}, S., {Morlino}, G., {Gabici}, S., \& {Aharonian}, F.~A. 2019, \mnras,
  487, 3199

\bibitem[{{Chatterjee} {et~al.}(2004){Chatterjee}, {Cordes}, {Vlemmings},
  {Arzoumanian}, {Goss}, \& {Lazio}}]{Chatterjee+2004}
{Chatterjee}, S., {Cordes}, J.~M., {Vlemmings}, W.~H.~T., {et~al.} 2004, \apj,
  604, 339

\bibitem[{{Chen} {et~al.}(2017{\natexlab{a}}){Chen}, {Liu}, {Ren}, {Yuan},
  {Huang}, {Yu}, {Xiang}, {Wang}, {Tian}, \& {Zhang}}]{Chenbq+2017}
{Chen}, B.~Q., {Liu}, X.~W., {Ren}, J.~J., {et~al.} 2017{\natexlab{a}}, \mnras,
  472, 3924

\bibitem[{{Chen} {et~al.}(2017{\natexlab{b}}){Chen}, {Xiong}, \&
  {Yang}}]{Chenxp+2017}
{Chen}, X., {Xiong}, F., \& {Yang}, J. 2017{\natexlab{b}}, \aap, 604, A13

\bibitem[{{Chen} {et~al.}(2014){Chen}, {Jiang}, {Zhou}, {Su}, {Zhou}, {Li}, \&
  {Zhang}}]{Chen+2014}
{Chen}, Y., {Jiang}, B., {Zhou}, P., {et~al.} 2014, in IAU Symposium, Vol. 296,
  Supernova Environmental Impacts, ed. A.~{Ray} \& R.~A. {McCray}, 170--177

\bibitem[{{Chen} {et~al.}(2013){Chen}, {Zhou}, \& {Chu}}]{Chen+2013}
{Chen}, Y., {Zhou}, P., \& {Chu}, Y.-H. 2013, \apjl, 769, L16

\bibitem[{{Chen} {et~al.}(2023){Chen}, {Sefako}, {Yang}, {Jiang}, {Su},
  {Zhang}, \& {Zhou}}]{Chenzw+2023}
{Chen}, Z., {Sefako}, R., {Yang}, Y., {et~al.} 2023, \mnras, 525, 107

\bibitem[{{Chevalier}(1999)}]{Chevalier1999}
{Chevalier}, R.~A. 1999, \apj, 511, 798

\bibitem[{{Cho} {et~al.}(1994){Cho}, {Kim}, \& {Fukui}}]{Cho+1994}
{Cho}, S.-H., {Kim}, K.~T., \& {Fukui}, Y. 1994, \aj, 108, 634

\bibitem[{{Claussen} {et~al.}(1997){Claussen}, {Frail}, {Goss}, \&
  {Gaume}}]{Claussen+1997}
{Claussen}, M.~J., {Frail}, D.~A., {Goss}, W.~M., \& {Gaume}, R.~A. 1997, \apj,
  489, 143

\bibitem[{{Condon} {et~al.}(1991){Condon}, {Broderick}, \&
  {Seielstad}}]{Condon+1991}
{Condon}, J.~J., {Broderick}, J.~J., \& {Seielstad}, G.~A. 1991, \aj, 102, 2041

\bibitem[{{Condon} {et~al.}(1994){Condon}, {Broderick}, {Seielstad}, {Douglas},
  \& {Gregory}}]{Condon+1994}
{Condon}, J.~J., {Broderick}, J.~J., {Seielstad}, G.~A., {Douglas}, K., \&
  {Gregory}, P.~C. 1994, \aj, 107, 1829

\bibitem[{{Condon} {et~al.}(1998){Condon}, {Cotton}, {Greisen}, {Yin},
  {Perley}, {Taylor}, \& {Broderick}}]{Condon+1998}
{Condon}, J.~J., {Cotton}, W.~D., {Greisen}, E.~W., {et~al.} 1998, \aj, 115,
  1693

\bibitem[{{Condon} {et~al.}(1993){Condon}, {Griffith}, \&
  {Wright}}]{Condon+1993}
{Condon}, J.~J., {Griffith}, M.~R., \& {Wright}, A.~E. 1993, \aj, 106, 1095

\bibitem[{{Cong}(1977)}]{Cong1977}
{Cong}, H.~I.~L. 1977, PhD thesis, Columbia University

\bibitem[{{Cordes} \& {Lazio}(2002)}]{CordesLazio2002}
{Cordes}, J.~M., \& {Lazio}, T.~J.~W. 2002, arXiv e-prints, arXiv:0207156

\bibitem[{{Cornett} {et~al.}(1977){Cornett}, {Chin}, \& {Knapp}}]{Cornett+1977}
{Cornett}, R.~H., {Chin}, G., \& {Knapp}, G.~R. 1977, \aap, 54, 889

\bibitem[{{Dame}(2011)}]{Dame2011}
{Dame}, T.~M. 2011, arXiv e-prints, arXiv:1101.1499

\bibitem[{{de O{\~n}a Wilhelmi} {et~al.}(2020){de O{\~n}a Wilhelmi}, {Sushch},
  {Brose}, {Mestre}, {Su}, \& {Zanin}}]{deOnaWilhelmi+2020}
{de O{\~n}a Wilhelmi}, E., {Sushch}, I., {Brose}, R., {et~al.} 2020, \mnras,
  497, 3581

\bibitem[{{de Wilt} {et~al.}(2017){de Wilt}, {Rowell}, {Walsh}, {Burton},
  {Rathborne}, {Fukui}, {Kawamura}, \& {Aharonian}}]{deWilt+2017}
{de Wilt}, P., {Rowell}, G., {Walsh}, A.~J., {et~al.} 2017, \mnras, 468, 2093

\bibitem[{{Dell'Ova} {et~al.}(2020){Dell'Ova}, {Gusdorf}, {Gerin}, {Riquelme},
  {G{\"u}sten}, {Noriega-Crespo}, {Tram}, {Houde}, {Guillard}, {Lehmann},
  {Lesaffre}, {Louvet}, {Marcowith}, \& {Padovani}}]{DellOva+2020}
{Dell'Ova}, P., {Gusdorf}, A., {Gerin}, M., {et~al.} 2020, \aap, 644, A64

\bibitem[{{Denoyer}(1979{\natexlab{a}})}]{Denoyer1979b}
{Denoyer}, L.~K. 1979{\natexlab{a}}, \apjl, 232, L165

\bibitem[{{Denoyer}(1979{\natexlab{b}})}]{Denoyer1979a}
---. 1979{\natexlab{b}}, \apjl, 228, L41

\bibitem[{{Dickman} {et~al.}(1992){Dickman}, {Snell}, {Ziurys}, \&
  {Huang}}]{Dickman+1992}
{Dickman}, R.~L., {Snell}, R.~L., {Ziurys}, L.~M., \& {Huang}, Y.-L. 1992,
  \apj, 400, 203

\bibitem[{{Dobashi} {et~al.}(2019){Dobashi}, {Shimoikura}, {Endo}, {Takagi},
  {Nakamura}, {Shimajiri}, \& {Bernard}}]{Dobashi+2019}
{Dobashi}, K., {Shimoikura}, T., {Endo}, N., {et~al.} 2019, \pasj, 71, S11

\bibitem[{{Dokara} {et~al.}(2021){Dokara}, {Brunthaler}, {Menten}, {Dzib},
  {Reich}, {Cotton}, {Anderson}, {Chen}, {Gong}, {Medina}, {Ortiz-Le{\'o}n},
  {Rugel}, {Urquhart}, {Wyrowski}, {Yang}, {Beuther}, {Billington}, {Csengeri},
  {Carrasco-Gonz{\'a}lez}, \& {Roy}}]{Dokara+2021}
{Dokara}, R., {Brunthaler}, A., {Menten}, K.~M., {et~al.} 2021, \aap, 651, A86

\bibitem[{{Downes} {et~al.}(1986){Downes}, {Pauls}, \& {Salter}}]{Downes+1986}
{Downes}, A.~J.~B., {Pauls}, T., \& {Salter}, C.~J. 1986, \mnras, 218, 393

\bibitem[{{Dubner} {et~al.}(1999){Dubner}, {Giacani}, {Reynoso}, {Goss},
  {Roth}, \& {Green}}]{Dubner+1999}
{Dubner}, G., {Giacani}, E., {Reynoso}, E., {et~al.} 1999, \aj, 118, 930

\bibitem[{{Dubner} {et~al.}(2004){Dubner}, {Giacani}, {Reynoso}, \&
  {Par{\'o}n}}]{Dubner+2004}
{Dubner}, G., {Giacani}, E., {Reynoso}, E., \& {Par{\'o}n}, S. 2004, \aap, 426,
  201

\bibitem[{{Dubner} {et~al.}(1998){Dubner}, {Holdaway}, {Goss}, \&
  {Mirabel}}]{Dubner+1998}
{Dubner}, G.~M., {Holdaway}, M., {Goss}, W.~M., \& {Mirabel}, I.~F. 1998, \aj,
  116, 1842

\bibitem[{{Duvidovich} {et~al.}(2020){Duvidovich}, {Petriella}, \&
  {Giacani}}]{Duvidovich+2020}
{Duvidovich}, L., {Petriella}, A., \& {Giacani}, E. 2020, \mnras, 491, 5732

\bibitem[{{Eger} {et~al.}(2011){Eger}, {Rowell}, {Kawamura}, {Fukui},
  {Rolland}, \& {Stegmann}}]{Eger+2011}
{Eger}, P., {Rowell}, G., {Kawamura}, A., {et~al.} 2011, \aap, 526, A82

\bibitem[{{Faherty} {et~al.}(2007){Faherty}, {Walter}, \&
  {Anderson}}]{Faherty+2007}
{Faherty}, J., {Walter}, F.~M., \& {Anderson}, J. 2007, \apss, 308, 225

\bibitem[{{Feldt} \& {Green}(1993)}]{FeldtGreen1993}
{Feldt}, C., \& {Green}, D.~A. 1993, \aap, 274, 421

\bibitem[{{Ferrand} \& {Safi-Harb}(2012)}]{FerrandSafi-Harb2012}
{Ferrand}, G., \& {Safi-Harb}, S. 2012, Advances in Space Research, 49, 1313

\bibitem[{{Foster} \& {Routledge}(2002)}]{FosterRoutledge2002}
{Foster}, T., \& {Routledge}, D. 2002, in Astronomical Society of the Pacific
  Conference Series, Vol. 276, Seeing Through the Dust: The Detection of HI and
  the Exploration of the ISM in Galaxies, ed. A.~R. {Taylor}, T.~L.
  {Landecker}, \& A.~G. {Willis}, 123

\bibitem[{{Frail} {et~al.}(2013){Frail}, {Claussen}, \&
  {M{\'e}hault}}]{Frail+2013}
{Frail}, D.~A., {Claussen}, M.~J., \& {M{\'e}hault}, J. 2013, \apjl, 773, L19

\bibitem[{{Frail} {et~al.}(1996){Frail}, {Goss}, {Reynoso}, {Giacani}, {Green},
  \& {Otrupcek}}]{Frail+1996}
{Frail}, D.~A., {Goss}, W.~M., {Reynoso}, E.~M., {et~al.} 1996, \aj, 111, 1651

\bibitem[{{Frail} {et~al.}(1994){Frail}, {Goss}, \& {Slysh}}]{Frail+1994}
{Frail}, D.~A., {Goss}, W.~M., \& {Slysh}, V.~I. 1994, \apjl, 424, L111

\bibitem[{{Fraser} \& {Boubert}(2019)}]{FraserBoubert2019}
{Fraser}, M., \& {Boubert}, D. 2019, \apj, 871, 92

\bibitem[{{Fukuda} {et~al.}(2014){Fukuda}, {Yoshiike}, {Sano}, {Torii},
  {Yamamoto}, {Acero}, \& {Fukui}}]{Fukuda+2014}
{Fukuda}, T., {Yoshiike}, S., {Sano}, H., {et~al.} 2014, \apj, 788, 94

\bibitem[{{Fukui} {et~al.}(2021){Fukui}, {Sano}, {Yamane}, {Hayakawa}, {Inoue},
  {Tachihara}, {Rowell}, \& {Einecke}}]{Fukui+2021}
{Fukui}, Y., {Sano}, H., {Yamane}, Y., {et~al.} 2021, \apj, 915, 84

\bibitem[{{Furst} {et~al.}(1989){Furst}, {Hummel}, {Reich}, {Sofue}, {Sieber},
  {Reif}, \& {Dettmar}}]{Furst+1989}
{Furst}, E., {Hummel}, E., {Reich}, W., {et~al.} 1989, \aap, 209, 361

\bibitem[{{Gaensler} {et~al.}(2003){Gaensler}, {Schulz}, {Kaspi}, {Pivovaroff},
  \& {Becker}}]{Gaensler+2003}
{Gaensler}, B.~M., {Schulz}, N.~S., {Kaspi}, V.~M., {Pivovaroff}, M.~J., \&
  {Becker}, W.~E. 2003, \apj, 588, 441

\bibitem[{{Gaensler} {et~al.}(2008){Gaensler}, {Tanna}, {Slane}, {Brogan},
  {Gelfand}, {McClure-Griffiths}, {Camilo}, {Ng}, \& {Miller}}]{Gaensler+2008}
{Gaensler}, B.~M., {Tanna}, A., {Slane}, P.~O., {et~al.} 2008, \apjl, 680, L37

\bibitem[{{Gao} {et~al.}(2011){Gao}, {Han}, {Reich}, {Reich}, {Sun}, \&
  {Xiao}}]{Gao+2011}
{Gao}, X.~Y., {Han}, J.~L., {Reich}, W., {et~al.} 2011, \aap, 529, A159

\bibitem[{{Gao} {et~al.}(2020){Gao}, {Reich}, {Reich}, {Hou}, \&
  {Han}}]{Gao+2020}
{Gao}, X.~Y., {Reich}, P., {Reich}, W., {Hou}, L.~G., \& {Han}, J.~L. 2020,
  \mnras, 493, 2188

\bibitem[{{Gerbrandt} {et~al.}(2014){Gerbrandt}, {Foster}, {Kothes},
  {Geisb{\"u}sch}, \& {Tung}}]{Gerbrandt+2014}
{Gerbrandt}, S., {Foster}, T.~J., {Kothes}, R., {Geisb{\"u}sch}, J., \& {Tung},
  A. 2014, \aap, 566, A76

\bibitem[{{Giacani} {et~al.}(2009){Giacani}, {Smith}, {Dubner}, {Loiseau},
  {Castelletti}, \& {Paron}}]{Giacani+2009}
{Giacani}, E., {Smith}, M.~J.~S., {Dubner}, G., {et~al.} 2009, \aap, 507, 841

\bibitem[{{Giacani} {et~al.}(1998){Giacani}, {Dubner}, {Cappa}, \&
  {Testori}}]{Giacani+1998}
{Giacani}, E.~B., {Dubner}, G., {Cappa}, C., \& {Testori}, J. 1998, \aaps, 133,
  61

\bibitem[{{Giuliani} {et~al.}(2011){Giuliani}, {Cardillo}, {Tavani}, {Fukui},
  {Yoshiike}, {Torii}, {Dubner}, {Castelletti}, {Barbiellini}, {Bulgarelli},
  {Caraveo}, {Costa}, {Cattaneo}, {Chen}, {Contessi}, {Del Monte},
  {Donnarumma}, {Evangelista}, {Feroci}, {Gianotti}, {Lazzarotto}, {Lucarelli},
  {Longo}, {Marisaldi}, {Mereghetti}, {Pacciani}, {Pellizzoni}, {Piano},
  {Picozza}, {Pittori}, {Pucella}, {Rapisarda}, {Rappoldi}, {Sabatini},
  {Soffitta}, {Striani}, {Trifoglio}, {Trois}, {Vercellone}, {Verrecchia},
  {Vittorini}, {Colafrancesco}, {Giommi}, \& {Bignami}}]{Giuliani+2011}
{Giuliani}, A., {Cardillo}, M., {Tavani}, M., {et~al.} 2011, \apjl, 742, L30

\bibitem[{{Gorham}(1990)}]{Gorham1990}
{Gorham}, P.~W. 1990, \apj, 364, 187

\bibitem[{{Goss} \& {Robinson}(1968)}]{GossRobinson1968}
{Goss}, W.~M., \& {Robinson}, B.~J. 1968, \aplett, 2, 81

\bibitem[{{Gotthelf} \& {Halpern}(2008)}]{GotthelfHalpern2008}
{Gotthelf}, E.~V., \& {Halpern}, J.~P. 2008, \apj, 681, 515

\bibitem[{{Gotthelf} {et~al.}(2011){Gotthelf}, {Halpern}, {Terrier}, \&
  {Mattana}}]{Gotthelf+2011}
{Gotthelf}, E.~V., {Halpern}, J.~P., {Terrier}, R., \& {Mattana}, F. 2011,
  \apjl, 729, L16

\bibitem[{{Gray}(1994)}]{Gray1994}
{Gray}, A.~D. 1994, \mnras, 270, 847

\bibitem[{{Green} {et~al.}(1997){Green}, {Frail}, {Goss}, \&
  {Otrupcek}}]{Green+1997}
{Green}, A.~J., {Frail}, D.~A., {Goss}, W.~M., \& {Otrupcek}, R. 1997, \aj,
  114, 2058

\bibitem[{{Green}(2019)}]{Green2019}
{Green}, D.~A. 2019, Journal of Astrophysics and Astronomy, 40, 36

\bibitem[{{Green} \& {Dewdney}(1992)}]{GreenDewdney1992}
{Green}, D.~A., \& {Dewdney}, P.~E. 1992, \mnras, 254, 686

\bibitem[{{Green} \& {Gull}(1982)}]{GreenGull1982}
{Green}, D.~A., \& {Gull}, S.~F. 1982, \nat, 299, 606

\bibitem[{{Green} {et~al.}(1988){Green}, {Gull}, {Tan}, \&
  {Simon}}]{Green+1988}
{Green}, D.~A., {Gull}, S.~F., {Tan}, S.~M., \& {Simon}, A.~J.~B. 1988, \mnras,
  231, 735

\bibitem[{{Gusdorf} {et~al.}(2012){Gusdorf}, {Anderl}, {G{\"u}sten}, {Stutzki},
  {H{\"u}bers}, {Hartogh}, {Heyminck}, \& {Okada}}]{Gusdorf+2012}
{Gusdorf}, A., {Anderl}, S., {G{\"u}sten}, R., {et~al.} 2012, \aap, 542, L19

\bibitem[{{Gusdorf} {et~al.}(2014){Gusdorf}, {G{\"u}sten}, {Anderl}, {Hezareh},
  \& {Wiesemeyer}}]{Gusdorf+2014}
{Gusdorf}, A., {G{\"u}sten}, R., {Anderl}, S., {Hezareh}, T., \& {Wiesemeyer},
  H. 2014, in Supernova Environmental Impacts, ed. A.~{Ray} \& R.~A. {McCray},
  Vol. 296, 178--182

\bibitem[{{Helfand} {et~al.}(2006){Helfand}, {Becker}, {White}, {Fallon}, \&
  {Tuttle}}]{Helfand+2006}
{Helfand}, D.~J., {Becker}, R.~H., {White}, R.~L., {Fallon}, A., \& {Tuttle},
  S. 2006, \aj, 131, 2525

\bibitem[{{Hessels} {et~al.}(2008){Hessels}, {Nice}, {Gaensler}, {Kaspi},
  {Lorimer}, {Champion}, {Lyne}, {Kramer}, {Cordes}, {Freire}, {Camilo},
  {Ransom}, {Deneva}, {Bhat}, {Cognard}, {Crawford}, {Jenet}, {Kasian},
  {Lazarus}, {van Leeuwen}, {McLaughlin}, {Stairs}, {Stappers}, \&
  {Venkataraman}}]{Hessels+2008}
{Hessels}, J.~W.~T., {Nice}, D.~J., {Gaensler}, B.~M., {et~al.} 2008, \apjl,
  682, L41

\bibitem[{{Hewitt} {et~al.}(2009){Hewitt}, {Rho}, {Andersen}, \&
  {Reach}}]{Hewitt+2009ir}
{Hewitt}, J.~W., {Rho}, J., {Andersen}, M., \& {Reach}, W.~T. 2009, \apj, 694,
  1266

\bibitem[{{Hewitt} \& {Yusef-Zadeh}(2009)}]{Hewitt+2009oh}
{Hewitt}, J.~W., \& {Yusef-Zadeh}, F. 2009, \apjl, 694, L16

\bibitem[{{Hewitt} {et~al.}(2008){Hewitt}, {Yusef-Zadeh}, \&
  {Wardle}}]{Hewitt+2008}
{Hewitt}, J.~W., {Yusef-Zadeh}, F., \& {Wardle}, M. 2008, \apj, 683, 189

\bibitem[{{Hewitt} {et~al.}(2006){Hewitt}, {Yusef-Zadeh}, {Wardle}, {Roberts},
  \& {Kassim}}]{Hewitt+2006}
{Hewitt}, J.~W., {Yusef-Zadeh}, F., {Wardle}, M., {Roberts}, D.~A., \&
  {Kassim}, N.~E. 2006, \apj, 652, 1288

\bibitem[{{Heyer} \& {Dame}(2015)}]{HeyerDame2015}
{Heyer}, M., \& {Dame}, T.~M. 2015, \araa, 53, 583

\bibitem[{{Hezareh} {et~al.}(2013){Hezareh}, {Wiesemeyer}, {Houde}, {Gusdorf},
  \& {Siringo}}]{Hezareh+2013}
{Hezareh}, T., {Wiesemeyer}, H., {Houde}, M., {Gusdorf}, A., \& {Siringo}, G.
  2013, \aap, 558, A45

\bibitem[{{Higgs} {et~al.}(1983){Higgs}, {Landecker}, \& {Roger}}]{Higgs+1983}
{Higgs}, L.~A., {Landecker}, T.~L., \& {Roger}, R.~S. 1983, \aj, 88, 97

\bibitem[{{Hoffman} {et~al.}(2005){Hoffman}, {Goss}, {Brogan}, \&
  {Claussen}}]{Hoffman+2005}
{Hoffman}, I.~M., {Goss}, W.~M., {Brogan}, C.~L., \& {Claussen}, M.~J. 2005,
  \apj, 627, 803

\bibitem[{{Huang} {et~al.}(2012){Huang}, {Kong}, {Takata}, {Hui}, {Lin}, \&
  {Cheng}}]{Huang+2012}
{Huang}, R.~H.~H., {Kong}, A.~K.~H., {Takata}, J., {et~al.} 2012, \apj, 760, 92

\bibitem[{{Huang} {et~al.}(1983){Huang}, {Dame}, \& {Thaddeus}}]{Huang+1983}
{Huang}, Y.~L., {Dame}, T.~M., \& {Thaddeus}, P. 1983, \apj, 272, 609

\bibitem[{{Huang} {et~al.}(1986){Huang}, {Dickman}, \& {Snell}}]{Huang+1986}
{Huang}, Y.~L., {Dickman}, R.~L., \& {Snell}, R.~L. 1986, \apjl, 302, L63

\bibitem[{{Huang} \& {Thaddeus}(1986)}]{HuangThaddeus1986}
{Huang}, Y.~L., \& {Thaddeus}, P. 1986, \apj, 309, 804

\bibitem[{{Hurley-Walker} {et~al.}(2017){Hurley-Walker}, {Callingham},
  {Hancock}, {Franzen}, {Hindson}, {Kapi{\'n}ska}, {Morgan}, {Offringa},
  {Wayth}, {Wu}, {Zheng}, {Murphy}, {Bell}, {Dwarakanath}, {For}, {Gaensler},
  {Johnston-Hollitt}, {Lenc}, {Procopio}, {Staveley-Smith}, {Ekers}, {Bowman},
  {Briggs}, {Cappallo}, {Deshpande}, {Greenhill}, {Hazelton}, {Kaplan},
  {Lonsdale}, {McWhirter}, {Mitchell}, {Morales}, {Morgan}, {Oberoi}, {Ord},
  {Prabu}, {Shankar}, {Srivani}, {Subrahmanyan}, {Tingay}, {Webster},
  {Williams}, \& {Williams}}]{Hurley-Walker+2017}
{Hurley-Walker}, N., {Callingham}, J.~R., {Hancock}, P.~J., {et~al.} 2017,
  \mnras, 464, 1146

\bibitem[{{Hurley-Walker} {et~al.}(2019{\natexlab{a}}){Hurley-Walker},
  {Gaensler}, {Leahy}, {Filipovi{\'c}}, {Hancock}, {Franzen}, {Offringa},
  {Callingham}, {Hindson}, {Wu}, {Bell}, {For}, {Johnston-Hollitt},
  {Kapi{\'n}ska}, {Morgan}, {Murphy}, {McKinley}, {Procopio}, {Staveley-Smith},
  {Wayth}, \& {Zheng}}]{Hurley-Walker+2019b}
{Hurley-Walker}, N., {Gaensler}, B.~M., {Leahy}, D.~A., {et~al.}
  2019{\natexlab{a}}, \pasa, 36, e048

\bibitem[{{Hurley-Walker} {et~al.}(2019{\natexlab{b}}){Hurley-Walker},
  {Hancock}, {Franzen}, {Callingham}, {Offringa}, {Hindson}, {Wu}, {Bell},
  {For}, {Gaensler}, {Johnston-Hollitt}, {Kapi{\'n}ska}, {Morgan}, {Murphy},
  {McKinley}, {Procopio}, {Staveley-Smith}, {Wayth}, \&
  {Zheng}}]{Hurley-Walker+2019a}
{Hurley-Walker}, N., {Hancock}, P.~J., {Franzen}, T.~M.~O., {et~al.}
  2019{\natexlab{b}}, \pasa, 36, e047

\bibitem[{{Inoue} {et~al.}(2012){Inoue}, {Yamazaki}, {Inutsuka}, \&
  {Fukui}}]{Inoue+2012}
{Inoue}, T., {Yamazaki}, R., {Inutsuka}, S.-i., \& {Fukui}, Y. 2012, \apj, 744,
  71

\bibitem[{{Inutsuka} {et~al.}(2015){Inutsuka}, {Inoue}, {Iwasaki}, \&
  {Hosokawa}}]{Inutsuka+2015}
{Inutsuka}, S.-i., {Inoue}, T., {Iwasaki}, K., \& {Hosokawa}, T. 2015, \aap,
  580, A49

\bibitem[{{Jeong} {et~al.}(2012){Jeong}, {Byun}, {Koo}, {Lee}, {Lee}, \&
  {Kang}}]{Jeong+2012}
{Jeong}, I.-G., {Byun}, D.-Y., {Koo}, B.-C., {et~al.} 2012, \apss, 342, 389

\bibitem[{{Jeong} {et~al.}(2013){Jeong}, {Koo}, {Cho}, {Kramer}, {Stutzki}, \&
  {Byun}}]{Jeong+2013}
{Jeong}, I.-G., {Koo}, B.-C., {Cho}, W.-K., {et~al.} 2013, \apj, 770, 105

\bibitem[{Jiang {et~al.}(2010)Jiang, Chen, Wang, Su, Zhou, Safi-Harb, \&
  DeLaney}]{Jiang+2010}
Jiang, B., Chen, Y., Wang, J., {et~al.} 2010, ApJ, 712, 1147

\bibitem[{{Johanson} \& {Kerton}(2009)}]{JohansonKerton2009}
{Johanson}, A.~K., \& {Kerton}, C.~R. 2009, \aj, 138, 1615

\bibitem[{{Junkes} {et~al.}(1992){Junkes}, {Fuerst}, \& {Reich}}]{Junkes+1992}
{Junkes}, N., {Fuerst}, E., \& {Reich}, W. 1992, \aaps, 96, 1

\bibitem[{{Kargaltsev} {et~al.}(2017){Kargaltsev}, {Pavlov}, {Klingler}, \&
  {Rangelov}}]{Kargaltsev+2017}
{Kargaltsev}, O., {Pavlov}, G.~G., {Klingler}, N., \& {Rangelov}, B. 2017,
  Journal of Plasma Physics, 83, 635830501

\bibitem[{{Kassim}(1988)}]{Kassim1988}
{Kassim}, N.~E. 1988, \apjl, 328, L55

\bibitem[{{Kazes} \& {Caswell}(1977)}]{KazesCaswell1977}
{Kazes}, I., \& {Caswell}, J.~L. 1977, \aap, 58, 449

\bibitem[{{Keohane} {et~al.}(1996){Keohane}, {Rudnick}, \&
  {Anderson}}]{Keohane+1996}
{Keohane}, J.~W., {Rudnick}, L., \& {Anderson}, M.~C. 1996, \apj, 466, 309

\bibitem[{{Kilpatrick} {et~al.}(2014){Kilpatrick}, {Bieging}, \&
  {Rieke}}]{Kilpatrick+2014}
{Kilpatrick}, C.~D., {Bieging}, J.~H., \& {Rieke}, G.~H. 2014, \apj, 796, 144

\bibitem[{{Kilpatrick} {et~al.}(2016){Kilpatrick}, {Bieging}, \&
  {Rieke}}]{Kilpatrick+2016}
---. 2016, \apj, 816, 1

\bibitem[{{Kirichenko} {et~al.}(2015){Kirichenko}, {Danilenko}, {Shternin},
  {Shibanov}, {Ryspaeva}, {Zyuzin}, {Durant}, {Kargaltsev}, {Pavlov}, \&
  {Cabrera-Lavers}}]{Kirichenko+2015}
{Kirichenko}, A., {Danilenko}, A., {Shternin}, P., {et~al.} 2015, \apj, 802, 17

\bibitem[{{Koo} {et~al.}(2006){Koo}, {Kang}, \& {Salter}}]{Koo+2006}
{Koo}, B.-C., {Kang}, J.-h., \& {Salter}, C.~J. 2006, \apjl, 643, L49

\bibitem[{{Koo} \& {Moon}(1997)}]{KooMoon1997}
{Koo}, B.-C., \& {Moon}, D.-S. 1997, \apj, 485, 263

\bibitem[{{Koo} {et~al.}(2001){Koo}, {Rho}, {Reach}, {Jung}, \&
  {Mangum}}]{Koo+2001}
{Koo}, B.-C., {Rho}, J., {Reach}, W.~T., {Jung}, J., \& {Mangum}, J.~G. 2001,
  \apj, 552, 175

\bibitem[{{Koo} {et~al.}(1993){Koo}, {Yun}, {Ho}, \& {Lee}}]{Koo+1993}
{Koo}, B.-C., {Yun}, M.-S., {Ho}, P. T.~P., \& {Lee}, Y. 1993, \apj, 417, 196

\bibitem[{{Koo} {et~al.}(2008){Koo}, {McKee}, {Lee}, {Lee}, {Lee}, {Moon},
  {Hong}, {Kaneda}, \& {Onaka}}]{Koo+2008}
{Koo}, B.-C., {McKee}, C.~F., {Lee}, J.-J., {et~al.} 2008, \apjl, 673, L147

\bibitem[{{Koralesky} {et~al.}(1998){Koralesky}, {Frail}, {Goss}, {Claussen},
  \& {Green}}]{Koralesky+1998}
{Koralesky}, B., {Frail}, D.~A., {Goss}, W.~M., {Claussen}, M.~J., \& {Green},
  A.~J. 1998, \aj, 116, 1323

\bibitem[{{Kothes}(2013)}]{Kothes2013}
{Kothes}, R. 2013, \aap, 560, A18

\bibitem[{{Kothes} {et~al.}(2003){Kothes}, {Reich}, {Foster}, \&
  {Byun}}]{Kothes+2003}
{Kothes}, R., {Reich}, W., {Foster}, T., \& {Byun}, D.-Y. 2003, \apj, 588, 852

\bibitem[{{Kothes} {et~al.}(2018){Kothes}, {Sun}, {Gaensler}, \&
  {Reich}}]{Kothes+2018}
{Kothes}, R., {Sun}, X., {Gaensler}, B., \& {Reich}, W. 2018, \apj, 852, 54

\bibitem[{{Kothes} {et~al.}(2014){Kothes}, {Sun}, {Reich}, \&
  {Foster}}]{Kothes+2014}
{Kothes}, R., {Sun}, X.~H., {Reich}, W., \& {Foster}, T.~J. 2014, \apjl, 784,
  L26

\bibitem[{{Kothes} {et~al.}(2001){Kothes}, {Uyaniker}, \&
  {Pineault}}]{Kothes+2001}
{Kothes}, R., {Uyaniker}, B., \& {Pineault}, S. 2001, \apj, 560, 236

\bibitem[{{Kothes} {et~al.}(2002){Kothes}, {Uyaniker}, \& {Yar}}]{Kothes+2002}
{Kothes}, R., {Uyaniker}, B., \& {Yar}, A. 2002, \apj, 576, 169

\bibitem[{{Kuriki} {et~al.}(2018){Kuriki}, {Sano}, {Kuno}, {Seta}, {Yamane},
  {Inaba}, {Nagaya}, {Yoshiike}, {Okawa}, {Tsutsumi}, {Hattori}, {Kohno},
  {Fujita}, {Nishimura}, {Ohama}, {Matsuo}, {Tsuda}, {Torii}, {Minamidani},
  {Umemoto}, {Rowell}, {Bamba}, {Tachihara}, \& {Fukui}}]{Kuriki+2018}
{Kuriki}, M., {Sano}, H., {Kuno}, N., {et~al.} 2018, \apj, 864, 161

\bibitem[{{Landecker} {et~al.}(1987){Landecker}, {Dewdney}, {Vaneldik}, \&
  {Routledge}}]{Landecker+1987}
{Landecker}, T.~L., {Dewdney}, P.~E., {Vaneldik}, J.~F., \& {Routledge}, D.
  1987, \aj, 94, 111

\bibitem[{{Landecker} {et~al.}(1989){Landecker}, {Pineault}, {Routledge}, \&
  {Vaneldik}}]{Landecker+1989}
{Landecker}, T.~L., {Pineault}, S., {Routledge}, D., \& {Vaneldik}, J.~F. 1989,
  \mnras, 237, 277

\bibitem[{{Lau} {et~al.}(2017){Lau}, {Rowell}, {Burton}, {Fukui}, {Aharonian},
  {Oya}, {Vink}, {Ohm}, \& {Casanova}}]{Lau+2017}
{Lau}, J.~C., {Rowell}, G., {Burton}, M.~G., {et~al.} 2017, \mnras, 464, 3757

\bibitem[{{Leahy} {et~al.}(2014){Leahy}, {Green}, \& {Tian}}]{Leahy+2014}
{Leahy}, D., {Green}, K., \& {Tian}, W. 2014, \mnras, 438, 1813

\bibitem[{{Leahy} \& {Ranasinghe}(2016)}]{LeahyRanasinghe2016}
{Leahy}, D.~A., \& {Ranasinghe}, S. 2016, \apj, 817, 74

\bibitem[{{Leahy} {et~al.}(2008){Leahy}, {Tian}, \& {Wang}}]{Leahy+2008}
{Leahy}, D.~A., {Tian}, W., \& {Wang}, Q.~D. 2008, \aj, 136, 1477

\bibitem[{{Leahy} \& {Tian}(2007)}]{LeahyTian2007}
{Leahy}, D.~A., \& {Tian}, W.~W. 2007, \aap, 461, 1013

\bibitem[{{Leahy} \& {Tian}(2008{\natexlab{a}})}]{LeahyTian2008a}
---. 2008{\natexlab{a}}, \aap, 480, L25

\bibitem[{{Leahy} \& {Tian}(2008{\natexlab{b}})}]{LeahyTian2008}
---. 2008{\natexlab{b}}, \aj, 135, 167

\bibitem[{{Lee} {et~al.}(2009){Lee}, {Moon}, {Koo}, {Lee}, \&
  {Matthews}}]{Lee+2009}
{Lee}, H.-G., {Moon}, D.-S., {Koo}, B.-C., {Lee}, J.-J., \& {Matthews}, K.
  2009, \apj, 691, 1042

\bibitem[{{Lee} {et~al.}(2012{\natexlab{a}}){Lee}, {Koo}, {Snell}, {Yun},
  {Heyer}, \& {Burton}}]{Lee+2012}
{Lee}, J.-J., {Koo}, B.-C., {Snell}, R.~L., {et~al.} 2012{\natexlab{a}}, \apj,
  749, 34

\bibitem[{{Lee} {et~al.}(2004){Lee}, {Koo}, \& {Tatematsu}}]{Lee+2004}
{Lee}, J.-J., {Koo}, B.-C., \& {Tatematsu}, K. 2004, \apjl, 605, L113

\bibitem[{{Lee} {et~al.}(2012{\natexlab{b}}){Lee}, {Koo}, \& {Lee}}]{Lee+2012b}
{Lee}, J.-W., {Koo}, B.-C., \& {Lee}, J.-E. 2012{\natexlab{b}}, Journal of
  Korean Astronomical Society, 45, 117

\bibitem[{{Lee} {et~al.}(2020){Lee}, {Koo}, \& {Lee}}]{Lee+2020}
{Lee}, Y.-H., {Koo}, B.-C., \& {Lee}, J.-J. 2020, \aj, 160, 263

\bibitem[{{Liszt}(2009)}]{Liszt2009}
{Liszt}, H.~S. 2009, \aap, 508, 1331

\bibitem[{{Liu} {et~al.}(2017){Liu}, {Chen}, {Zhang}, {Liu}, {He}, {Zhou},
  {Zhou}, \& {Su}}]{Liu+2017}
{Liu}, B., {Chen}, Y., {Zhang}, X., {et~al.} 2017, \apj, 851, 37

\bibitem[{{Liu} {et~al.}(2018){Liu}, {Chen}, {Chen}, {Zhou}, {Wang}, \&
  {Su}}]{Liu+2018}
{Liu}, Q.-C., {Chen}, Y., {Chen}, B.-Q., {et~al.} 2018, \apj, 859, 173

\bibitem[{{Liu} {et~al.}(2020){Liu}, {Chen}, {Zhou}, {Zhang}, \&
  {Jiang}}]{Liu+2020}
{Liu}, Q.-C., {Chen}, Y., {Zhou}, P., {Zhang}, X., \& {Jiang}, B. 2020, \apj,
  892, 143

\bibitem[{{Lockett} {et~al.}(1999){Lockett}, {Gauthier}, \&
  {Elitzur}}]{Lockett+1999}
{Lockett}, P., {Gauthier}, E., \& {Elitzur}, M. 1999, \apj, 511, 235

\bibitem[{{Ma} {et~al.}(2019){Ma}, {Wang}, {Zhang}, {Li}, \& {Yang}}]{Ma+2019}
{Ma}, Y., {Wang}, H., {Zhang}, M., {Li}, C., \& {Yang}, J. 2019, \apj, 878, 44

\bibitem[{{Manchester} {et~al.}(2005){Manchester}, {Hobbs}, {Teoh}, \&
  {Hobbs}}]{Manchester+2005}
{Manchester}, R.~N., {Hobbs}, G.~B., {Teoh}, A., \& {Hobbs}, M. 2005, \aj, 129,
  1993

\bibitem[{{Matheson} {et~al.}(2016){Matheson}, {Safi-Harb}, \&
  {Kothes}}]{Matheson+2016}
{Matheson}, H., {Safi-Harb}, S., \& {Kothes}, R. 2016, \apj, 825, 134

\bibitem[{{Maxted} {et~al.}(2018){Maxted}, {Burton}, {Braiding}, {Rowell},
  {Sano}, {Voisin}, {Capasso}, {P{\"u}hlhofer}, \& {Fukui}}]{Maxted+2018}
{Maxted}, N., {Burton}, M., {Braiding}, C., {et~al.} 2018, \mnras, 474, 662

\bibitem[{{McDonnell} {et~al.}(2008){McDonnell}, {Wardle}, \&
  {Vaughan}}]{McDonnell+2008}
{McDonnell}, K.~E., {Wardle}, M., \& {Vaughan}, A.~E. 2008, \mnras, 390, 49

\bibitem[{{McEwen}(2016)}]{McEwen2016}
{McEwen}, B.~C. 2016, PhD thesis, The University of New Mexico

\bibitem[{{Nicholas} {et~al.}(2012){Nicholas}, {Rowell}, {Burton}, {Walsh},
  {Fukui}, {Kawamura}, \& {Maxted}}]{Nicholas+2012}
{Nicholas}, B.~P., {Rowell}, G., {Burton}, M.~G., {et~al.} 2012, \mnras, 419,
  251

\bibitem[{{Oliver} {et~al.}(1996){Oliver}, {Masheder}, \&
  {Thaddeus}}]{Oliver+1996}
{Oliver}, R.~J., {Masheder}, M.~R.~W., \& {Thaddeus}, P. 1996, \aap, 315, 578

\bibitem[{{Paron} {et~al.}(2015){Paron}, {Celis Pe{\~n}a}, {Ortega},
  {Petriella}, {Rubio}, {Dubner}, \& {Giacani}}]{Paron+2015}
{Paron}, S., {Celis Pe{\~n}a}, M., {Ortega}, M.~E., {et~al.} 2015, \aap, 580,
  A51

\bibitem[{{Paron} \& {Giacani}(2010)}]{ParonGiacani2010}
{Paron}, S., \& {Giacani}, E. 2010, \aap, 509, L4

\bibitem[{{Paron} {et~al.}(2012){Paron}, {Ortega}, {Petriella}, {Rubio},
  {Dubner}, \& {Giacani}}]{Paron+2012}
{Paron}, S., {Ortega}, M.~E., {Petriella}, A., {et~al.} 2012, \aap, 547, A60

\bibitem[{{Paron} {et~al.}(2013){Paron}, {Weidmann}, {Ortega}, {Albacete
  Colombo}, \& {Pichel}}]{Paron+2013}
{Paron}, S., {Weidmann}, W., {Ortega}, M.~E., {Albacete Colombo}, J.~F., \&
  {Pichel}, A. 2013, \mnras, 433, 1619

\bibitem[{{Petriella} {et~al.}(2013){Petriella}, {Paron}, \&
  {Giacani}}]{Petriella+2013}
{Petriella}, A., {Paron}, S.~A., \& {Giacani}, E.~B. 2013, \aap, 554, A73

\bibitem[{{Pihlstr{\"o}m} {et~al.}(2014){Pihlstr{\"o}m}, {Sjouwerman}, {Frail},
  {Claussen}, {Mesler}, \& {McEwen}}]{Pihlstrom+2014}
{Pihlstr{\"o}m}, Y.~M., {Sjouwerman}, L.~O., {Frail}, D.~A., {et~al.} 2014,
  \aj, 147, 73

\bibitem[{{Pollock}(1985)}]{Pollock1985}
{Pollock}, A.~M.~T. 1985, \aap, 150, 339

\bibitem[{{Ranasinghe} \& {Leahy}(2022)}]{RanasingheLeahy2022}
{Ranasinghe}, S., \& {Leahy}, D. 2022, \apj, 940, 63

\bibitem[{{Ranasinghe} {et~al.}(2021){Ranasinghe}, {Leahy}, \&
  {Stil}}]{Ranasinghe+2021}
{Ranasinghe}, S., {Leahy}, D., \& {Stil}, J. 2021, Universe, 7, 338

\bibitem[{{Ranasinghe} {et~al.}(2020){Ranasinghe}, {Leahy}, \&
  {Tian}}]{Ranasinghe+2020}
{Ranasinghe}, S., {Leahy}, D., \& {Tian}, W.~W. 2020, J. High Energy Phys.,
  Gravitation Cosmol., 6, 9

\bibitem[{{Ranasinghe} \& {Leahy}(2017)}]{RanasingheLeahy2017}
{Ranasinghe}, S., \& {Leahy}, D.~A. 2017, \apj, 843, 119

\bibitem[{{Ranasinghe} \& {Leahy}(2018{\natexlab{a}})}]{RanasingheLeahy2018b}
---. 2018{\natexlab{a}}, \mnras, 477, 2243

\bibitem[{{Ranasinghe} \& {Leahy}(2018{\natexlab{b}})}]{RanasingheLeahy2018a}
---. 2018{\natexlab{b}}, \aj, 155, 204

\bibitem[{{Ranasinghe} {et~al.}(2018){Ranasinghe}, {Leahy}, \&
  {Tian}}]{Ranasinghe+2018}
{Ranasinghe}, S., {Leahy}, D.~A., \& {Tian}, W. 2018, Open Physics Journal, 4,
  1

\bibitem[{{Reach} \& {Rho}(1999)}]{ReachRho1999}
{Reach}, W.~T., \& {Rho}, J. 1999, \apj, 511, 836

\bibitem[{{Reach} {et~al.}(2005){Reach}, {Rho}, \& {Jarrett}}]{Reach+2005}
{Reach}, W.~T., {Rho}, J., \& {Jarrett}, T.~H. 2005, \apj, 618, 297

\bibitem[{{Reach} {et~al.}(2006){Reach}, {Rho}, {Tappe}, {Pannuti}, {Brogan},
  {Churchwell}, {Meade}, {Babler}, {Indebetouw}, \& {Whitney}}]{Reach+2006}
{Reach}, W.~T., {Rho}, J., {Tappe}, A., {et~al.} 2006, \aj, 131, 1479

\bibitem[{{Reich} \& {Sun}(2019)}]{ReichSun2019}
{Reich}, W., \& {Sun}, X.-H. 2019, Research in Astronomy and Astrophysics, 19,
  045

\bibitem[{{Reid} {et~al.}(2016){Reid}, {Dame}, {Menten}, \&
  {Brunthaler}}]{Reid+2016}
{Reid}, M.~J., {Dame}, T.~M., {Menten}, K.~M., \& {Brunthaler}, A. 2016, \apj,
  823, 77

\bibitem[{{Reid} {et~al.}(2019){Reid}, {Menten}, {Brunthaler}, {Zheng}, {Dame},
  {Xu}, {Li}, {Sakai}, {Wu}, {Immer}, {Zhang}, {Sanna}, {Moscadelli}, {Rygl},
  {Bartkiewicz}, {Hu}, {Quiroga-Nu{\~n}ez}, \& {van Langevelde}}]{Reid+2019}
{Reid}, M.~J., {Menten}, K.~M., {Brunthaler}, A., {et~al.} 2019, \apj, 885, 131

\bibitem[{{Reynolds} \& {Moffett}(1993)}]{ReynoldsMoffett1993}
{Reynolds}, S.~P., \& {Moffett}, D.~A. 1993, \aj, 105, 2226

\bibitem[{{Reynoso} \& {Goss}(2002)}]{ReynosoGoss2002}
{Reynoso}, E.~M., \& {Goss}, W.~M. 2002, \apj, 575, 871

\bibitem[{{Reynoso} \& {Mangum}(2000)}]{ReynosoMangum2000}
{Reynoso}, E.~M., \& {Mangum}, J.~G. 2000, \apj, 545, 874

\bibitem[{{Reynoso} \& {Mangum}(2001)}]{ReynosoMangum2001}
---. 2001, \aj, 121, 347

\bibitem[{{Reynoso} {et~al.}(1999){Reynoso}, {Vel{\'a}zquez}, {Dubner}, \&
  {Goss}}]{Reynoso+1999}
{Reynoso}, E.~M., {Vel{\'a}zquez}, P.~F., {Dubner}, G.~M., \& {Goss}, W.~M.
  1999, \aj, 117, 1827

\bibitem[{{Rho} {et~al.}(2021){Rho}, {Jarrett}, {Tram}, {Lim}, {Reach},
  {Bieging}, {Lee}, {Koo}, \& {Whitney}}]{Rho+2021}
{Rho}, J., {Jarrett}, T.~H., {Tram}, L.~N., {et~al.} 2021, \apj, 917, 47

\bibitem[{{Roberts} {et~al.}(1993){Roberts}, {Goss}, {Kalberla}, {Herbstmeier},
  \& {Schwarz}}]{Roberts+1993}
{Roberts}, D.~A., {Goss}, W.~M., {Kalberla}, P.~M.~W., {Herbstmeier}, U., \&
  {Schwarz}, U.~J. 1993, \aap, 274, 427

\bibitem[{{Roberts} {et~al.}(2002){Roberts}, {Hessels}, {Ransom}, {Kaspi},
  {Freire}, {Crawford}, \& {Lorimer}}]{Roberts+2002}
{Roberts}, M. S.~E., {Hessels}, J. W.~T., {Ransom}, S.~M., {et~al.} 2002,
  \apjl, 577, L19

\bibitem[{{Rosado} {et~al.}(2007){Rosado}, {Arias}, \&
  {Ambrocio-Cruz}}]{Rosado+2007}
{Rosado}, M., {Arias}, L., \& {Ambrocio-Cruz}, P. 2007, \aj, 133, 89

\bibitem[{{Rousseau} {et~al.}(2012){Rousseau}, {Grondin}, {Van Etten},
  {Lemoine-Goumard}, {Bogdanov}, {Hessels}, {Kaspi}, {Arzoumanian}, {Camilo},
  {Casandjian}, {Espinoza}, {Johnston}, {Lyne}, {Smith}, {Stappers}, \&
  {Caliandro}}]{Rousseau+2012}
{Rousseau}, R., {Grondin}, M.~H., {Van Etten}, A., {et~al.} 2012, \aap, 544, A3

\bibitem[{{Routledge} {et~al.}(1991){Routledge}, {Dewdney}, {Landecker}, \&
  {Vaneldik}}]{Routledge+1991}
{Routledge}, D., {Dewdney}, P.~E., {Landecker}, T.~L., \& {Vaneldik}, J.~F.
  1991, \aap, 247, 529

\bibitem[{{Safi-Harb} {et~al.}(2005){Safi-Harb}, {Dubner}, {Petre}, {Holt}, \&
  {Durouchoux}}]{Safi-Harb+2005}
{Safi-Harb}, S., {Dubner}, G., {Petre}, R., {Holt}, S.~S., \& {Durouchoux}, P.
  2005, \apj, 618, 321

\bibitem[{{Saken} {et~al.}(1992){Saken}, {Fesen}, \& {Shull}}]{Saken+1992}
{Saken}, J.~M., {Fesen}, R.~A., \& {Shull}, J.~M. 1992, \apjs, 81, 715

\bibitem[{{Salpeter}(1955)}]{Salpeter1955}
{Salpeter}, E.~E. 1955, \apj, 121, 161

\bibitem[{{Sano} {et~al.}(2020){Sano}, {Inoue}, {Tokuda}, {Tanaka}, {Yamazaki},
  {Inutsuka}, {Aharonian}, {Rowell}, {Filipovi{\'c}}, {Yamane}, {Yoshiike},
  {Maxted}, {Uchida}, {Hayakawa}, {Tachihara}, {Uchiyama}, \&
  {Fukui}}]{Sano+2020}
{Sano}, H., {Inoue}, T., {Tokuda}, K., {et~al.} 2020, \apjl, 904, L24

\bibitem[{{Sano} {et~al.}(2021){Sano}, {Yoshiike}, {Yamane}, {Hayashi},
  {Enokiya}, {Tokuda}, {Tachihara}, {Rowell}, {Filipovi{\'c}}, \&
  {Fukui}}]{Sano+2021}
{Sano}, H., {Yoshiike}, S., {Yamane}, Y., {et~al.} 2021, \apj, 919, 123

\bibitem[{{Sasaki} {et~al.}(2006){Sasaki}, {Kothes}, {Plucinsky}, {Gaetz}, \&
  {Brunt}}]{Sasaki+2006}
{Sasaki}, M., {Kothes}, R., {Plucinsky}, P.~P., {Gaetz}, T.~J., \& {Brunt},
  C.~M. 2006, \apjl, 642, L149

\bibitem[{{Sashida} {et~al.}(2013){Sashida}, {Oka}, {Tanaka}, {Aono},
  {Matsumura}, {Nagai}, \& {Seta}}]{Sashida+2013}
{Sashida}, T., {Oka}, T., {Tanaka}, K., {et~al.} 2013, \apj, 774, 10

\bibitem[{{Sato}(1979)}]{Sato1979}
{Sato}, F. 1979, \aplett, 20, 43

\bibitem[{{Saz Parkinson} {et~al.}(2010){Saz Parkinson}, {Dormody}, {Ziegler},
  {Ray}, {Abdo}, {Ballet}, {Baring}, {Belfiore}, {Burnett}, {Caliandro},
  {Camilo}, {Caraveo}, {de Luca}, {Ferrara}, {Freire}, {Grove}, {Gwon},
  {Harding}, {Johnson}, {Johnson}, {Johnston}, {Keith}, {Kerr},
  {Kn{\"o}dlseder}, {Makeev}, {Marelli}, {Michelson}, {Parent}, {Ransom},
  {Reimer}, {Romani}, {Smith}, {Thompson}, {Watters}, {Weltevrede}, {Wolff}, \&
  {Wood}}]{SazParkinson+2010}
{Saz Parkinson}, P.~M., {Dormody}, M., {Ziegler}, M., {et~al.} 2010, \apj, 725,
  571

\bibitem[{{Seta} {et~al.}(2004){Seta}, {Hasegawa}, {Sakamoto}, {Oka}, {Sawada},
  {Inutsuka}, {Koyama}, \& {Hayashi}}]{Seta+2004}
{Seta}, M., {Hasegawa}, T., {Sakamoto}, S., {et~al.} 2004, \aj, 127, 1098

\bibitem[{{Seta} {et~al.}(1998{\natexlab{a}}){Seta}, {Winnewisser}, {Hasegawa},
  {White}, \& {Oka}}]{Seta1998iau}
{Seta}, M., {Winnewisser}, G., {Hasegawa}, T., {White}, G.~J., \& {Oka}, T.
  1998{\natexlab{a}}, in The Central Regions of the Galaxy and Galaxies, ed.
  Y.~{Sofue}, Vol. 184, 195

\bibitem[{{Seta} {et~al.}(1998{\natexlab{b}}){Seta}, {Hasegawa}, {Dame},
  {Sakamoto}, {Oka}, {Handa}, {Hayashi}, {Morino}, {Sorai}, \&
  {Usuda}}]{Seta+1998}
{Seta}, M., {Hasegawa}, T., {Dame}, T.~M., {et~al.} 1998{\natexlab{b}}, \apj,
  505, 286

\bibitem[{{Shan} {et~al.}(2018){Shan}, {Zhu}, {Tian}, {Zhang}, {Zhang}, {Wu},
  \& {Yang}}]{Shan+2018}
{Shan}, S.~S., {Zhu}, H., {Tian}, W.~W., {et~al.} 2018, \apjs, 238, 35

\bibitem[{{Shan} {et~al.}(2012){Shan}, {Yang}, {Shi}, {Yao}, {Zuo}, {Lin},
  {Chen}, {Zhang}, {Duan}, {Cao}, {Li}, {Li}, {Liu}, \& {Zhong}}]{Shan+2012}
{Shan}, W.~L., {Yang}, J., {Shi}, S.~C., {et~al.} 2012, IEEE Transactions on
  Terahertz Science and Technology, 2, 593

\bibitem[{{Sheidaei}(2011)}]{Sheidaei2011}
{Sheidaei}, F. 2011, in International Cosmic Ray Conference, Vol.~7,
  International Cosmic Ray Conference, 244

\bibitem[{{Sofue} {et~al.}(2021){Sofue}, {Kohno}, \& {Umemoto}}]{Sofue+2021}
{Sofue}, Y., {Kohno}, M., \& {Umemoto}, T. 2021, \apjs, 253, 17

\bibitem[{{Stanimirovi{\'c}} {et~al.}(2003){Stanimirovi{\'c}}, {Weisberg},
  {Dickey}, {de la Fuente}, {Devine}, {Hedden}, \&
  {Anderson}}]{Stanimirovic+2003}
{Stanimirovi{\'c}}, S., {Weisberg}, J.~M., {Dickey}, J.~M., {et~al.} 2003,
  \apj, 592, 953

\bibitem[{{Su} {et~al.}(2009){Su}, {Chen}, {Yang}, {Koo}, {Zhou}, {Jeong}, \&
  {Zhang}}]{Su+2009}
{Su}, Y., {Chen}, Y., {Yang}, J., {et~al.} 2009, \apj, 694, 376

\bibitem[{{Su} {et~al.}(2011){Su}, {Chen}, {Yang}, {Koo}, {Zhou}, {Lu},
  {Jeong}, \& {DeLaney}}]{Su+2011}
---. 2011, \apj, 727, 43

\bibitem[{{Su} {et~al.}(2014{\natexlab{a}}){Su}, {Fang}, {Yang}, {Zhou}, \&
  {Chen}}]{Su+2014b}
{Su}, Y., {Fang}, M., {Yang}, J., {Zhou}, P., \& {Chen}, Y. 2014{\natexlab{a}},
  \apj, 788, 122

\bibitem[{{Su} {et~al.}(2014{\natexlab{b}}){Su}, {Yang}, {Zhou}, {Zhou}, \&
  {Chen}}]{Su+2014}
{Su}, Y., {Yang}, J., {Zhou}, X., {Zhou}, P., \& {Chen}, Y. 2014{\natexlab{b}},
  \apj, 796, 122

\bibitem[{{Su} {et~al.}(2015){Su}, {Zhang}, {Shao}, \& {Yang}}]{Su+2015}
{Su}, Y., {Zhang}, S., {Shao}, X., \& {Yang}, J. 2015, \apj, 811, 134

\bibitem[{{Su} {et~al.}(2017{\natexlab{a}}){Su}, {Zhou}, {Yang}, {Chen},
  {Chen}, {Gong}, \& {Zhang}}]{Su+2017b}
{Su}, Y., {Zhou}, X., {Yang}, J., {et~al.} 2017{\natexlab{a}}, \apj, 845, 48

\bibitem[{{Su} {et~al.}(2018){Su}, {Zhou}, {Yang}, {Chen}, {Chen}, \&
  {Zhang}}]{Su+2018}
---. 2018, \apj, 863, 103

\bibitem[{{Su} {et~al.}(2017{\natexlab{b}}){Su}, {Zhou}, {Yang}, {Chen},
  {Chen}, {Liu}, {Wang}, {Li}, \& {Zhang}}]{Su+2017a}
---. 2017{\natexlab{b}}, \apj, 836, 211

\bibitem[{{Su} {et~al.}(2019){Su}, {Yang}, {Zhang}, {Gong}, {Wang}, {Zhou},
  {Wang}, {Chen}, {Sun}, {Chen}, {Xu}, \& {Jiang}}]{Su+2019}
{Su}, Y., {Yang}, J., {Zhang}, S., {et~al.} 2019, \apjs, 240, 9

\bibitem[{{Sun} \& {Chen}(2019)}]{SunChen2019}
{Sun}, L., \& {Chen}, Y. 2019, \apj, 872, 45

\bibitem[{{Sun} {et~al.}(2020){Sun}, {Yang}, {Xu}, {Zhang}, {Su}, {Wang},
  {Chen}, {Lu}, {Sun}, {Ju}, {Zhang}, {Zhou}, \& {Jiang}}]{Sun+2020}
{Sun}, Y., {Yang}, J., {Xu}, Y., {et~al.} 2020, \apjs, 246, 7

\bibitem[{{Supan} {et~al.}(2015){Supan}, {Castelletti}, {Joshi}, {Surnis}, \&
  {Supanitsky}}]{Supan+2015}
{Supan}, L., {Castelletti}, G., {Joshi}, B.~C., {Surnis}, M.~P., \&
  {Supanitsky}, D. 2015, \aap, 576, A81

\bibitem[{{Supan} {et~al.}(2022){Supan}, {Fischetto}, \&
  {Castelletti}}]{Supan+2022}
{Supan}, L., {Fischetto}, G., \& {Castelletti}, G. 2022, \aap, 664, A89

\bibitem[{{Tatematsu} {et~al.}(1990{\natexlab{a}}){Tatematsu}, {Fukui},
  {Iwata}, {Seward}, \& {Nakano}}]{Tatematsu+1990a}
{Tatematsu}, K., {Fukui}, Y., {Iwata}, T., {Seward}, F.~D., \& {Nakano}, M.
  1990{\natexlab{a}}, \apj, 351, 157

\bibitem[{{Tatematsu} {et~al.}(1990{\natexlab{b}}){Tatematsu}, {Fukui},
  {Landecker}, \& {Roger}}]{Tatematsu+1990}
{Tatematsu}, K., {Fukui}, Y., {Landecker}, T.~L., \& {Roger}, R.~S.
  1990{\natexlab{b}}, \aap, 237, 189

\bibitem[{{Tatematsu} {et~al.}(1987){Tatematsu}, {Fukui}, {Nakano}, {Kogure},
  {Ogawa}, \& {Kawabata}}]{Tatematsu+1987}
{Tatematsu}, K., {Fukui}, Y., {Nakano}, M., {et~al.} 1987, \aap, 184, 279

\bibitem[{{Tian} \& {Leahy}(2008)}]{TianLeahy2008b}
{Tian}, W.~W., \& {Leahy}, D.~A. 2008, \mnras, 391, L54

\bibitem[{{Tian} \& {Leahy}(2011)}]{TianLeahy2011}
---. 2011, \apjl, 729, L15

\bibitem[{{Tian} \& {Leahy}(2013)}]{TianLeahy2013}
---. 2013, \apjl, 769, L17

\bibitem[{{Tian} {et~al.}(2007{\natexlab{a}}){Tian}, {Leahy}, \&
  {Wang}}]{Tian+2007}
{Tian}, W.~W., {Leahy}, D.~A., \& {Wang}, Q.~D. 2007{\natexlab{a}}, \aap, 474,
  541

\bibitem[{{Tian} {et~al.}(2007{\natexlab{b}}){Tian}, {Li}, {Leahy}, \&
  {Wang}}]{Tian+2007b}
{Tian}, W.~W., {Li}, Z., {Leahy}, D.~A., \& {Wang}, Q.~D. 2007{\natexlab{b}},
  \apjl, 657, L25

\bibitem[{{Tian} {et~al.}(2010){Tian}, {Li}, {Leahy}, {Yang}, {Yang},
  {Yamazaki}, \& {Lu}}]{Tian+2010}
{Tian}, W.~W., {Li}, Z., {Leahy}, D.~A., {et~al.} 2010, \apj, 712, 790

\bibitem[{{Traverso} {et~al.}(1999){Traverso}, {Reynoso}, \&
  {Dubner}}]{Traverso+1999}
{Traverso}, P.~H., {Reynoso}, E.~M., \& {Dubner}, G.~M. 1999, Boletin de la
  Asociacion Argentina de Astronomia La Plata Argentina, 43, 50

\bibitem[{{Trimble}(1968)}]{Trimble1968}
{Trimble}, V. 1968, \aj, 73, 535

\bibitem[{{Trushkin}(2001)}]{Trushkin2001}
{Trushkin}, S.~A. 2001, in ESA Special Publication, Vol. 459, Exploring the
  Gamma-Ray Universe, ed. A.~{Gimenez}, V.~{Reglero}, \& C.~{Winkler}, 109--111

\bibitem[{{Turner} {et~al.}(1992){Turner}, {Chan}, {Green}, \&
  {Lubowich}}]{Turner+1992}
{Turner}, B.~E., {Chan}, K.-W., {Green}, S., \& {Lubowich}, D.~A. 1992, \apj,
  399, 114

\bibitem[{{Uyaniker} {et~al.}(2002){Uyaniker}, {Kothes}, \&
  {Brunt}}]{Uyaniker+2002}
{Uyaniker}, B., {Kothes}, R., \& {Brunt}, C.~M. 2002, \apj, 565, 1022

\bibitem[{{van Dishoeck} {et~al.}(1993){van Dishoeck}, {Jansen}, \&
  {Phillips}}]{vanDishoeck+1993}
{van Dishoeck}, E.~F., {Jansen}, D.~J., \& {Phillips}, T.~G. 1993, \aap, 279,
  541

\bibitem[{{Van Etten} {et~al.}(2008){Van Etten}, {Romani}, \&
  {Ng}}]{VanEtten+2008}
{Van Etten}, A., {Romani}, R.~W., \& {Ng}, C.~Y. 2008, \apj, 680, 1417

\bibitem[{{Voisin} {et~al.}(2016){Voisin}, {Rowell}, {Burton}, {Walsh},
  {Fukui}, \& {Aharonian}}]{Voisin+2016}
{Voisin}, F., {Rowell}, G., {Burton}, M.~G., {et~al.} 2016, \mnras, 458, 2813

\bibitem[{{Voisin} {et~al.}(2019){Voisin}, {Rowell}, {Burton}, {Fukui}, {Sano},
  {Aharonian}, {Maxted}, {Braiding}, {Blackwell}, \& {Lau}}]{Voisin+2019}
{Voisin}, F.~J., {Rowell}, G.~P., {Burton}, M.~G., {et~al.} 2019, \pasa, 36,
  e014

\bibitem[{{Wallace} {et~al.}(1994){Wallace}, {Landecker}, \&
  {Taylor}}]{Wallace+1994}
{Wallace}, B.~J., {Landecker}, T.~L., \& {Taylor}, A.~R. 1994, \aap, 286, 565

\bibitem[{{Wallace} {et~al.}(1997){Wallace}, {Landecker}, \&
  {Taylor}}]{Wallace+1997}
---. 1997, \aj, 114, 2068

\bibitem[{{Wang}(2011)}]{Wang2011}
{Wang}, W. 2011, Research in Astronomy and Astrophysics, 11, 824

\bibitem[{{Wang} {et~al.}(2020){Wang}, {Beuther}, {Rugel}, {Soler}, {Stil},
  {Ott}, {Bihr}, {McClure-Griffiths}, {Anderson}, {Klessen}, {Goldsmith},
  {Roy}, {Glover}, {Urquhart}, {Heyer}, {Linz}, {Smith}, {Bigiel}, {Dempsey},
  \& {Henning}}]{Wang+2020}
{Wang}, Y., {Beuther}, H., {Rugel}, M.~R., {et~al.} 2020, \aap, 634, A83

\bibitem[{{Wardle} \& {Yusef-Zadeh}(2002)}]{WardleYusef-Zadeh2002}
{Wardle}, M., \& {Yusef-Zadeh}, F. 2002, Science, 296, 2350

\bibitem[{{Wayth} {et~al.}(2015){Wayth}, {Lenc}, {Bell}, {Callingham},
  {Dwarakanath}, {Franzen}, {For}, {Gaensler}, {Hancock}, {Hindson},
  {Hurley-Walker}, {Jackson}, {Johnston-Hollitt}, {Kapi{\'n}ska}, {McKinley},
  {Morgan}, {Offringa}, {Procopio}, {Staveley-Smith}, {Wu}, {Zheng}, {Trott},
  {Bernardi}, {Bowman}, {Briggs}, {Cappallo}, {Corey}, {Deshpande}, {Emrich},
  {Goeke}, {Greenhill}, {Hazelton}, {Kaplan}, {Kasper}, {Kratzenberg},
  {Lonsdale}, {Lynch}, {McWhirter}, {Mitchell}, {Morales}, {Morgan}, {Oberoi},
  {Ord}, {Prabu}, {Rogers}, {Roshi}, {Shankar}, {Srivani}, {Subrahmanyan},
  {Tingay}, {Waterson}, {Webster}, {Whitney}, {Williams}, \&
  {Williams}}]{Wayth+2015}
{Wayth}, R.~B., {Lenc}, E., {Bell}, M.~E., {et~al.} 2015, \pasa, 32, e025

\bibitem[{{Weisberg} {et~al.}(1995){Weisberg}, {Siegel}, {Frail}, \&
  {Johnston}}]{Weisberg+1995}
{Weisberg}, J.~M., {Siegel}, M.~H., {Frail}, D.~A., \& {Johnston}, S. 1995,
  \apj, 447, 204

\bibitem[{{White} {et~al.}(1987){White}, {Rainey}, {Hayashi}, \&
  {Kaifu}}]{White+1987}
{White}, G.~J., {Rainey}, R., {Hayashi}, S.~S., \& {Kaifu}, N. 1987, \aap, 173,
  337

\bibitem[{{Willis} \& {Dickel}(1971)}]{WillisDickel1971}
{Willis}, A.~G., \& {Dickel}, J.~R. 1971, \aplett, 8, 203

\bibitem[{{Wilner} {et~al.}(1998){Wilner}, {Reynolds}, \&
  {Moffett}}]{Wilner+1998}
{Wilner}, D.~J., {Reynolds}, S.~P., \& {Moffett}, D.~A. 1998, \aj, 115, 247

\bibitem[{{Wootten}(1981)}]{Wootten1981}
{Wootten}, A. 1981, \apj, 245, 105

\bibitem[{{Wootten}(1977)}]{Wootten1977}
{Wootten}, H.~A. 1977, \apj, 216, 440

\bibitem[{{Xu} \& {Wang}(2012)}]{XuWang2012}
{Xu}, J.-L., \& {Wang}, J.-J. 2012, \aap, 543, A24

\bibitem[{{Xu} {et~al.}(2011){Xu}, {Wang}, \& {Miller}}]{Xu+2011}
{Xu}, J.-L., {Wang}, J.-J., \& {Miller}, M. 2011, \apj, 727, 81

\bibitem[{{Yamada} {et~al.}(2017){Yamada}, {Oka}, {Takekawa}, {Iwata},
  {Tsujimoto}, {Tokuyama}, {Furusawa}, {Tanabe}, \& {Nomura}}]{Yamada+2017}
{Yamada}, M., {Oka}, T., {Takekawa}, S., {et~al.} 2017, \apjl, 834, L3

\bibitem[{{Yang} {et~al.}(2006){Yang}, {Zhang}, {Cai}, {Lu}, \&
  {Tan}}]{Yang+2006}
{Yang}, J., {Zhang}, J.-L., {Cai}, Z.-Y., {Lu}, D.-R., \& {Tan}, Y.-H. 2006,
  \cjaa, 6, 210

\bibitem[{{Yoshiike} {et~al.}(2013){Yoshiike}, {Fukuda}, {Sano}, {Ohama},
  {Moribe}, {Torii}, {Hayakawa}, {Okuda}, {Yamamoto}, {Tajima}, {Mizuno},
  {Nishimura}, {Kimura}, {Maezawa}, {Onishi}, {Mizuno}, {Ogawa}, {Giuliani},
  {Koo}, \& {Fukui}}]{Yoshiike+2013}
{Yoshiike}, S., {Fukuda}, T., {Sano}, H., {et~al.} 2013, \apj, 768, 179

\bibitem[{{Yu} {et~al.}(2019){Yu}, {Chen}, {Jiang}, \& {Zijlstra}}]{Yu+2019}
{Yu}, B., {Chen}, B.~Q., {Jiang}, B.~W., \& {Zijlstra}, A. 2019, \mnras, 488,
  3129

\bibitem[{{Yusef-Zadeh} {et~al.}(1999){Yusef-Zadeh}, {Goss}, {Roberts},
  {Robinson}, \& {Frail}}]{Yusef-Zadeh+1999}
{Yusef-Zadeh}, F., {Goss}, W.~M., {Roberts}, D.~A., {Robinson}, B., \& {Frail},
  D.~A. 1999, \apj, 527, 172

\bibitem[{{Yusef-Zadeh} {et~al.}(2003){Yusef-Zadeh}, {Wardle}, {Rho}, \&
  {Sakano}}]{Yusef-Zadeh+2003}
{Yusef-Zadeh}, F., {Wardle}, M., {Rho}, J., \& {Sakano}, M. 2003, \apj, 585,
  319

\bibitem[{{Zhang} {et~al.}(2015){Zhang}, {Chen}, {Su}, {Zhou}, {Pannuti}, \&
  {Zhou}}]{Zhang+2015}
{Zhang}, G.-Y., {Chen}, Y., {Su}, Y., {et~al.} 2015, \apj, 799, 103

\bibitem[{{Zhang} {et~al.}(2013){Zhang}, {Chen}, {Li}, \& {Zhou}}]{Zhang+2013}
{Zhang}, X., {Chen}, Y., {Li}, H., \& {Zhou}, X. 2013, \mnras, 429, L25

\bibitem[{{Zhang} {et~al.}(2010){Zhang}, {Gao}, \& {Wang}}]{Zhang+2010}
{Zhang}, Z., {Gao}, Y., \& {Wang}, J. 2010, Science China Physics, Mechanics,
  and Astronomy, 53, 1357

\bibitem[{{Zhou} \& {Chen}(2011)}]{ZhoupChen2011}
{Zhou}, P., \& {Chen}, Y. 2011, \apj, 743, 4

\bibitem[{{Zhou} {et~al.}(2016{\natexlab{a}}){Zhou}, {Chen}, {Safi-Harb},
  {Zhou}, {Sun}, {Zhang}, \& {Zhang}}]{Zhoup+2016a}
{Zhou}, P., {Chen}, Y., {Safi-Harb}, S., {et~al.} 2016{\natexlab{a}}, \apj,
  831, 192

\bibitem[{{Zhou} {et~al.}(2016{\natexlab{b}}){Zhou}, {Chen}, {Zhang}, {Li},
  {Safi-Harb}, {Zhou}, \& {Zhang}}]{Zhoup+2016}
{Zhou}, P., {Chen}, Y., {Zhang}, Z.-Y., {et~al.} 2016{\natexlab{b}}, \apj, 826,
  34

\bibitem[{{Zhou} {et~al.}(2018){Zhou}, {Li}, {Zhang}, {Vink}, {Chen}, {Arias},
  {Patnaude}, \& {Bregman}}]{Zhoup+2018}
{Zhou}, P., {Li}, J.-T., {Zhang}, Z.-Y., {et~al.} 2018, \apj, 865, 6

\bibitem[{{Zhou} {et~al.}(2020{\natexlab{a}}){Zhou}, {Zhou}, {Chen}, {Wang},
  {Vink}, \& {Wang}}]{Zhoup+2020}
{Zhou}, P., {Zhou}, X., {Chen}, Y., {et~al.} 2020{\natexlab{a}}, \apj, 905, 99

\bibitem[{{Zhou} {et~al.}(2022){Zhou}, {Zhang}, {Zhou}, {Arias}, {Koo}, {Vink},
  {Zhang}, {Sun}, {Du}, {Zhu}, {Chen}, {Bovino}, \& {Lee}}]{Zhoup+2022}
{Zhou}, P., {Zhang}, G.-Y., {Zhou}, X., {et~al.} 2022, \apj, 931, 144

\bibitem[{{Zhou} {et~al.}(2009){Zhou}, {Chen}, {Su}, \& {Yang}}]{Zhou+2009}
{Zhou}, X., {Chen}, Y., {Su}, Y., \& {Yang}, J. 2009, \apj, 691, 516

\bibitem[{{Zhou} {et~al.}(2020{\natexlab{b}}){Zhou}, {Su}, {Yang}, {Chen},
  {Xu}, {Chen}, \& {Zhang}}]{Zhou+2020}
{Zhou}, X., {Su}, Y., {Yang}, J., {et~al.} 2020{\natexlab{b}}, \apj, 900, 155

\bibitem[{{Zhou} {et~al.}(2014){Zhou}, {Yang}, {Fang}, \& {Su}}]{Zhou+2014}
{Zhou}, X., {Yang}, J., {Fang}, M., \& {Su}, Y. 2014, \apj, 791, 109

\bibitem[{{Zhou} {et~al.}(2016{\natexlab{c}}){Zhou}, {Yang}, {Fang}, {Su},
  {Sun}, \& {Chen}}]{Zhou+2016}
{Zhou}, X., {Yang}, J., {Fang}, M., {et~al.} 2016{\natexlab{c}}, \apj, 833, 4

\bibitem[{Zhou {et~al.}(2023)Zhou, Su, Yang, Chen, Sun, Jiang, Wang, Wang,
  Zhang, Xu, Yan, Yuan, Chen, Ao, \& Ma}]{Zhou+2023data}
Zhou, X., Su, Y., Yang, J., {et~al.} 2023, {A Systematic Study of Associations
  between Supernova Remnants and Molecular Clouds -- Figures of Individual
  Objects}

\bibitem[{{Zhu} {et~al.}(2013){Zhu}, {Tian}, {Torres}, {Pedaletti}, \&
  {Su}}]{Zhu+2013}
{Zhu}, H., {Tian}, W.~W., {Torres}, D.~F., {Pedaletti}, G., \& {Su}, H.~Q.
  2013, \apj, 775, 95

\bibitem[{{Zhu} {et~al.}(2014){Zhu}, {Tian}, \& {Zuo}}]{Zhu+2014}
{Zhu}, H., {Tian}, W.~W., \& {Zuo}, P. 2014, \apj, 793, 95

\bibitem[{{Zuo} {et~al.}(2011){Zuo}, {Li}, {Sun}, {Yang}, {Li}, {Xu}, {He},
  {Fan}, \& {Fan}}]{Zuo+2011}
{Zuo}, Y.~X., {Li}, Y., {Sun}, J.~X., {et~al.} 2011, Acta Astronomica Sinica,
  52, 152

\end{thebibliography}
\end{document}